\documentclass{article}
\usepackage{graphicx} 
\usepackage[margin=1in]{geometry}
\usepackage{natbib}
\usepackage{hyperref}
\usepackage{xcolor}
\usepackage{enumitem}
\usepackage{tikz}
\usepackage{quiver}
\usepackage{amsmath,amsthm,amsfonts,amssymb}
\usepackage{cleveref}
\usepackage{algorithm}
\usepackage{algpseudocode}
\usepackage{diagbox}
\usepackage{multirow}
\usepackage{authblk}

\newcommand{\remove}[1]{}
\newcommand{\mc}[1]{\ensuremath{\mathcal{#1}}}
\newcommand{\Var}{\mathrm{Var}}
\newcommand{\p}[1]{\left( #1 \right)}
\renewcommand{\b}[1]{\left[ #1 \right]}
\DeclareMathOperator*{\EE}{\mathbb{E}}
\newcommand{\E}[1]{\EE \b{#1}}

\def\bx{\mathbf{x}}

\def\bX{\mathbf{X}}

\def\bZ{\mathbf{Z}}

\title{Evaluating the Impacts of Swapping on the US Decennial
Census{\thanks{This work was supported in part by Sloan Foundation grant
G-2017-9890, a Google Research Scholar award, and the Center for Research on
Computation and Society (CRCS), Harvard.}}}
\author[1]{Mar{\'i}a Ballesteros}
\author[2]{Cynthia Dwork}
\author[1]{Gary King}
\author[3]{Conlan Olson}
\author[4*]{Manish Raghavan}

\affil[1]{Institute for Quantitative Social Science, Harvard University}
\affil[2]{Department of Computer Science, Harvard University}
\affil[3]{Department of Computer Science, Columbia University}
\affil[4]{Sloan School of Management and Department of Electrical Engineering and Computer Science, MIT}

\affil[*]{Author names appear in alphabetical order}

\date{}

\begin{document}

\def\diffp{Differential Privacy}
\def\truthy{verisimilar}

\maketitle
\begin{abstract}

To meet its dual burdens of providing useful statistics and ensuring privacy of individual respondents, the US Census Bureau has for decades introduced some form of ``noise'' into published statistics. Initially, they used a method known as ``swapping'' (1990--2010).
In 2020, they switched to an algorithm called TopDown that ensures a form of \diffp.
While the TopDown algorithm has been made public, no implementation of swapping has been released and many details of the deployed swapping methodology deployed have been kept secret.
Further, the Bureau has not published (even a synthetic) ``original'' dataset and its swapped version.
It is therefore difficult to evaluate the effects of swapping, and to compare these effects to those of other privacy technologies.
To address these difficulties we describe and implement a parameterized swapping algorithm based on Census publications, court documents, and informal interviews with Census employees.
With this implementation, we characterize the impacts of swapping on a range of statistical quantities of interest.
We provide intuition for the types of shifts induced by swapping and compare against those introduced by TopDown.
We find that even when swapping and TopDown introduce errors of similar magnitude, the direction in which statistics are biased need not be the same across the two techniques.
More broadly, our implementation provides researchers with the tools to analyze and potentially correct for the impacts of disclosure avoidance systems on the quantities they study.

\end{abstract}

\section{Introduction}
Census statistics are used for a wide variety of planning, administrative, and redistricting purposes.  In order to meet its dual burdens of providing useful statistics and protecting the privacy of individual respondents, the US Census Bureau has for decades introduced noise into published statistics.
From 1990--2010, the decennial census used a form of ``swapping''~\citep{dalenius1982data,mckenna-rmd}.
The details of the method are kept confidential.

In 2010, the Bureau determined that the existing disclosure avoidance system
(DAS) was vulnerable to reconstruction attacks (see \citet{abowd_attack} for a detailed description of how the {B}ureau launched a linkage-assisted reconstruction attack on the 2010 data).
In 2020, the Decennial Census deployed a new DAS known as TopDown, which seeks to provide formal differential privacy guarantees~\citep{abowd2018us}.
Unlike swapping, the TopDown algorithm is public.\footnote{\url{https://github.com/uscensusbureau/DAS_2020_Redistricting_Production_Code}}

At a high level, swapping works by choosing households that are at a particularly high disclosure risk---say, a household that is racially unique in its neighborhood---and ``swapping'' each such household with another household in a different region, ``hiding'' the unique household in an area where it may be less identifiable. The precise instantiation of this framework used by the Bureau is kept secret, and indeed, we do not even know whether any single algorithm was used across all states in the 1990--2010 decennial censuses.

Prior to 2020, although it was generally known that swapping was deployed,  consumers of {C}ensus statistics tended to treat published data as ground truth.  This became impossible with the announcement of the deployment of differential privacy, the publication of demonstration data and, eventually, the TopDown algorithm itself.\footnote{Differential privacy does not rely on secrecy of the algorithm to maintain privacy for respondents.}  In a nutshell, consumers were forced to confront the ``ground truth'' fiction. Each DAS, swapping and TopDown, has both privacy properties and impacts on data analysis. We know that the Bureau chose TopDown because it can provide better privacy protection than that obtained with the level of swapping used in the 2010 census. In this work, we turn to the impact of these disclosure avoidance systems on data analysis.

Studying the impacts of disclosure avoidance systems is challenging due to a lack of both data and information.
In the case of TopDown, the Census Bureau has provided the details of the algorithm and a series of ``before and after'' demonstration data releases to facilitate research (see Section~\ref{sec:related} for details).
No analogous releases exist for swapping.
In this work, we facilitate research into the effects of swapping through a careful implementation of a swapping algorithm based on information released by the Census Bureau (see Section~\ref{sec:swapping}); specifically, we drawn from a variety of publications, presentations, and statements by Bureau employees, as well as court documents,
to create an approximation to the Bureau's practices.
For details that remain unknown, we conduct robustness checks to verify that our conclusions hold across a range of reasonable choices.

We use this implementation to study a variety of impacts that swapping has on the Census data. In Section~\ref{sec:effects}, to investigate the basic statistical properties of swapping, we measure the errors that swapping introduces to population counts by race, and compare them to those of TopDown and a simple variant, ToyDown, created by \citet{cohen2021census}, at a specific setting and apportioning of the privacy parameter~$\varepsilon$.  We further examine detailed properties of swapping with respect to household size, mixed-race households, and racial entropy across geographies. Surprisingly, despite the fact that swapping targets households that are racially unique in their current location and ``moves'' them elsewhere, we find that swapping \textit{increases} racial entropy. We provide a detailed analysis of why this is the case.

We next consider the impact of swapping on downstream uses of Census data for both public policy and social science (Section~\ref{sec:downstream}).
Prior work has found that differential privacy biases statistics derived from ecological regression (ER) analyses, which are used in cases involving the Voting Rights Act~\citep{cohen2021census}.
We find that the same is true for swapping, but, crucially, swapping appears to bias these statistics \textit{in the opposite direction}.
We again provide intuition for why this might be the case.
Thus, viewing swapping as a similarly noisy predecessor to differential privacy may obscure key differences in the direction of effects of that noise.

We further compare swapping to TopDown in terms of impact on social science applications, where recent work has shown that the noise introduced by the TopDown algorithm 
can lead to substantial variance in downstream analyses~\citep{MuellerJ.Tom2022T2UC,hauer2021differential,ruggles2019differential,santos2020differential,winkler2021differential}.
In particular, \citet{MuellerJ.Tom2022T2UC} show that TopDown can introduce large errors in rural areas.
We find that the same is true for swapping.
We conclude by proposing a more general framework for using our methods to estimate the impacts of swapping on downstream statistics of interest.

To summarize, the key contributions of this work are as follows:
\begin{itemize}
    \item We implement a swapping algorithm that we believe closely resembles the process used by the US Census Bureau based on a careful analysis of public-domain documents. Code for this implementation can be found at \url{https://github.com/ceolson/census-swapping}.
    \item Using this implementation, we provide a detailed description of the impacts of swapping and how those impacts compare to the impacts of TopDown and ToyDown.
    \item We highlight several important findings, including an increase in racial entropy due to swapping, biases in opposite directions to those induced by differential privacy, and the ability to replicate studies describing errors from differential privacy.
    \item More broadly, we offer a framework to estimate and correct for the effects of disclosure avoidance systems in downstream analyses.
\end{itemize}

\section{Related Work}
\label{sec:related}
\paragraph*{Disclosure Avoidance in the US Decennial Census.}

The US Census Bureau has a statutory obligation to protect the privacy of respondents under Title 13 of the U.S. Code.
To do so, they have deployed a number of Disclosure Avoidance Systems (DASs) over the years \citep{mckenna-rmd}.
Of note, for the purposes of this work, are two methods in particular: data swapping, used in 1990, 2000, and 2010; and TopDown, the Bureau's implementation of differential privacy, used in 2020.

Data swapping, introduced by \citet{dalenius1982data}, works by exchanging the values contained in particular cells in a dataset while leaving aggregate statistics unchanged.
For a survey on variants of swapping over the years, see \citet{fienberg2004data}.
At a high level, the swapping algorithm used by the Census Bureau swapped the records associated with one household with those of another household in another location.
While the exact details of this procedure are unknown, we compile a combination of reports and statements issued by the Census Bureau and related personnel \citep{abowd-declaration,griffin1989disclosure,mckenna-rmd,ramanayake2010balancing,steel-zayatz,wright2021empirical,zayatz-sdc,zayatz-jos,zayatz-2010,hawala2003microdata} that collectively provide enough information to implement a close approximation to the original Census procedure.
We provide further details in \Cref{sec:swapping}.

Beginning in 2020, the Census Bureau began using a new DAS based on differential privacy \citep{dwork2006calibrating,abowd2018us}.
Instead of swapping records, the Bureau's algorithm (known as TopDown) adds noise to tabulated counts, providing formal privacy guarantees, and then, in a {\em post-processing} phase, solves a constrained optimization problem to find ``nearby'' noisy counts satisfying hierarchical consistency and other requirements (\textit{e.g.}, non-negativity) \citep{abowd2019census}. 
Unlike swapping, the algorithm itself is public knowledge; the realized values of the added noise are kept secret to protect privacy.

Our focus in this work is on the Decennial Census.
Disclosure avoidance is also used in the American Community Survey (ACS), which collects more detailed information on a sample of the population (see \citet{zayatz-2010}).
The Census Bureau uses a variety of disclosure avoidance techniques to release
ACS data, including swapping \citep[Section 13.8]{acs_design}, \citep{pums_accuracy,rodriguez_acm}.

\paragraph*{Analyzing Census DASs.}

Research on the effects of swapping in the Census, especially in comparison with differential privacy, is closely related to our work \citep{kim2015effect,christ2022differential,bailie2023can}.
\citet{kim2015effect} implements a version of swapping for the 2007--2011 American Community Survey in Alleghany County, PA.
\citet{kim2015effect} provides results using both synthetic data at the tract level and privileged access to the true (unswapped) microdata through the Census Bureau.
In contrast, our methods make analyses like this possible \emph{without} privileged access, which may benefit future researchers, and they are sufficiently general to be used across the country.
\citet{christ2022differential} compare the effects of swapping and differential privacy on both utility and privacy.
To do so, they create synthetic data by combining blocks from nine counties across the country from the 2010 Census.
They provide two plausible implementations of swapping which, like \citet{kim2015effect}, operate by swapping individual records with an external pool of candidates that is representative of the US population.
In contrast, our swapping implementation operates on \textit{households}, as opposed to \textit{individuals}, which is consistent with practices in the 1990--2010 Decennial Censuses.

\citet{bailie2023can} give a swapping algorithm that provides a differential privacy guarantee and analyze it on Census data.
In contrast, the swapping procedure used by the Census, which we aim to reconstruct here, has no formal differential privacy guarantees.
Because swapping requires access to household-level data at the block level, \citet{bailie2023can} evaluate their approach on the 1940 Census, for which the original microdata have been made public.
Our approach instead generates synthetic microdata for the 2010 Census \citep{syn-data}, to which we then apply our swapping procedure.

While it is still fairly new, the TopDown algorithm and its close variants have been the subject of a number of studies \citep{wright2021empirical,cohen2021census,kenny2021impact,kenny2023evaluating}.
Both \citet{cohen2021census} and \citet{kenny2021impact} consider how TopDown will impact analyses related to voting rights.
Because the original TopDown algorithm is difficult to run reliably, we adopt a simpler variant (``ToyDown'') introduced by \citet{cohen2021census}, similar to the ``Reasonably Good Algorithm'' of \citet{long-jason-report}.
\citet{wright2021empirical} analyze the reliability of TopDown outputs for redistricting applications.
\citet{kenny2023evaluating} characterize the amount of noise introduced by the overall TopDown process.
We build upon some of their techniques to calibrate a key parameter of our algorithm against released TopDown data.
Beyond the uncertainty introduced by the DAS, \citet{steed2022policy} and \citet{boyd2022differential} detail how other sources of error lead to inaccurate data and downstream policy impacts.

\paragraph*{Synthetic data generation.}

Our swapping implementation requires access to household-level microdata, which are in general only made public by the Census Bureau after 72 years.
Instead, we rely on \textit{synthetic} microdata for the 2010 Census, derived from a combination of Census data products using techniques from \citet{syn-data}.
Synthetic data are not perfect: while they match tabular data released by the Census, they will differ in distribution and locations of households.
Moreover, disclosure avoidance techniques have already been used prior to the release of the data products on which they are based, meaning that when we run our swapping implementation, it will effectively be on \textit{already-swapped} data.
Still, prior work has had to rely on a variety of heuristics to obtain household-level datasets \citep{cohen2021census,christ2022differential}, and the implementation provided by \citet{syn-data} provides a principled way to sample synthetic data.
We provide further details on this approach in \Cref{fig:databases}.

\section{Swapping Implementation}
\label{sec:swapping}

Swapping works by exchanging the details of pairs of households within a state. 
To a first approximation, the swapping ``algorithm'' is as follows: for each household in the state, with some probability, swap the details of that household with a randomly selected household in the same state.
The exact implementation of this process is unknown; the Census Bureau has not publicly released the details of their swapping implementation.
Indeed, the ``swapping algorithm'' is not necessarily an entirely algorithmic procedure; the level of human involvement is unknown.

Nevertheless, we provide an implementation of swapping that is faithful to public disclosures made by the Census Bureau and its representatives \citep{abowd-declaration,mckenna-rmd,zayatz-2010,zayatz-sdc,zayatz-jos,ramanayake2010balancing,hawala2003microdata}. All of these sources confirm the following framework for swapping:
\begin{enumerate}
    \item Within a state, households are assigned a probability of being swapped based on their local uniqueness, which is determined by ``flagging'' variables \citep{steel-zayatz}.
    \item If selected to be swapped, a household is paired with another household that matches on certain ``key'' variables \citep{steel-zayatz}. The matching household is chosen based on its geographic proximity: neither too close nor too far \citep{mckenna-rmd}.
    \item The maximum number of swaps is determined by the ``swap rate'' \citep{steel-zayatz,abowd-ces}, which is the fraction of households in the state that are swapped.
\end{enumerate}
Within this framework, ``[t]he decision variables in the swapping program are the swap rate, the flagging variables, the matching key variables and the ranking of pairs'' \citep{ramanayake2010balancing}.
We consider these four decisions, plus additional implementation details, in what follows.

\paragraph{Swap rate}
The swap rate is not public, but we have some information about reasonable ranges. \citet{bailie2023can} claim that the swap rate is 2--4\%. John Abowd, the former Chief Scientist at the U.S. Census Bureau, states that the swap rates that the Bureau considered when studying the feasibility of using swapping in 2020 were higher than the swap rates used in 2010 \citep{abowd-declaration}. An example of the swap rates that the Bureau considered for 2020 appears in \citet{hawes21determining} (also written by Bureau officials), in which the authors present swap rates in the range of 5--50\%. Based on this information, we use 2\% as a reasonable low swap rate and and 10\% as a reasonable high swap rate. In some places, we also present results for higher swap rates.

\paragraph{Flagging variables} Multiple sources \citep{zayatz-2010,ramanayake2010balancing} highlight the importance of demographic uniqueness within a small geographic region for classifying households into these tiers.
While the precise set of flagging variables used to determine reidentification risk are not made public, sources indicate that ``key demographic variables'' serve as flagging variables~\citep{zayatz-2010}. In this work, we choose flagging variables to be counts of individuals by race, the number of Hispanic individuals, the total number of persons, and the number of adults. To sort households by reidentification risk, we sort them by the number of other households in the same block that have the same values across all of the flagging variables. 

\paragraph{Matching key variables} Documentation on matching variables is inconsistent. According to \citet{zayatz-sdc}, ``The key variables were number of people in the household of each race by Hispanic/Non-Hispanic by age group (\textless18,18+), number of units in building, rent/value of home, and tenure.''
\citet{griffin1989disclosure} states that the planned swapping methodology matches on ``number of persons in household, population characteristics of race, Hispanic origin and age, and on housing characteristics of units at building, rent/value and tenure.''
If true, however, this would guarantee that block-level counts by race and ethnicity are invariant to swapping, which goes against assumptions commonly held in the literature~\citep[e.g.,~][]{christ2022differential,kenny2023evaluating}.
Moreover, \citet{steel-zayatz} appear to contradict this in their evaluation of the impacts of swapping, writing ``Our procedure targeted households with unique characteristics, and this increased the number of zero cells (for these characteristics) because it tended to draw people from cells with values 1, 2, and 3 (presumably where the entire contribution to the cell is from one household and hence selected for high disclosure risk) and swap them into cells with larger values.''

\citet{abowd-declaration} (the chief scientist at the Census Bureau) offers an even more clear contradiction in a sworn declaration: ``For the 2000 and 2010 censuses, those key matching characteristics were (1) the whole number of persons in the household, and (2) the whole number of persons aged 18 or older in the household.''
\citet{wright2021empirical} corroborate this, writing, ``the SWA [swapped] TOTAL counts and the corresponding CEF [unswapped] TOTAL counts at the block level were the same in 2010. The same is true for TOTAL18 counts for the 18 years and over population at the block level.''
\citet{kenny2023comment} draw the same conclusion from \citet{wright2021empirical}.
One potential explanation for this apparent contradiction is that the extended set of matching characteristics described by \citet{zayatz-sdc} and \citet{griffin1989disclosure} were used for the 1990 Census, and a smaller set of matching characteristics were used in 2000 and 2010 as per \citet{abowd-declaration}.
In this work, we take Abowd's definition of key matching characteristics: number of persons and number of persons aged 18 or older.
\paragraph{Ranking of pairs} According to \citet{mckenna-rmd}, ``A household record is typically swapped with another household within a large area but in a different smaller area within the larger one, for example, across tracts but within the same county.'' We were unable to find more specific information than this on how match partners were chosen. In our implementation, we require households to be swapped with a household from a different tract in the same state. Combining this requirement with the requirement that swap partners match on key variables, we can define the set of potential swap partners to be the households that match key variables, are in a different tract, and are in the same state. Among these potential swap partners, we prioritize the geographically closest households.\footnote{Each household is assigned a geographic location based on the block in which it appears.}

\paragraph{Additional implementation details}
There are more minor implementation design decisions that we describe in
\Cref{appendix:swapping_details}. Since we have no way of knowing which
decisions the Census Bureau made, we study two variants of our swapping
implementation. We use one ``standard'' implementation of swapping to conduct
the experiments reported in this work. We also make an alternative
implementation of swapping, which we call our ``high variance'' implementation
because it is tuned to have a higher variance than our standard implementation.
The details of the differences between these variants are in
\Cref{appendix:swapping_details}.

We run all of our experiments using the high variance implementation of swapping and find that our conclusions are largely unchanged. Results for the high variance implementation are not included in this work (because they are similar to the results for the standard implementation), with the exception of the variance estimation results, which we report for both our standard and high variance implementations.

Many of the choices we make are heuristics.
As such, the analyses we provide in the remainder of the paper should not be viewed as authoritative claims about the exact impacts of swapping.
Instead, we view our implementation as approximately correct in its overall approach, and the analyses we produce are reasonable estimates in terms of both magnitude (as a function of the swap rate) and direction.

\section{Characterizing the Effects of Swapping}
\label{sec:effects}

With this implementation, we are ready to analyze the effects of swapping.
In this section, we focus on basic descriptive statistics of the changes induced by swapping to the microdata.
In particular, we focus on race and household size.
Section~\ref{sec:downstream} considers the use of Census data in specific downstream applications.

\subsection{Setup and notation}

We begin by introducing our notation and set up. We view a database as a list of households $\bX=(\bx_1,\ldots,\bx_n)$.
From the perspective of our swapping implementation, each household $\bx_i$ is described by the count of people of each race, the count of Hispanic people, the count of people over 18, and the Census block where the household is located.

Let $\bX_{\text{CEF}}$ be the Census Edited File (CEF), which is the unprotected Census microdata (which we never observe). Let $\bX_{\text{Released}}$ be the released Census data (which is the result of applying swapping and other disclosure avoidance methods to the CEF). Let $\bX_{\text{Synthetic}}$ be the unmodified synthetic database. Note that $\bX_{\text{Released}}$ and $\bX_{\text{Synthetic}}$ match on any statistic aggregated to the Census block level by construction~\citep{syn-data}.

By running our swapping implementation on $\bX_{\text{Synthetic}}$, we can produce swapped databases. In this section, we consider swapping at a 2\% swap rate and a 10\% swap rate (representing the limits of the range of what we believe are realistic swap rates). Let $\bX_{\text{Swapped2}}$ be the result of applying swapping at a 2\% swap rate to $\bX_{\text{Synthetic}}$ and let $\bX_{\text{Swapped10}}$ be the result of applying swapping at a 10\% swap rate to $\bX_{\text{Synthetic}}$. 

By running ToyDown, the simplified version of TopDown implemented by \citet{cohen2021census}, on our synthetic data, we can also produce $\bX_{\text{ToyDown}}$. To produce this database, we run ToyDown with $\epsilon=3.26$ and an equal split of privacy budget across geographic levels. This value of $\epsilon$ was chosen by calibrating the variance of ToyDown against the variance of TopDown in demonstration data (see Section \ref{sec:variance} for details).
Technically, our implementation of ToyDown differs from the one used in \citep{cohen2021census} because we do not treat Hispanic as a race category to be consistent with our swapping algorithm.

Finally, we have data produced by TopDown, the Census Bureau's real differential privacy-based disclosure avoidance system, from demonstration data released to help consumers prepare for the new DAS. This demonstration data included P.L. 94-171 data from two independent runs of the TopDown algorithm under production settings, one released in 2021~\citep{june21topdown} and the other released in 2023~\citep{april23topdown}. We call these two databases $\bX_{\text{TopDown2021}}$ and $\bX_{\text{TopDown2023}}$. Figure \ref{fig:databases} shows the databases that we consider and the relationships between them.

\begin{figure*}
    \centering
    \begin{tikzcd}[cells={nodes={draw=black}}]
        & {\mathbf{X}_{\text{CEF}}} \\
        \\
        {\mathbf{X}_{\text{TopDown}}} && {\mathbf{X}_\text{Released}} && {\mathbf{X}_\text{Synthetic}} \\
        \\
        &&& {\mathbf{X}_\text{ToyDown}} && {\mathbf{X}_\text{Swapped}}
        \arrow["{\text{\scriptsize TopDown}}"'{pos=0.6}, from=1-2, to=3-1]
        \arrow["{\parbox{5cm}{\scriptsize Swapping (Bureau's)\\ and other DAS}}"{pos=0.6}, from=1-2, to=3-3]
        \arrow["{\text{\scriptsize Swapping (ours)}}"{pos=0.6}, from=3-5, to=5-6]
        \arrow["{\text{\scriptsize ToyDown}}"'{pos=0.7}, from=3-5, to=5-4]
        \arrow["{\text{\scriptsize Equal at block level}}", dashed, tail reversed, from=3-3, to=3-5]
        \arrow["\text{Panel D}"', curve={height=30pt}, tail reversed, from=3-1, to=3-3, color=red]
        \arrow["\text{Panels A and B}", curve={height=-50pt}, tail reversed, from=3-5, to=5-6, color=red]
        \arrow["\text{Panel C}", curve={height=-30pt}, tail reversed, from=3-5, to=5-4, color=red]
        \arrow["\text{Quantity of Interest}", curve={height=-40pt}, tail reversed, from=1-2, to=3-3, color=blue]
    \end{tikzcd}


        
    \caption{Various versions of the Census data and the relationships between them. Note that there are two releases of $\bX_\textnormal{TopDown}$, which we call $\bX_\textnormal{TopDown2021}$ and $\bX_\textnormal{TopDown2023}$. We also produce multiple versions of $\bX_{\textnormal{Swapped}}$: $\bX_{\textnormal{Swapped2}}$ is produced by swapping at a 2\% swap rate and $\bX_{\textnormal{Swapped10}}$ is produced by swapping at a 10\% swap rate. The red arrows show the comparisons made in the panels in Figures \ref{fig:swapping_errors} and \ref{fig:relative_swapping_errors}. The blue arrow shows the comparison of interest (which we cannot directly observe).}
    \label{fig:databases}
\end{figure*}

For a race $r\in\mathcal R$ and a geographic level $\Gamma$ (like the set of Census blocks in a state $s$, which we denote $\mathcal B_s$, or the set of counties in a state $s$, which we denote $\mathcal C_s$) we can aggregate households to obtain counts of that race for each geographic unit in that geographic level. For $r\in\mathcal R$, $\gamma\in\Gamma$, and a database $\bX$, we write the number of people of race $r$ in region $\gamma$ according to the database $\bX$ as $c_\gamma^r(\bX)$.

In our experiments, we study five states: Alabama, Wisconsin, Texas, Vermont,
and Nevada. These states were selected to cover a wide range of sizes,
demographic characteristics, and geographic locations. When showing detailed
results about one state, we use Alabama. In many cases, corresponding results
for other states are in \Cref{appendix:additional_tables_figures}.

\subsection{Quantifying the Errors Introduced by Swapping}\label{sec:errors}

We begin by studying the bias of swapping on race counts. It is important to remember that the \emph{state-wide average} bias of the count of any race is zero, since swapping (unlike TopDown) does not change the statewide counts of each race.
Therefore, we study the error in the individual counts at lower geographic levels. These errors will average to zero across the entire state, but we are interested in their magnitudes.

For a race $r$ and a geographic region $\gamma$, we can define the error
\[\textsf{Error}_\gamma^r(\bX_{1},\bX_{2})\triangleq c_\gamma^r(\bX_{1})-c_\gamma^r(\bX_{2}).\]
Ideally, we would estimate quantities $\textsf{Error}_\gamma^r(\bX_{\text{CEF}},\bX_{\text{Released}})$, which represent the error in race counts due to the Bureau's implementation of swapping applied to the underlying Census data (shown in blue in Figure~\ref{fig:databases}).
This, however, is impossible without privileged access to confidential Census microdata.
Instead, we study the quantity $\textsf{Error}_\gamma^r(\bX_{\text{Synthetic}},\bX_{\text{Swapped}})$, which we can define for swapping at a 2\% swap rate or a 10\% swap rate.
Note that this, too, is the result of applying the swapping to an underlying dataset.
However, it differs from our ideal quantity of interest $\textsf{Error}_\gamma^r(\bX_{\text{CEF}},\bX_{\text{Released}})$ in two key ways: (1) the ``unswapped'' data we use are synthetic microdata, and (2) the swapping algorithm used is our own implementation as opposed to the Bureau's.
Our conclusions are valid insofar as these two components (the synthetic data and the swapping algorithm) are sufficiently close to their real-world analogues (the true microdata and the Bureau's swapping procedure).

For comparison, we also consider two baselines using differential privacy.
The first is a direct comparison with ToyDown, which we apply to the same synthetic data:
$\textsf{Error}_\gamma^r(\bX_{\text{Synthetic}},\bX_{\text{ToyDown}})$.
In addition, we compare the TopDown demonstration data against the released 2010 Census data, yielding
$\textsf{Error}_\gamma^r(\bX_{\text{Released}},\bX_{\text{TopDown2021}})$.
We caution that $\textsf{Error}_\gamma^r(\bX_{\text{Released}},\bX_{\text{TopDown2021}})$ should not be directly compared with the three other quantities $\textsf{Error}_\gamma^r(\bX_{\text{Synthetic}},$ $\bX_{\text{Swapped2}})$, $\textsf{Error}_\gamma^r(\bX_{\text{Synthetic}},\bX_{\text{Swapped10}})$, and $\textsf{Error}_\gamma^r(\bX_{\text{Synthetic}},$ $\bX_{\text{ToyDown}})$; these three quantities measure the error introduced by applying a single process (swapping or ToyDown, respectively) to a dataset, while $\textsf{Error}_\gamma^r(\bX_{\text{Released}},\bX_{\text{TopDown2021}})$ measures the error between the results of \emph{two different} processes---the Census Bureau's swapping method and TopDown---being applied to the same underlying dataset.\footnote{As an illustrative note, a direct comparison to $\textsf{Error}_\gamma^r(\bX_{\text{Released}},\bX_{\text{TopDown2021}})$ using synthetic data would be the quantity $\textsf{Error}_\gamma^r(\bX_{\text{Swapped}},\bX_{\text{ToyDown}})$.}
See Figure~\ref{fig:databases} for further intuition for these comparisons.

\begin{figure*}
    \centering
    \begin{tabular}{rccl}
        \multirow{-9}{*}{A}&\includegraphics[scale=0.3]{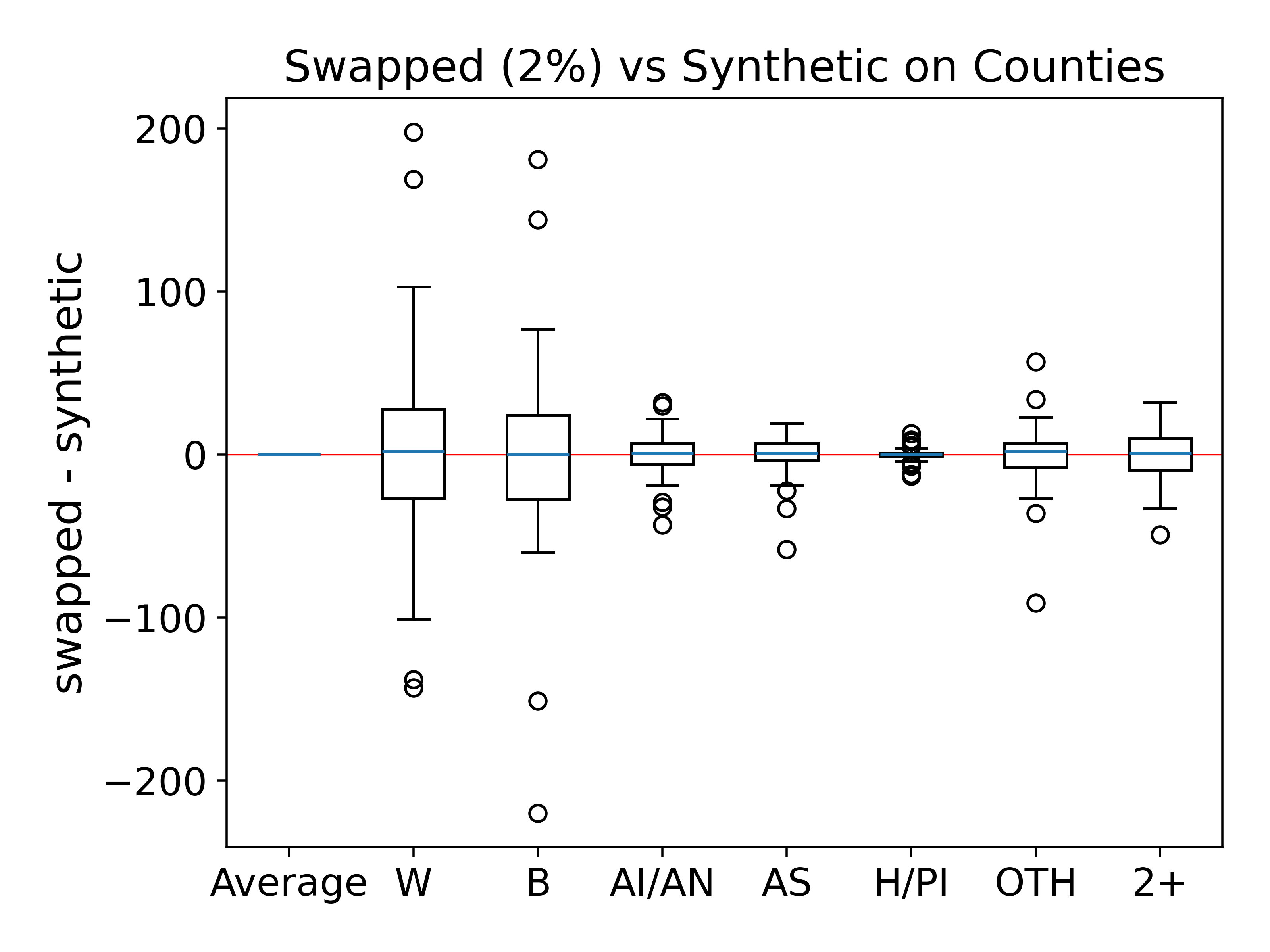}&
        \includegraphics[scale=0.3]{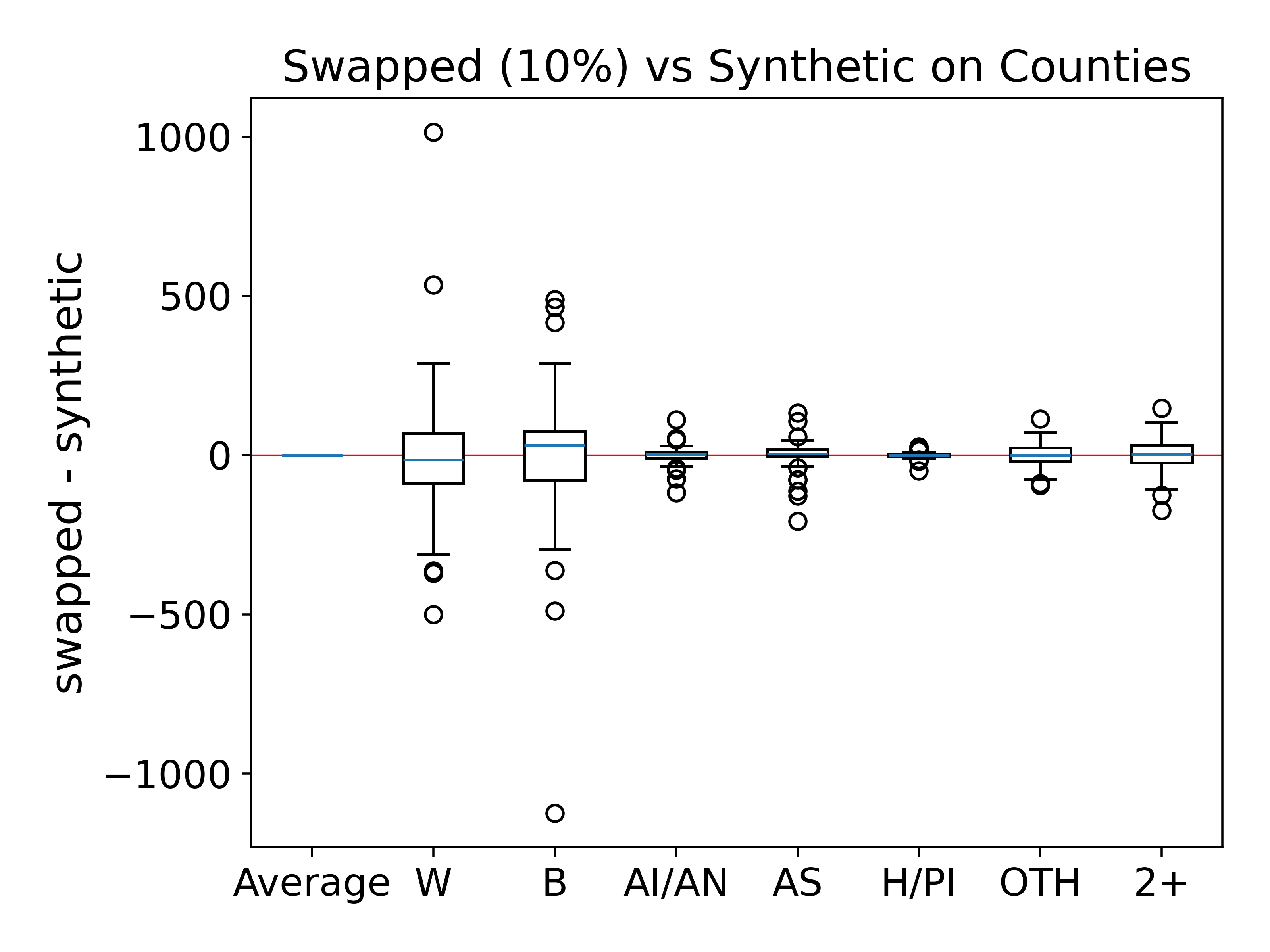}&\multirow{-9}{*}{B}\\
        \multirow{-9}{*}{C}&\includegraphics[scale=0.3]{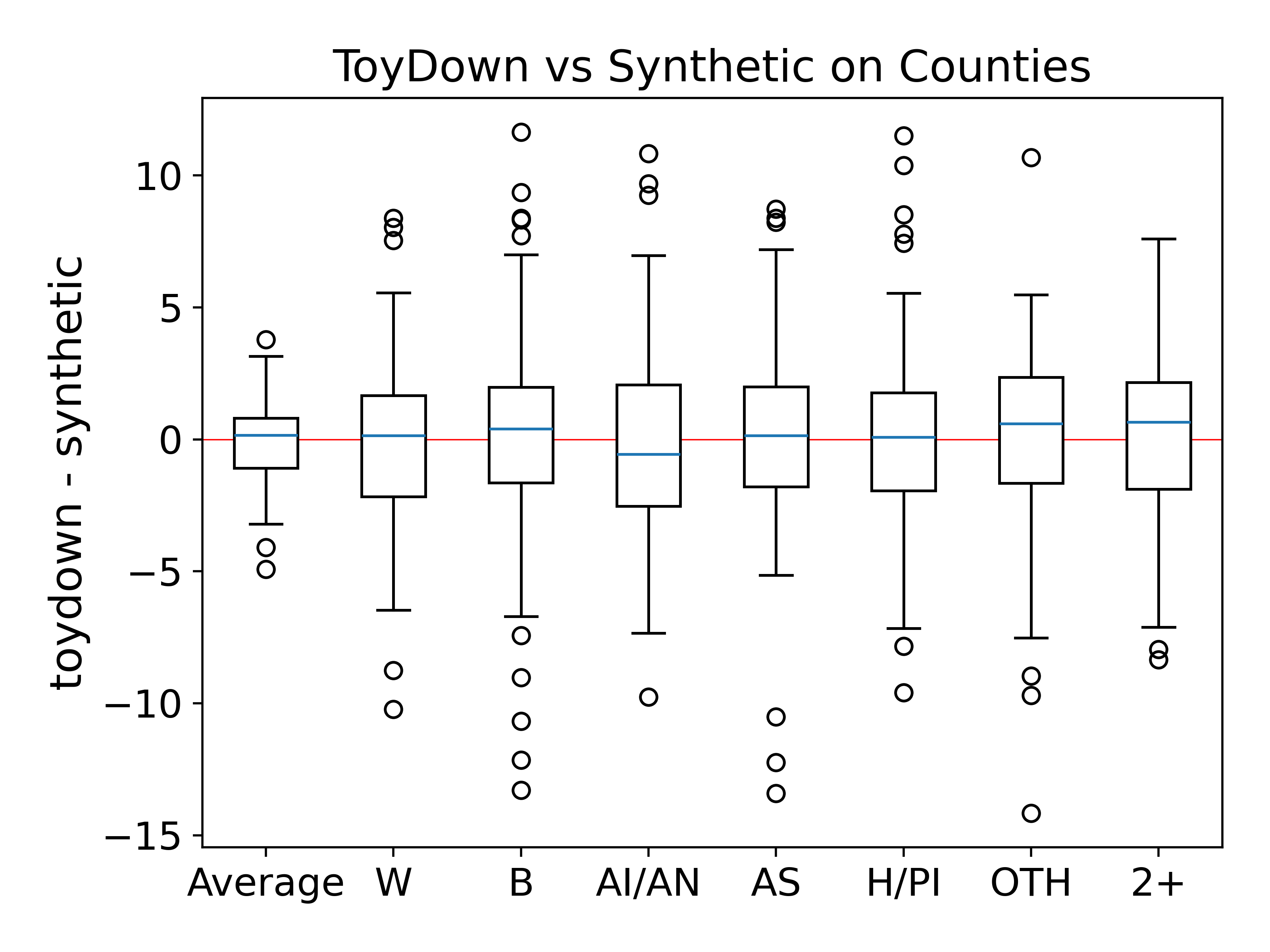}&
        \includegraphics[scale=0.3]{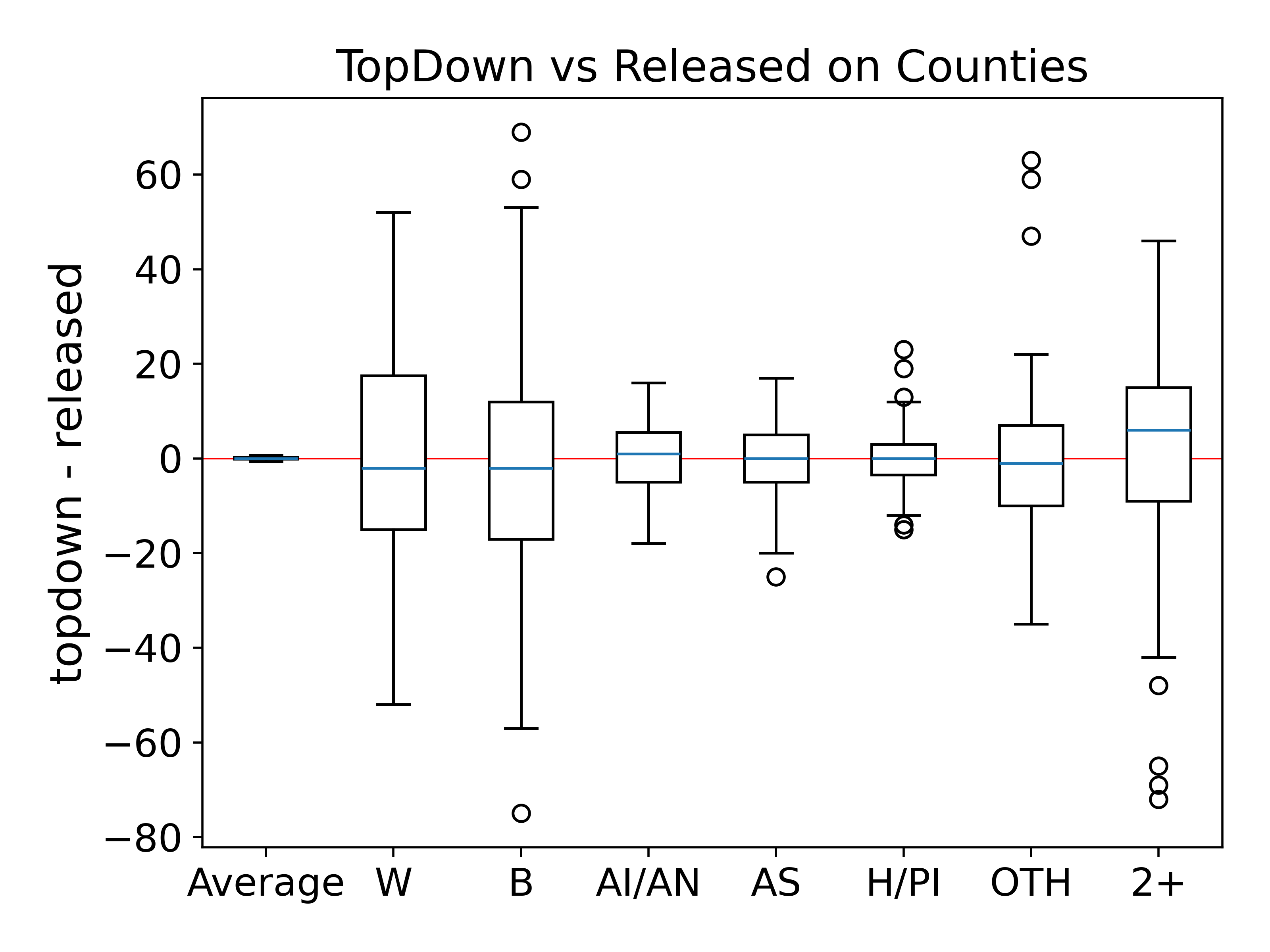}&\multirow{-9}{*}{D}
    \end{tabular}
    \caption{Plots of the four quantities $\textnormal{\textsf{Error}}_\gamma^r(\bX_{\textnormal{Synthetic}},\bX_{\textnormal{Swapped2}})$, $\textnormal{\textsf{Error}}_\gamma^r(\bX_{\textnormal{Synthetic}},\bX_{\textnormal{Swapped10}})$, $\textnormal{\textsf{Error}}_\gamma^r(\bX_{\textnormal{Synthetic}},\bX_{\textnormal{ToyDown}})$, $\textnormal{\textsf{Error}}_\gamma^r(\bX_{\textnormal{Released}},\bX_{\textnormal{TopDown2021}})$ for all $\gamma\in\mathcal C_{\textnormal{Alabama}}$, where $\mathcal C_{\textnormal{Alabama}}$ is the set of all counties in Alabama. To generate the ``Average'' plot, for each county, the errors for each race were averaged. Note that the scale of the $y$-axes are different in different plots.}
    \label{fig:swapping_errors}
\end{figure*}

Figure~\ref{fig:swapping_errors} shows $\textsf{Error}_\gamma^r(\bX_{\text{Synthetic}},\bX_{\text{Swapped2}})$, $\textsf{Error}_\gamma^r(\bX_{\text{Synthetic}},$ $\bX_{\text{Swapped10}})$, $\textsf{Error}_\gamma^r(\bX_{\text{Synthetic}},\bX_{\text{ToyDown}})$, and $\textsf{Error}_\gamma^r(\bX_{\text{Released}},$
\newline $\bX_{\text{TopDown2021}})$ at the county level in Alabama.
As noted earlier, the errors for swapping (at both 2\% and 10\%) are both mean zero.\footnote{The swap errors do not necessarily have \emph{median} zero; the blue lines in the boxplots are the medians.}
We observe that errors due to swapping (Panels A and B) are generally greater than errors due to ToyDown (Panel C). Also, swapping results in higher errors on large racial groups, while ToyDown distributes errors evenly across racial groups. The picture for TopDown (Panel D) may superimpose the errors due to swapping and TopDown, which is consistent with the fact that both swapping and TopDown stand between $\bX_{\text{TopDown}}$ and $\bX_{\text{Released}}$.

For downstream applications (for example, any application using the proportion of each race) errors relative to population size are also important. To evaluate relative errors, we use the definition
\[\textsf{RelativeError}_\gamma^r(\bX_{1},\bX_{2})\triangleq \cfrac{2}{1+\cfrac{c_\gamma^r(\bX_{1})}{c_\gamma^r(\bX_{2})}}\] as a measure of the effect of swapping on the count of race $r$ in region $\gamma$, normalized by the size of the count. If the count increases by a relatively large amount from $\bX_1$ to $\bX_2$, $\textsf{RelativeError}_\gamma^r(\bX_{1},\bX_{2})$ will be close to 2. If the count decreases by a relatively large amount, $\textsf{RelativeError}_\gamma^r(\bX_{1},\bX_{2})$ will be close to 0. If the count is unchanged, $\textsf{RelativeError}_\gamma^r(\bX_{1},\bX_{2})=1$. If both counts are 0, we treat the relative error as 1. In practice, the way we treat zeros is not very important because the only county-level race counts that are zero are the number of Hawaiian or Pacific Islander people in five counties (before swapping).

\begin{figure*}
    \centering
    \begin{tabular}{rccl}
        \multirow{-9}{*}{A}&\includegraphics[scale=0.3]{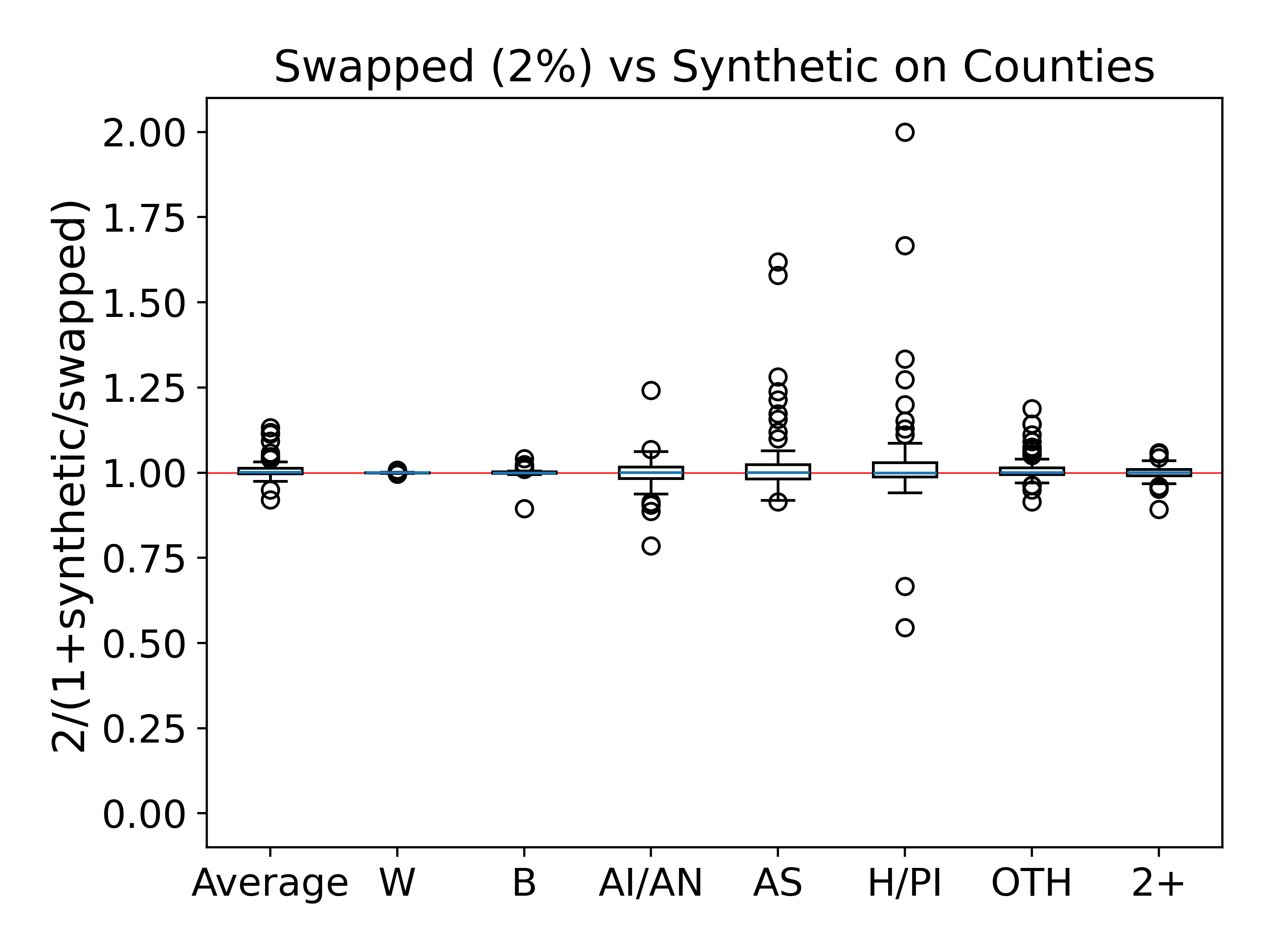}&
        \includegraphics[scale=0.3]{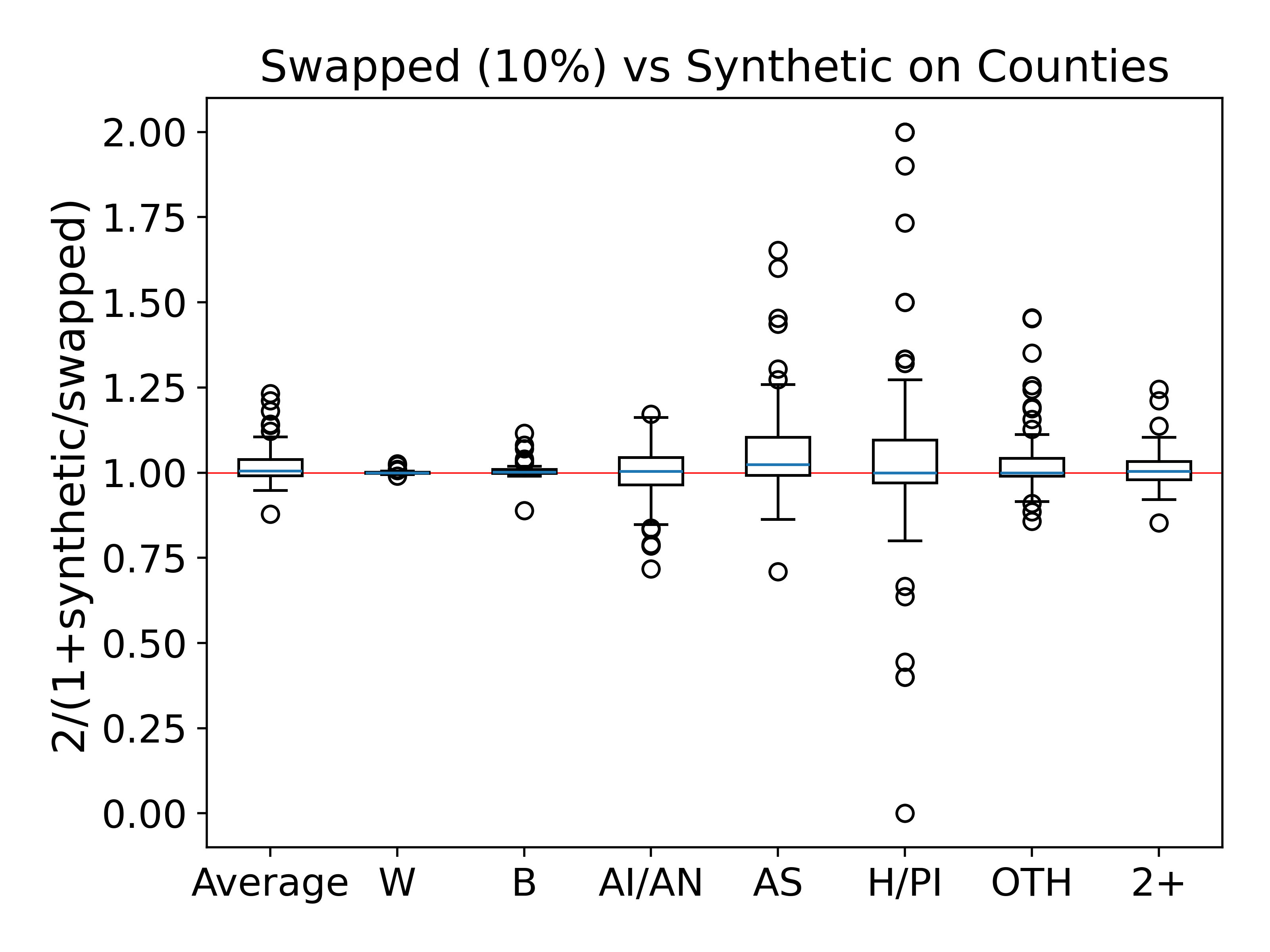}&\multirow{-9}{*}{B}\\
        \multirow{-9}{*}{C}&\includegraphics[scale=0.3]{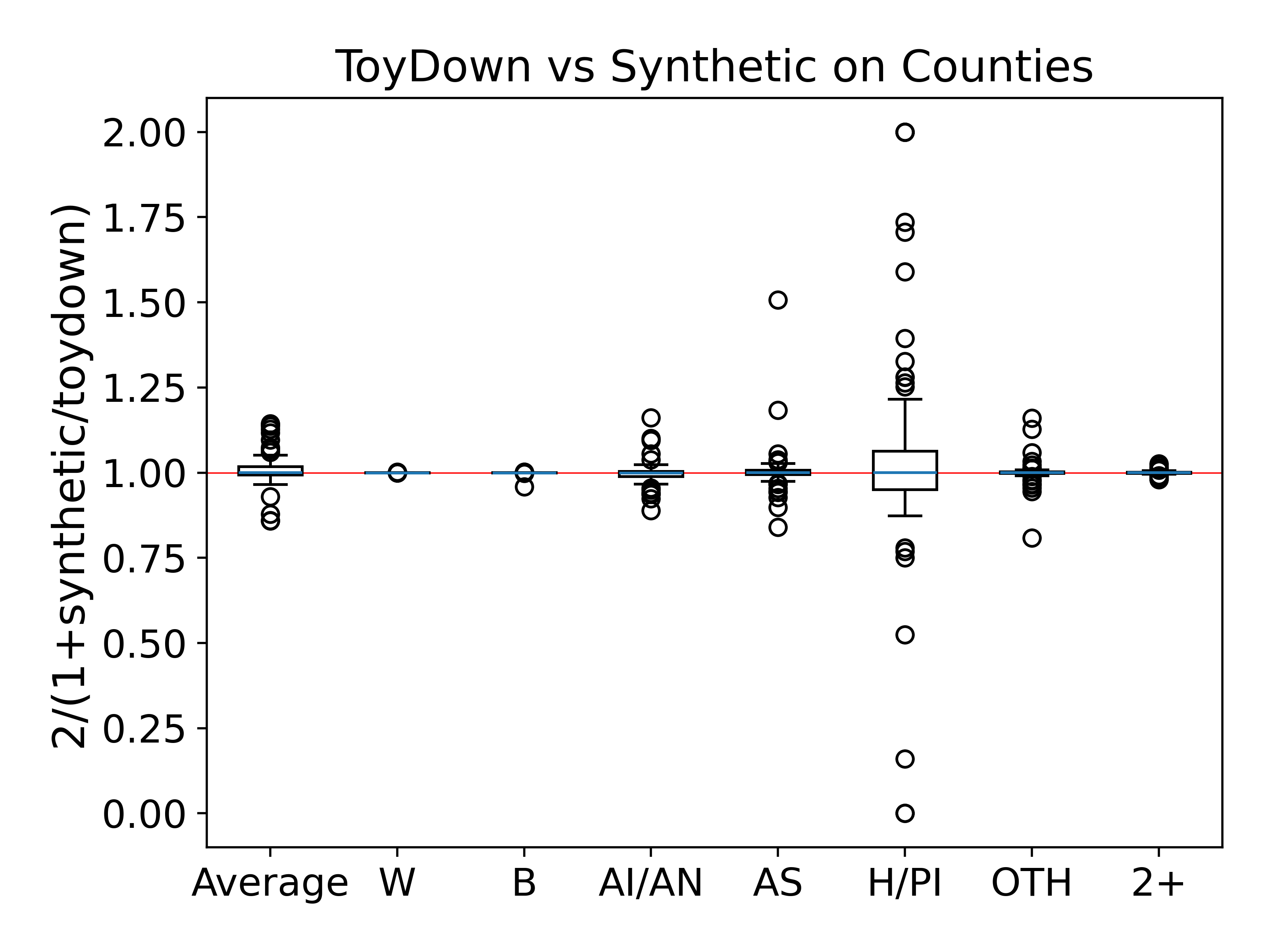}&
        \includegraphics[scale=0.3]{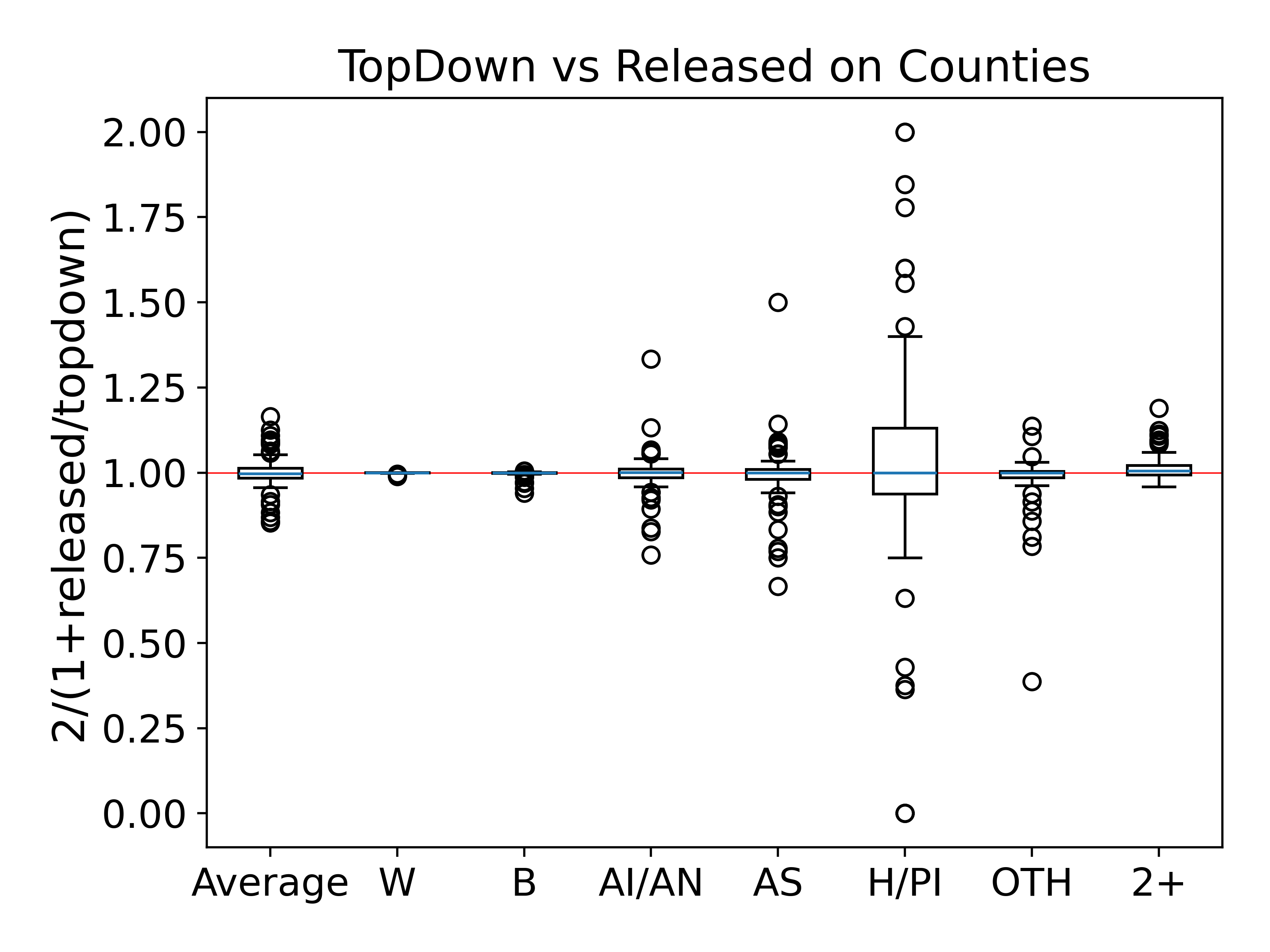}&\multirow{-9}{*}{D}
    \end{tabular}
    \caption{Plots of the four quantities $\textnormal{\textsf{RelativeError}}_\gamma^r(\bX_{\textnormal{Synthetic}},\bX_{\textnormal{Swapped2}})$, $\textnormal{\textsf{RelativeError}}_\gamma^r(\bX_{\textnormal{Synthetic}},\bX_{\textnormal{Swapped10}})$, $\textnormal{\textsf{RelativeError}}_\gamma^r(\bX_{\textnormal{Synthetic}},\bX_{\textnormal{ToyDown}})$, and $\textnormal{\textsf{RelativeError}}_\gamma^r(\bX_{\textnormal{Released}},\bX_{\textnormal{TopDown2021}})$ for all $\gamma\in\mathcal C_{\textnormal{Alabama}}$, where $\mathcal C_{\textnormal{Alabama}}$ is the set of all counties in Alabama. To generate the ``Average'' plot, for each county, the relative errors for each race were averaged.}
    \label{fig:relative_swapping_errors}
\end{figure*}

Again, our ideal comparison is $\textsf{RelativeError}_\gamma^r(\bX_{\text{CEF}},\bX_{\text{Released}})$. However, since we do not observe $\bX_{\text{CEF}}$, we instead measure \\$\textsf{RelativeError}_\gamma^r(\bX_{\text{Synthetic}},\bX_{\text{Swapped}})$ (defined for swapping at a 2\% swap rate and a 10\% swap rate), as well as $\textsf{RelativeError}_\gamma^r(\bX_{\text{Synthetic}},$ $\bX_{\text{ToyDown}})$ and $\textsf{RelativeError}_\gamma^r(\bX_{\text{Released}},\bX_{\text{TopDown2021}})$. These are shown in Figure~\ref{fig:relative_swapping_errors}. We observe that 2\% swapping has smaller relative errors than ToyDown, but 10\% swapping has comparable relative errors to ToyDown. Both swapping and ToyDown exhibit very low relative error on large racial groups and much larger relative error on small racial groups (especially with a 10\% swap rate, but often with a 2\% swap rate as well). In the case of ToyDown, this is consistent with the fact that ToyDown distributes \emph{absolute} errors relatively evenly across racial groups, so groups with smaller sizes have larger \emph{relative} errors. In the case of swapping, it appears that, while swapping has larger absolute errors on larger racial groups, this is more than canceled out by the effect of the group sizes. The overall effect is that swapping, at least at a 10\% swap rate, introduces larger relative errors to smaller groups.

\subsection{Quantifying the Variance of Swapping}\label{sec:variance}

In addition to biasing quantities of interest, because they are randomized algorithms, both swapping and TopDown introduce variance into their outputs.
Here, we compare the variance they introduce into block-level counts by race.
Our techniques draw inspiration from \citet{kenny2023evaluating}, who establish estimators for the bias and variance introduced by TopDown.
\citet{bailie2023can} also compare swapping at different rates against TopDown,
but with one key difference: while we compare the data \textit{accuracy} of
swapping and TopDown, they compare the formal \textit{privacy guarantees} of
swapping and ToyDown. Though we do not study formal privacy guarantees, our
study of variance is motivated by its connection to privacy. Noise is necessary
to ensure privacy \citep{dinur_nissim}, and the variance of a statistical process is a measure of how much noise it introduces. 

\def\vartd{\ensuremath{V_s^{\text{TD}}}}
\def\varsw{\ensuremath{V_s^{\text{swap}}}}
\def\ctd{\ensuremath{C_s^\text{TD}}}
\def\ctda{\ensuremath{C_s^{\text{TD}_1}}}
\def\ctdb{\ensuremath{C_s^{\text{TD}_2}}}

\def\vartdhat{\ensuremath{\widehat{V}_s^{\text{TD}}}}
\def\varswhat{\ensuremath{\widehat{V}_s^{\text{swap}}}}

\def\epstd{\ensuremath{\varepsilon_s^\text{TD}}(b, r)}
\def\epstda{\ensuremath{\varepsilon_s^{\text{TD}_1}}(b, r)}
\def\epstdb{\ensuremath{\varepsilon_s^{\text{TD}_2}}(b, r)}

\def\epssw{\ensuremath{\varepsilon_s^\text{swap}}(b, r)}

\def\cswa{\ensuremath{C_s^{\text{swap}_1}}}
\def\cswb{\ensuremath{C_s^{\text{swap}_2}}}

Similarly to \citet{kenny2023evaluating}, define
\begin{align*}
    \varepsilon_\gamma^r(\bX_{\text{Swapped}})\triangleq  c_\gamma^r(\bX_{\text{Swapped}})-\EE_{\bX\sim\textsc{Swapping}}[c_\gamma^r(\bX)],
\end{align*} where $\bX\sim\textsc{Swapping}$ indicates that $\bX$ is generated by running swapping with fresh randomness.
With this, the variance of the counts by race at the block level can be written
\begin{align*}
    V_{\text{swap}} &\triangleq \frac{1}{|\mc B_s| |\mc R|} \sum_{\gamma \in \mc B_s} \sum_{r \in \mc R} \Var(\varepsilon_\gamma^r(\bX_{\text{Swapped}})),
\end{align*}
where $\mathcal B_s$ is the set of blocks in a state $s$ and $\mc R$ is the set of Census race categories.

We do not observe the expected counts $\EE_{\bX\sim\textsc{Swapping}}[c_\gamma^r(\bX)]$, so we can't directly compute $\varepsilon_\gamma^r(\bX_{\text{Swapped}})$.
However, we can produce an unbiased estimate of $V_{\text{Swapped}}$ by running swapping twice to produce $\bX_{\textnormal{Swapped}}$ and $\bX_{\textnormal{Swapped}}'$. This yields the estimator
\begin{align*}
    \widehat V_\text{Swapped} &\triangleq \frac{1}{2 |\mc B_s| |\mc R|}  \sum_{\gamma \in \mc B_s} \sum_{r \in \mc R} \p{c_\gamma^r(\bX_{\text{Swapped}})-c_\gamma^r(\bX_{\text{Swapped}}')}^2.
\end{align*}
Note that $\E{\widehat V_\text{Swapped}} = V_\text{Swapped}$ since
\begin{align*}
    &\hspace{-10mm}\E{\p{c_\gamma^r(\bX_{\text{Swapped}})-c_\gamma^r(\bX_{\text{Swapped}}')}^2} \\
    &= \E{\p{\varepsilon_\gamma^r(\bX_{\text{Swapped}})-\varepsilon_\gamma^r(\bX_{\text{Swapped}}')}^2} \\
    &= 2\E{\varepsilon_\gamma^r(\bX_{\text{Swapped}})^2} - 2\E{\varepsilon_\gamma^r(\bX_{\text{Swapped}})}^2 \tag{$\varepsilon_\gamma^r(\bX_{\text{Swapped}})$, $\varepsilon_\gamma^r(\bX_{\text{Swapped}}')$ are i.i.d.} \\
    &= 2\Var(\varepsilon_\gamma^r(\bX_{\text{Swapped}})).
\end{align*}

Following a similar approach for TopDown, we can define \[\varepsilon_\gamma^r(\bX_{\text{TopDown}})\triangleq c_\gamma^r(\bX_{\text{TopDown}}) - \EE_{\bX\sim\textsc{TopDown}}[c_\gamma^r(\bX)],\] where $\bX\sim\textsc{TopDown}$ indicates that $\bX$ is generated by running TopDown with fresh randomness. 
Then, we have
\begin{align*}
    V_\text{TopDown}
    &\triangleq \frac{1}{|\mc B_s| |\mc R|} \sum_{\gamma \in \mc B_s} \sum_{r \in \mc R} \Var(\varepsilon_\gamma^r(\bX_{\text{TopDown}}))
\end{align*}
Following \citep{kenny2023evaluating}, we use the two publicly available TopDown protected datasets $\bX_\text{TopDown2021}$ and $\bX_\text{TopDown2023}$ to form the estimator
\begin{align*}\label{eq:variance_estimator}
    &\hspace{-1mm}\widehat V_\text{TopDown} \\
    &\triangleq \frac{1}{2 |\mc B_s| |\mc R|}  \sum_{\gamma \in \mc B_s} \sum_{r \in \mc R} \p{c_\gamma^r(\bX_{\text{TopDown2021}})-c_\gamma^r(\bX_{\text{TopDown2023}})}^2.
\end{align*}

\begin{figure*}
    \centering
    \includegraphics[scale=0.35]{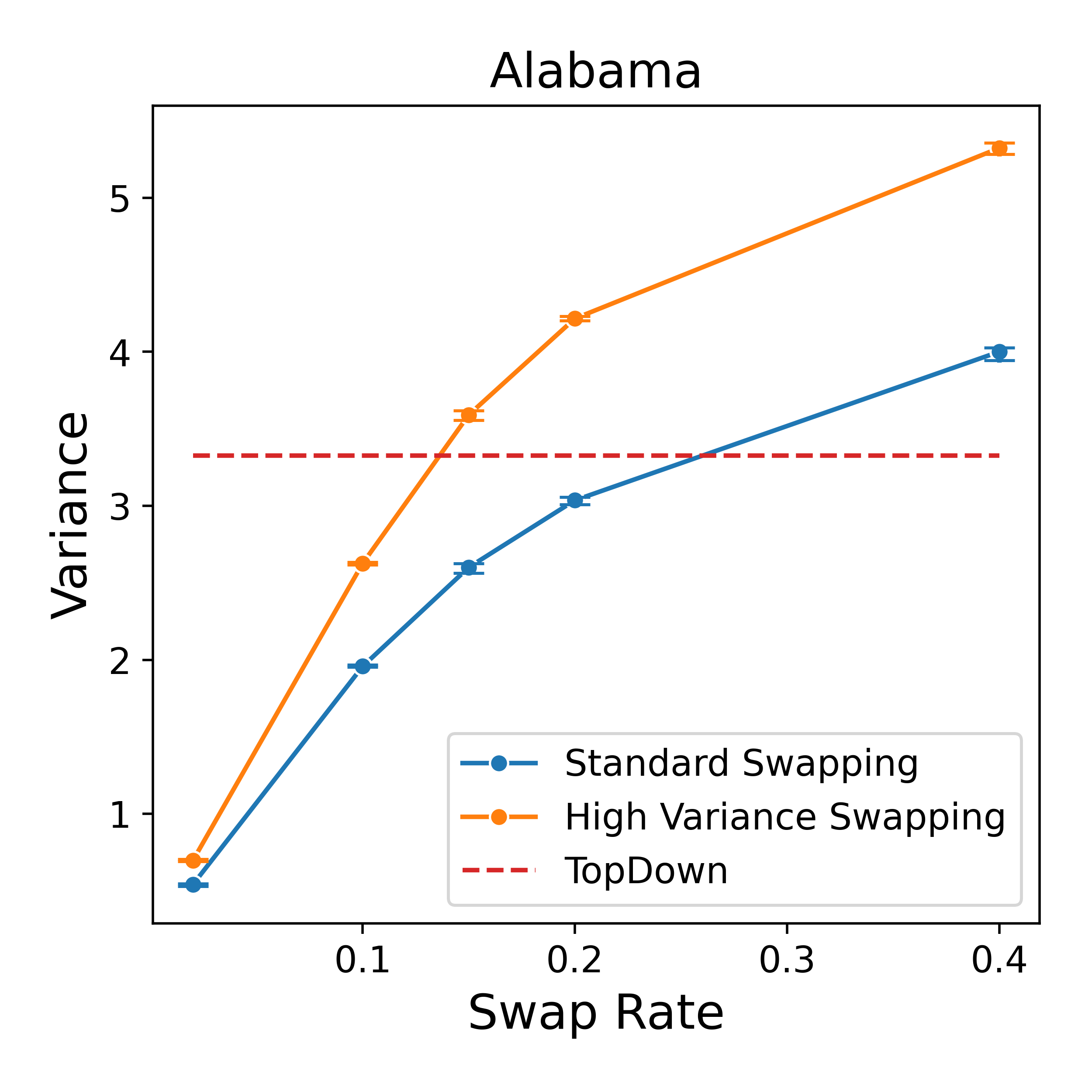}\includegraphics[scale=0.35]{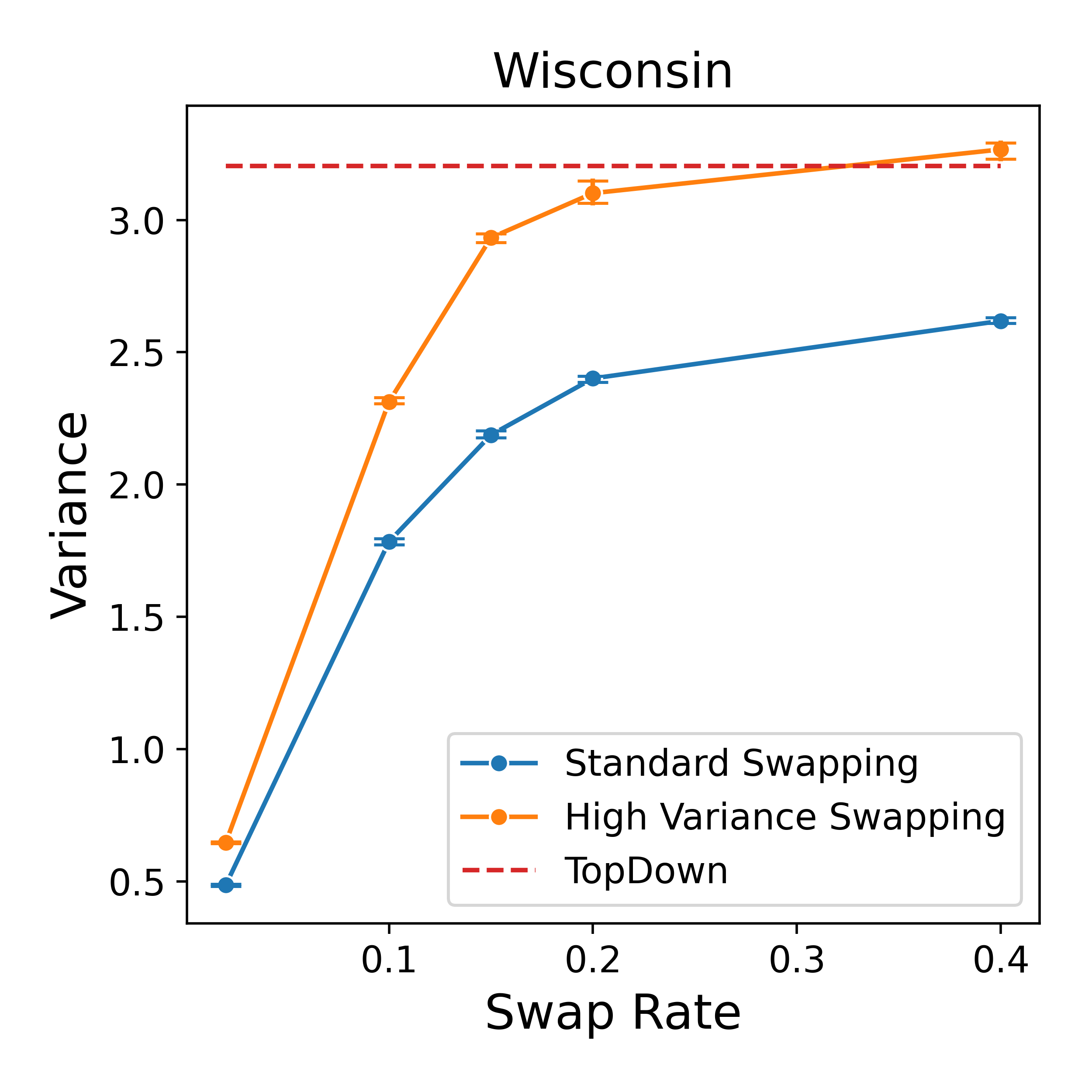}\includegraphics[scale=0.35]{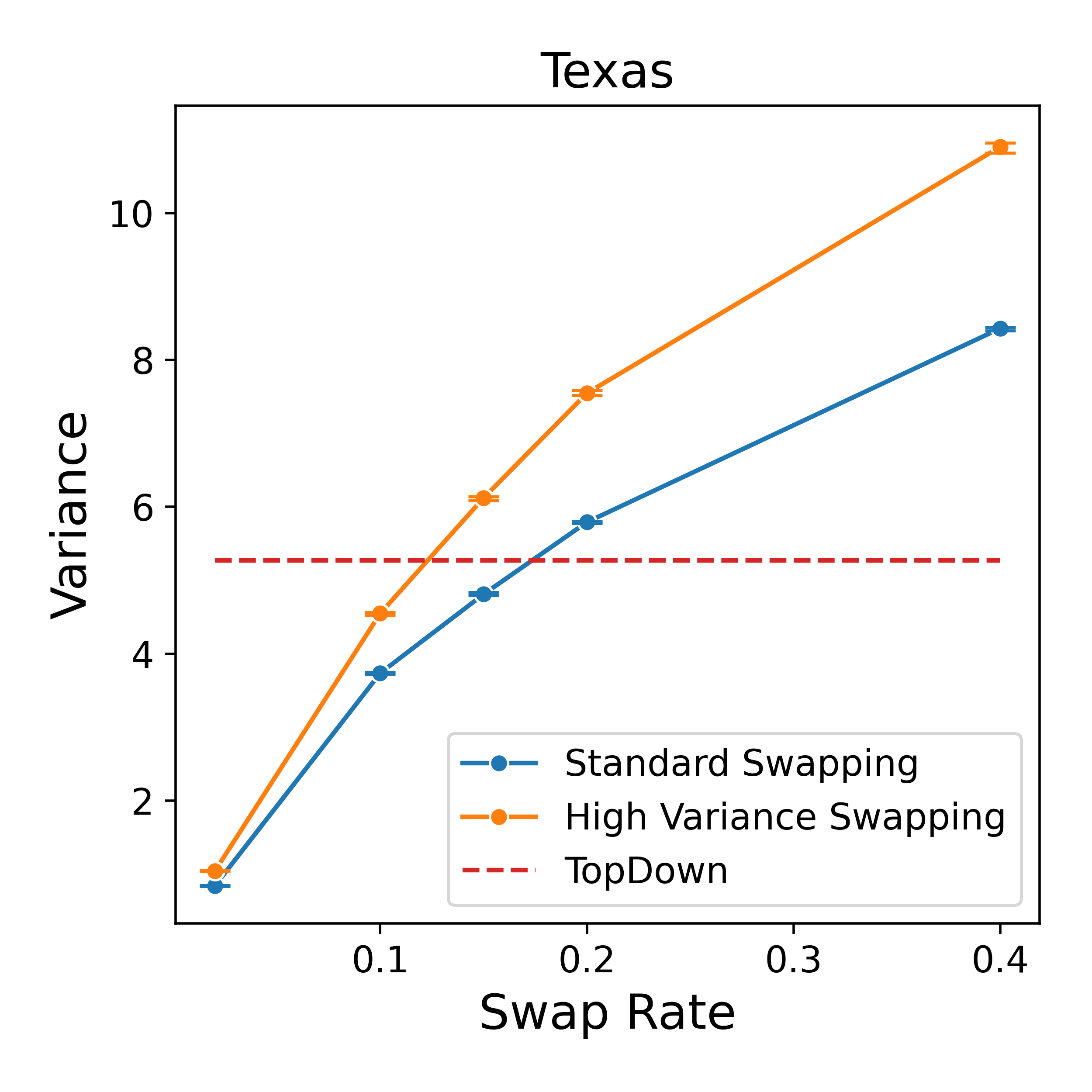}
    \caption{The variance of swapping at different swap rates. Error bars show
    the minimum and maximum of 5 runs of our estimator. The horizontal dotted
  lines show the variance estimate for TopDown. Note that we do not characterize
the error in our estimate for TopDown. Plots for two more states appear in
\Cref{fig:variance_plots_vt_nv}.}
    \label{fig:variance_plots}
\end{figure*}

While we are unable to precisely characterize the variance of the TopDown estimator $\widehat V_\text{TopDown}$ (since we only observe two samples), we can estimate $\Var(\widehat V_\text{Swapped})$ since we can sample it as many times as we like (to obtain each sample, we run swapping twice to generate two independent swapped datasets).
In Figure \ref{fig:variance_plots}, we show the estimated variance of swapping for different swap rates.
Because the variance of swapping is a function of many parameters of the swapping algorithm, we report results for both our standard implementation of swapping and our high variance implementation of swapping.
While we believe these two versions of swapping are realistic, we emphasize that our variance estimates are not necessarily close to the variance of the Bureau's swapping method. 

For comparison, we also plot the estimated variance of TopDown as a horizontal line. First, we observe that the variance estimates for both processes are different across different states. Generally, all variance estimates are higher in larger states. Second, in all states, it takes a high swap rate to reach the same variance as TopDown (in some states, it appears there is no swap rate resulting in variance as high as TopDown). Swapping at the rates that we believe the Census realistically used has lower variance than TopDown. We caution that we do not claim that the swap rate should be the rate that yields equal variance to that of TopDown. This is because (1) it is not clear that the Bureau was interested in equalizing the variances of the 2010 DAS and the 2020 DAS, (2) our variance estimates of swapping do not necessarily estimate the variance of the Bureau's swapping algorithm, and (3) our two versions of swapping (standard and high variance) yield significantly different swap rates at which  swapping and TopDown have comparable variances.

Given the connection between variance and privacy described above, these observations suggest that our swapping algorithm with realistic swap rates likely has lower variance, and therefore lower privacy, than TopDown.
This observation is consistent with statements made by \citet{abowd-declaration}.

We also use the above method for variance estimation to calibrate $\epsilon$ for the ToyDown data used in Section \ref{sec:errors} to match the variance of TopDown.
For different values of $\epsilon$, we run ToyDown twice, producing $\bX_{\text{ToyDown},\epsilon}$ and 
$\bX_{\text{ToyDown},\epsilon}'$. Then, we measure
\begin{align*}
&\hspace{-2mm}\widehat{V}_{\text{ToyDown}}(\epsilon) \\
&\triangleq\frac{1}{2|\mathcal B_s||\mathcal R|}\sum_{\gamma\in\mathcal B_s}\sum_{r\in\mathcal R}(c_\gamma^r(\bX_{\text{ToyDown},\epsilon})-c_\gamma^r(\bX_{\text{ToyDown},\epsilon}'))^2.
\end{align*}
We find that $\widehat{V}_{\text{ToyDown}}(3.26)\approx \widehat{V}_{\text{TopDown}}$, suggesting that ToyDown with $\epsilon=3.26$ provides roughly the same amount of variance as the TopDown implementation run by the Census Bureau.

\subsection{Effect of Swapping on Racial Entropy}\label{sec:racial_entropy}

Here, we consider how swapping impacts the racial distribution of the population.
We use entropy as our primary measure of interest here.
Higher racial entropy would indicate that populations are more geographically segregated.
Here, we focus our analysis on Alabama.
Formally, define the racial entropy of a geographic area to be the entropy of the empirical distribution of race in that area; i.e., for a geographic area $\gamma$,
\[\textsf{RacialEntropy}_\gamma(\bX)=-\sum_{r\in\mathcal R}\frac{c_\gamma^r(\bX)}{\sum_{r'\in\mathcal R} c_\gamma^{r'}(\bX)}\log\left(\frac{c_\gamma^r(\bX)}{\sum_{r'\in\mathcal R} c_\gamma^{r'}(\bX)}\right).\]
Because swapping specifically targets rare households and moves them to a different tract, we might expect it to have a homogenizing effect, resulting in reduced tract-level racial entropy.
Surprisingly, we find that, on average, swapping increases the racial entropy of each tract. The average entropy increases from unswapped data to swapped data at a 2\% rate and again from swapped data at a 2\% rate to swapped data at a 10\% rate. \Cref{tab:entropies} shows the average entropy of tracts before and after swapping.

\begin{table*}
\centering
\begin{tabular}{|l|l|l|l|}
\hline
State & \begin{tabular}[c]{@{}l@{}}Average tract entropy\\ before swapping\end{tabular} & \begin{tabular}[c]{@{}l@{}}Average tract entropy\\ after 2\% swapping\end{tabular} & \begin{tabular}[c]{@{}l@{}}Average tract entropy\\ after 10\% swapping\end{tabular} \\ \hline
AL    & 0.598                                                                           & 0.603                                                                              & 0.624                                                                               \\ \hline
WI    & 0.431                                                                           & 0.434                                                                              & 0.447                                                                               \\ \hline
TX    & 0.807                                                                           & 0.810                                                                              & 0.819                                                                               \\ \hline
NV    & 0.997                                                                           & 0.999                                                                              & 1.006                                                                               \\ \hline
VT    & 0.232                                                                           & 0.233                                                                              & 0.241                                                                               \\ \hline
\end{tabular}
\caption{Average racial entropy at the tract level before and after swapping.}
\label{tab:entropies}
\end{table*}

To see why this is the case, observe that a household can be targeted for swapping on the basis of things besides having uncommon races.
    \begin{enumerate}
        \item Household size is a non-racial factor to swapping, and it turns out to be quite influential in determining which households are swapped.
        \item Households may be swapped not because they have uncommon races, but because their exact composition of races is uncommon. One reflection of this is that mixed race households are over-represented among the households involved in swapping. 
    \end{enumerate}
The more independently of race that swapping happens, the more it increases racial entropy instead of decreasing it. Race is still important in swapping, but does not fully determine who is swapped. The following statistic illustrates this: in Alabama, white people make up 69\% of the total population and 39\% of the people targeted in swapping at a 10\% swap rate. We see that white people (i.e. people with the statewide majority race) are indeed underrepresented among the population targeted for swapping, but they are still swapped often. This is because they may live in households of unusual sizes or with unusual racial compositions.

\begin{figure*}
    \centering
    \begin{tabular}{cc}
        \includegraphics[scale=0.25]{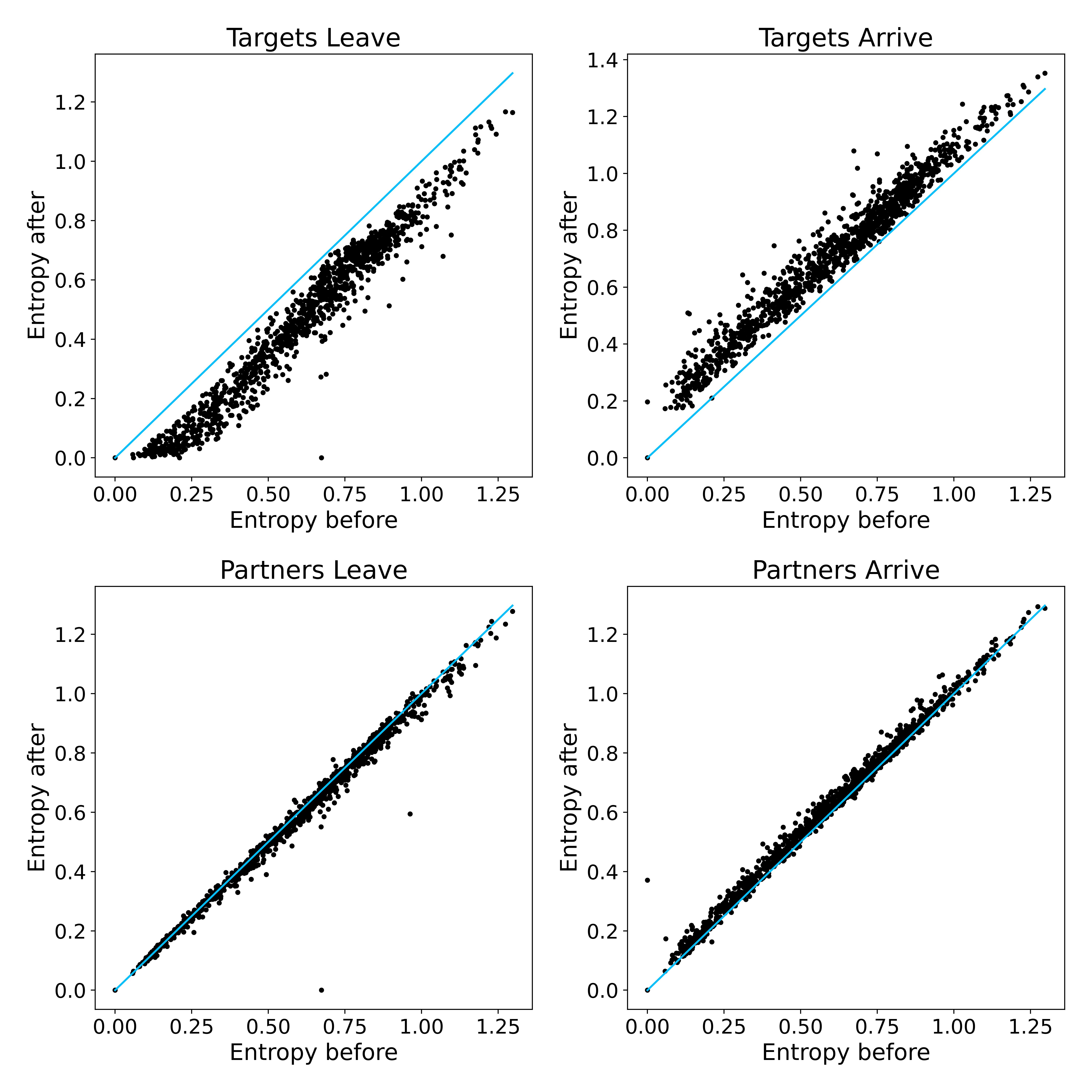}&\includegraphics[scale=0.3]{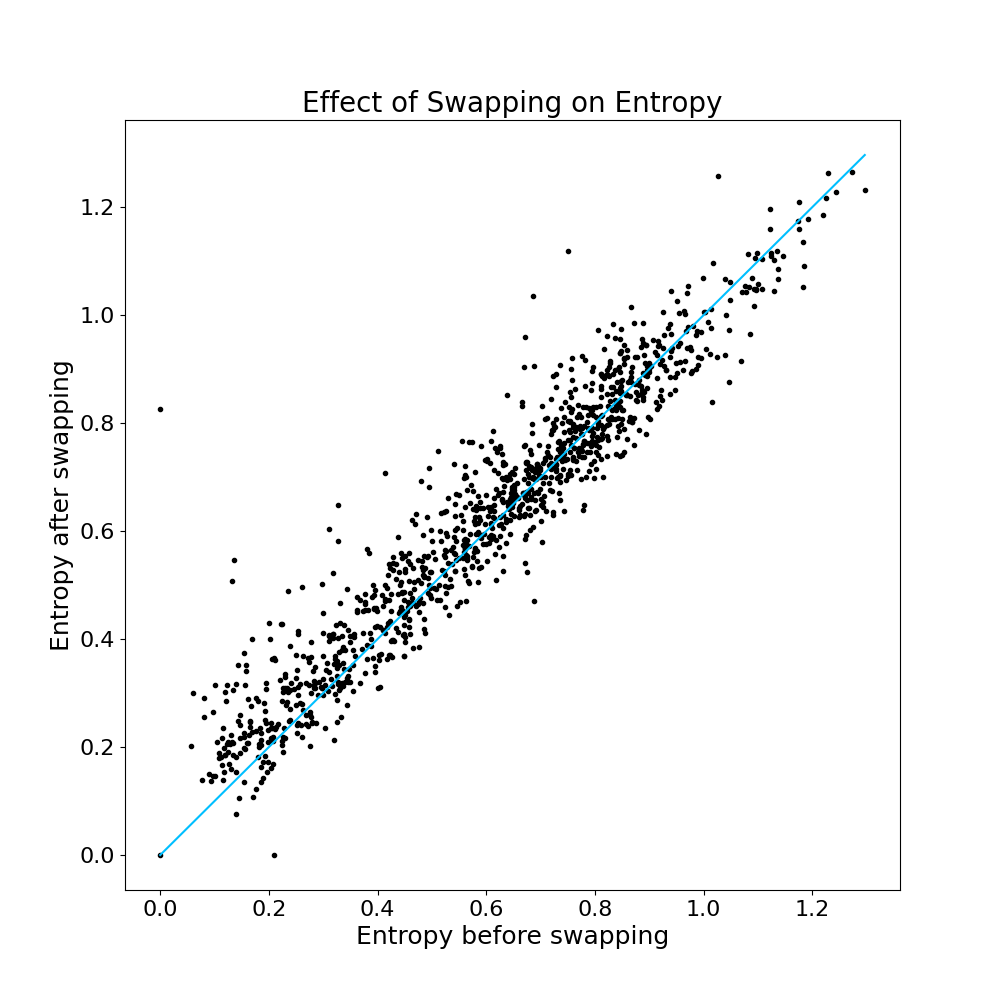}
    \end{tabular}
    \caption{The effect of swapping on racial entropy at the tract level for a
    10\% swap rate in Alabama. The left panel shows the effect of each step of
  swapping on entropy. The right panel shows the overall effect. Corresponding
plots for 2\% and 10\% swap rates in other states appear in
\Cref{fig:entropy_al,fig:entropy_wi,fig:entropy_tx,fig:entropy_nv,fig:entropy_vt}
in \Cref{appendix:additional_tables_figures}.}
    \label{fig:entropy_parts}
\end{figure*}

Another reason that swapping increases entropy, on average, is because many components of swapping are targeted only minimally. Conceptually, we can break each swap down into four steps: (1) the targeted household leaving its original location, (2) the targeted household arriving at its new location, (3) the partnered household leaving its original location, and (4) the partnered household arriving at its new location. The intuition that swapping decreases entropy by removing uncommon households only holds for the first step. This is illustrated in Figure~\ref{fig:entropy_parts}, which shows the effect on entropy of each step of swapping. The effect of step (1) (removing all the targeted households) indeed decreases racial entropy. Step (2) (removing all the partner households) also slightly decreases racial entropy. However, steps (3) and (4) (adding all the targeted households to their partners' tracts and adding all the partner households to the tracts of the targets) both have the effect of increasing racial entropy. The overall effect is to increase racial entropy.

We study several other interesting impacts of swapping in
\Cref{appendix:additional_swapping_analyses}. These include the rates of swapping on different household sizes, the effects of swapping on the distribution of people by race, and statistics describing which races are swapped for which.

\section{Downstream Impacts of Swapping}
\label{sec:downstream}

Having established basic characteristics of swapping, we next turn to its impacts on downstream analyses that rely on Census data.
We consider two prototypical analyses in public policy (Section~\ref{sec:voting}) and social science (Section~\ref{sec:rural}) settings.
Our results suggest that swapping can be a significant source of error in downstream analyses, and we suggest a general framework using our swapping implementation to estimate how swapping will impact other analyses in Section~\ref{sec:framework}.

\subsection{Voting}
\label{sec:voting}
The estimation of voting patterns by race is crucial for applications such as enforcing the Voting Rights Act requirement that voting districts are drawn in a way that allows minority voters to elect their preferred candidates \citep{ecological_inference_freedman}.
Given many sets of marginal counts of votes and racial groups (one for each voting district or precinct), two common methods are used to infer voting preferences: ecological regression (ER) and ecological inference (EI). Following \citet{cohen2021census}, we focus on ER.

In ecological regression, the support of each racial group for each candidate is learned separately. To learn the support of racial group $r$ for candidate $c$, ecological regression is a simple linear regression with the percent of race $r$ in the voting area as the explanatory variable and the percent of votes for $c$ as the response variable. This regression line is used to predict the percent of votes for candidate $c$ in a hypothetical voting district composed of all people of race $r$, which represents the support of race $r$ for candidate $c$.

In Section~\ref{sec:racial_entropy}, we observed that swapping has the effect of increasing racial entropy. If we compare ecological regression before and after swapping, swapping will tend to move the values for percent of race $r$ away from 0 and 1 and towards the middle. This increases the magnitude of the slope of the regression line, resulting in greater apparent polarization (i.e., the support of different racial groups for a given candidate will vary more from the overall support for that candidate).

\begin{figure*}
    \centering
    \includegraphics[scale=0.4]{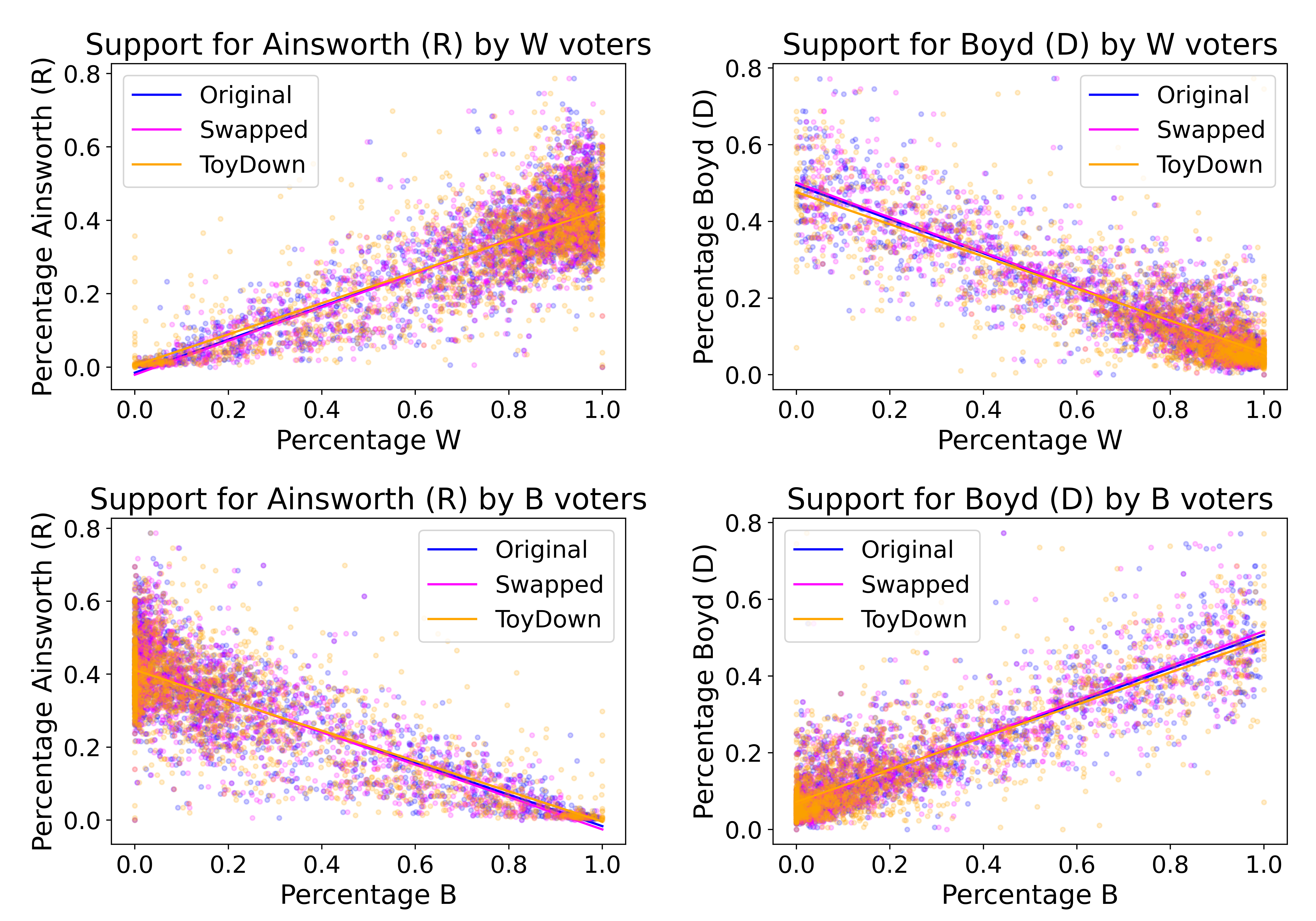}
    \caption{Ecological regression run on unswapped (in blue), swapped (in magenta), and ToyDown protected (in orange) data. The outputs of ER (the estimates of supports of racial group $r$ for candidate $c$) are the intersections of the regression lines with the line $x=1$. In this figure, ER \emph{is not} weighted by population.}
    \label{fig:er_small}
\end{figure*}

This effect is illustrated in Figure~\ref{fig:er_small}, which is the result of running ER on Alabama using election data from the 2018 lieutenant gubernatorial general election between Will Ainsworth, a white Republican candidate, and Will Boyd, a Black Democratic candidate. 
Since ER requires extrapolating to a hypothetical voting district with 100\% of race $r$, it is only reliable for racial groups $r$ that have reasonably high percentages in many of the real voting districts. Therefore, Figure~\ref{fig:er_small} only shows white and Black voter support for the Democratic and Republican candidates. We include the result of ER for all combinations of races and candidates in Figure~\ref{fig:er} in Appendix~\ref{appendix:additional_tables_figures}.

The regression lines based on swapped data (magenta) have slopes of greater magnitude than the regression lines based on unswapped data (blue). This means that estimates of racial support based on swapped data are slightly more polarized than estimates of racial support based on unswapped data. For Figure~\ref{fig:er_small}, the data was swapped at a 10\% swap rate.

On the other hand, the regression lines based on ToyDown protected data (orange) have slopes of slightly smaller magnitude than those based on unswapped data. This replicates the effect found by \citet{cohen2021census}, and occurs because the addition of noise to the covariates reduces the correlation between the covariates and the outcome variable, resulting in a regression line with a slope of smaller magnitude. To create Figure~\ref{fig:er_small}, we used ToyDown with $\epsilon=0.25$ to exaggerate the effect of ToyDown. Since $\epsilon$ and the swap rate affect the magnitude of the effects, it is not meaningful to compare the relative magnitudes of the effect due to swapping versus the effect due to ToyDown. Our key observation is simply that the effect of swapping on the slope of ER regression lines is in the opposite direction from the effect of ToyDown on the slope of ER regression lines.\footnote{We compare swapping to ToyDown to align our results with those of \citep{cohen2021census}, but we also ran ER with swapping compared to TopDown and observed qualitatively similar effects.}

Like \citet{cohen2021census}, we observe that weighting ER by the population of the voting districts reduces the effect of ToyDown to almost nothing. In contrast, weighting does not appear to reduce the effect of swapping in the same way. The result of ER with weighting is shown in Figure~\ref{fig:er_weighted} in Appendix \ref{appendix:additional_tables_figures}.
This suggests that interventions that mitigate errors due to differential privacy may not hold for swapping, and perhaps vice versa.
Our results indicate that the effects of swapping can differ substantially from those of differential privacy.

\subsection{Social science}
\label{sec:rural}
The use of DP in the 2020 Census disclosure avoidance system has caused concern about the impact of TopDown on important analyses in social sciences \citep{MuellerJ.Tom2022T2UC,ruggles2019differential,hauer2021differential,santos2020differential,winkler2021differential}.
Here, we choose one such study and investigate whether the concerns it raises also hold for swapping.
In particular, we build upon a study by \citet{MuellerJ.Tom2022T2UC}, who observe that TopDown introduces significant errors in population and race counts in rural areas. 

\begin{figure*}
    \centering
    \includegraphics[width=\textwidth]{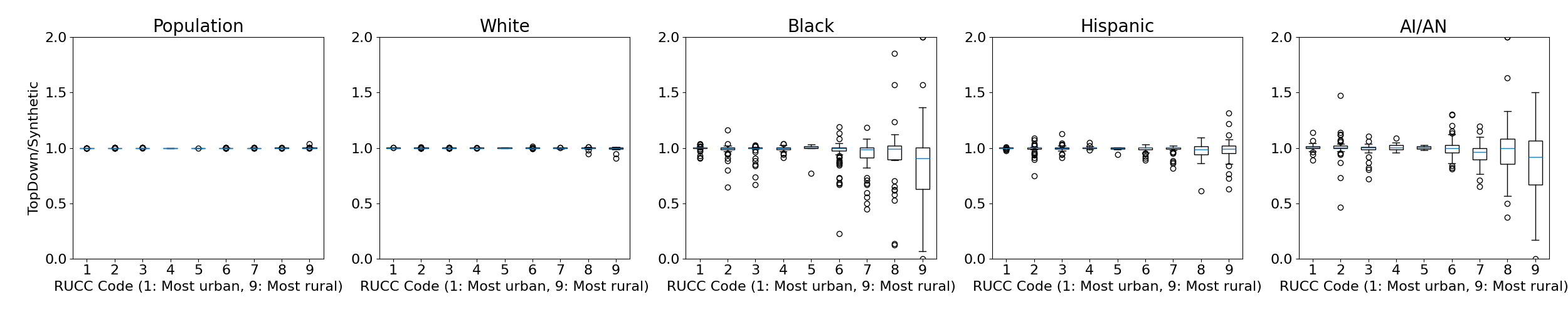}

    \includegraphics[width=\textwidth]{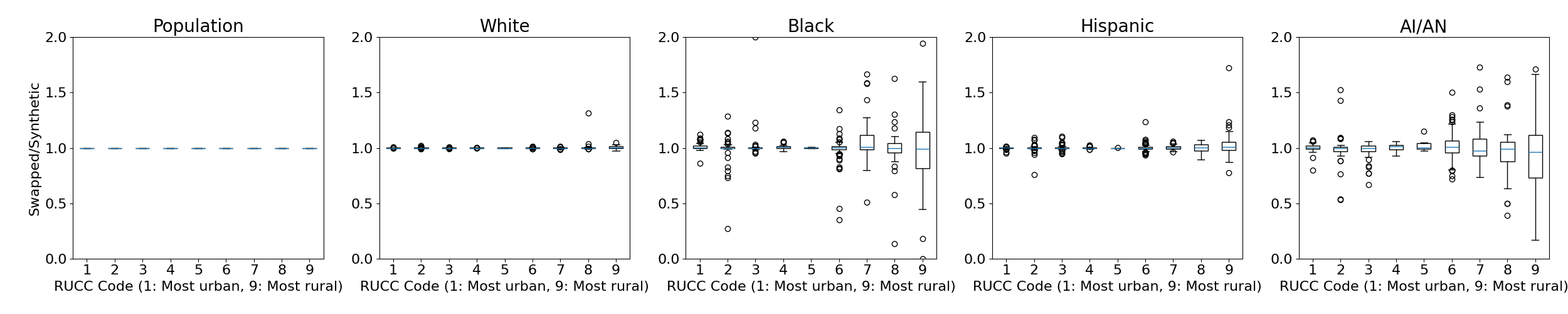}
    \caption{In the top panel, we create a similar figure to \cite[Figure 2]{MuellerJ.Tom2022T2UC}. The vertical axis is the TopDown count divided by the count in the released summary data for 2010. The horizontal axis groups each county by its ROCC number, which represents how urban or rural it is (1 is very urban, 9 is very rural). Infinite or $0/0$ values are left off of the plot. In the bottom panel, we create the analogous figure for swapping. Note that all vertical axes are scaled from 0 to 2 for readability (some outliers are cut out of the graph).}
    \label{fig:rural_errors}
\end{figure*}

With our methods, we can carry out an analogous analysis on the impact of swapping, shown in Figure~\ref{fig:rural_errors}. We replicate the qualitative observations of \citet{MuellerJ.Tom2022T2UC} and make new observations about swapping.\footnote{Our results are based only on Alabama, Nevada, Texas, Vermont, and Wisconsin, while the results in \citet{MuellerJ.Tom2022T2UC} are based on the whole U.S. While we limit our analyses to those states for computational reasons, in principle, it would be possible to use our methods to swap the whole country.} We conclude that, while differential privacy indeed introduces problematic relative errors in small racial groups in rural areas, so does swapping. Even at a 2\% swap rate (which we believe to be conservative), the relative errors introduced by swapping in counts of small racial groups in rural areas are as great or greater than those introduced by TopDown.\footnote{Note that, to match \citet{MuellerJ.Tom2022T2UC}, we actually compare TopDown to the released 2010 data, which is swapped (this corresponds to the arrow marked ``Panel D'' in Figure \ref{fig:databases}). As mentioned before, this is a comparison between two different processes applied to the same underlying data, not a direct comparison between the underlying data and the processed data.} Note that total population counts are not affected by swapping, so the errors introduced by swapping are always zero (compared to small, but nonzero, errors introduced by TopDown).

Our results have broader implications for social science that relies on Census data.
In general, these data already contain a variety of sources of error or noise including nonresponse and misreporting \citep{steed2022policy,boyd2022differential}.
While differential privacy is an additional source of noise for 2020 and beyond, swapping may have been an equally significant (albeit harder to quantify) source of error for analyses based on 1990--2010 censuses.
We conclude with a discussion of how researchers might seek to account for these errors using our implementation.

\subsection{A General Framework}
\label{sec:framework}

\def\bZ{\mathbf{Z}}

Here, we offer a general recipe for researchers seeking to characterize the effects of swapping on a particular statistic.
Suppose one is interested in a statistic $s$ that depends on Census data $\bX_{\text{Released}}$ and some auxiliary information $\mathbf Z$.
Note that for Census years 1990--2010, $\bX_{\text{Released}}$ has already been swapped by the Census Bureau prior to its release.
A researcher might be interested in $s(\bX_{\text{CEF}}, \bZ)$, but instead can only compute $s(\bX_{\text{Released}}, \bZ)$.
To investigate the effects of swapping on $s$, one could follow our recipe:
\begin{enumerate}
    \item Generate a synthetic database $\bX_{\text{Synthetic}}$ (for example, using the methods of \citet{syn-data}).
    \item Use our implementation to generate multiple datasets $\bX_{\text{Swapped}}$.
    \item Estimate the bias and variance of the random variable $\Delta \triangleq s(\bX_{\text{Swapped}},\mathbf Z)-s(\bX_{\text{Synthetic}},\mathbf Z))$ over the randomness of our swapping algorithm.
    This will require multiple independent runs of swapping.
\end{enumerate}
Under the assumption that our implementation of swapping behaves similarly to the Census Bureau's true swapping procedure, $s(\bX_{\text{Released}},\mathbf Z)-s(\bX_{\text{CEF}},\mathbf Z))$ behaves like a draw from the distribution of $\Delta$.
If $\Delta$ has high variance, we cannot reliably measure our statistic of interest $s$ using a single Census release.
If, on the other hand, $\Delta$ has nonzero mean and low variance, we might ``de-bias'' $s$ as $s(\bX_{\text{CEF}}, \bZ) \approx s(\bX_{\text{Released}}, \bZ) - \mathbb{E}[\Delta]$.
We suggest reporting $\Delta$ across a range of parameter values (in particular, the swap rate).
Using the same method with ToyDown instead of swapping, one can study the effect of a TopDown-like process.
In this way, researchers can characterize the effects of both swapping-based and DP-based disclosure avoidance systems.

\section{Discussion and Limitations}

Reliable Census statistics are essential to a range of research and public policy applications.
With the introduction of differential privacy into published US Census data, stakeholders are understandably concerned about noisier estimates and potentially worse decisions.
However, prior Census releases have been subject to swapping, a different (and more opaque) form of disclosure avoidance that has been hard to quantify thus far.
Our work seeks to close this gap by characterizing the effects of swapping and providing others with the tools to analyze its effects in their own contexts.

We find that swapping can have similarly significant impacts on downstream analyses as differential privacy.
In a sense, this is to be expected: if swapping
seeks to provide comparable privacy protections to differential privacy, it must inject randomness on a similar scale.
More surprisingly, we find that swapping cannot simply be treated as a like-for-like replacement for differential privacy in terms of the \textit{types} of errors it introduces.
Swapping can bias estimates in opposite directions as differential privacy, and remedies to make analyses robust to differential privacy may not work for swapping.
We hope that the methods and tools we provide will give researchers a principled way to reason about and potentially mitigate errors introduced by swapping.

Our work carries a number of significant limitations.
While we give our best effort to provide an implementation of swapping that is consistent with the Bureau's, there will undoubtedly be differences.
Our robustness checks seek to address this, but do not rule out the possibility that some of the effects we describe are artifacts of our implementation.
Similarly, the synthetic data we generate will in general differ from the true microdata, and our conclusions are only valid insofar as the effects of swapping on this synthetic data are comparable to its effects on the true microdata.
The magnitudes of the effects we observe are in general hard to quantify without more knowledge about the actual parameter values (such as the swap rate) used by the Bureau.
And finally, swapping is not the only disclosure avoidance technique used in the 1990--2010 censuses.
Some entries were censored, rounded, top-coded, or otherwise altered to meet confidentiality requirements; we are unable to evaluate these in this work.

Our work is intended to be a starting point for future research on the impacts of disclosure avoidance systems on Census data and downstream analyses.
When analyzing Census data, researchers can and should use simulations like ours to determine how privacy preserving techniques are likely to affect their estimates.
We seek to do this for swapping; future work could do the same TopDown.
Moreover, future work could extend our methods to other Census data products like the American Community Survey.
Our hope is that this line of work ultimately enables better research and decision-making based on data generated via fully- or partially-known random processes.

\bibliographystyle{plainnat}
\bibliography{refs}

\appendix

 \section{Additional Swapping Implementation Details}\label{appendix:swapping_details}Beyond the main swapping implementation choices described in Section \ref{sec:swapping}, there are a few more minor choices we must make. Among these choices, we conduct a robustness check by creating two implementations of swapping, one standard implementation and one ``high-variance'' implementation.
\begin{itemize}
    \item To pick a swap partner, we do not deterministically pick the geographically closest potential swap partner; instead, we take the $k$ closest potential swap partners and randomly chose one. In our standard implementation of swapping, $k=10$. In our high variance swapping implementation, $k=100$.
    \item We know that the Census Bureau assigns each household to a tier from 1 (least risk) to 4 (greatest risk), where tier 4 households are swapped with probability 1 \citep{steel-zayatz}. To assign these tiers, we sort the households by reidentification risk (using the method above to determine reidentification risk) and split the list into tiers. We set the distribution of tier assignments such that (1) there are twice as many tier 3 households as tier 4 households, and three times as many tier 2 households as tier 4 households and (2) in expectation, we will swap all of the tier 4 households and half of the tier 3 households by the time we make the desired number of swaps (assuming the standard swap probabilities described below). For reference, using a $10\%$ swap rate, this means that $6.25\%$ of the households are in tier 4, $12.5\%$ are in tier 3, $18.75\%$ are in tier 2, and $62.5\%$ are in tier 1.\footnote{Since we iterate through households in order of their tier, households in tier 1 or 2 are very unlikely to be \emph{targeted} to be swapped, but still have a chance of being involved in a swap as the swap \emph{partner}.}
    \item The decision to swap a household is randomized, where the probability of a house being swapped depends on its risk tier. Concretely, we sort the households by tier from higher tier to lower tier, randomly shuffling the households within each tier. Then, we iterate through the households. Tier 4 households are swapped with probability 1 (following \citet{steel-zayatz}); tier 3 households are swapping with probability $p_3$, tier 2 households with probability $p_2$, and tier 1 households with probability $p_1$. We continue swapping until we have swapped a total number of households equal to the swap rate multiplied by the number of households in the state.
    Our standard implementation uses $p_3 = 0.6$, $p_2 = 0.3$, and $p_1 = 0.1$. Our high variance swapping implementation uses $p_3'=0.3$, $p_2'=p_2$, and $p_1'=p_1$. This change increases the expected number of households considered as swap targets before the swap rate is exhausted.
\end{itemize}

 \section{Additional Analyses of Swapping}\label{appendix:additional_swapping_analyses}\paragraph{Household Size} Household size is an important factor in determining which households are swapped. This is because the inclusion of race counts as flagging variables for determining reidentification risk implies that total household size is also a flagging variable (because total household size is the sum of race counts). Consistent with this, we observe that larger (more uncommon) household sizes are over-represented among the swapped households. For example, 2 person households are the most common household size in Alabama, making up $35.59\%$ of households. However, they make up only $15.69\%$ of the swapped households under a $2\%$ swap rate and $22.12\%$ of the swapped households under a $10\%$ swap rate. In contrast, $3.41\%$ of all households in Alabama are of size 6 or greater, but these households make up $21.85\%$ of the households swapped at a $2\%$ swap rate and $20.39\%$ of the households swapped at a $10\%$ swap rate. The full table of the distribution over household sizes for the state, the swapped households under a $2\%$ swap rate, and the swapped households under a $10\%$ swap rate is shown in Table~\ref{table:hh_dist} in Appendix~\ref{appendix:additional_tables_figures}.

\begin{figure}
    \centering
    \includegraphics[scale=0.22]{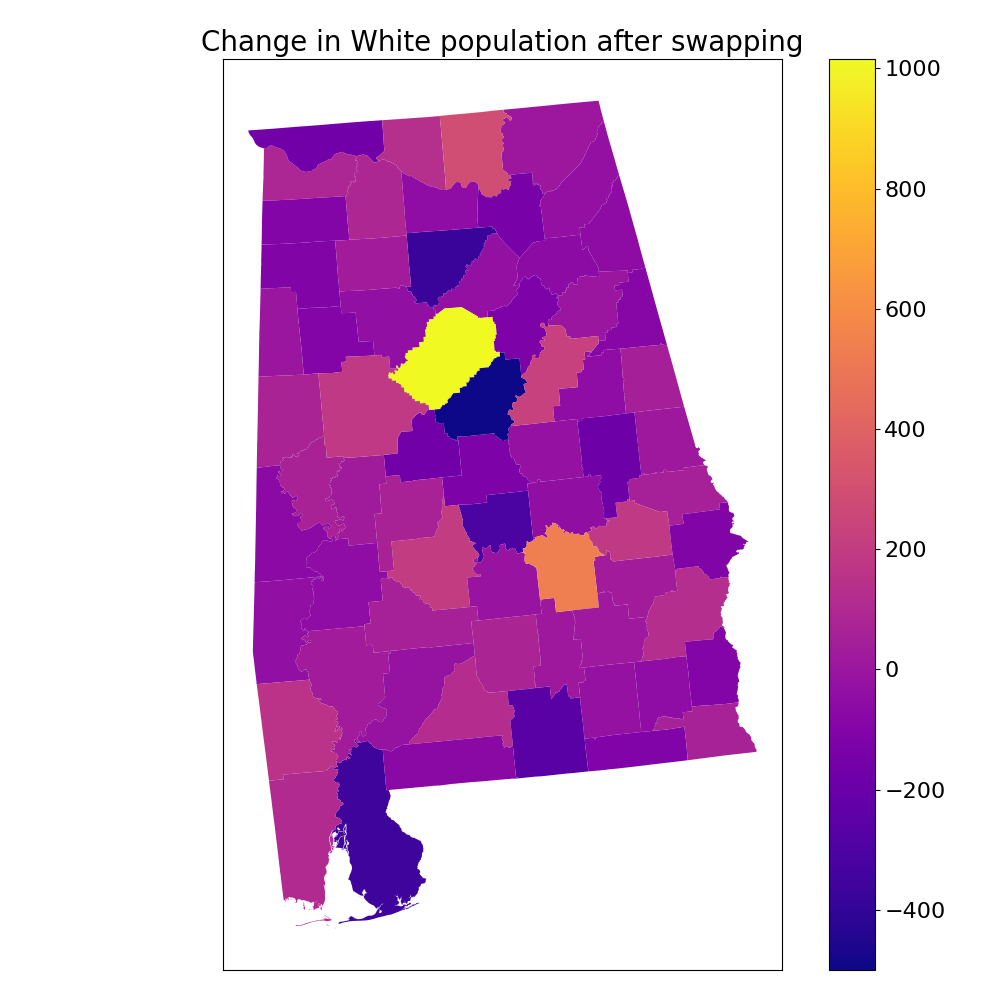}\includegraphics[scale=0.22]{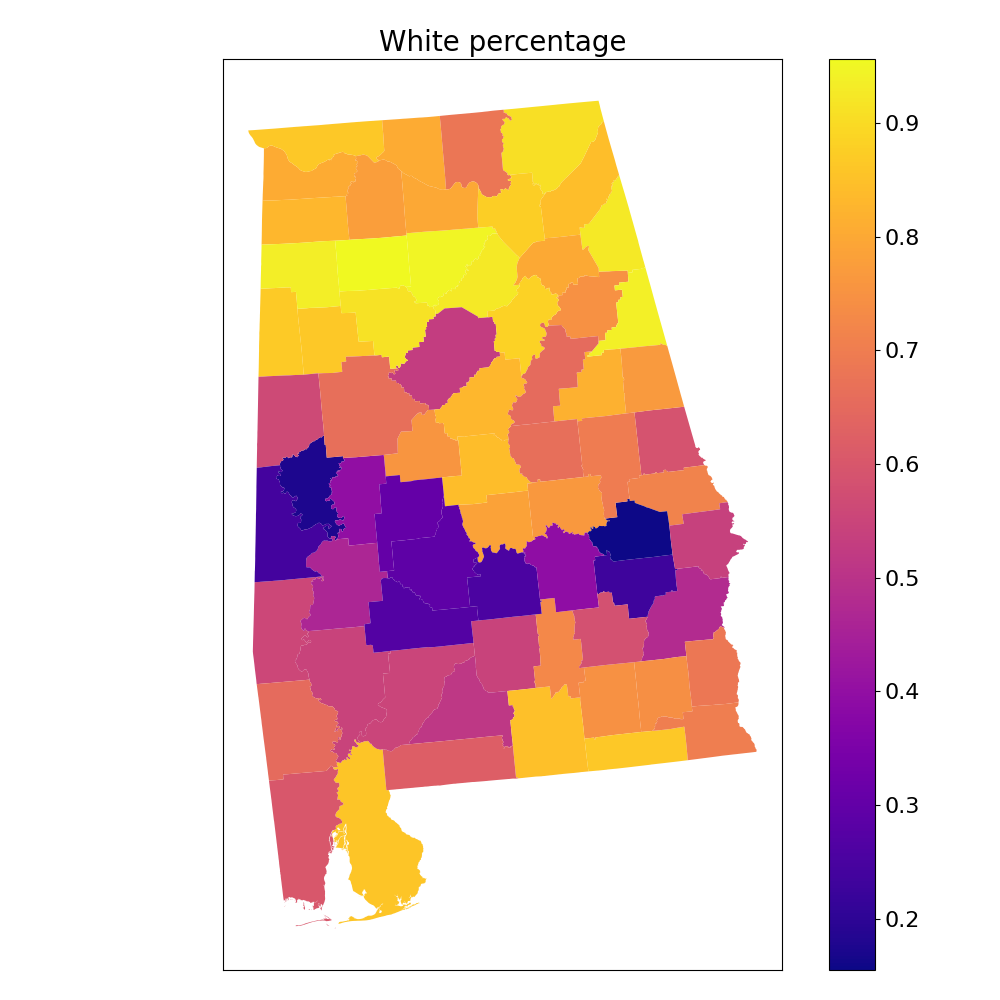}\hspace{5mm}  \includegraphics[scale=0.22]{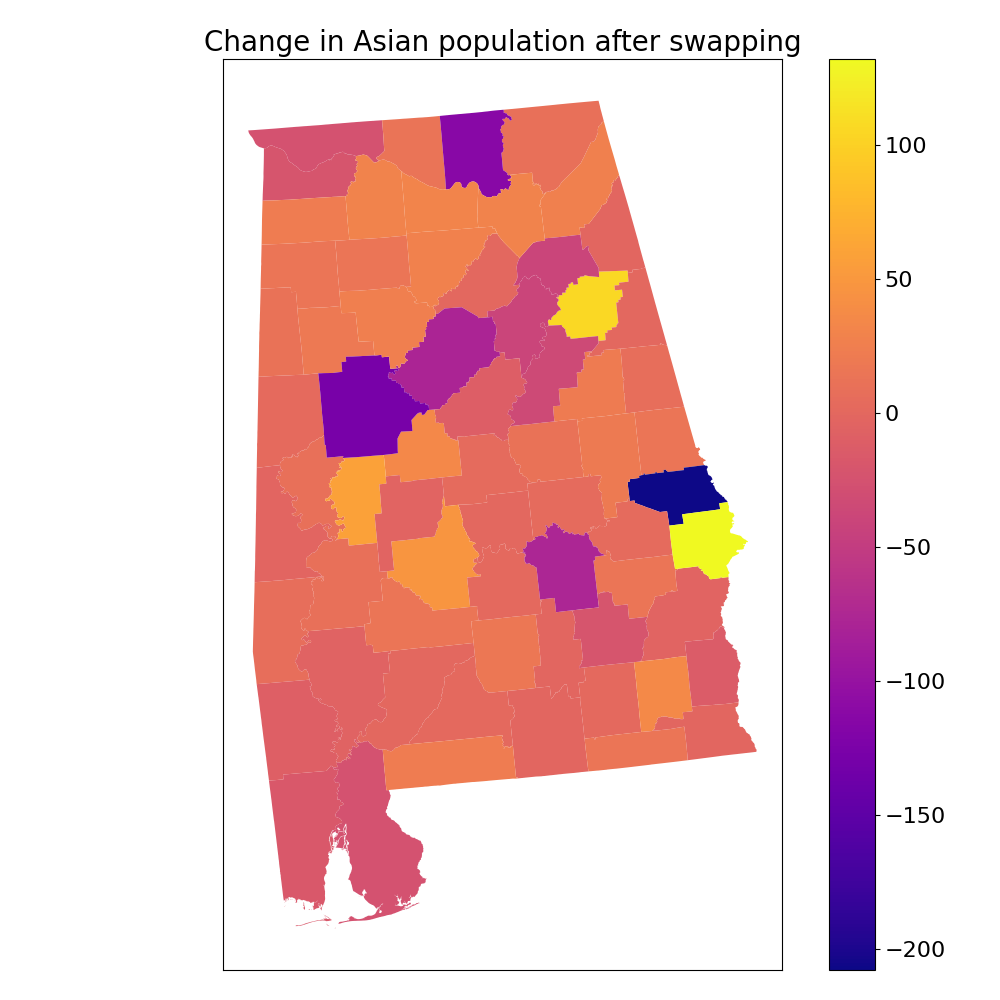}\includegraphics[scale=0.22]{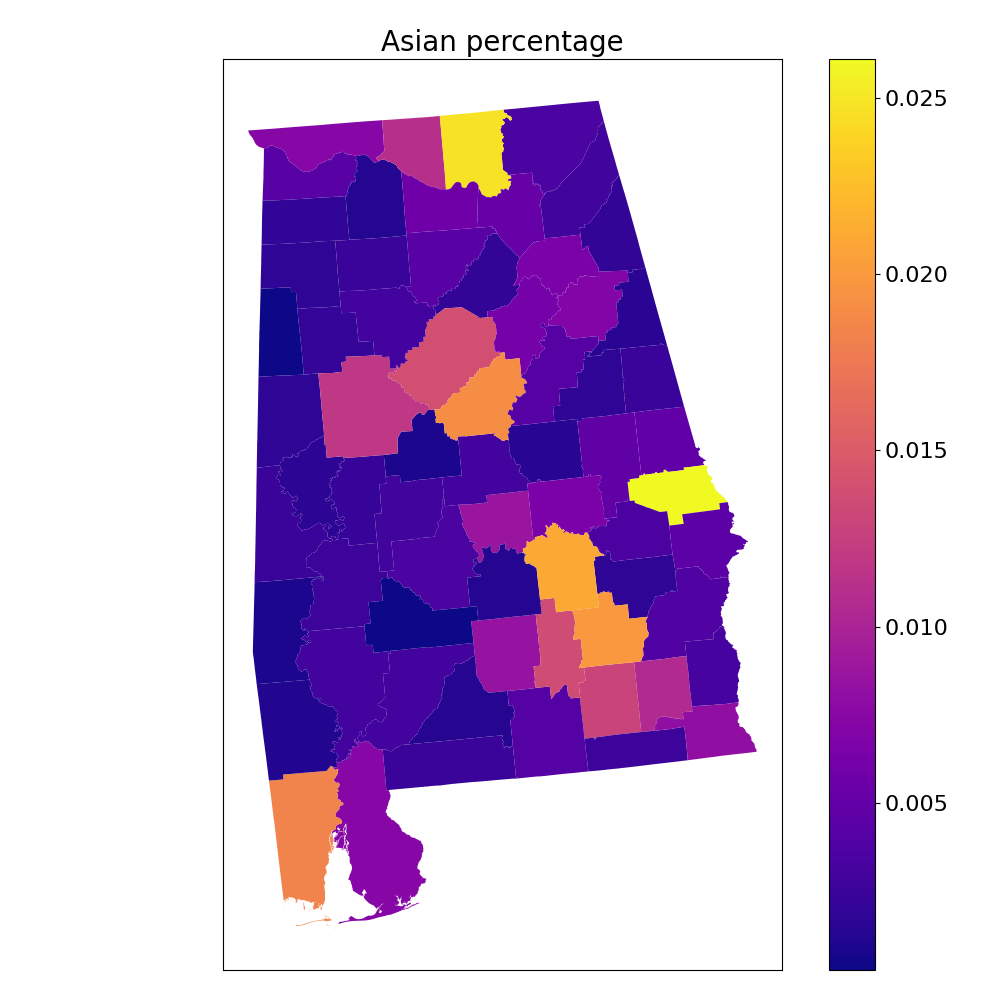}
    \caption{The effect of swapping on white and Asian population by county in Alabama. For reference, the percent white and percent Asian for each county is also shown.}
    \label{fig:map_w_al}
\end{figure}

\begin{figure}
    \centering
    \begin{tabular}{lcccc}
        \rotatebox{90}{\hspace{20mm}White}&\includegraphics[width=0.22\textwidth]{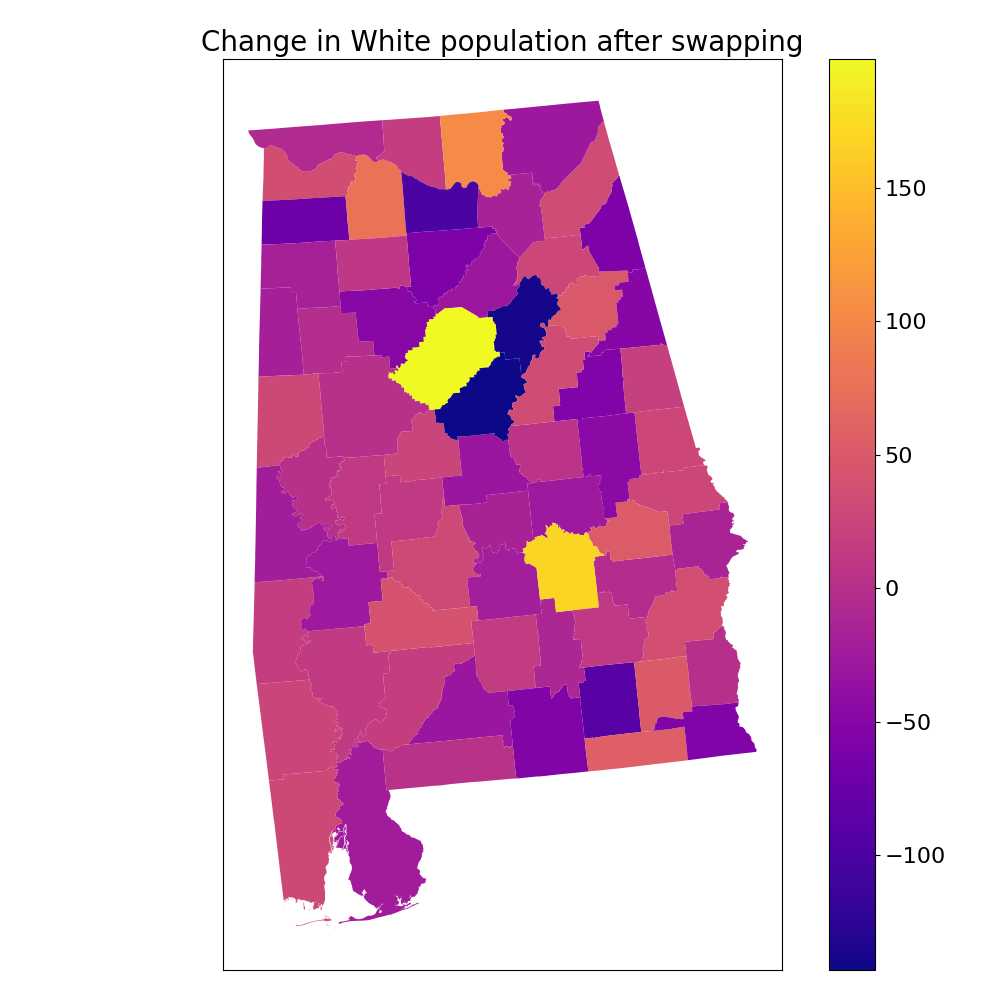}
                                          &\includegraphics[width=0.22\textwidth]{AL_figures/w_pop_change_0.1_map.png}&\includegraphics[width=0.22\textwidth]{AL_figures/w_pct_map.png}&\includegraphics[width=0.22\textwidth]{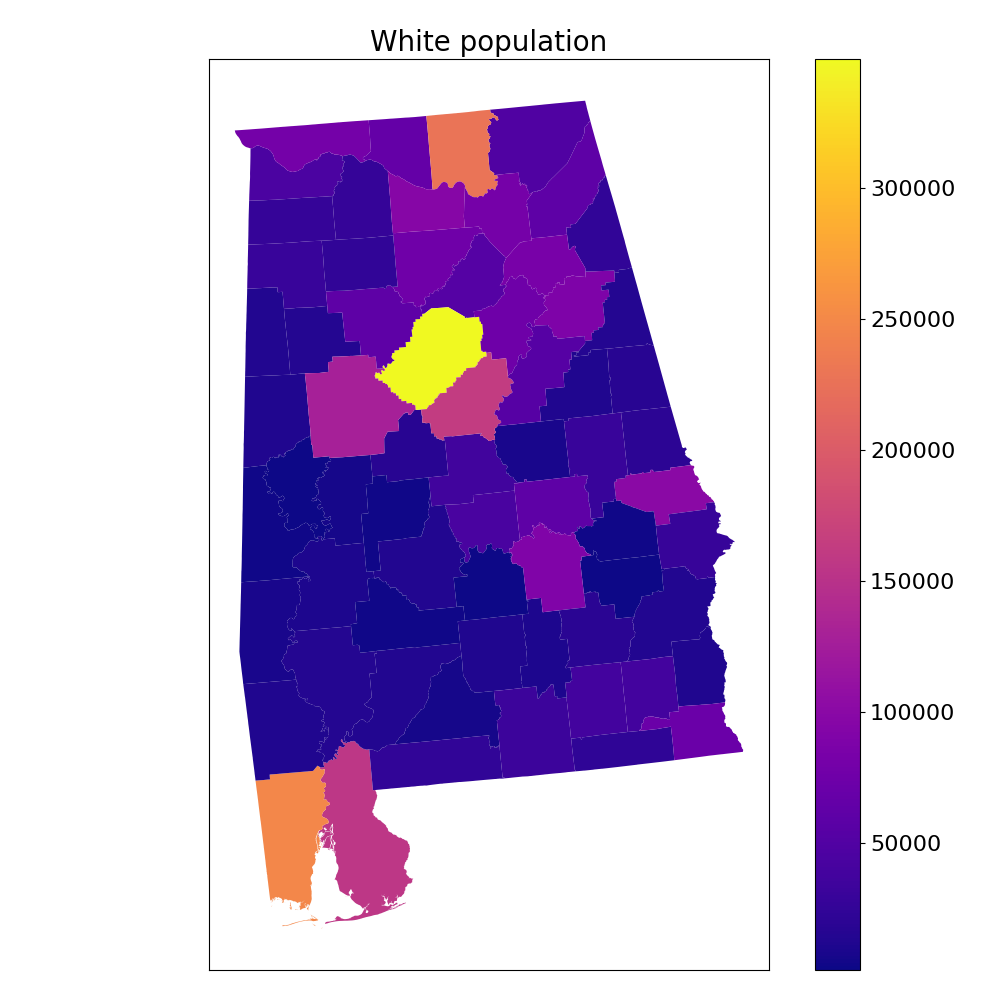}\\
        
        \rotatebox{90}{\hspace{20mm}Black}&\includegraphics[width=0.22\textwidth]{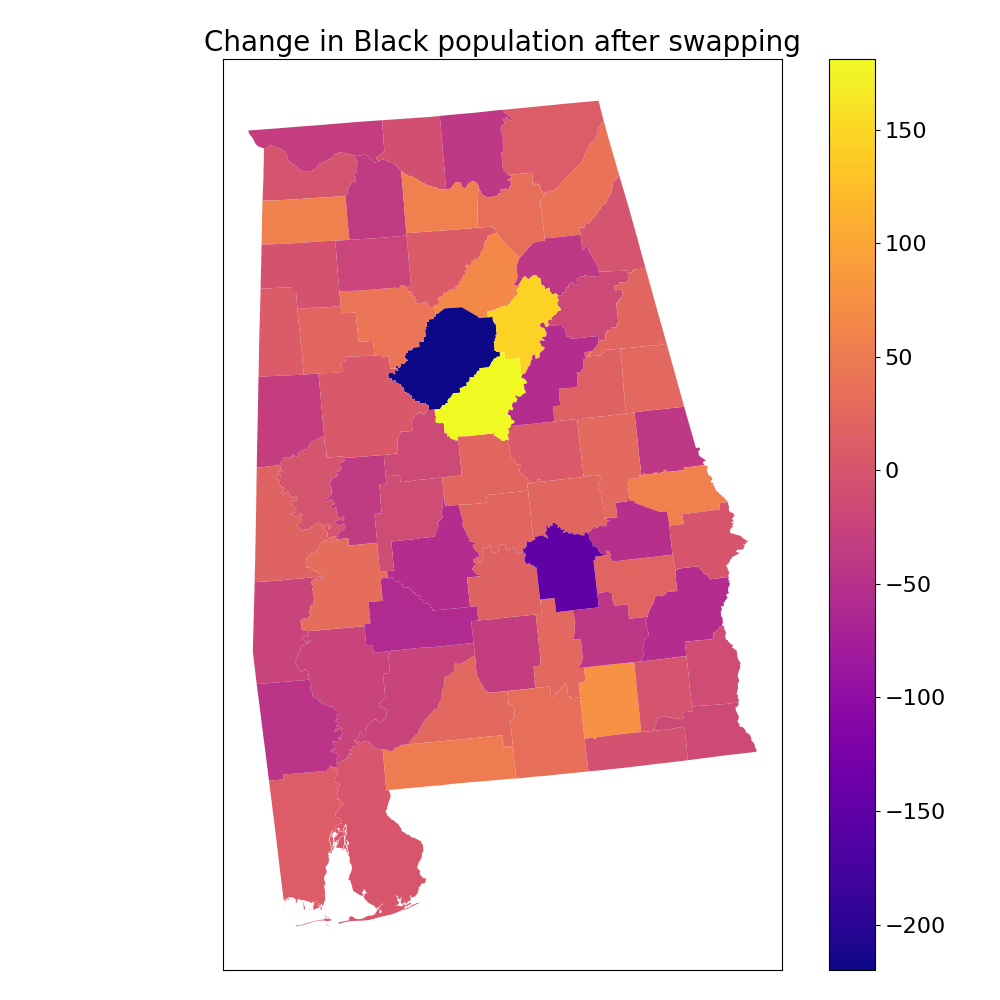}
                                          &\includegraphics[width=0.22\textwidth]{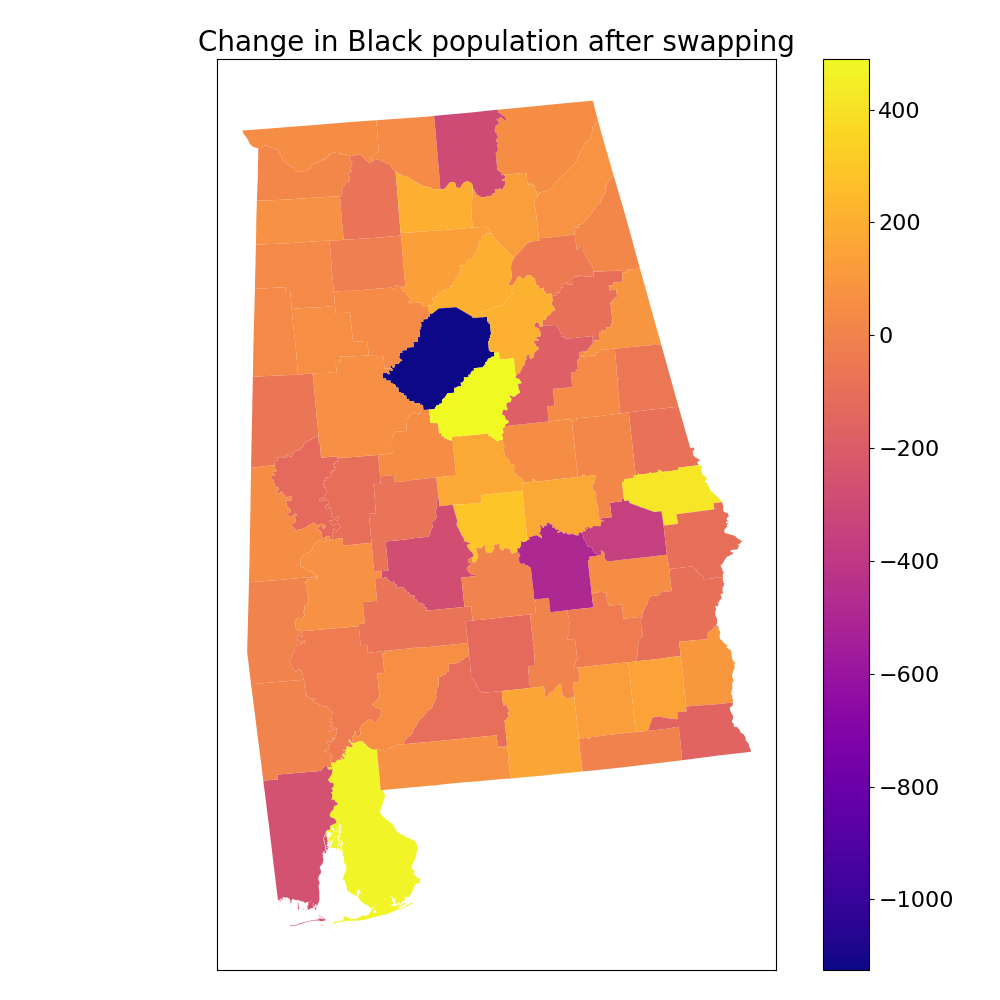}&\includegraphics[width=0.22\textwidth]{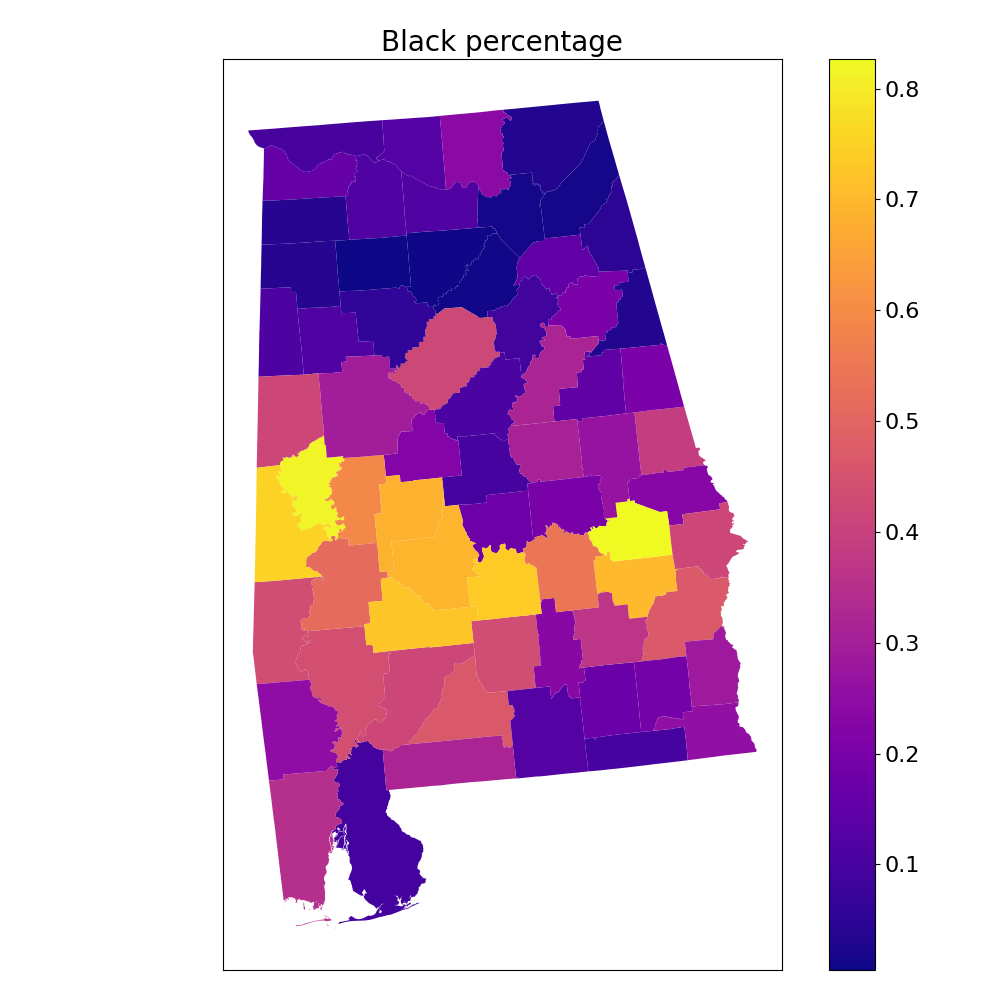}&\includegraphics[width=0.22\textwidth]{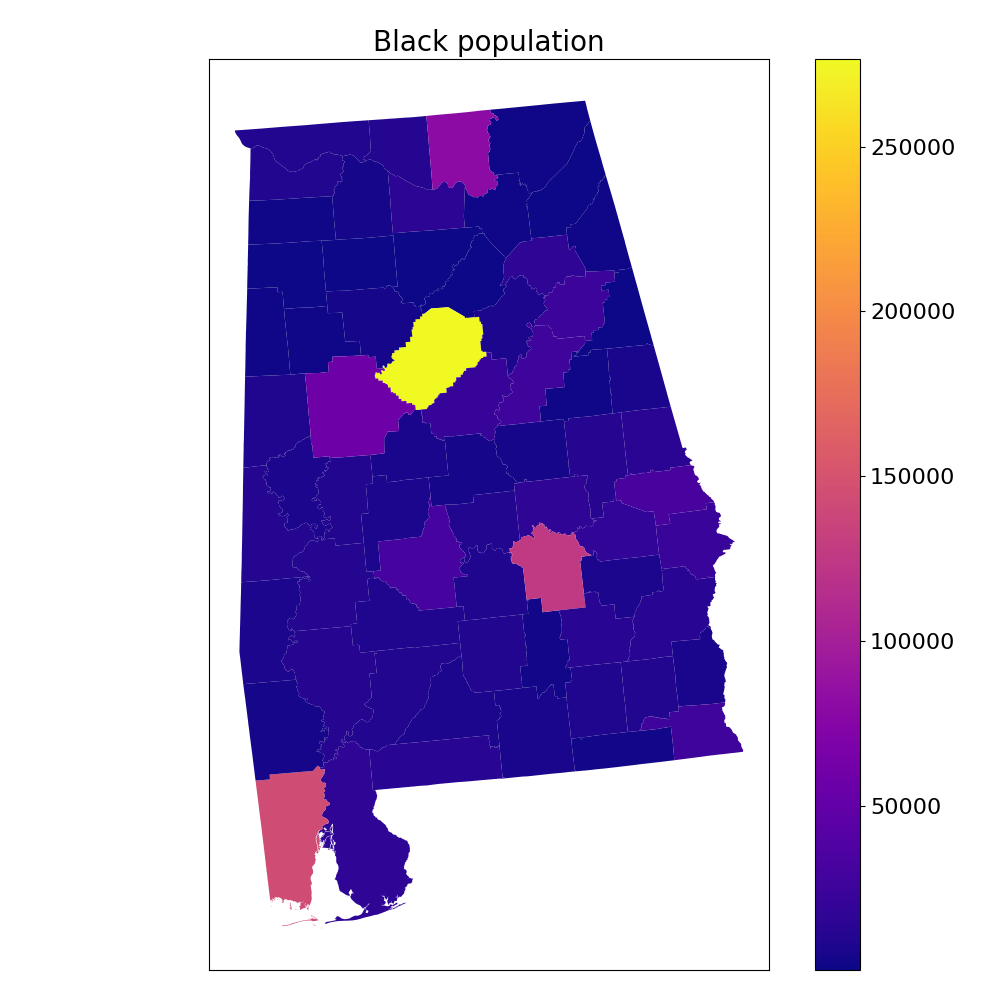}\\

        \rotatebox{90}{\hspace{20mm}Asian}&\includegraphics[width=0.22\textwidth]{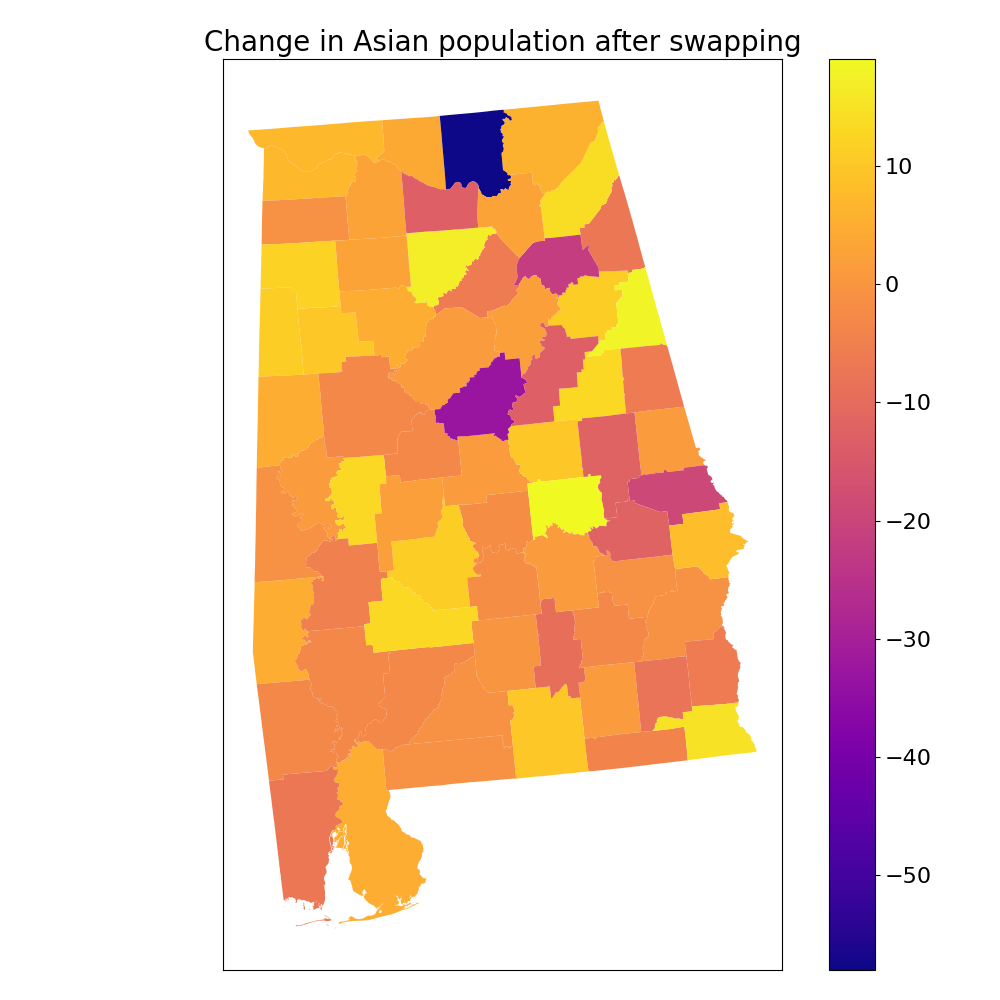}
                                          &\includegraphics[width=0.22\textwidth]{AL_figures/as_pop_change_0.1_map.png}&\includegraphics[width=0.22\textwidth]{AL_figures/as_pct_map.png}&\includegraphics[width=0.22\textwidth]{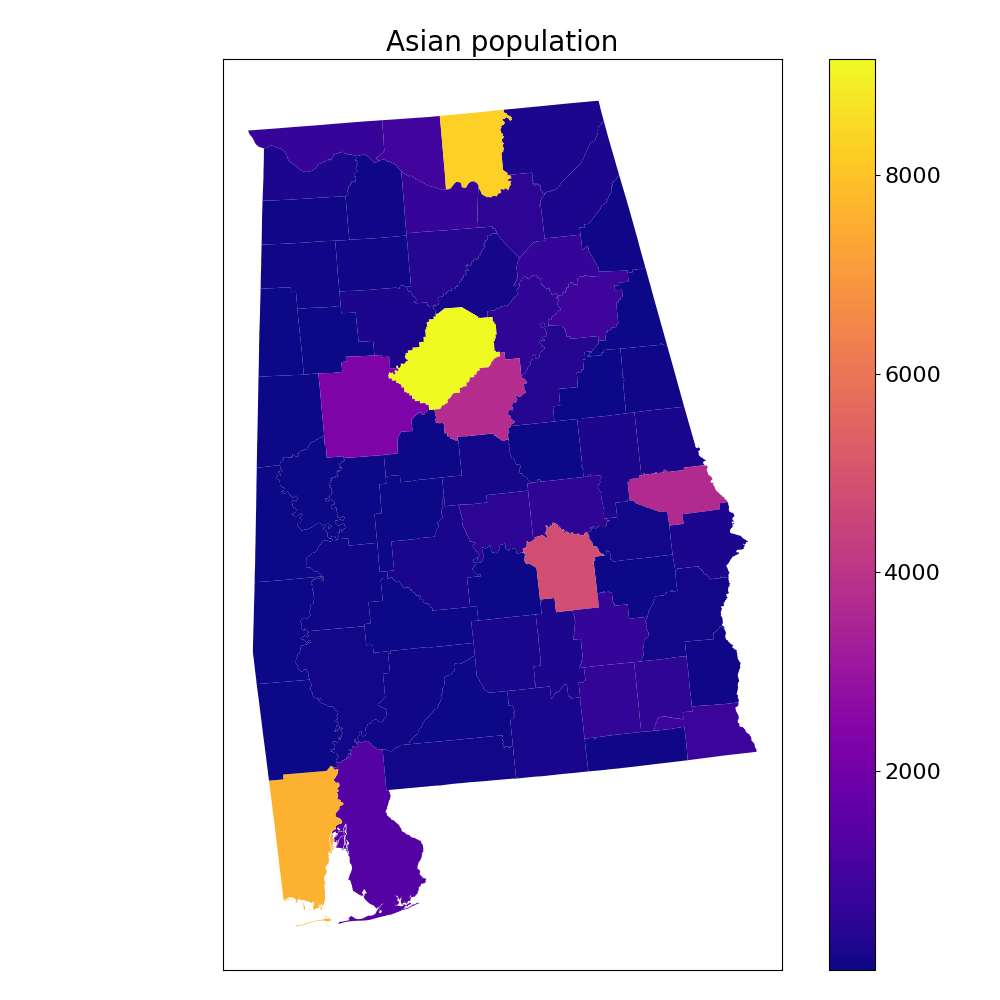}\\

        \rotatebox{90}{\hspace{20mm}Hispanic}&\includegraphics[width=0.22\textwidth]{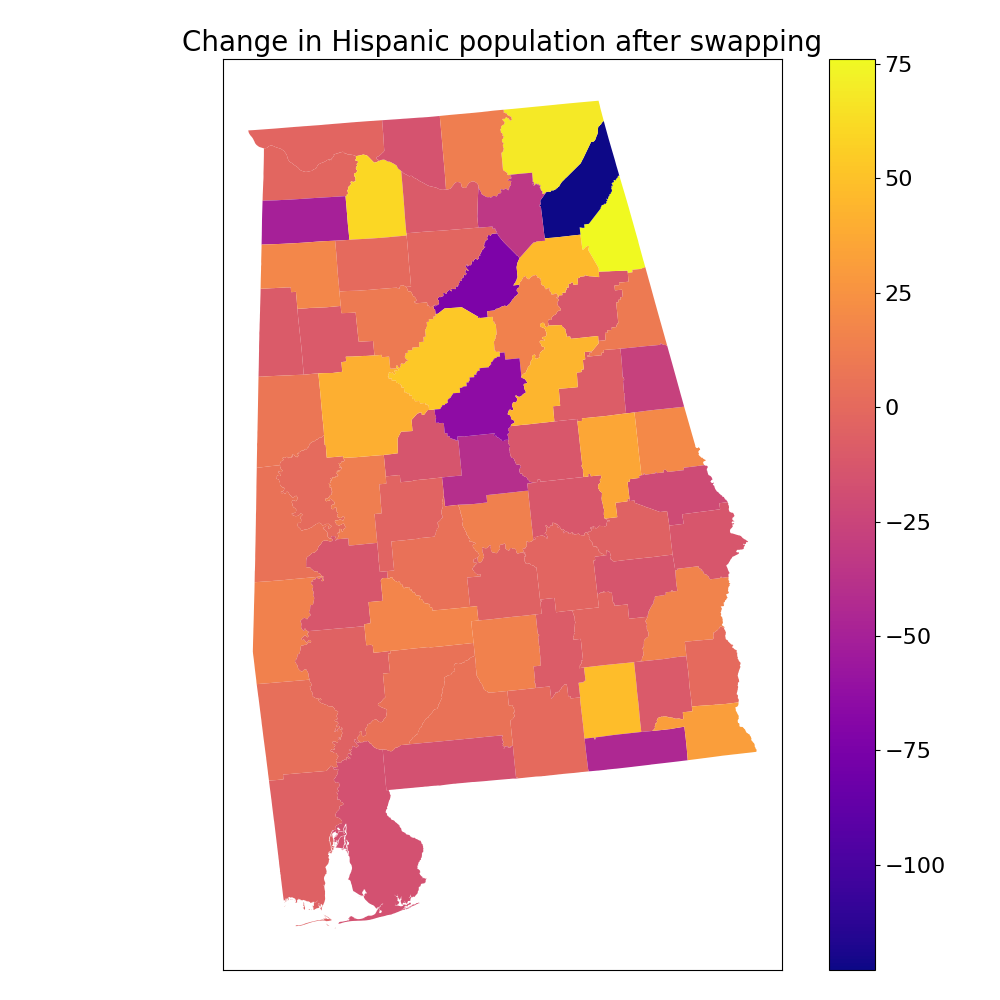}
                                             &\includegraphics[width=0.22\textwidth]{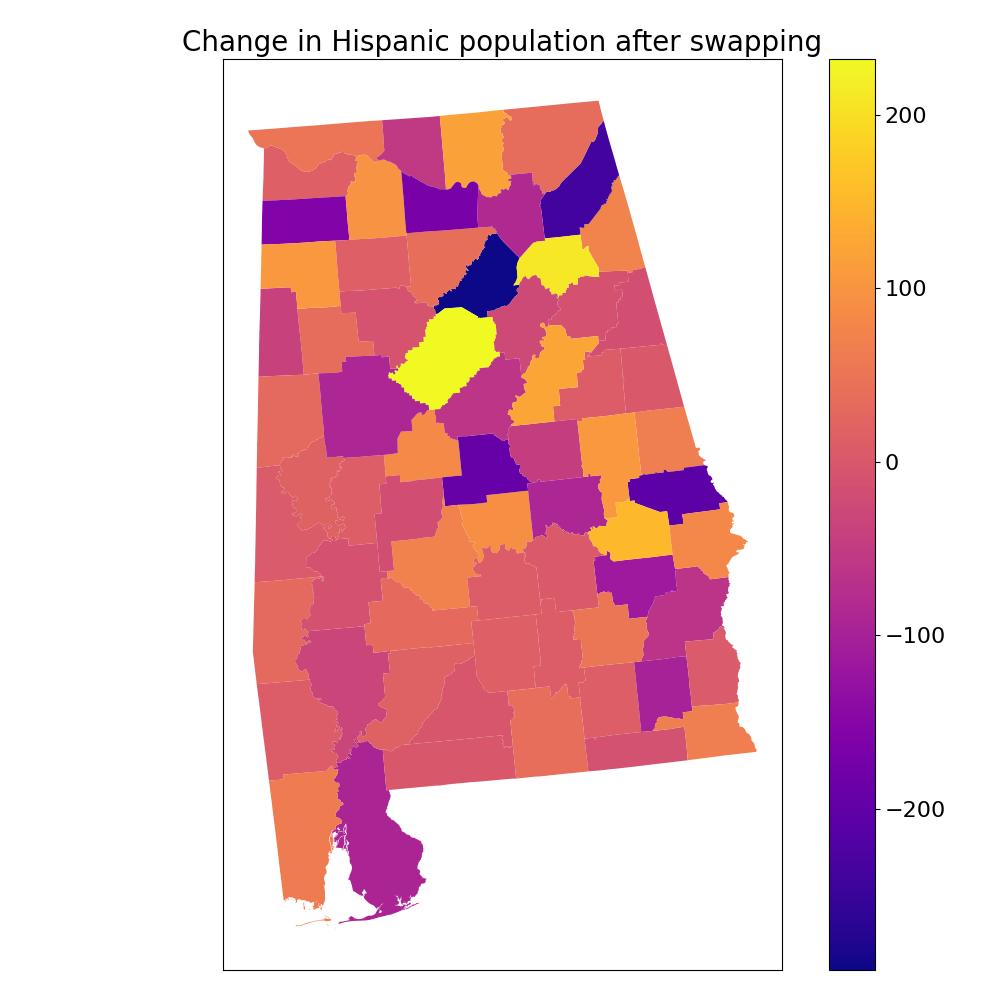}&\includegraphics[width=0.22\textwidth]{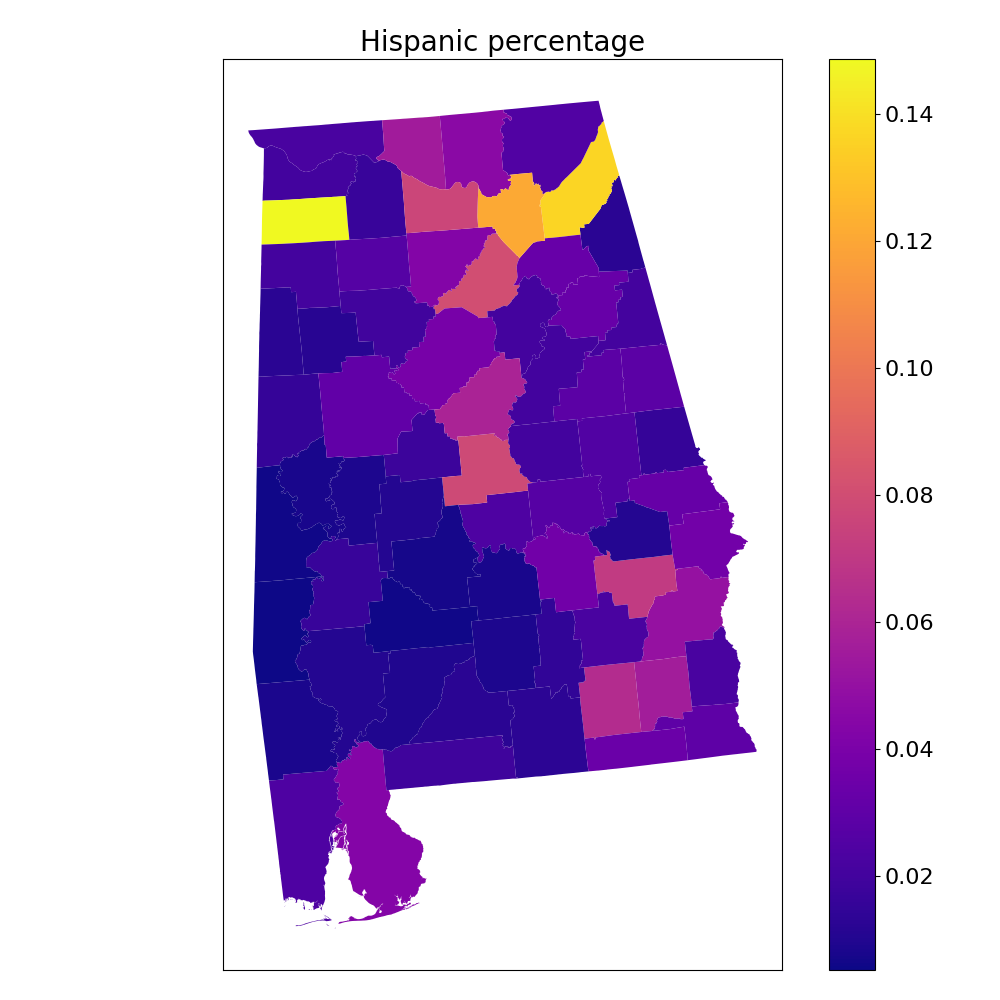}&\includegraphics[width=0.22\textwidth]{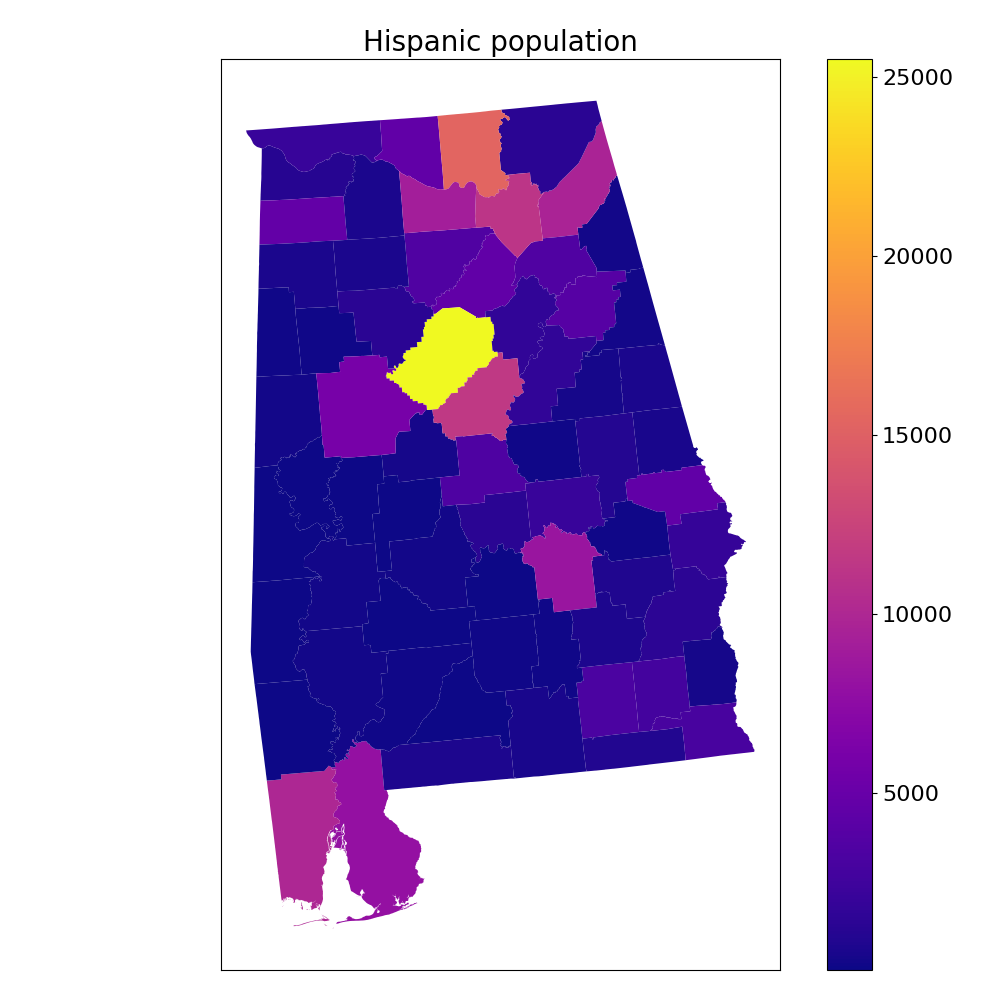}\\
        &2\% Swap Rate Effect & 10\% Swap Rate Effect & Percentage & Population
    \end{tabular}
    \caption{The effect of swapping on White and Hispanic population in Alabama, with the White/Hispanic percentage and White/Hispanic population shown for reference.}
    \label{fig:al_full_maps}
\end{figure}

\paragraph{Distribution of People by Race} Although swapping preserves household sizes and voting age counts, it often changes race and ethnicity counts. Therefore, swapping causes changes in the distribution of people by race or ethnicity across the state.
Figure~\ref{fig:map_w_al} illustrates the effects of swapping on white households by county in Alabama. Maps showing white, Black, Asian, and Hispanic population effects in Alabama, Wisconsin, Texas, Nevada, and Vermont are in Figures \ref{fig:al_full_maps}, \ref{fig:wi_full_maps}, \ref{fig:tx_full_maps}, \ref{fig:nv_full_maps}, \ref{fig:vt_full_maps} in Appendix~\ref{appendix:additional_tables_figures}.

We observe that counties that \emph{gain} significant numbers of a racial group are usually counties that have a relatively \emph{small} number of residents in that group before swapping, and counties that \emph{lose} significant numbers of a racial group are usually counties that have a relatively \emph{large} number of residents in that group before swapping. This is consistent with a view in which swapping effectively mixes nearby geographic regions, causing regression to the mean. We caution that this is only a loose rule of thumb and is sometimes violated. Another reason we stop short of finding a general pattern is that some of the maps showing which counties lose and gain people after swapping change noticeably depending on whether we use our standard implementation of swapping or our high variance implementation of swapping (these maps are the only results we generate that sometimes change meaningfully between the two swapping implementations).

Note that, while we observe that counties experiencing significant changes in swapping have particularly high or low populations, it is not the case that any county with a particularly high or low population experiences a significant change after swapping. For example, counties that have a large number of a certain group do not necessarily lose people in that group due to swapping. 

\paragraph{Which Races are Being Swapped for Which?}
Each swap involves two households: the household that was targeted for swapping (because of its relatively high risk for reidentification) and the household that is selected as the target's partner. In Table \ref{tab:target_distribution}, we show the races of the households that are chosen as swap targets with a 10\% swap rate in Alabama. In Table \ref{tab:partner_distribution}, we show the distribution of the race of the partner household, for a given type of target household. Corresponding tables for a 2\% swap rate in Alabama are Tables \ref{tab:target_distribution_2pct} and \ref{tab:partner_distribution_2pct} in Appendix \ref{appendix:additional_tables_figures}.

We make three main observations:
\begin{enumerate}
    \item Looking at Table \ref{tab:target_distribution}, households with multiple races are strongly over-represented among the swap targets. Also, households with uncommon races are slightly over-represented among the swap targets.
    \item Looking at Table \ref{tab:partner_distribution}, the distribution of partners for any particular type of target is similar to the race distribution among all partner households, which in turn is similar to the overall race distribution across all households. We hypothesize this is because partners are selected largely randomly (except for the optimization for geographic proximity) from households of the same size and voting age count. Ignoring for a moment the slight dependencies between race and household size and age distribution, the observed distribution of partners is to be expected.
    \item After considering the first two effects, we observe households have an elevated likelihood of being swapped with a household with the same race. This is visible in the relatively large numbers on the diagonal of Table \ref{tab:partner_distribution}. This may be because there is geographic clustering of households with the same race. This may also be because there is a correlation between household size and household race. Such a correlation would mean that swap partners, which necessarily have the same size, are more likely to have the same race. Table~\ref{tab:household_size_race_corr} in Appendix~\ref{appendix:additional_tables_figures} illustrates this correlation.
\end{enumerate}

\begin{table}
    \centering
    Distribution of Races of Target Households for a 10\% Swap Rate
    \begin{tabular}{|r|c|c|c|c|c|c|c|c|}
    \hline
                  & W    & B    & AI/AN & AS  & H/PI & OTH  & 2+  & Multiple Races \\ \hline
    \% overall    & 70.8 & 23.9 & 0.3   & 0.7 & 0.0  & 1.1  & 0.6 & 2.5            \\ \hline
    \% of targets & 21.2 & 10.4 & 4.3   & 5.4 & 0.6  & 10.3 & 8.8 & 39.0           \\ \hline
    \end{tabular}

    \caption{This table shows the distribution of the race among targeted households, compared to the distribution of the race of all households.}
    \label{tab:target_distribution}
\end{table}

\begin{table}

\setlength\doublerulesep{5mm} 
    \centering
    Partner Distribution for a 10\% Swap Rate
    \begin{tabular}{|r|c|c|c|c|c|c|c|c|}
    \hline
    \diagbox[width=\dimexpr \textwidth/8+5\tabcolsep\relax, height=1cm]{Target}{Partner}
                   & \multicolumn{1}{c|}{W} & \multicolumn{1}{c|}{B} & \multicolumn{1}{c|}{AI/AN} & \multicolumn{1}{c|}{AS} & \multicolumn{1}{c|}{H/PI} & \multicolumn{1}{c|}{OTH} & \multicolumn{1}{c|}{2+} & \multicolumn{1}{c|}{Multiple Races} \\ \hline
    W              & 67.6                 & 27.1                 & 0.2                      & 1.2                   & 0.1                     & 3.1                    & 0.6                   & 5.5                               \\ \hline
    B              & 49.6                 & 46.2                 & 0.2                      & 1.0                   & 0.1                     & 2.3                    & 0.6                   & 6.6                               \\ \hline
    AI/AN          & 77.0                 & 19.9                 & 0.6                      & 0.5                   & 0.0                     & 1.1                    & 0.9                   & 2.0                               \\ \hline
    AS             & 69.2                 & 23.1                 & 0.1                      & 5.5                   & 0.0                     & 1.5                    & 0.6                   & 3.2                               \\ \hline
    H/PI           & 70.7                 & 24.9                 & 0.7                      & 0.5                   & 0.3                     & 2.8                    & 0.0                   & 1.4                               \\ \hline
    OTH            & 64.4                 & 26.1                 & 0.3                      & 0.7                   & 0.0                     & 7.8                    & 0.6                   & 2.9                               \\ \hline
    2+             & 71.3                 & 25.6                 & 0.5                      & 0.7                   & 0.0                     & 1.0                    & 0.8                   & 2.0                               \\ \hline
    Multiple Races & 70.3                 & 26.1                 & 0.3                      & 1.0                   & 0.0                     & 1.8                    & 0.5                   & 5.3                               \\ \hline\hline
    \% overall     & 70.8                 & 23.9                 & 0.3                      & 0.7                   & 0.0                     & 1.1                    & 0.6                   & 2.5                               \\ \hline
    \% of partners & 64.3                 & 26.6                 & 0.3                      & 1.1                   & 0.0                     & 2.5                    & 0.6                   & 4.4                               \\ \hline
    \end{tabular}
    \caption{For a swap target of a particular race, this table shows the distribution of the race of the partner household. The row labels are the race of the target household. Each row shows the distribution of the race of the partner household for that particular type of target household (each row is normalized to sum to 100). For reference, below is distribution of the household races among all households and among the households selected as partners.}
    \label{tab:partner_distribution}
\end{table}

 \section{Additional Tables and Figures}\label{appendix:additional_tables_figures}\begin{figure}
    \centering
    \includegraphics[width=0.45\textwidth]{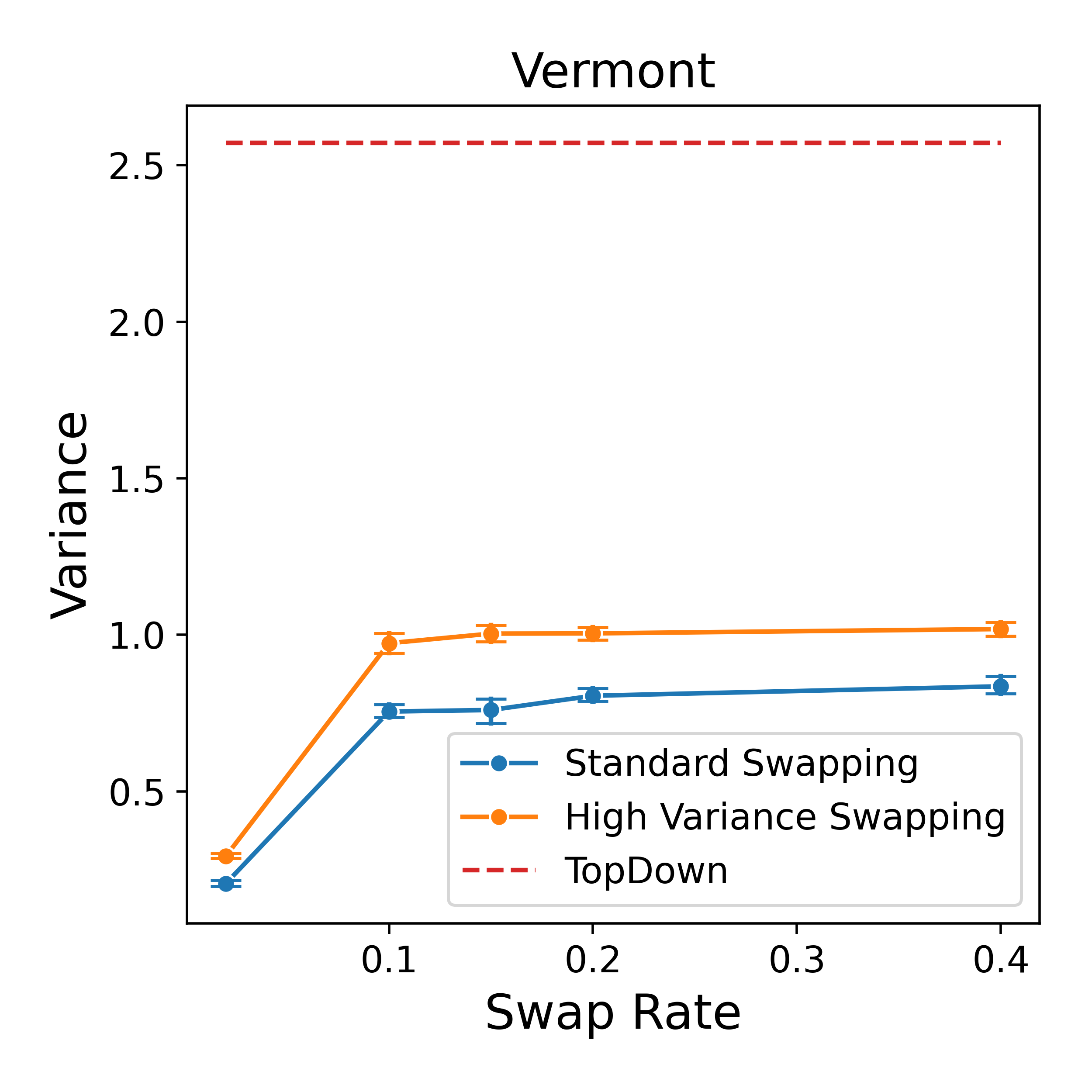}
    \includegraphics[width=0.45\textwidth]{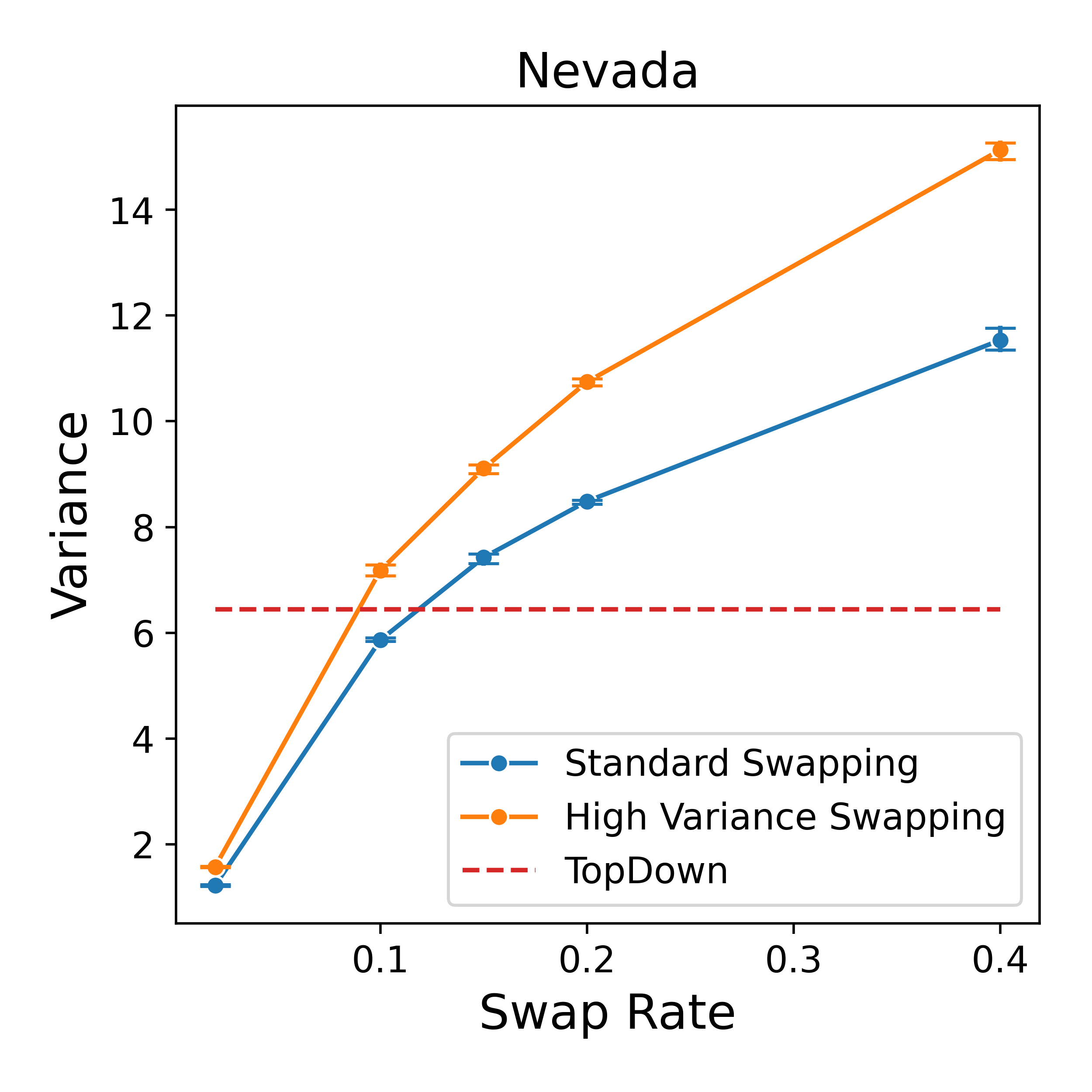}
    \caption{The variance of swapping at different swap rates. Error bars show the minimum and maximum of 5 runs of our estimator. The horizontal dotted lines show the variance estimate for TopDown. Note that we do not characterize the error in our estimate for TopDown.}
    \label{fig:variance_plots_vt_nv}
\end{figure}

\begin{table}
    \centering
\begin{tabular}{|l|l|l|l|l|l|l|l|l|}
\hline
State & TopDown & \begin{tabular}[c]{@{}l@{}}2\% \\ Swapping\end{tabular} & \begin{tabular}[c]{@{}l@{}}10\% \\ Swapping\end{tabular} & \begin{tabular}[c]{@{}l@{}}15\% \\ Swapping\end{tabular} & \begin{tabular}[c]{@{}l@{}}20\% \\ Swapping\end{tabular} & \begin{tabular}[c]{@{}l@{}}40\% \\ Swapping\end{tabular} & Blocks & \begin{tabular}[c]{@{}l@{}}AI/AN\\ Areas\end{tabular} \\ \hline
AL    & 1.15536 & 0.21892      & 0.62777       & 0.76887       & 0.8588        & 1.03619       & 135439 & 9           \\ \hline
TX    & 1.63866 & 0.32916      & 1.04552       & 1.24336       & 1.40136       & 1.74628       & 454658 & 3           \\ \hline
WI    & 1.06293 & 0.19732      & 0.57016       & 0.65949       & 0.69759       & 0.74886       & 152756 & 12          \\ \hline
VT    & 0.87533 & 0.12685      & 0.34582       & 0.34762       & 0.35103       & 0.38097       & 17541  & 0           \\ \hline
NV    & 2.01754 & 0.4715       & 1.49084       & 1.7347        & 1.90895       & 2.29154       & 35617  & 28          \\ \hline
\end{tabular}
    \caption{Mean absolute error of TopDown and swapping at various swap rates, in different states.}
    \label{tab:mean_abs_errors}
\end{table}

\begin{table}
\centering
\begin{tabular}{|l|l|l|l|}
\hline
\begin{tabular}[c]{@{}l@{}}Household\\ Size\end{tabular} & \begin{tabular}[c]{@{}l@{}}Total\\ Percentage\end{tabular} & \begin{tabular}[c]{@{}l@{}}Percentage of Swapped Households \\ (2\% Swap Rate)\end{tabular} & \begin{tabular}[c]{@{}l@{}}Percentage of Swapped Households \\ (10\% Swap Rate)\end{tabular} \\ \hline
1                                                        & 26.32\%                                                    & 5.69\%                                                                                       & 12.26\%                                                                                       \\ \hline
2                                                        & 35.59\%                                                    & 15.69\%                                                                                      & 22.12\%                                                                                       \\ \hline
3                                                        & 15.20\%                                                    & 20.73\%                                                                                      & 17.65\%                                                                                       \\ \hline
4                                                        & 14.14\%                                                    & 21.73\%                                                                                      & 16.44\%                                                                                       \\ \hline
5                                                        & 5.34\%                                                     & 14.32\%                                                                                      & 11.13\%                                                                                       \\ \hline
6                                                        & 1.84\%                                                     & 8.12\%                                                                                       & 8.12\%                                                                                        \\ \hline
7                                                        & 0.63\%                                                     & 5.30\%                                                                                       & 5.31\%                                                                                        \\ \hline
8                                                        & 0.24\%                                                     & 2.71\%                                                                                       & 2.34\%                                                                                        \\ \hline
9                                                        & 0.26\%                                                     & 2.27\%                                                                                       & 1.65\%                                                                                        \\ \hline
10                                                       & 0.07\%                                                     & 1.33\%                                                                                       & 0.68\%                                                                                        \\ \hline
11                                                       & 0.05\%                                                     & 0.90\%                                                                                       & 0.45\%                                                                                        \\ \hline
12                                                       & 0.04\%                                                     & 0.66\%                                                                                       & 0.43\%                                                                                        \\ \hline
13                                                       & 0.01\%                                                     & 0.14\%                                                                                       & 0.10\%                                                                                        \\ \hline
14                                                       & 0.26\%                                                     & 0.42\%                                                                                       & 1.32\%                                                                                        \\ \hline
15                                                       & 0.00\%                                                     & 0.00\%                                                                                       & 0.00\%                                                                                        \\ \hline
16                                                       & 0.00\%                                                     & 0.00\%                                                                                       & 0.00\%                                                                                        \\ \hline
36                                                       & 0.00\%                                                     & 0.00\%                                                                                       & 0.00\%                                                                                        \\ \hline
\end{tabular}
\caption{The distribution over household sizes for the state, the swapped households under a $2\%$ swap rate, and the swapped households under a $10\%$ swap rate.}
\label{table:hh_dist}
\end{table}

\begin{figure}
    \centering
    \begin{tabular}{lcccc}
        \rotatebox{90}{\hspace{20mm}White}&\includegraphics[width=0.22\textwidth]{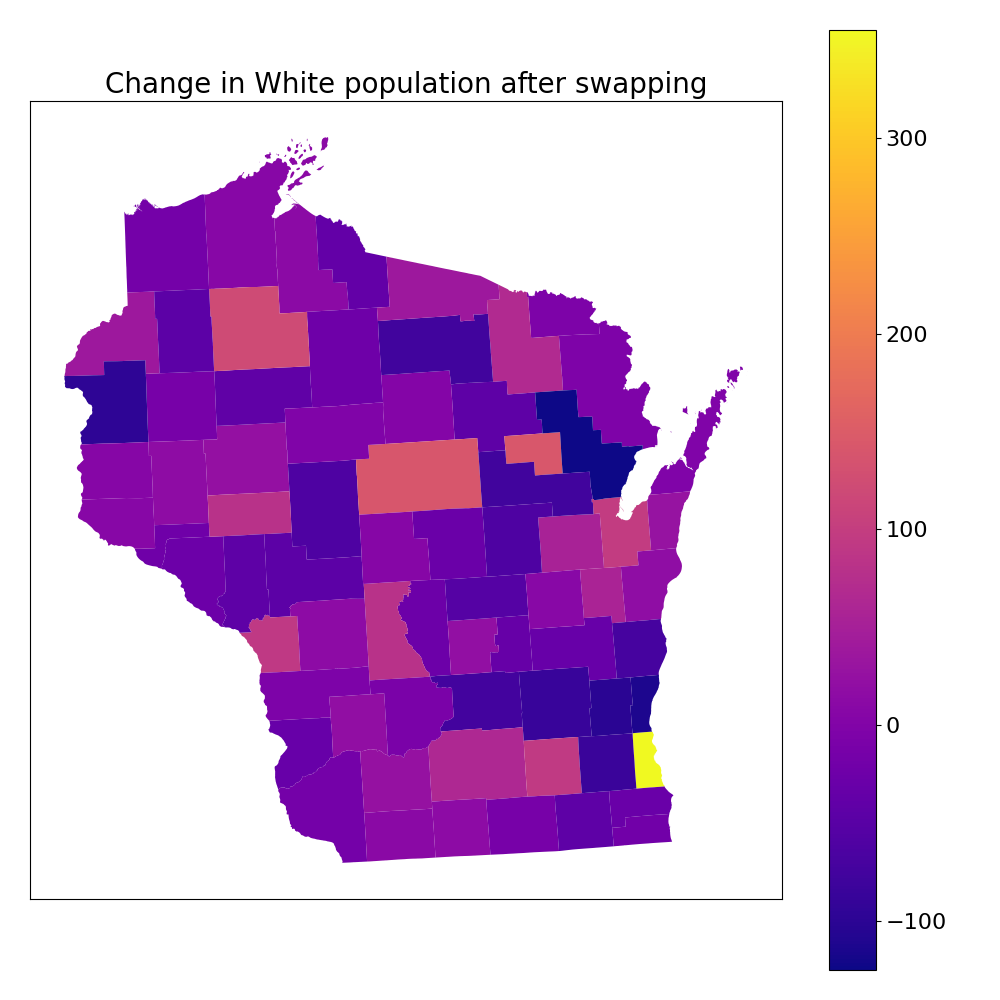}
                                          &\includegraphics[width=0.22\textwidth]{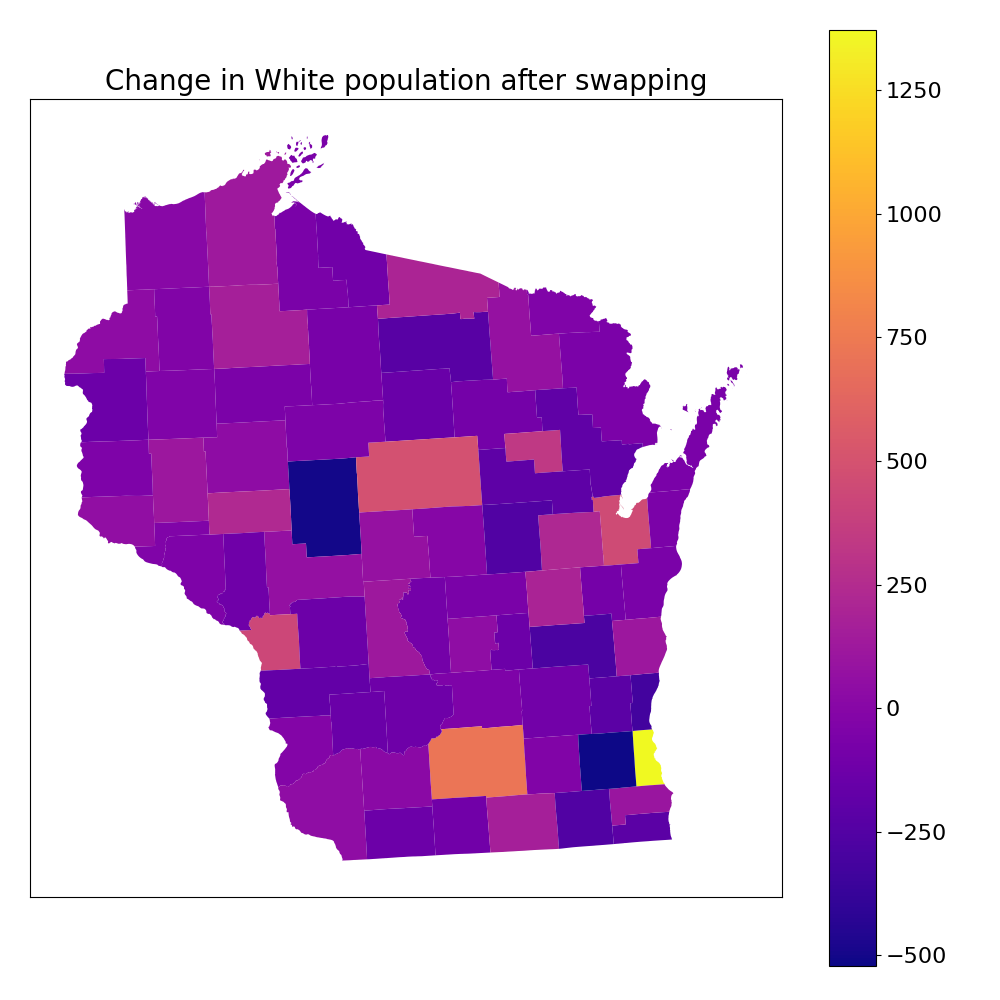}&\includegraphics[width=0.22\textwidth]{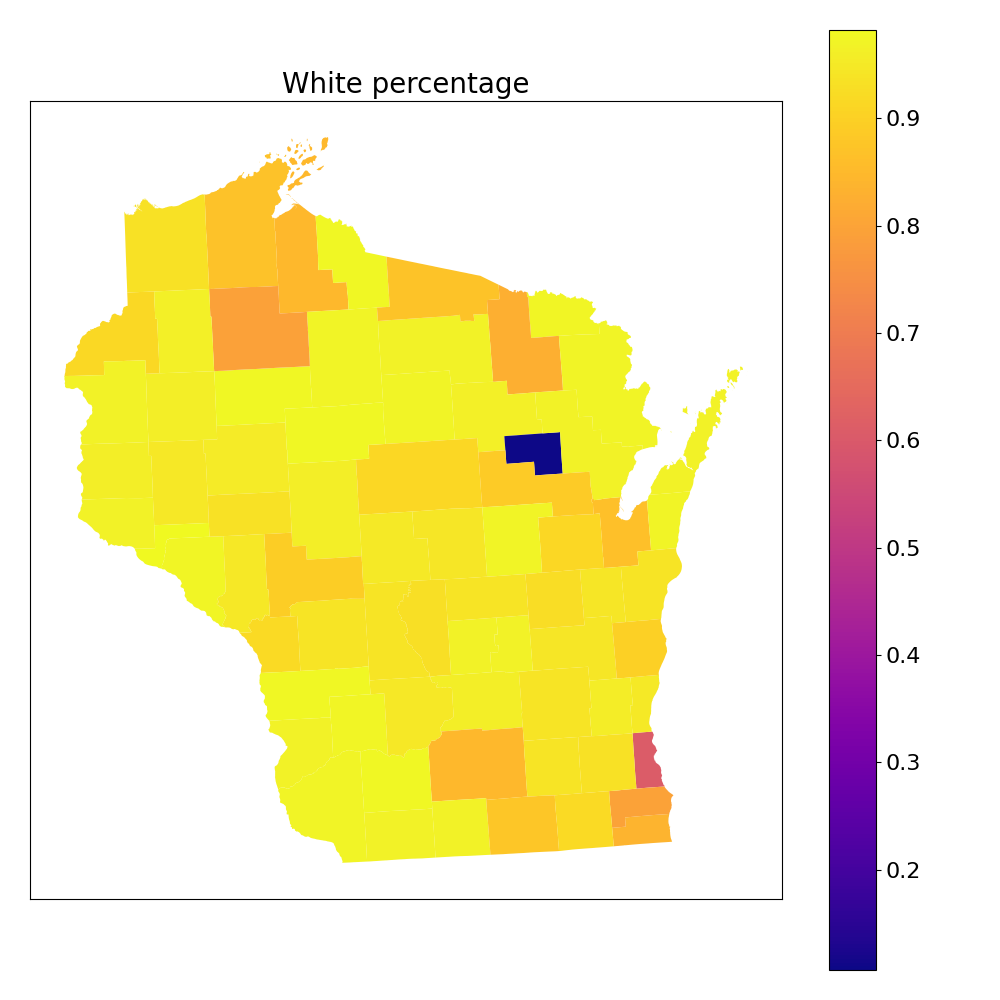}&\includegraphics[width=0.22\textwidth]{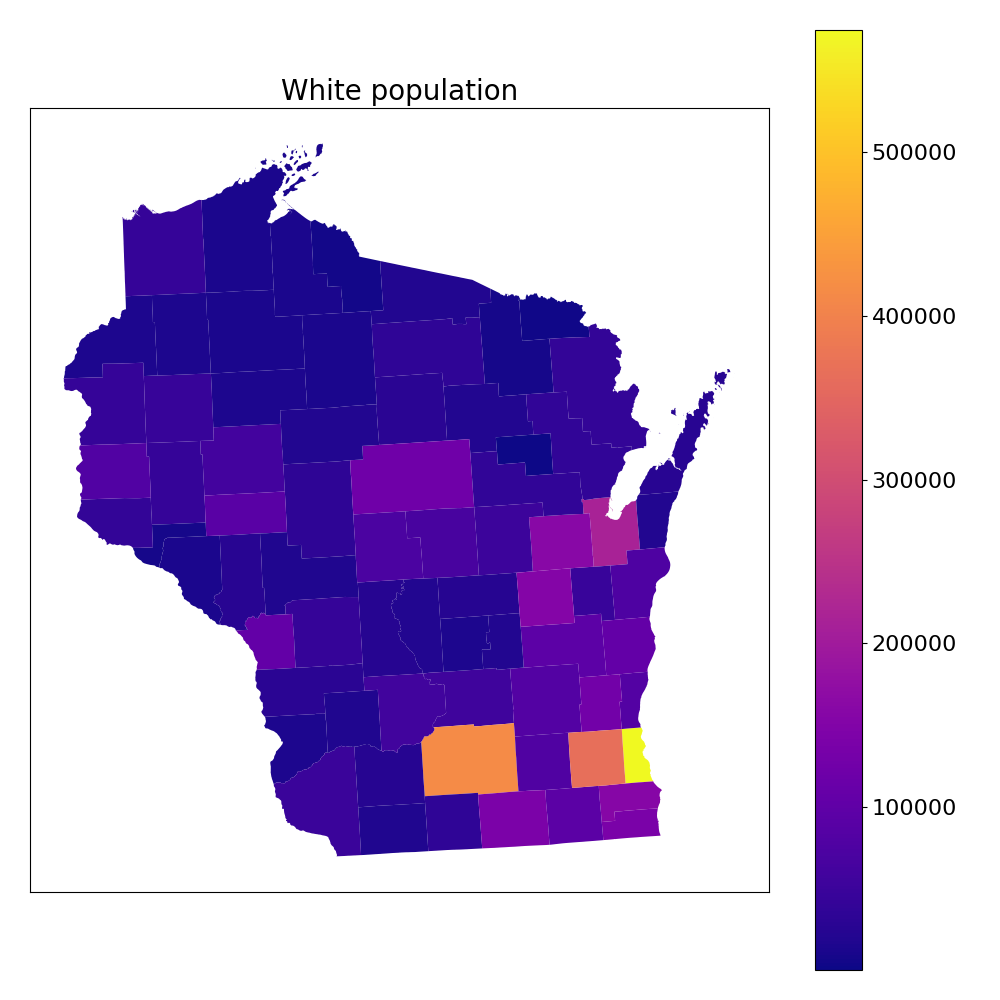}\\

        \rotatebox{90}{\hspace{20mm}Black}&\includegraphics[width=0.22\textwidth]{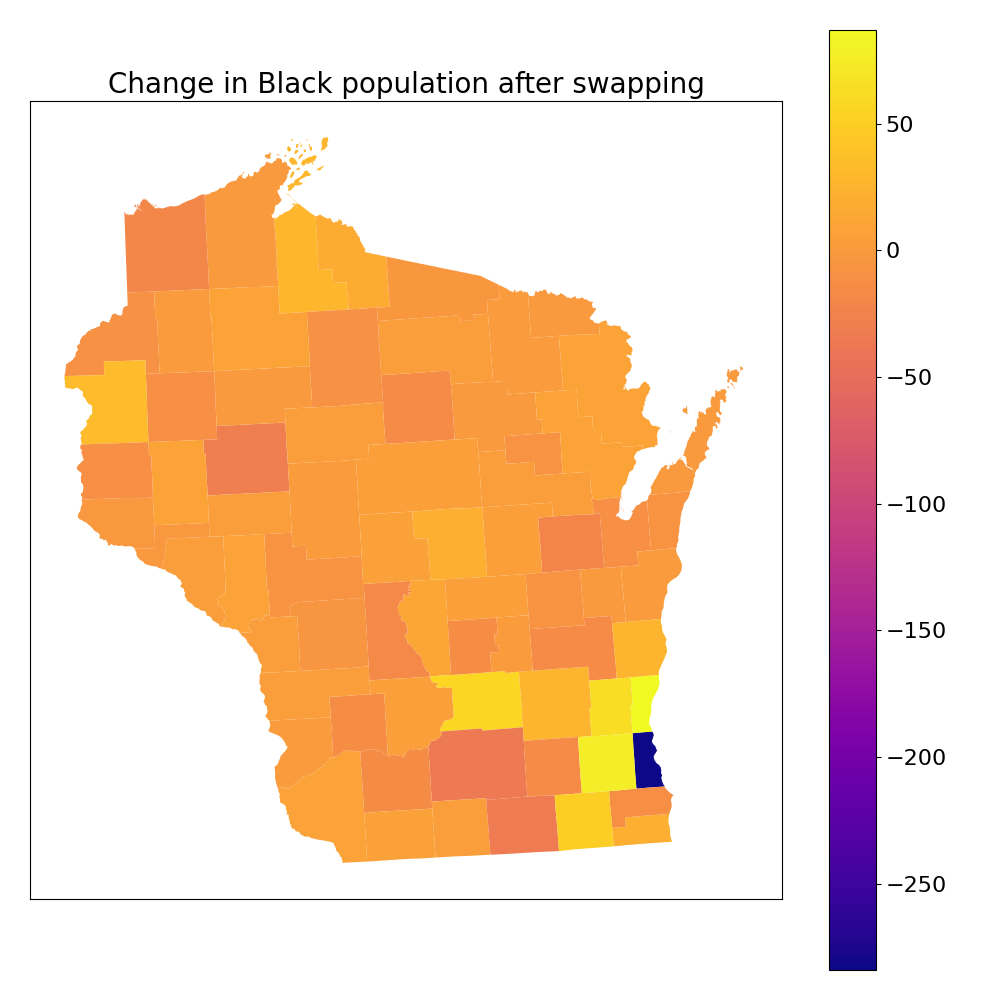}
                                          &\includegraphics[width=0.22\textwidth]{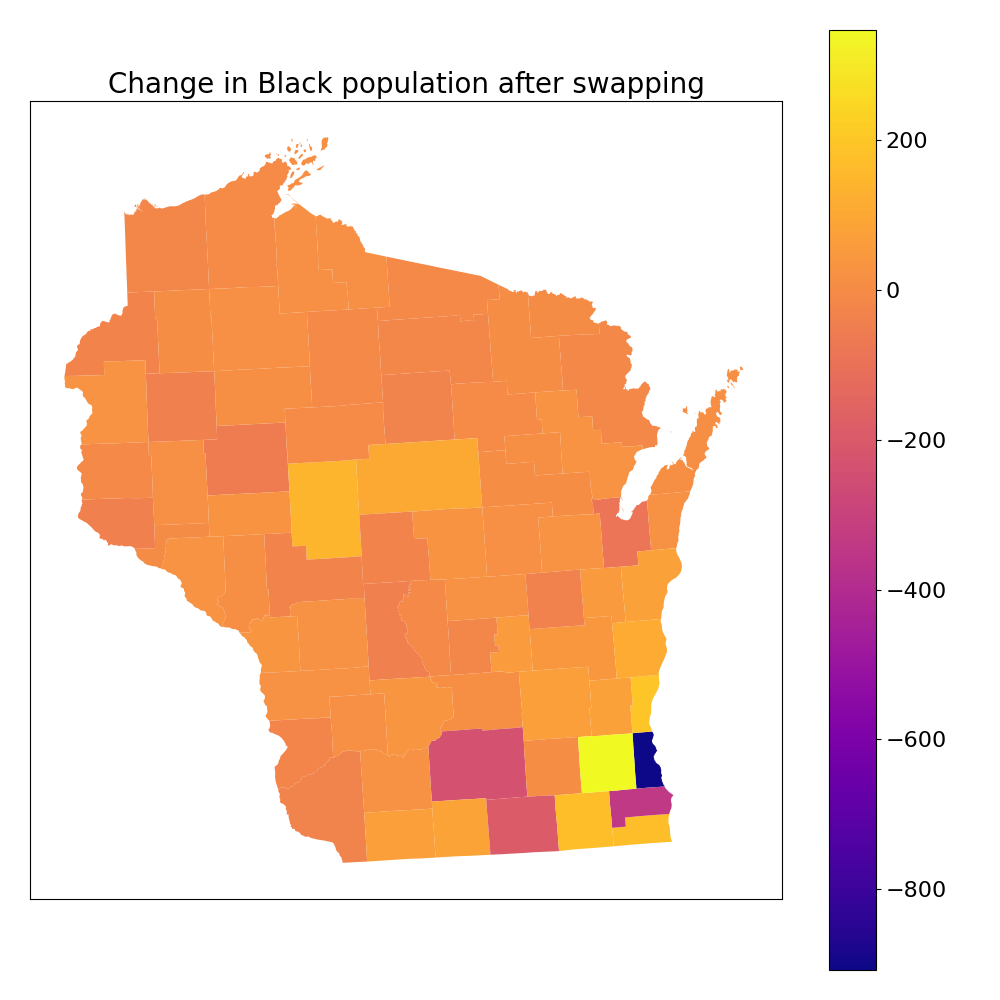}&\includegraphics[width=0.22\textwidth]{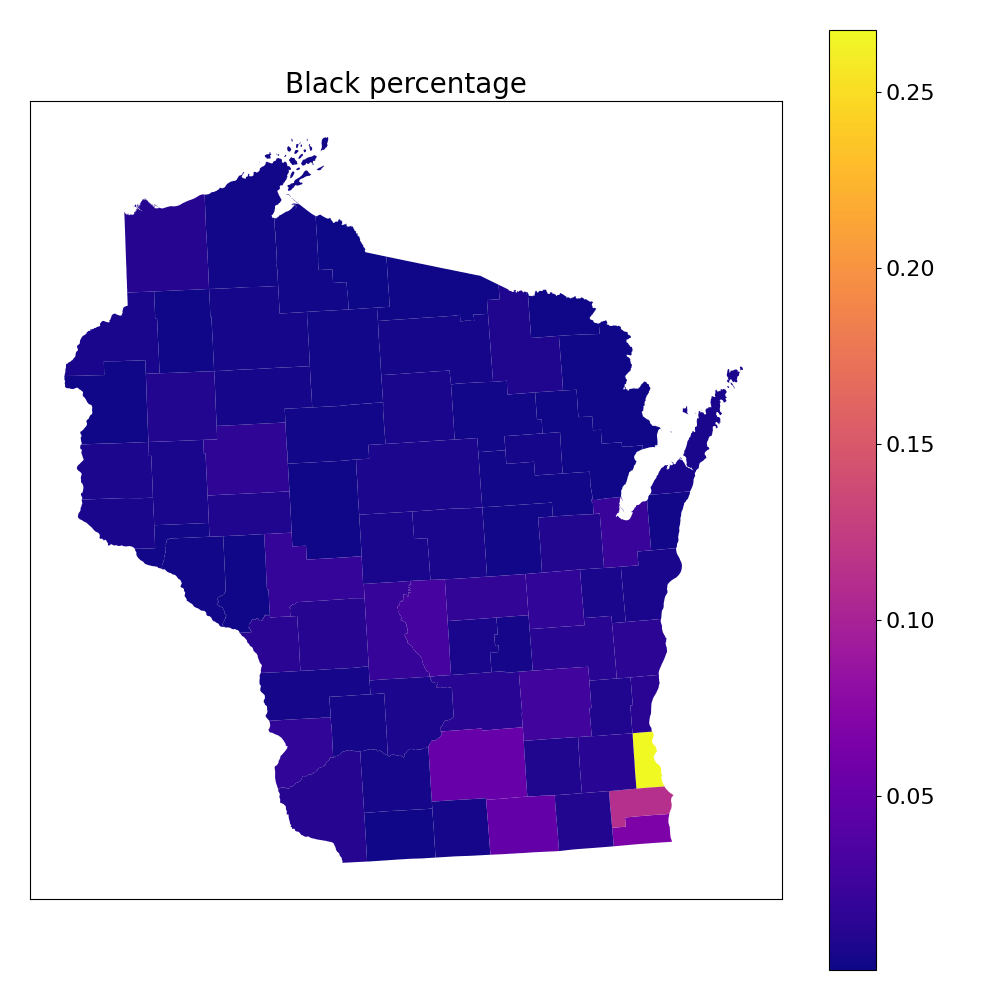}&\includegraphics[width=0.22\textwidth]{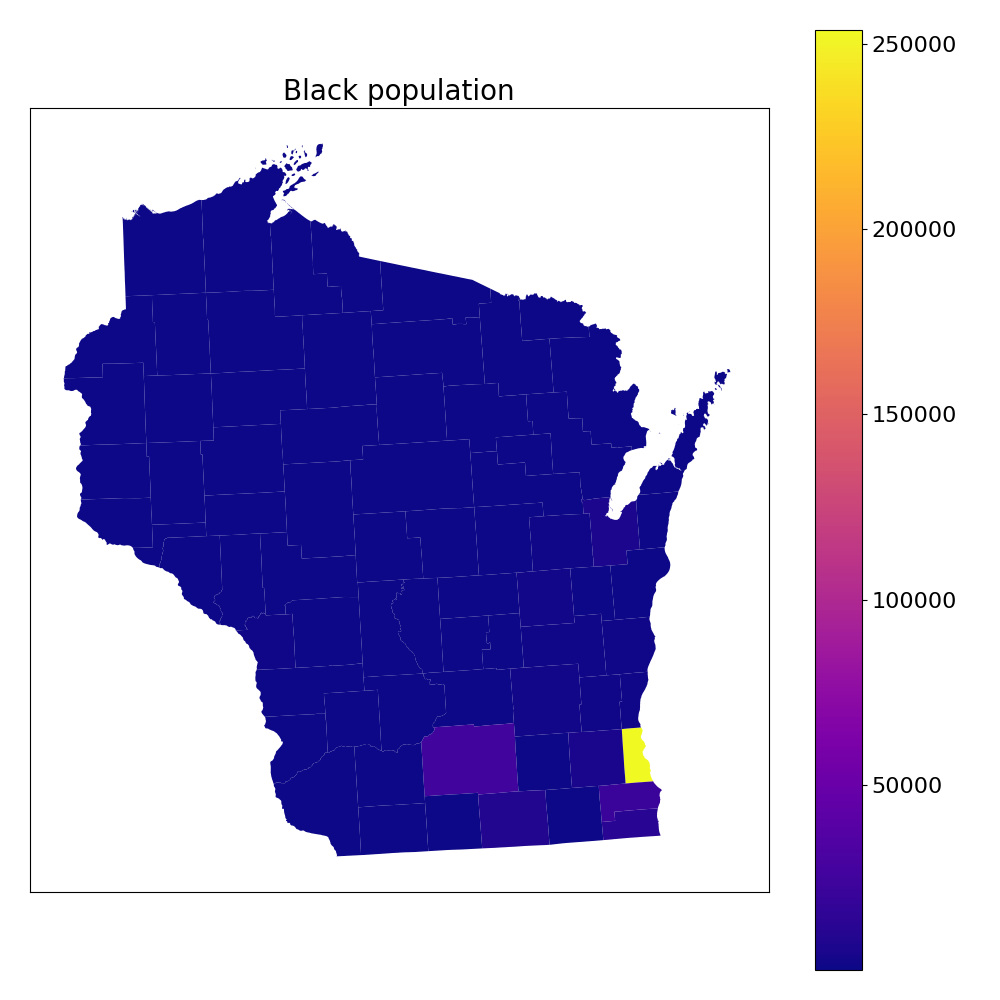}\\

        \rotatebox{90}{\hspace{20mm}Asian}&\includegraphics[width=0.22\textwidth]{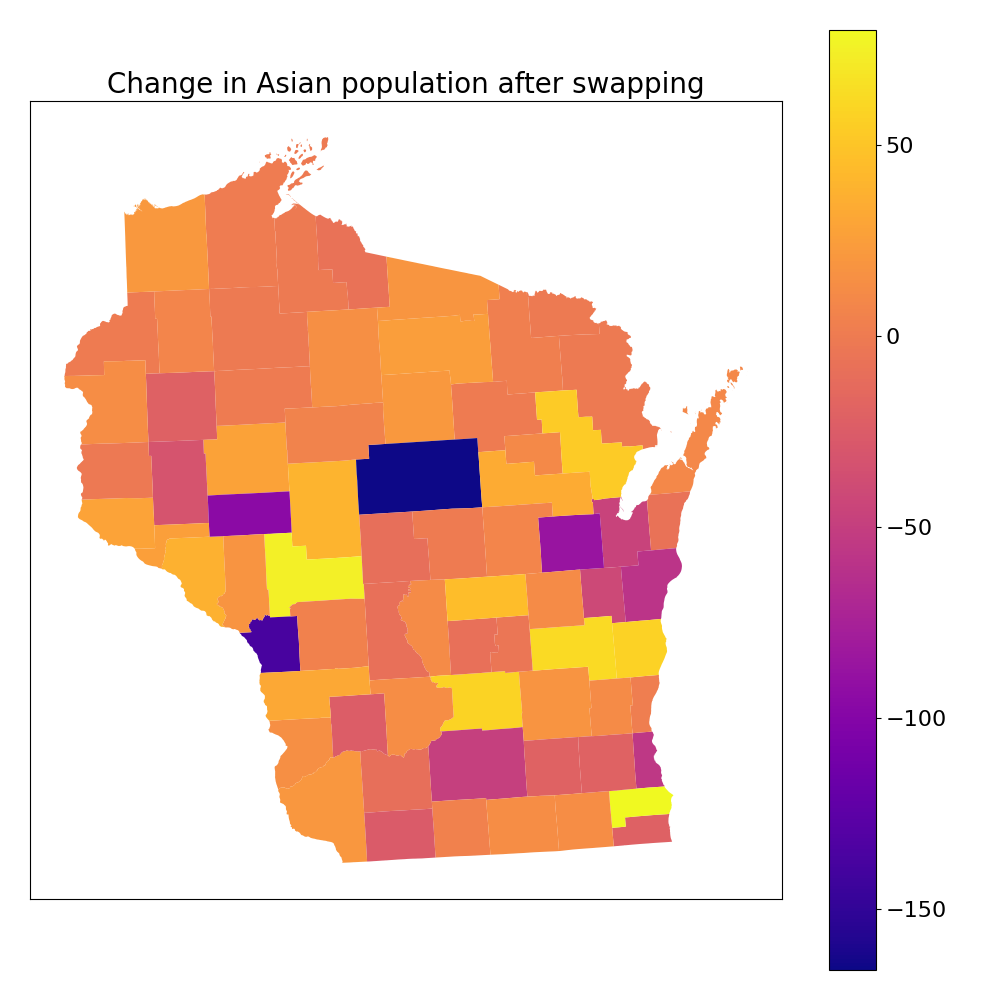}
                                          &\includegraphics[width=0.22\textwidth]{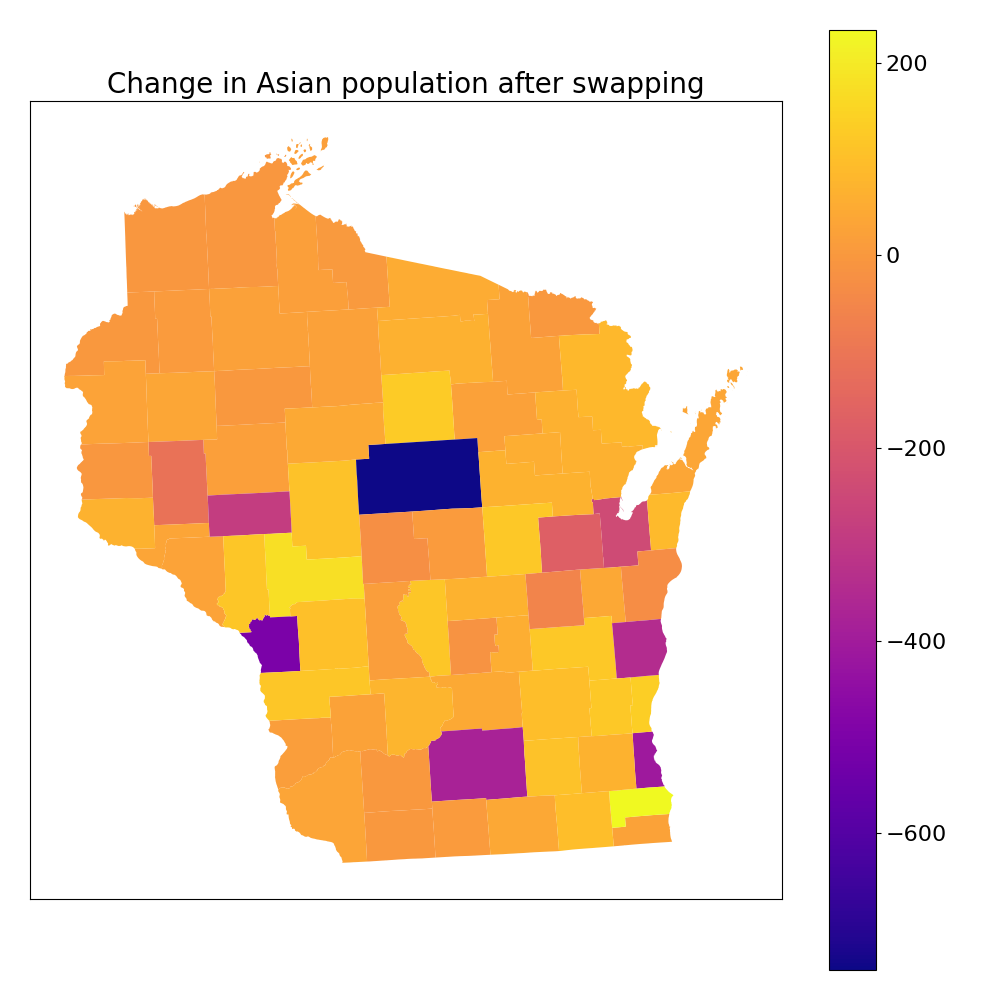}&\includegraphics[width=0.22\textwidth]{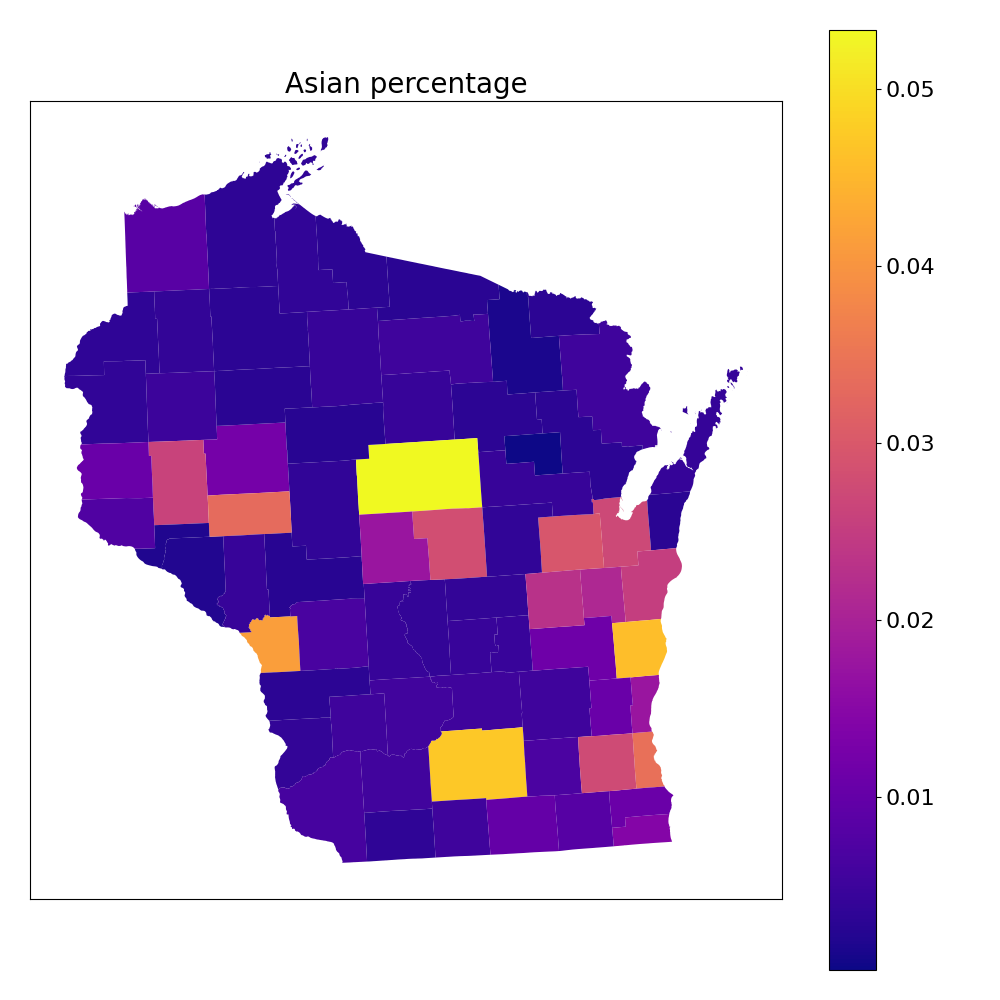}&\includegraphics[width=0.22\textwidth]{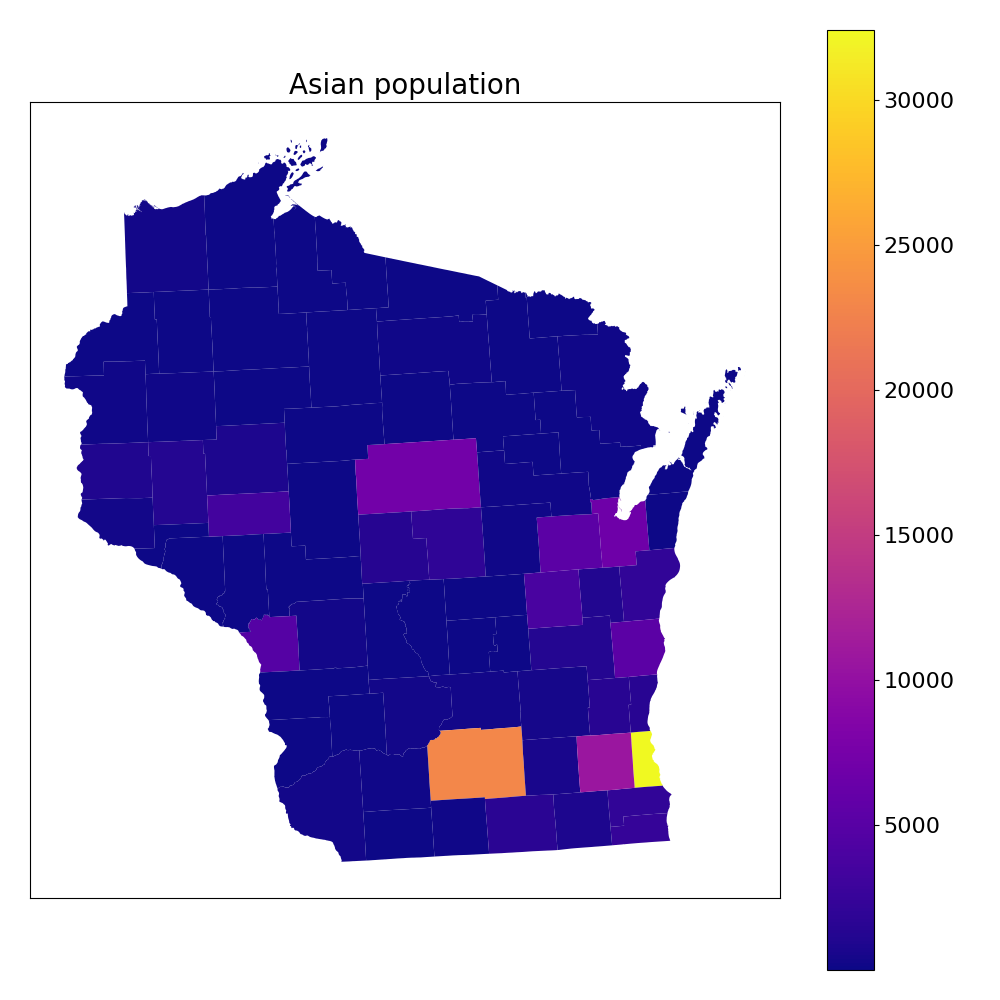}\\
        
        \rotatebox{90}{\hspace{20mm}Hispanic}&\includegraphics[width=0.22\textwidth]{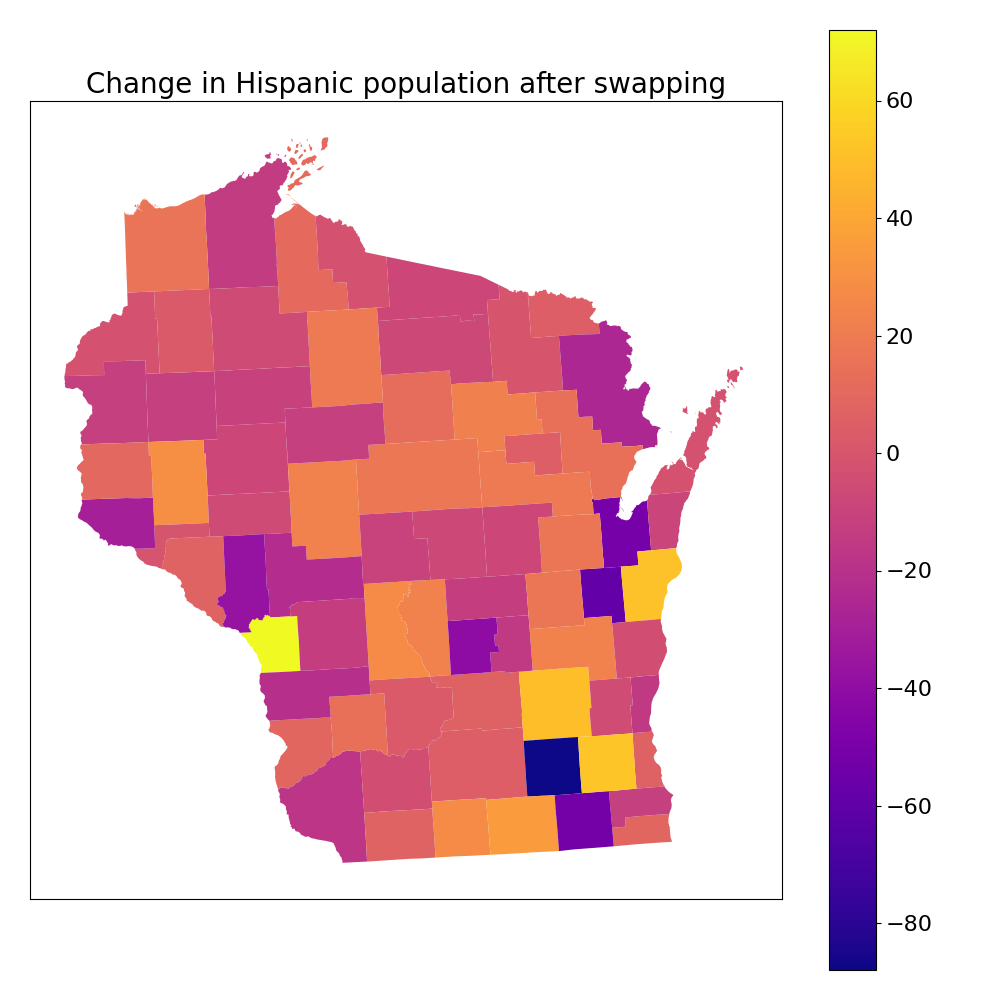}
                                             &\includegraphics[width=0.22\textwidth]{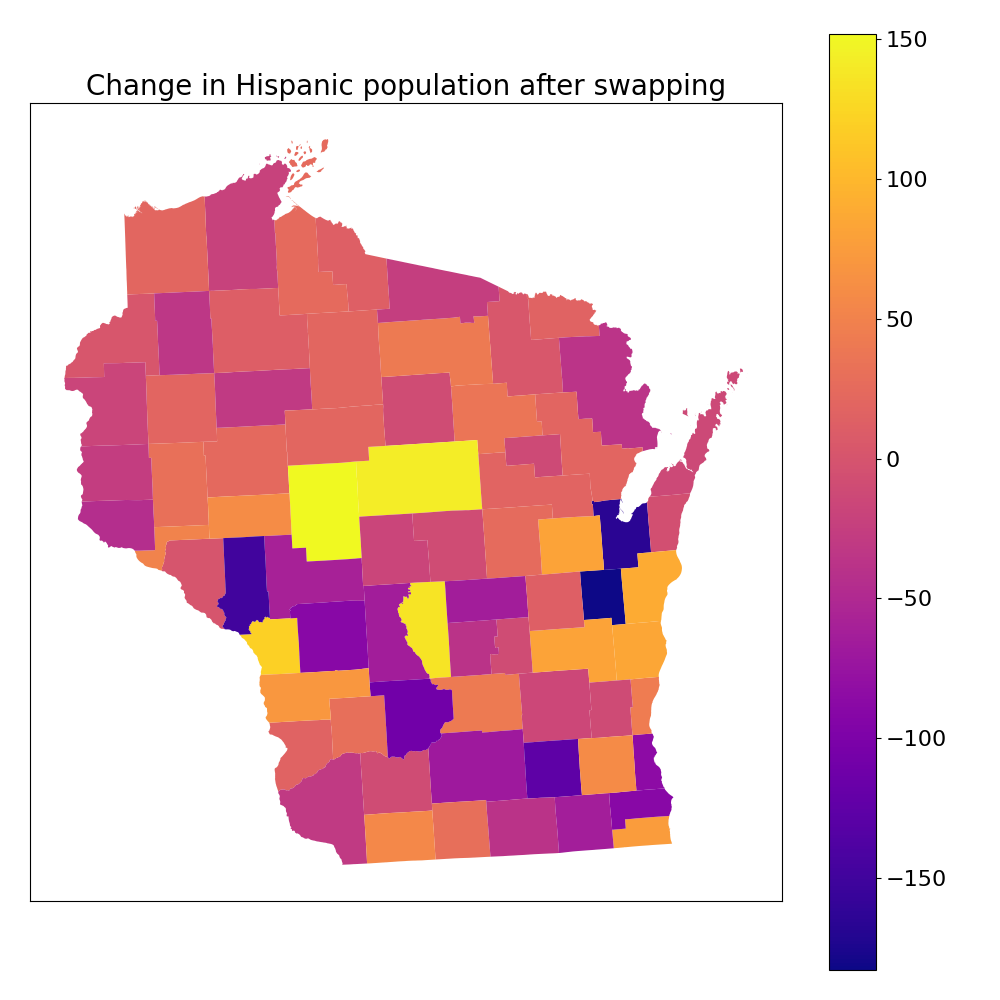}&\includegraphics[width=0.22\textwidth]{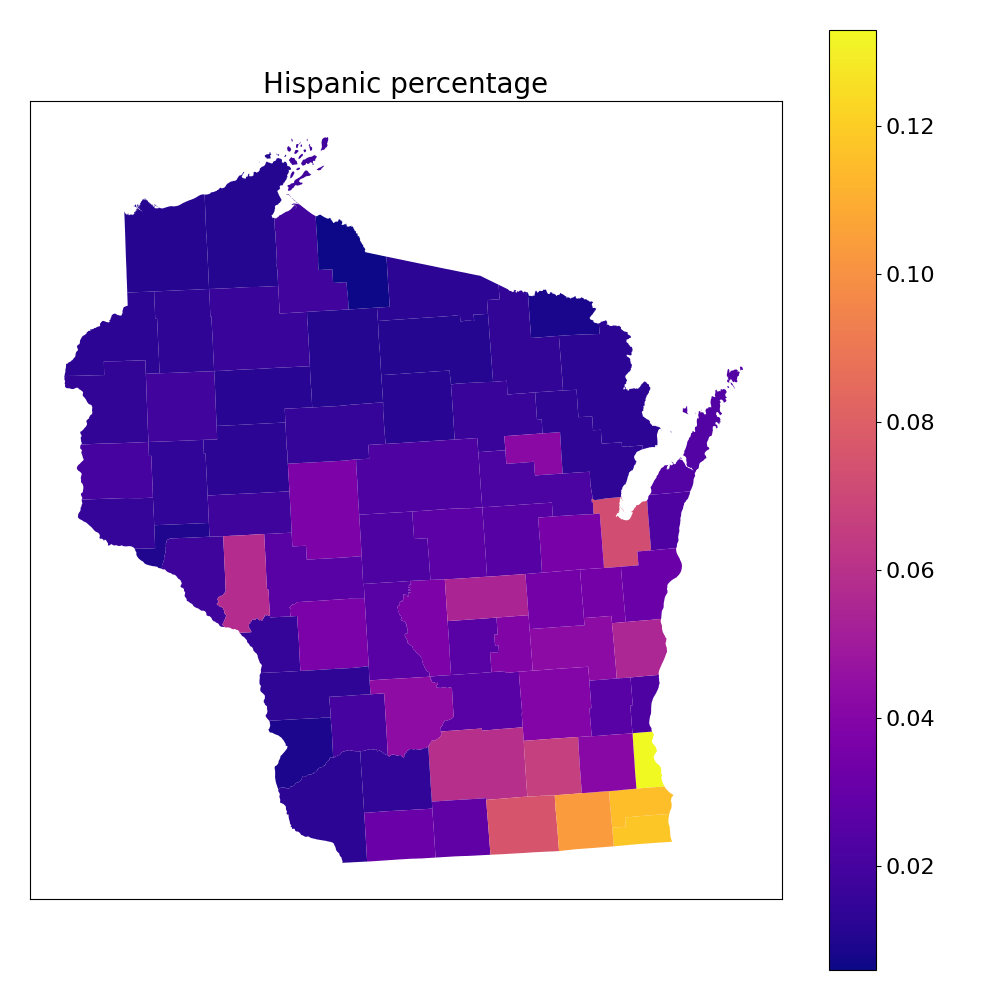}&\includegraphics[width=0.22\textwidth]{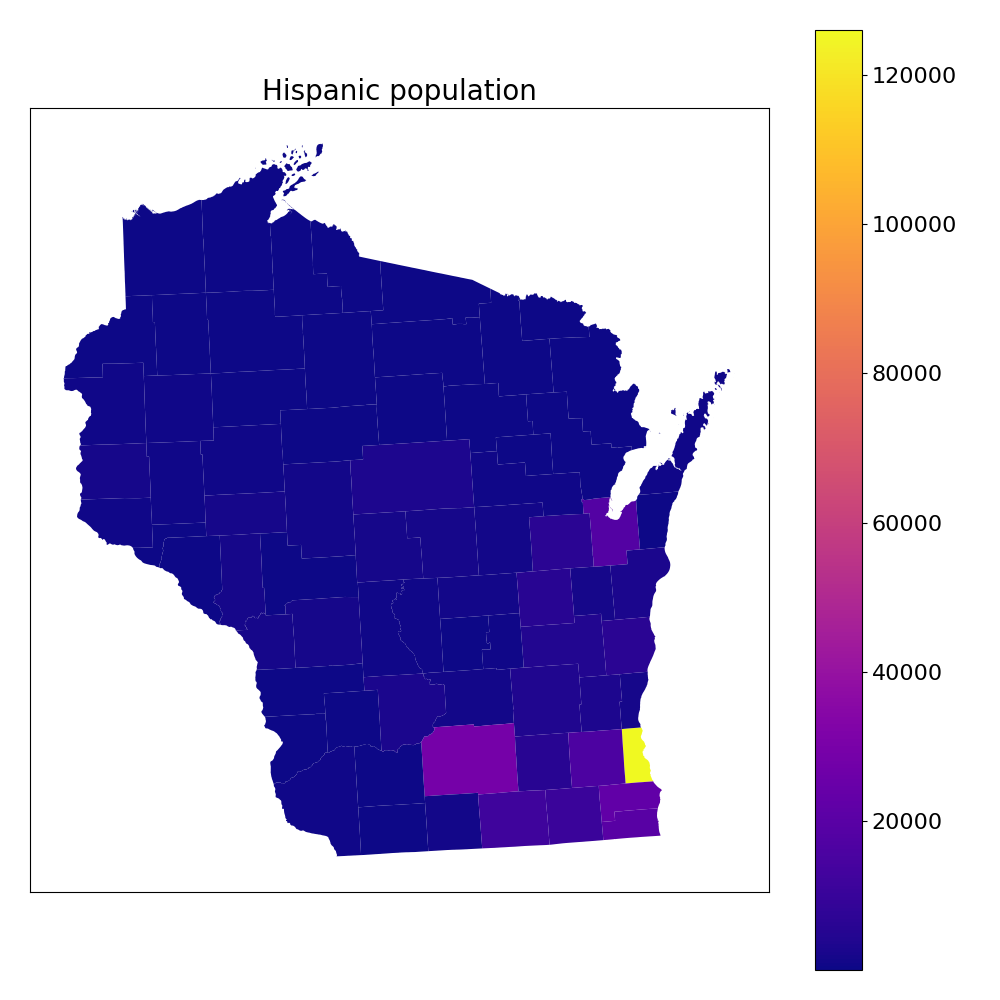}\\
        &2\% Swap Rate Effect & 10\% Swap Rate Effect & Percentage & Population
    \end{tabular}
    \caption{The effect of swapping on White and Hispanic population in Wisconsin, with the White/Hispanic percentage and White/Hispanic population shown for reference.}
    \label{fig:wi_full_maps}
\end{figure}

\begin{figure}
    \centering
    \begin{tabular}{lcccc}
        \rotatebox{90}{\hspace{20mm}White}&\includegraphics[width=0.22\textwidth]{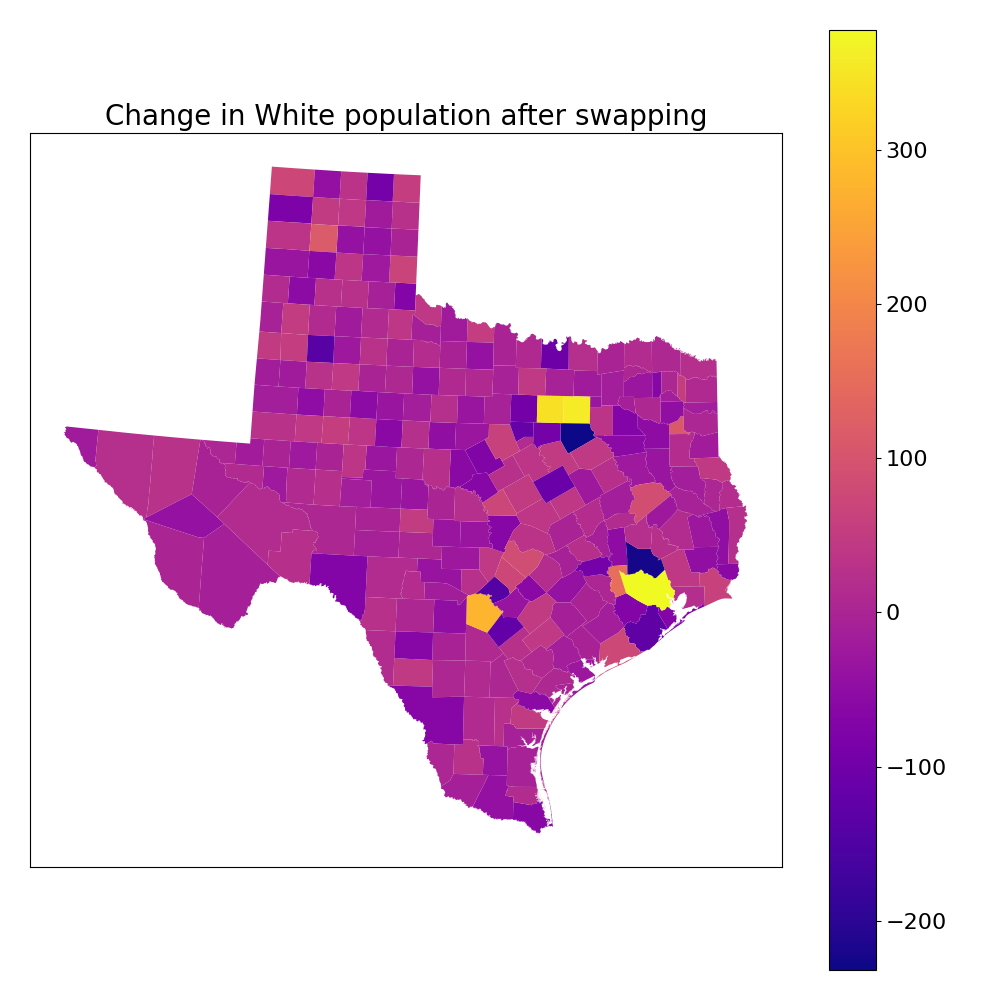}
                                          &\includegraphics[width=0.22\textwidth]{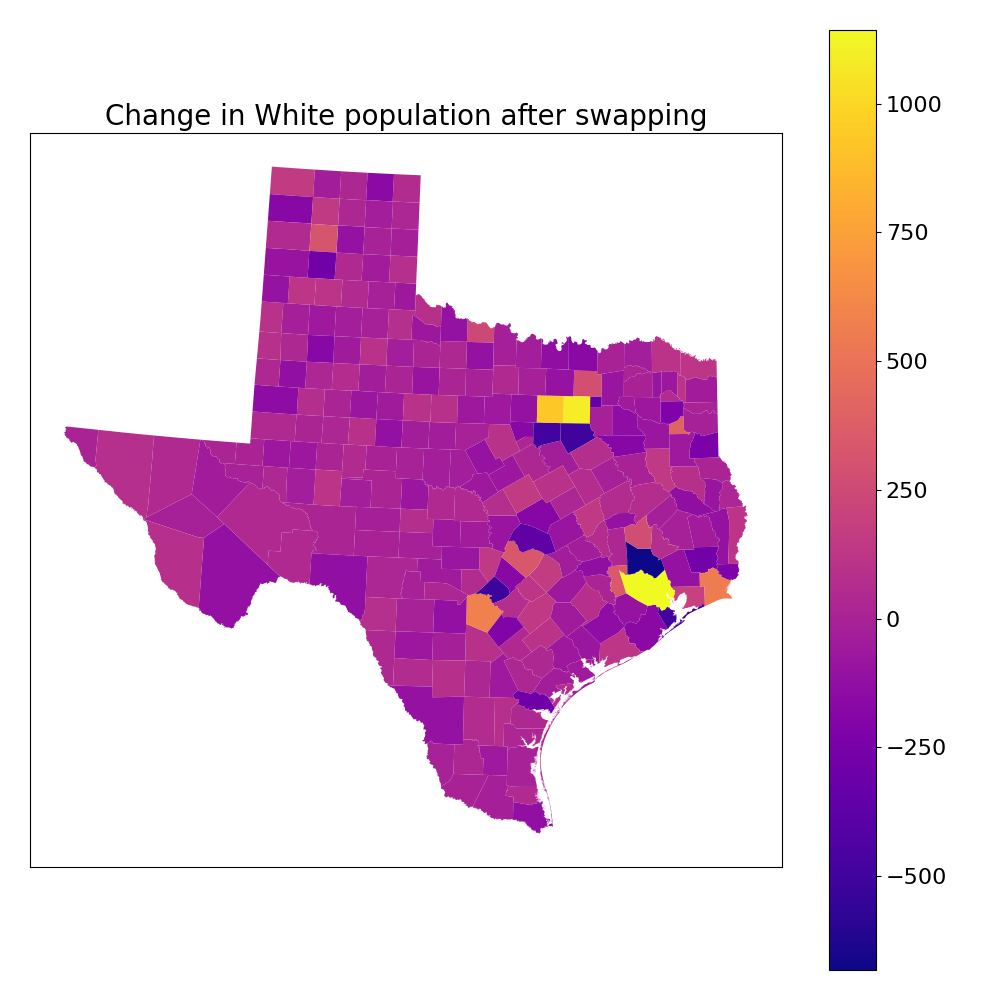}&\includegraphics[width=0.22\textwidth]{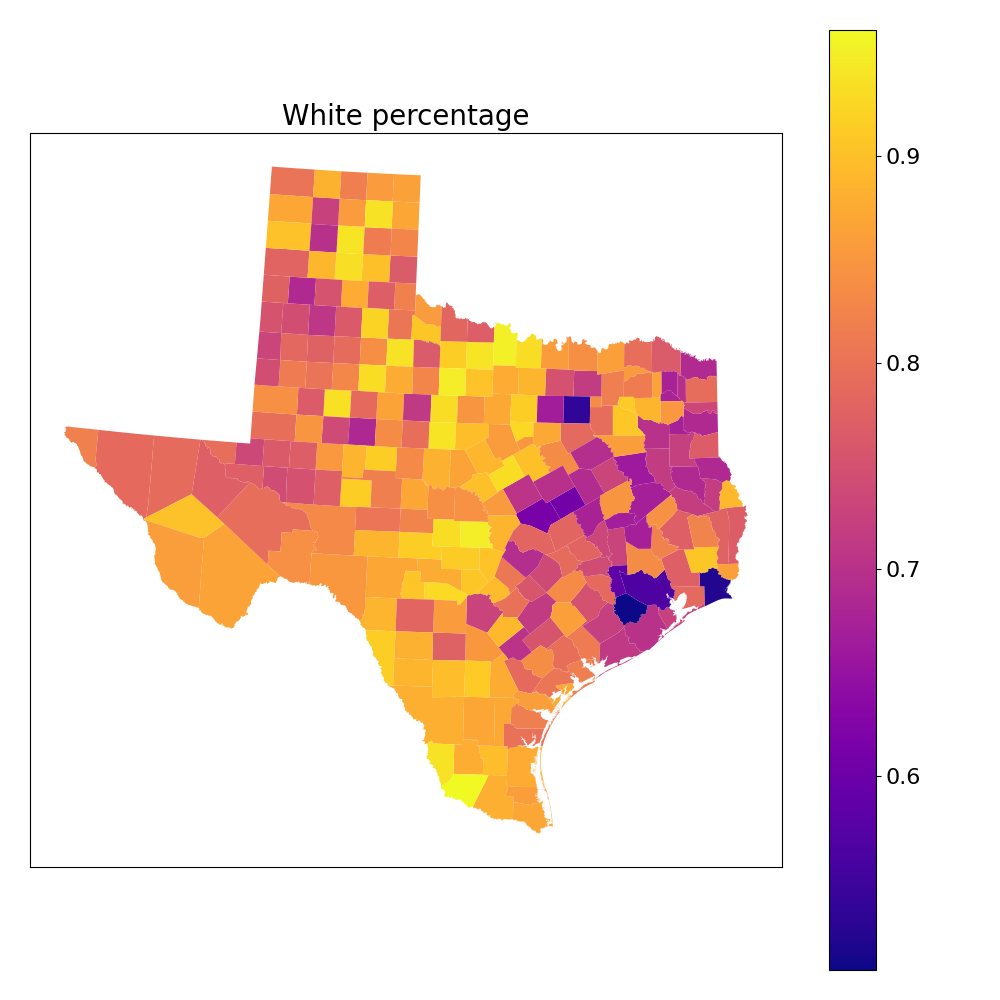}&\includegraphics[width=0.22\textwidth]{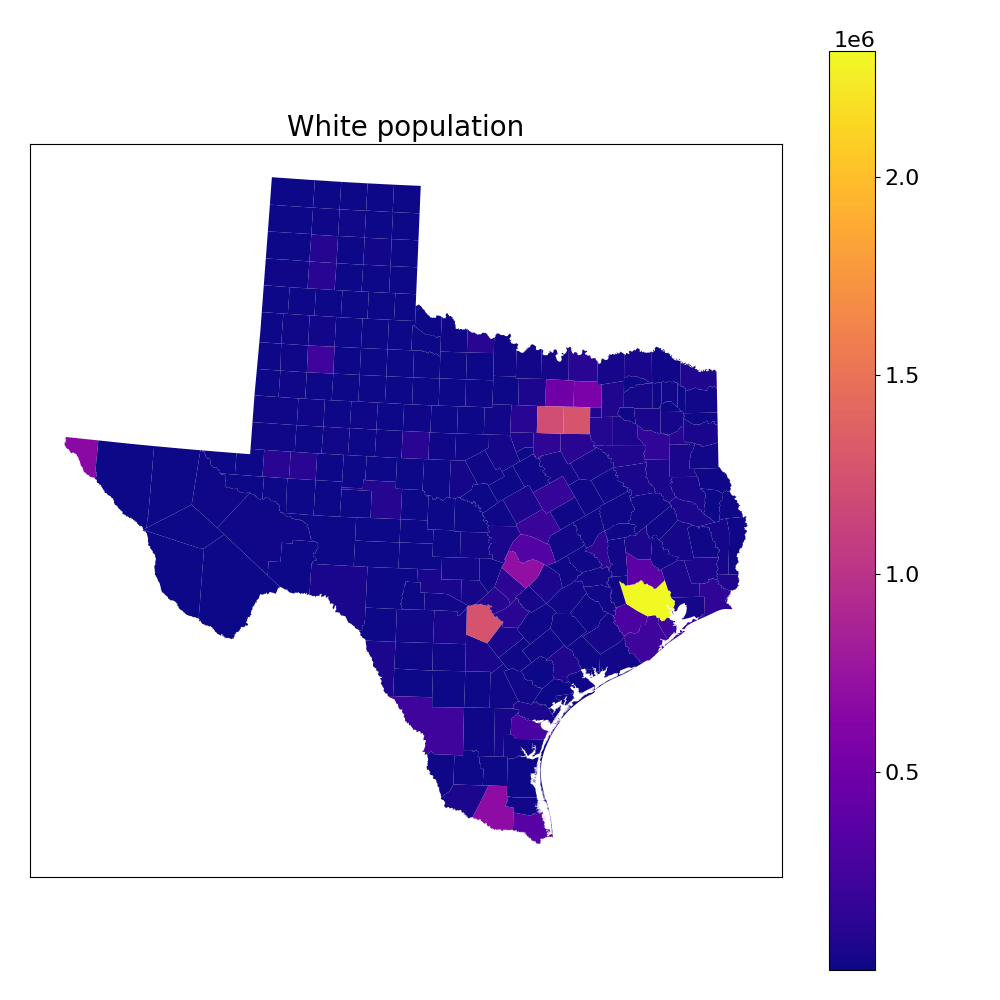}\\

        \rotatebox{90}{\hspace{20mm}Black}&\includegraphics[width=0.22\textwidth]{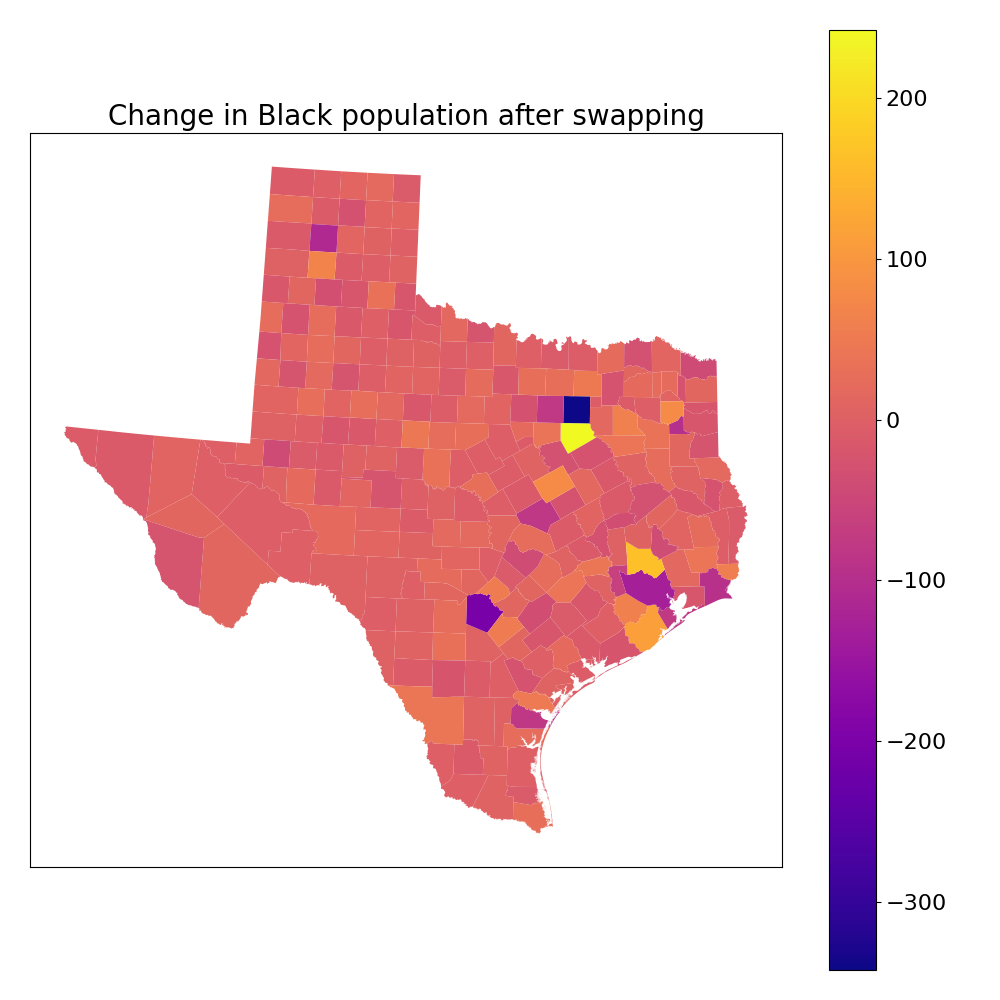}
                                          &\includegraphics[width=0.22\textwidth]{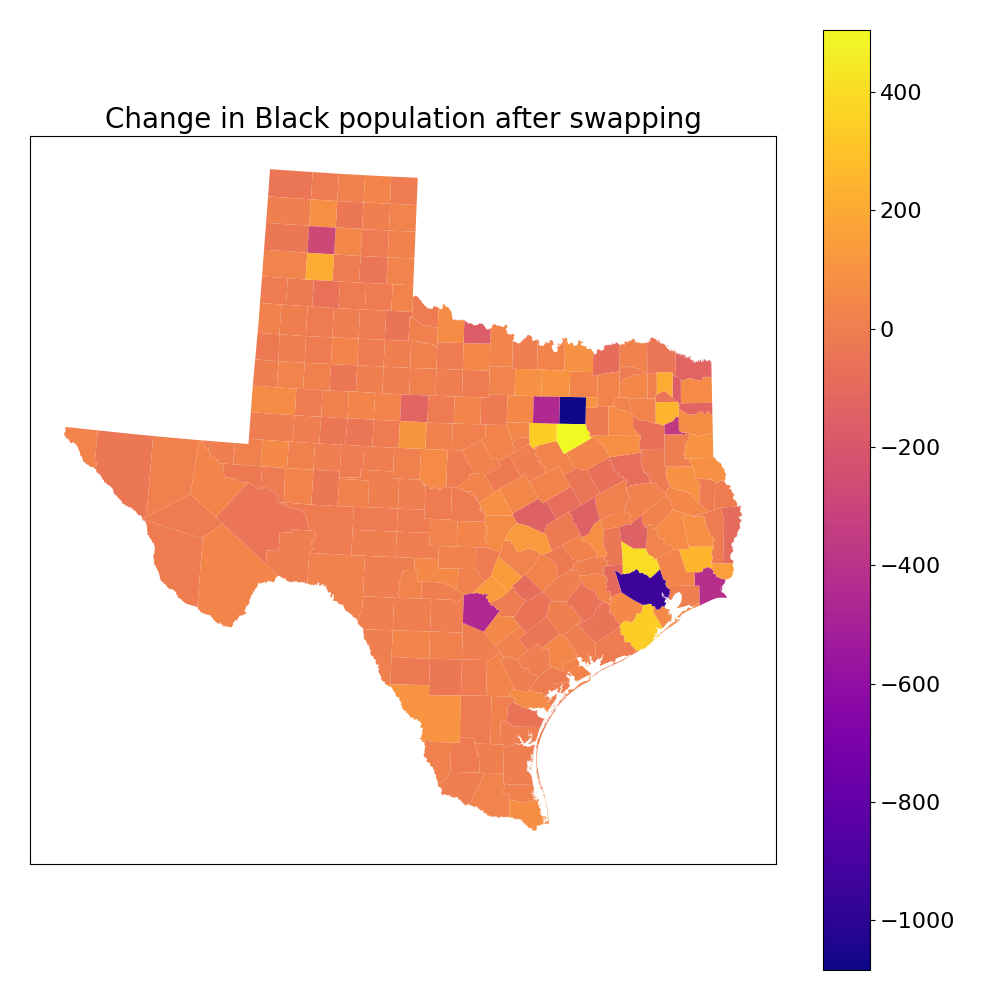}&\includegraphics[width=0.22\textwidth]{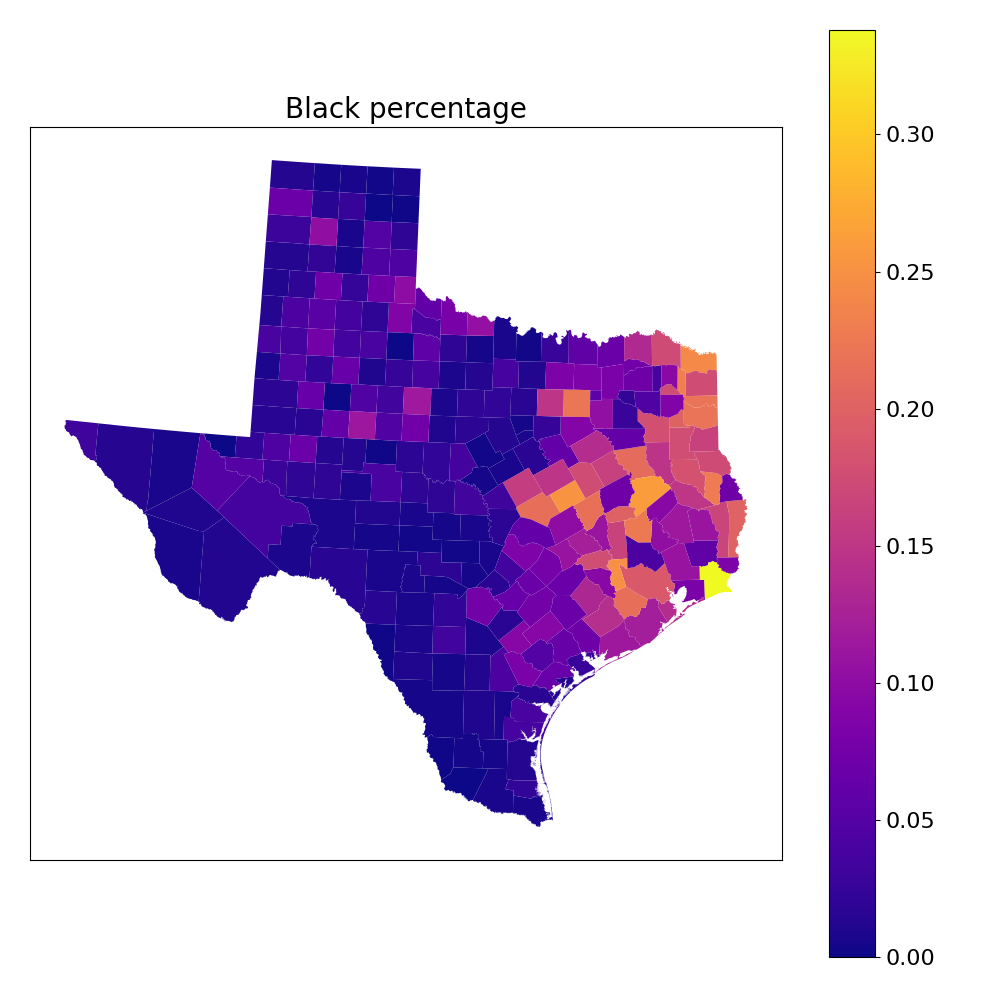}&\includegraphics[width=0.22\textwidth]{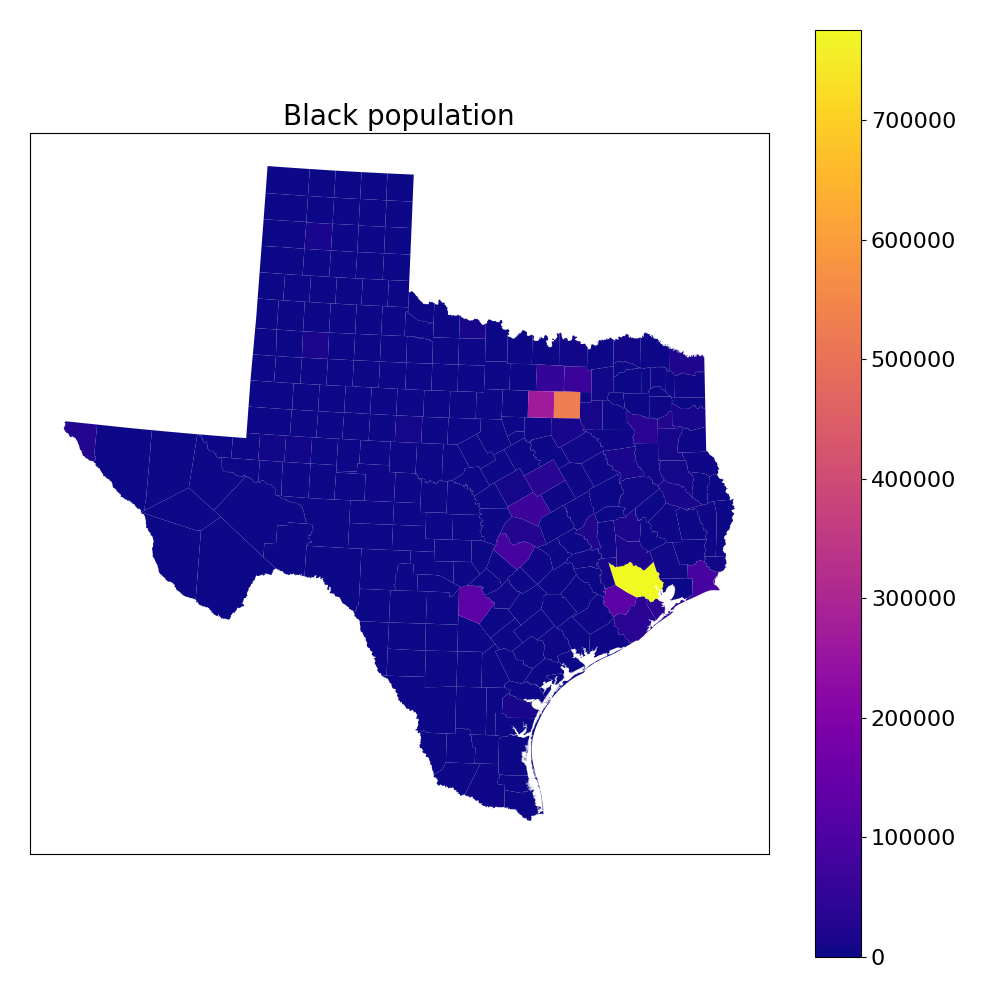}\\

        \rotatebox{90}{\hspace{20mm}Asian}&\includegraphics[width=0.22\textwidth]{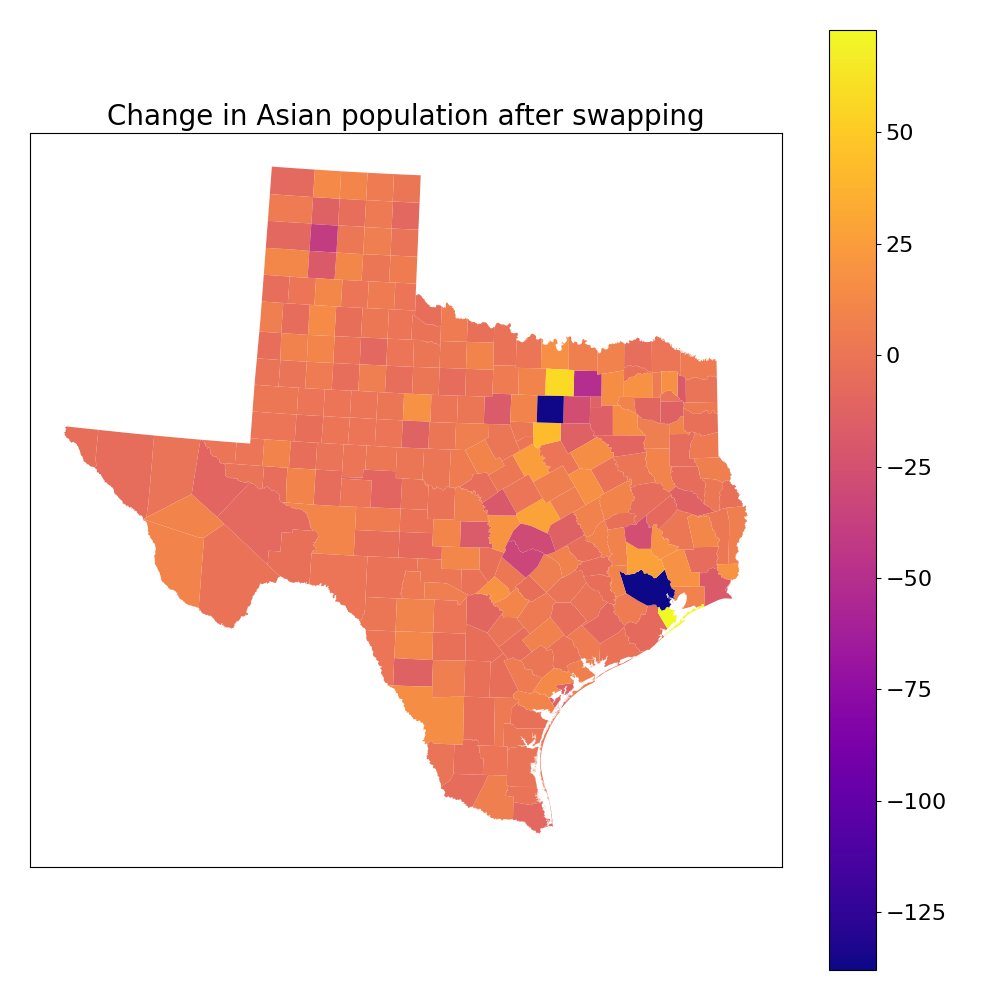}
                                          &\includegraphics[width=0.22\textwidth]{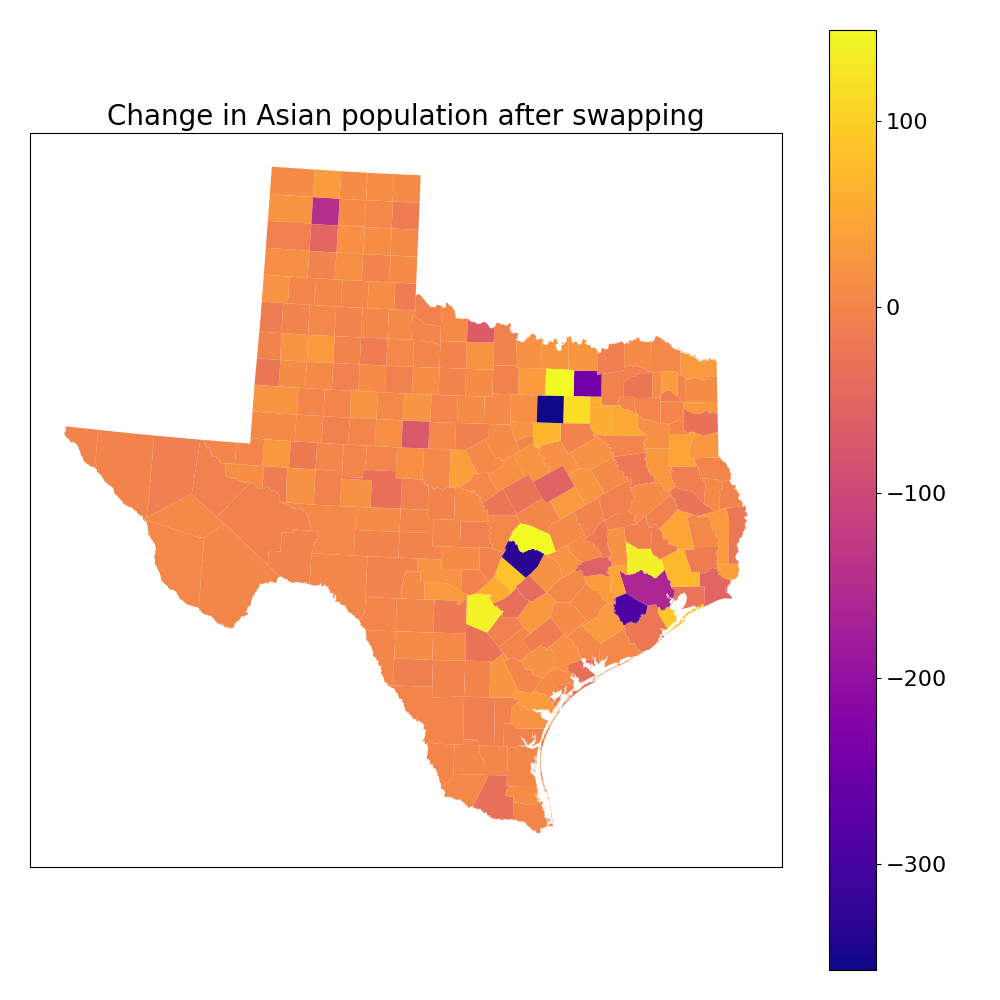}&\includegraphics[width=0.22\textwidth]{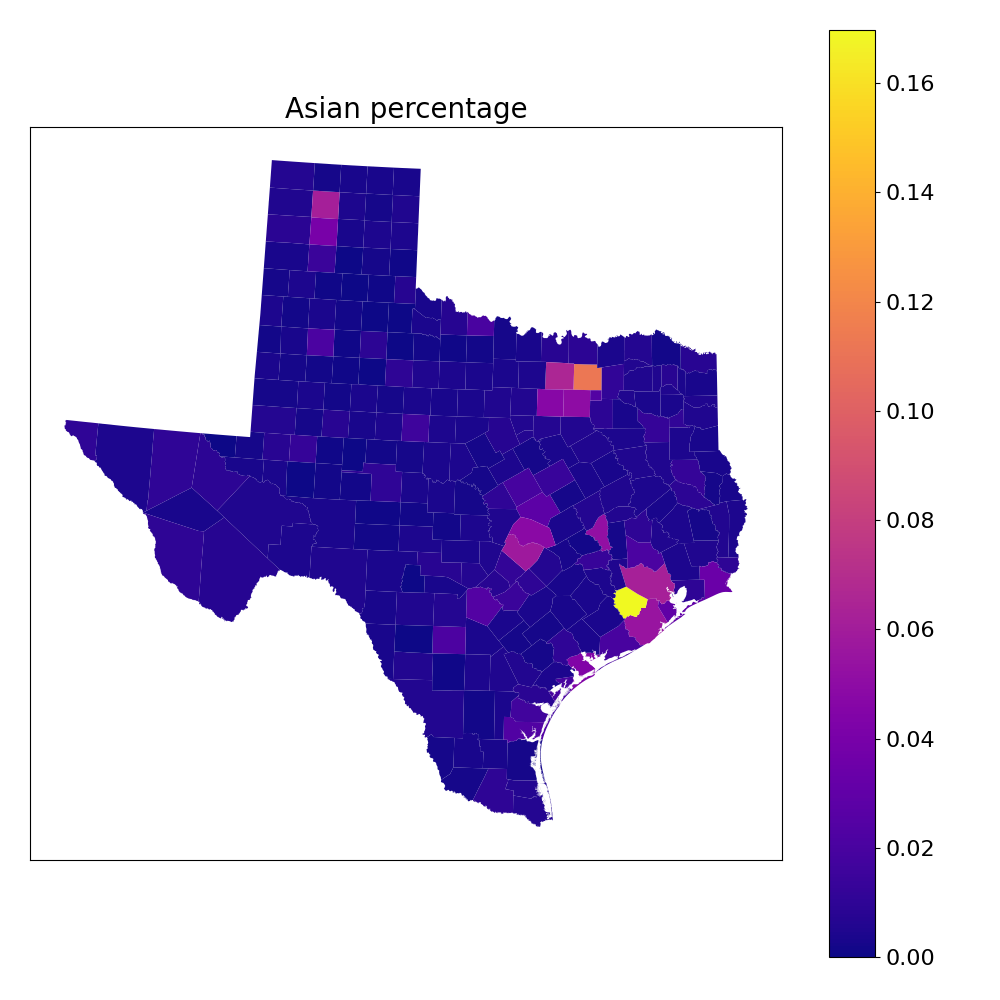}&\includegraphics[width=0.22\textwidth]{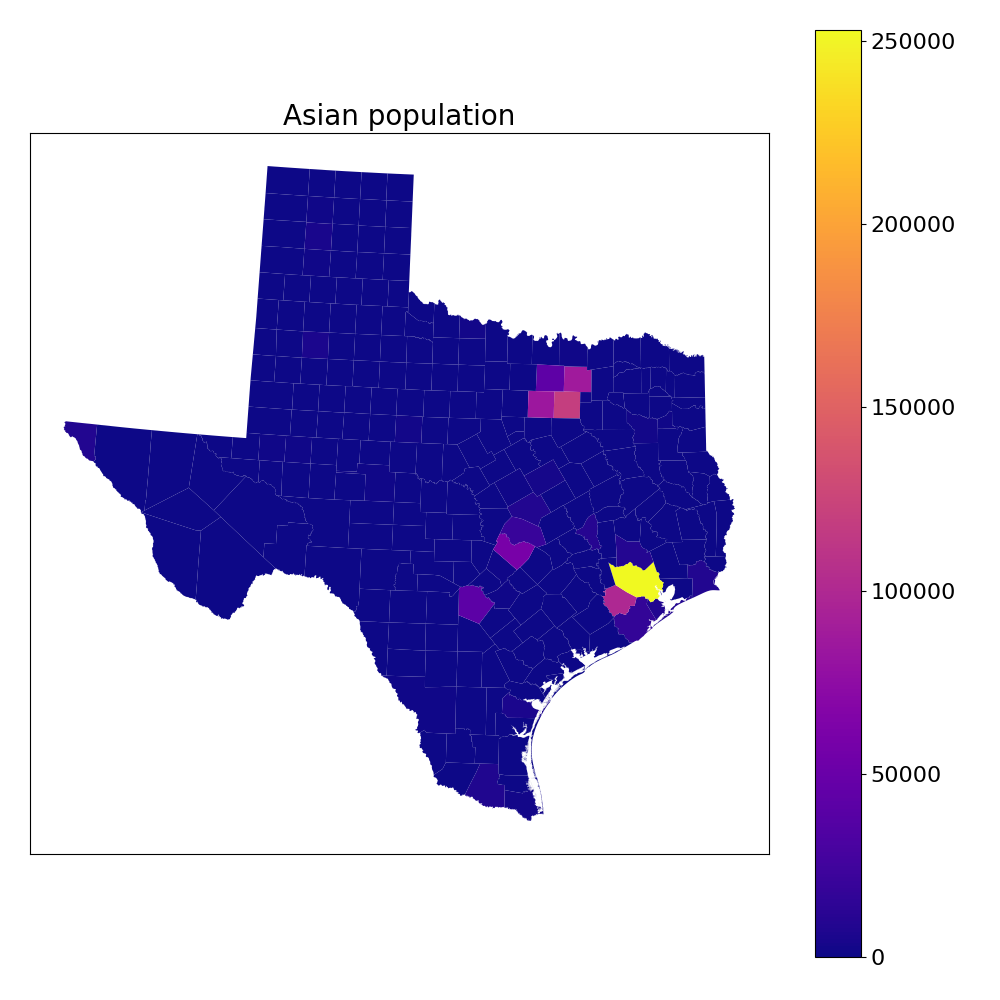}\\
        
        \rotatebox{90}{\hspace{20mm}Hispanic}&\includegraphics[width=0.22\textwidth]{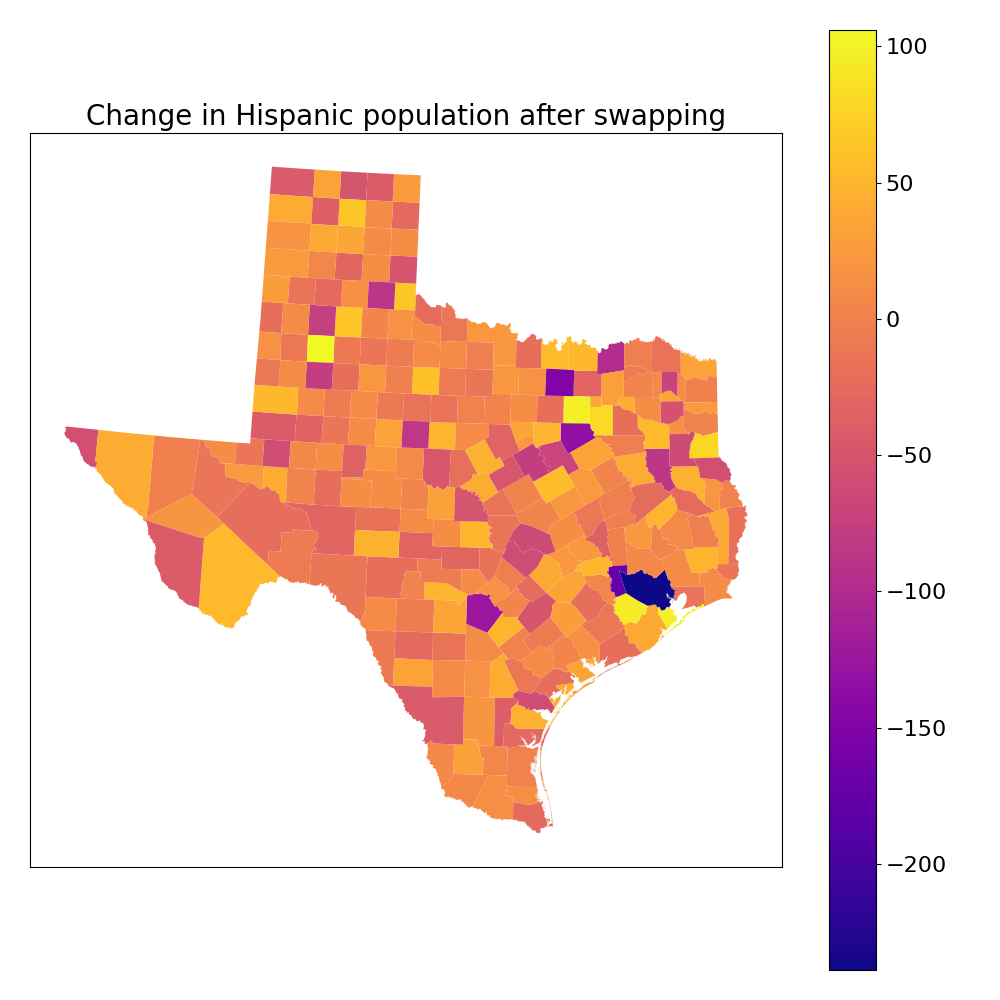}
                                             &\includegraphics[width=0.22\textwidth]{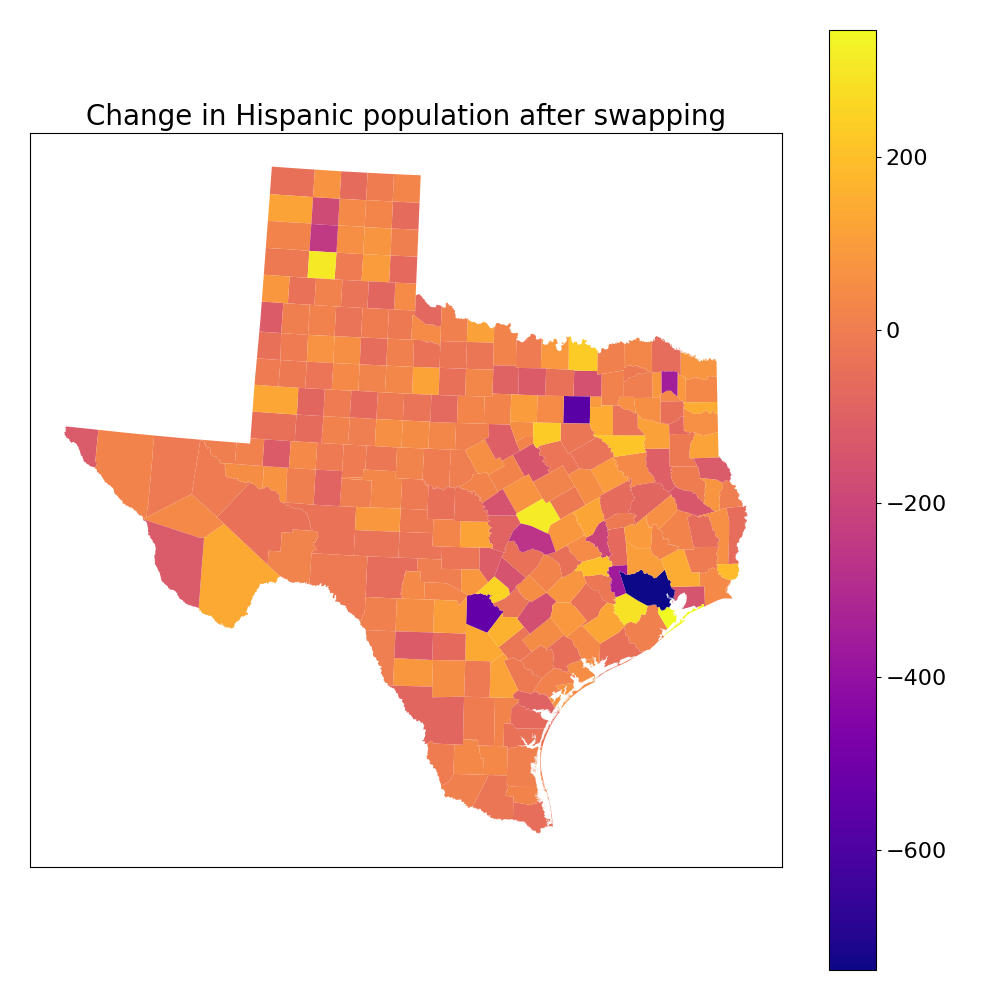}&\includegraphics[width=0.22\textwidth]{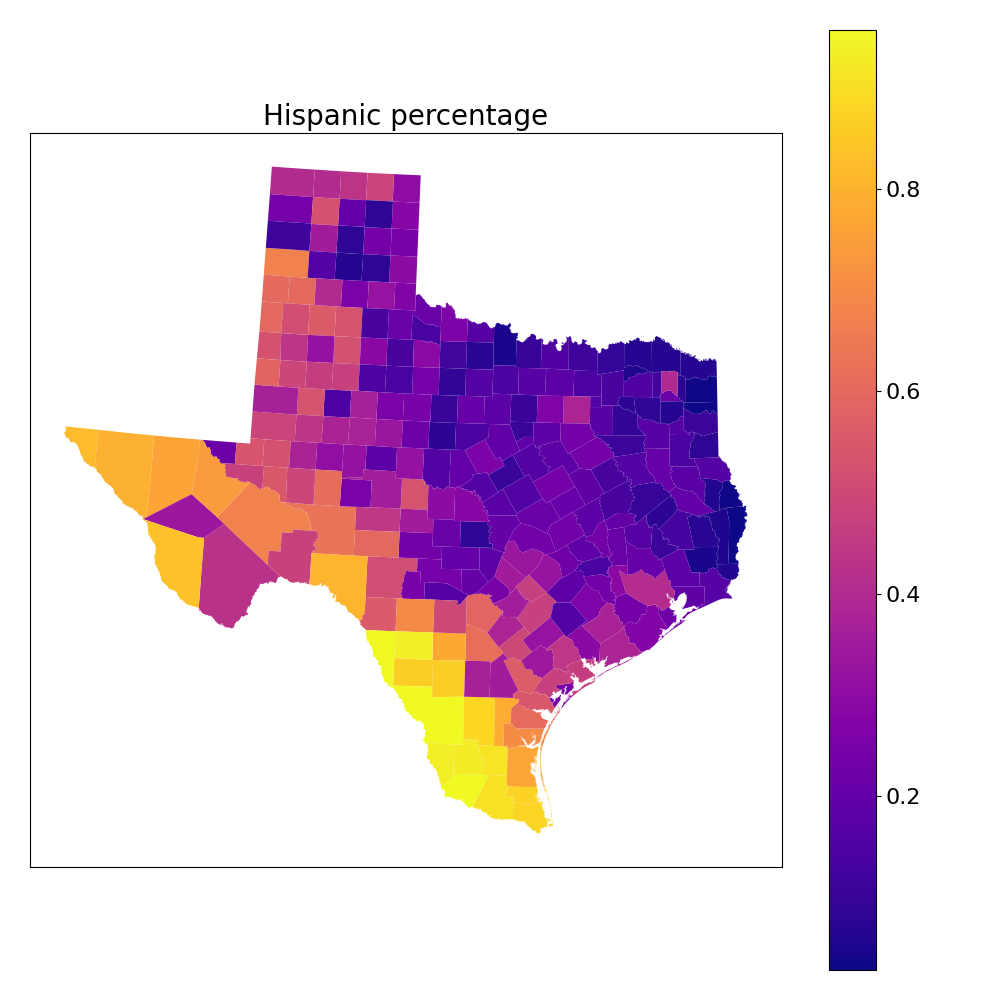}&\includegraphics[width=0.22\textwidth]{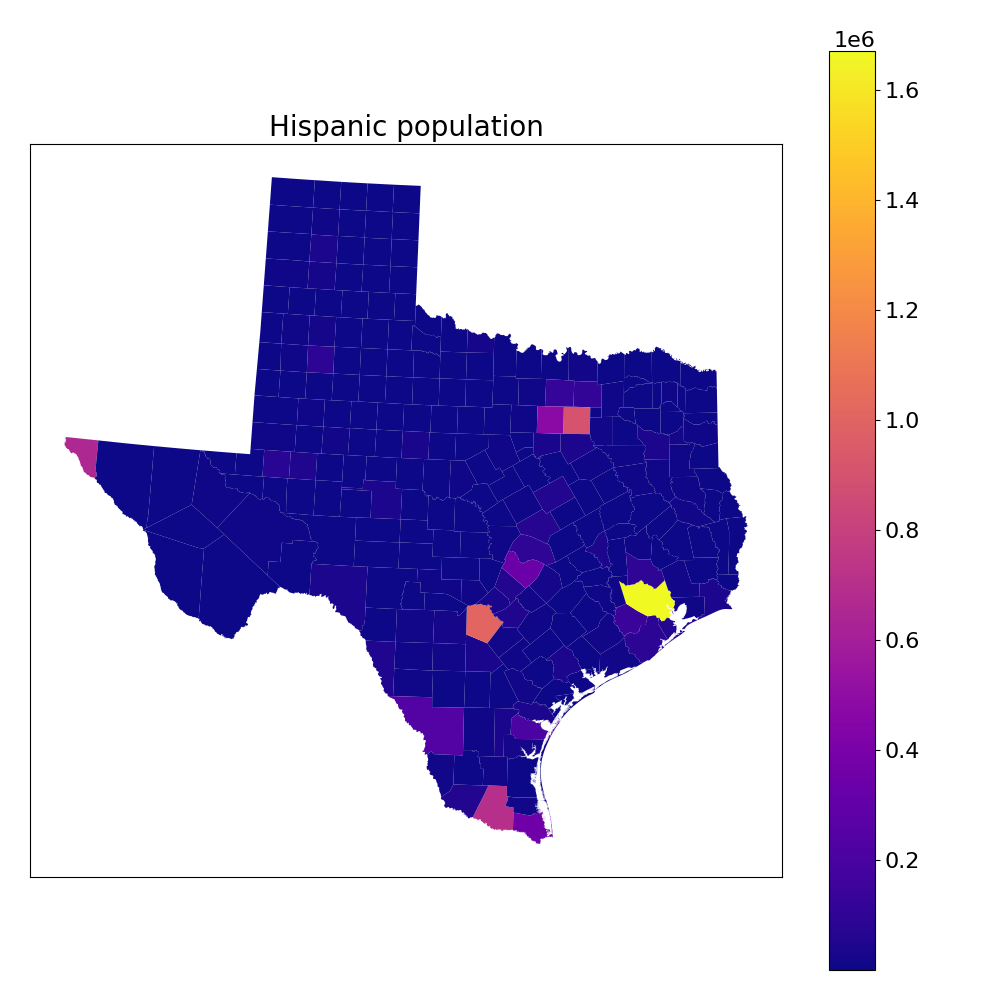}\\
        &2\% Swap Rate Effect & 10\% Swap Rate Effect & Percentage & Population
    \end{tabular}
    \caption{The effect of swapping on White and Hispanic population in Texas, with the White/Hispanic percentage and White/Hispanic population shown for reference.}
    \label{fig:tx_full_maps}
\end{figure}

\begin{figure}
    \centering
    \begin{tabular}{lcccc}
        \rotatebox{90}{\hspace{20mm}White}&\includegraphics[width=0.22\textwidth]{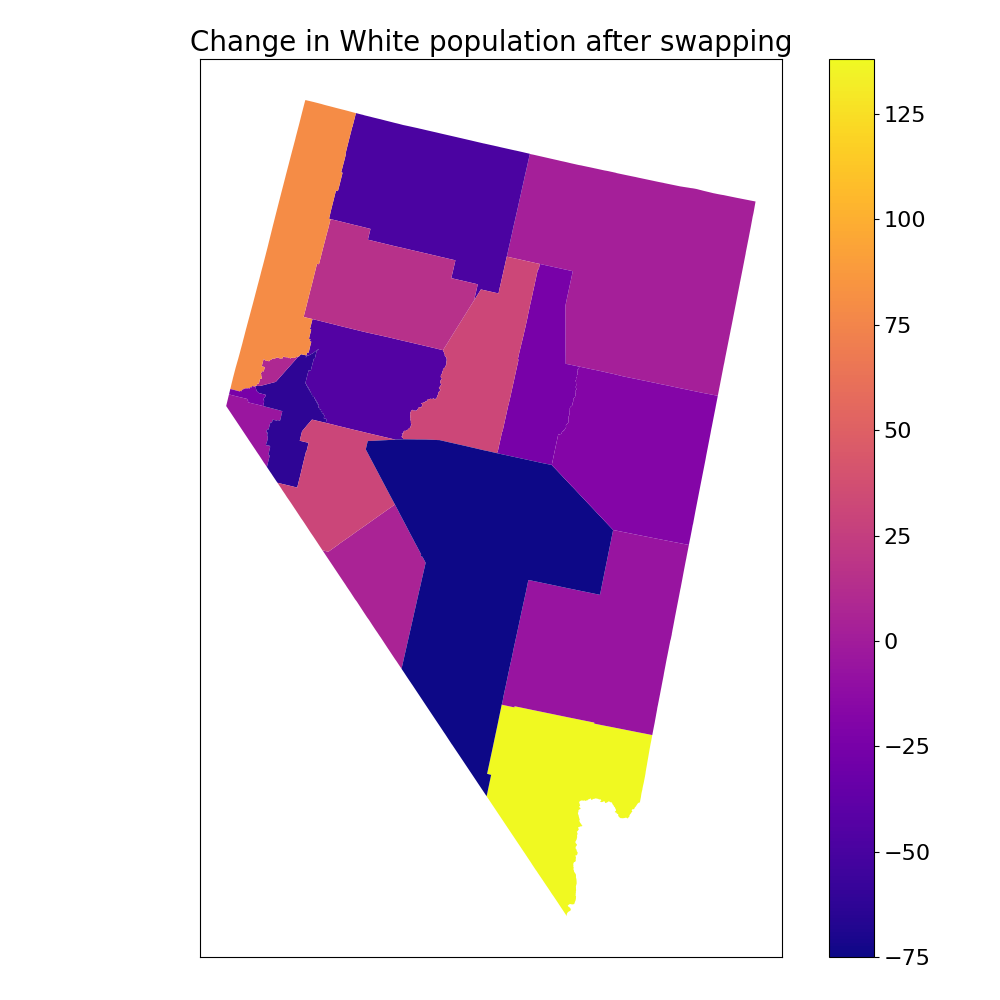}
                                          &\includegraphics[width=0.22\textwidth]{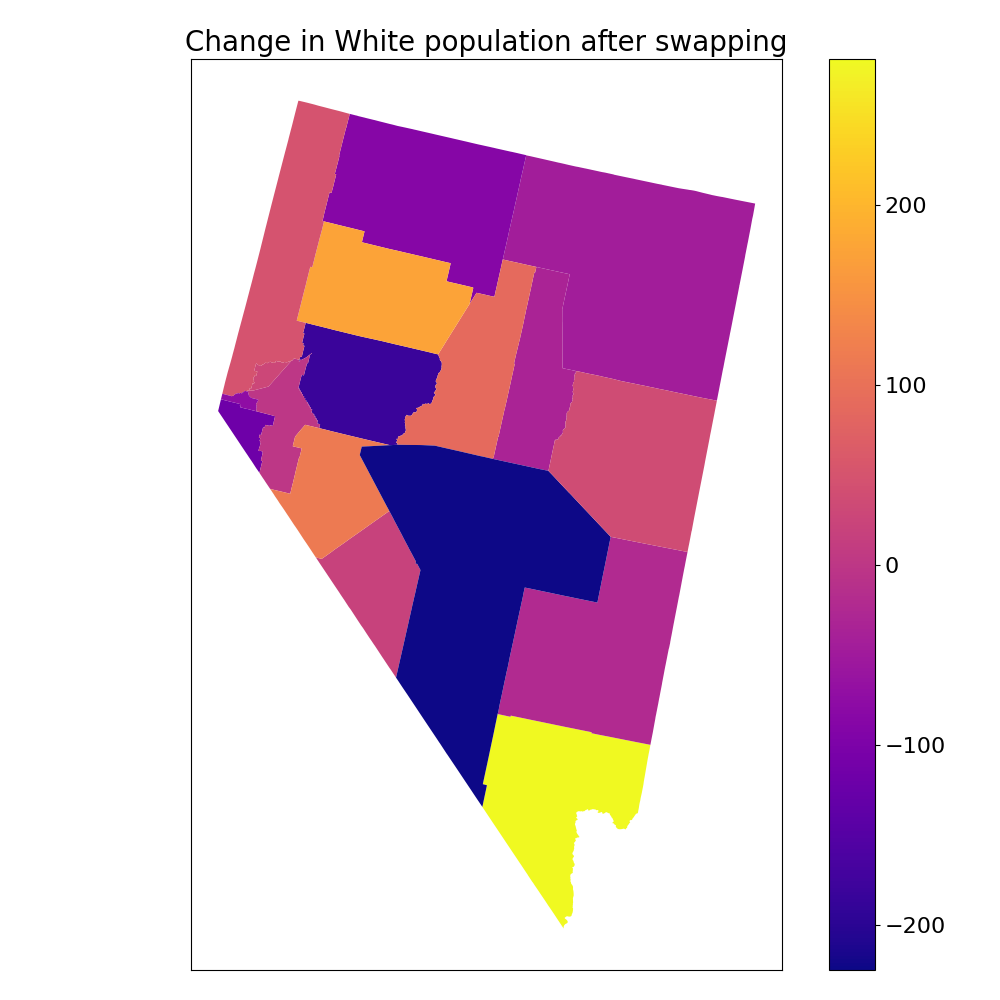}&\includegraphics[width=0.22\textwidth]{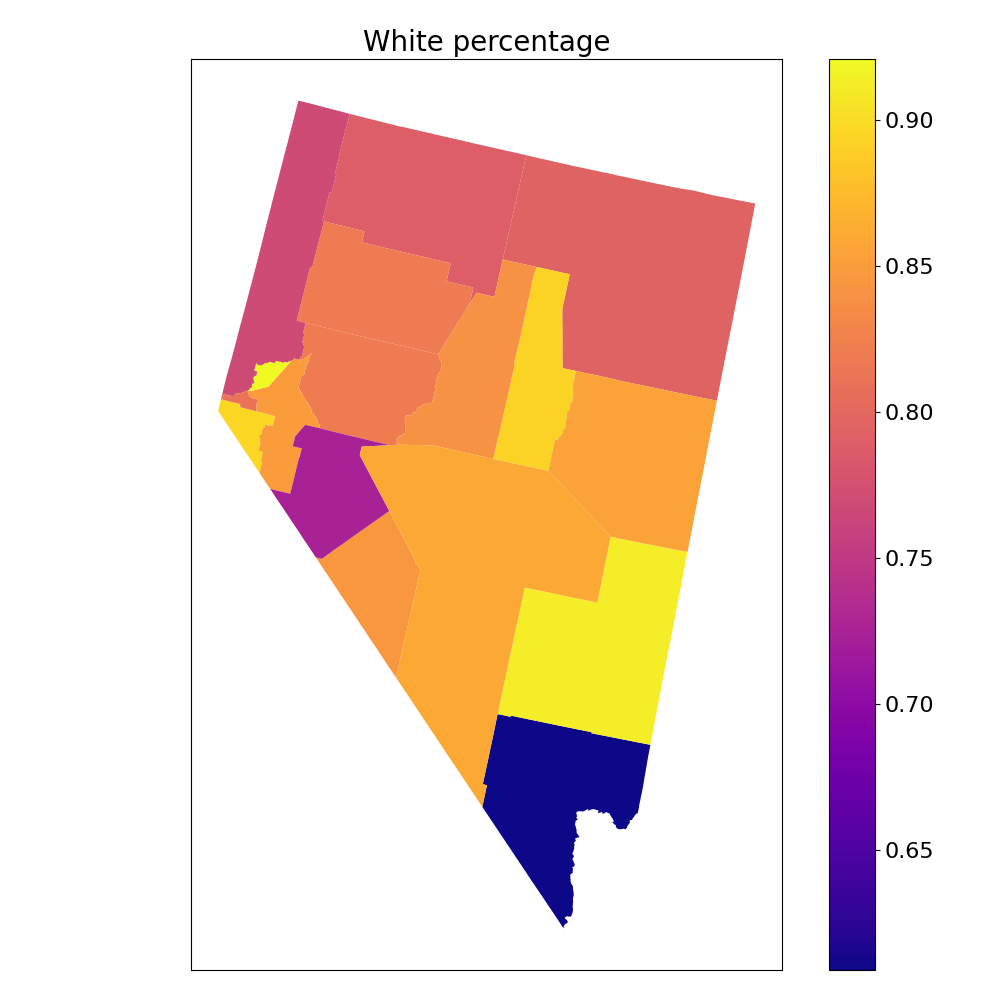}&\includegraphics[width=0.22\textwidth]{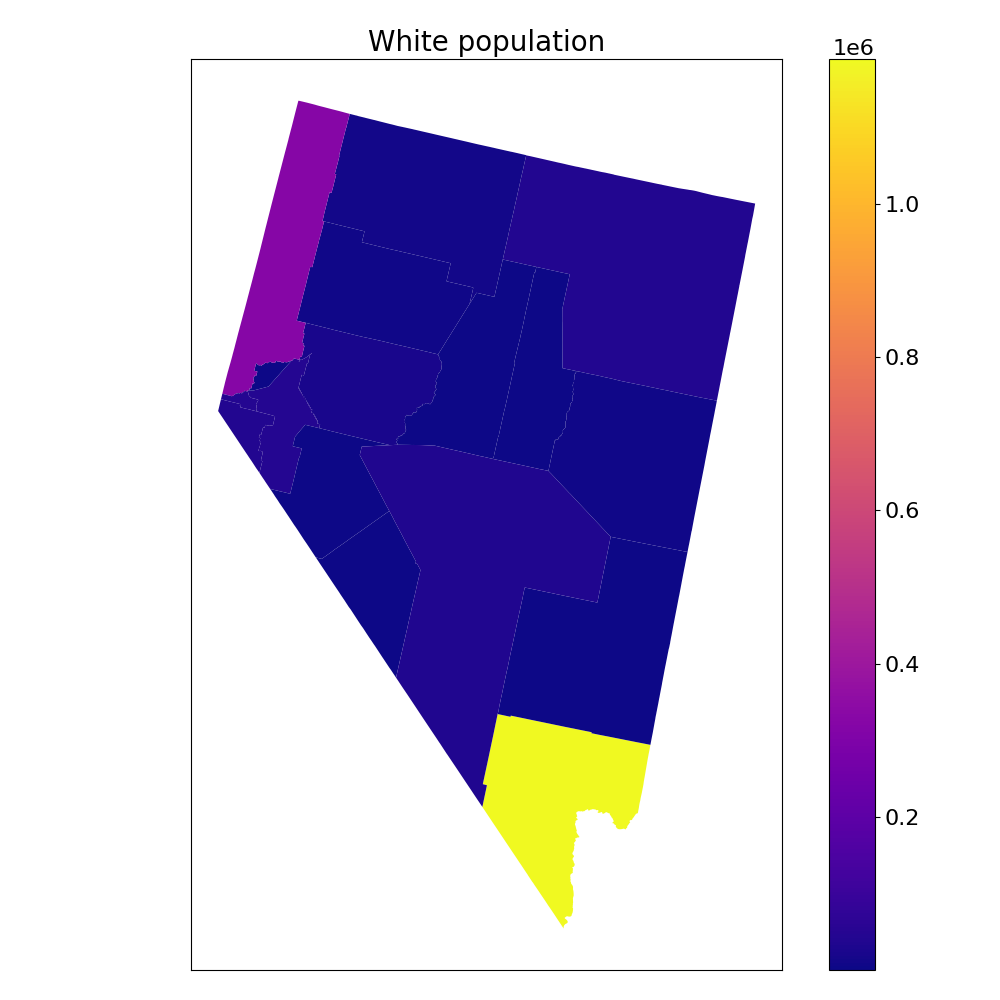}\\

        \rotatebox{90}{\hspace{20mm}Black}&\includegraphics[width=0.22\textwidth]{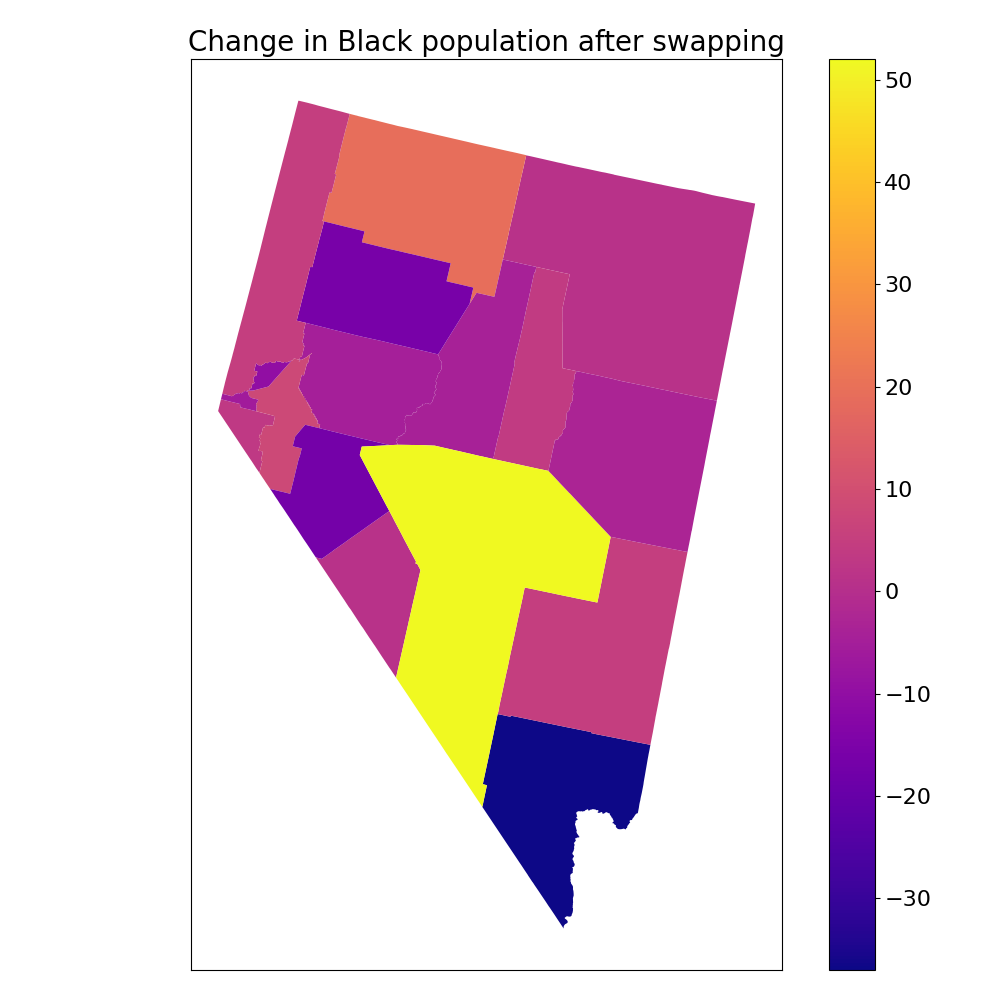}
                                          &\includegraphics[width=0.22\textwidth]{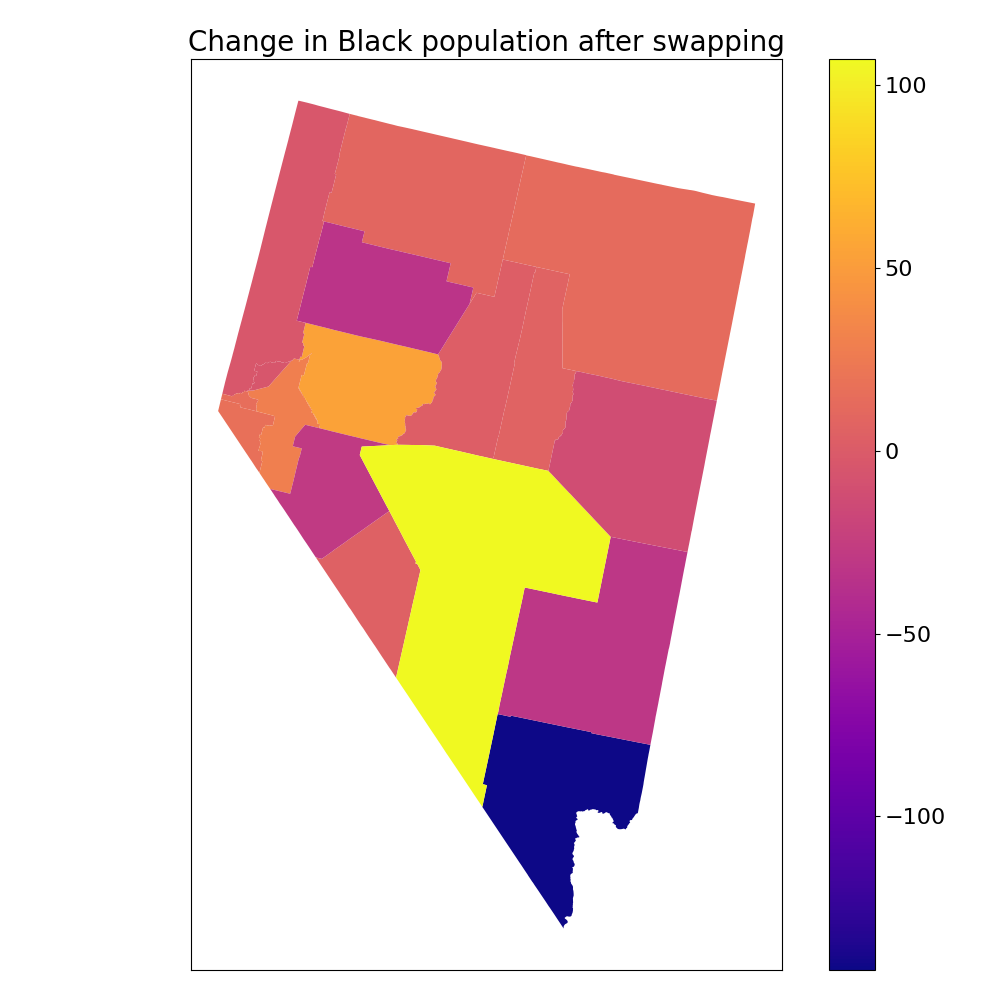}&\includegraphics[width=0.22\textwidth]{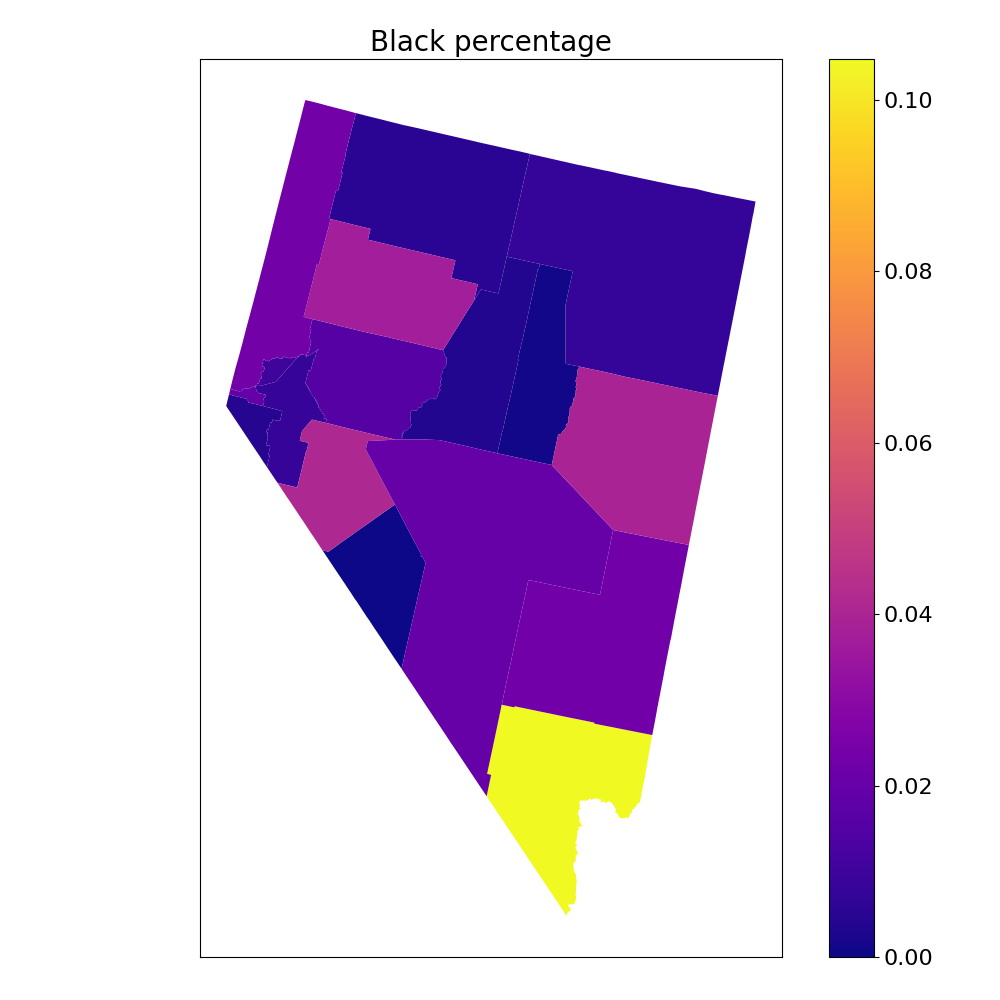}&\includegraphics[width=0.22\textwidth]{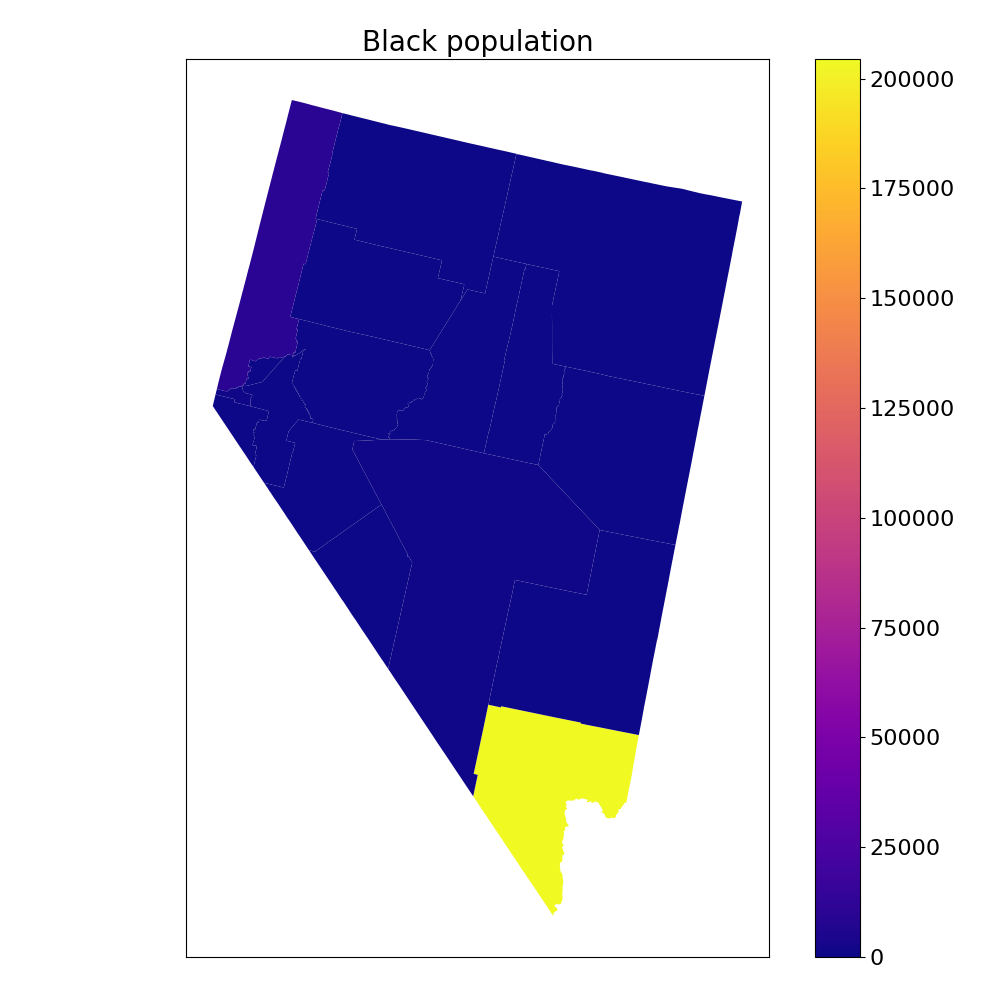}\\

        \rotatebox{90}{\hspace{20mm}Asian}&\includegraphics[width=0.22\textwidth]{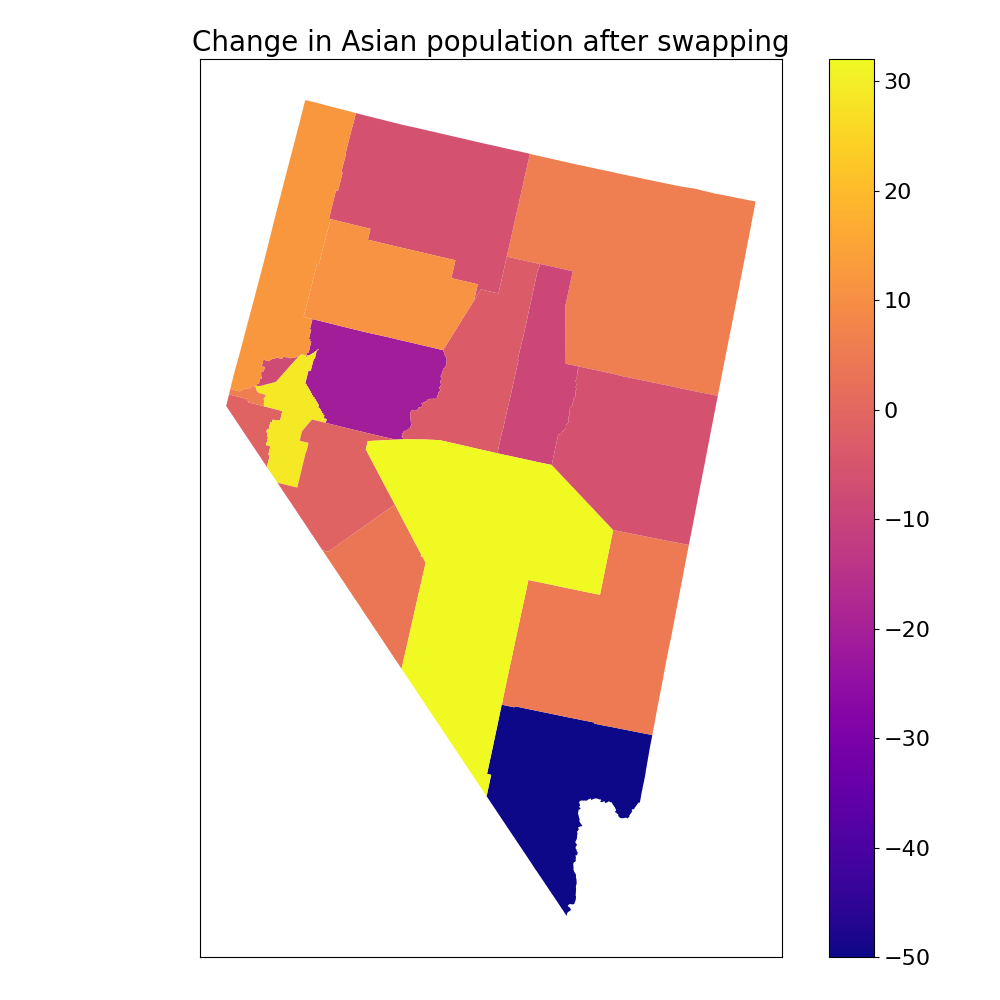}
                                          &\includegraphics[width=0.22\textwidth]{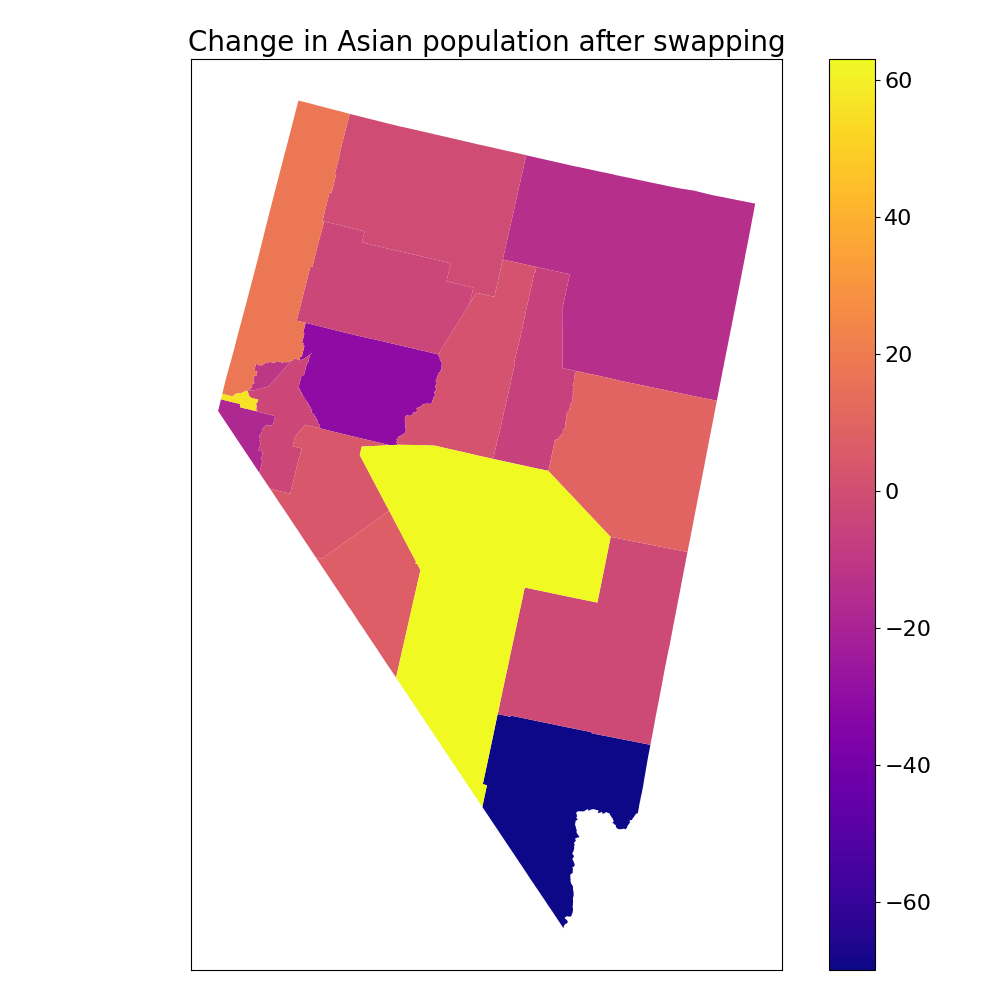}&\includegraphics[width=0.22\textwidth]{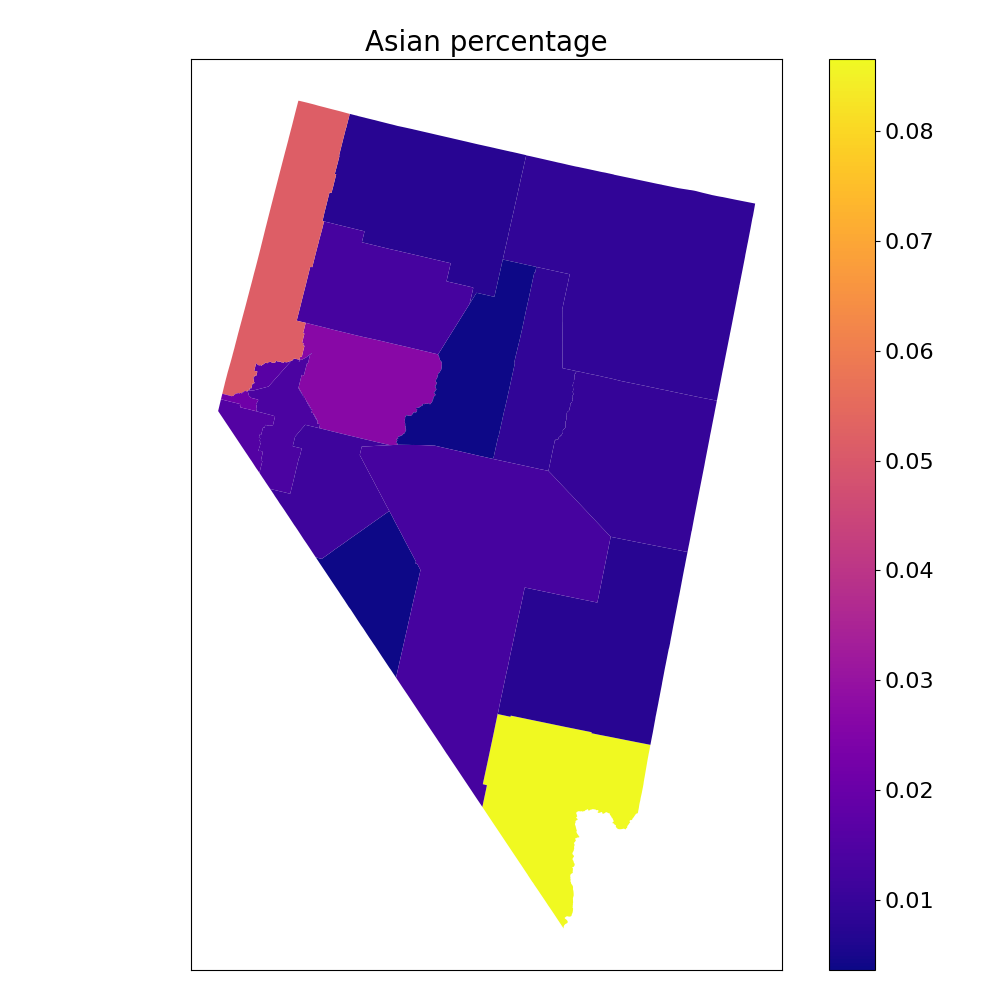}&\includegraphics[width=0.22\textwidth]{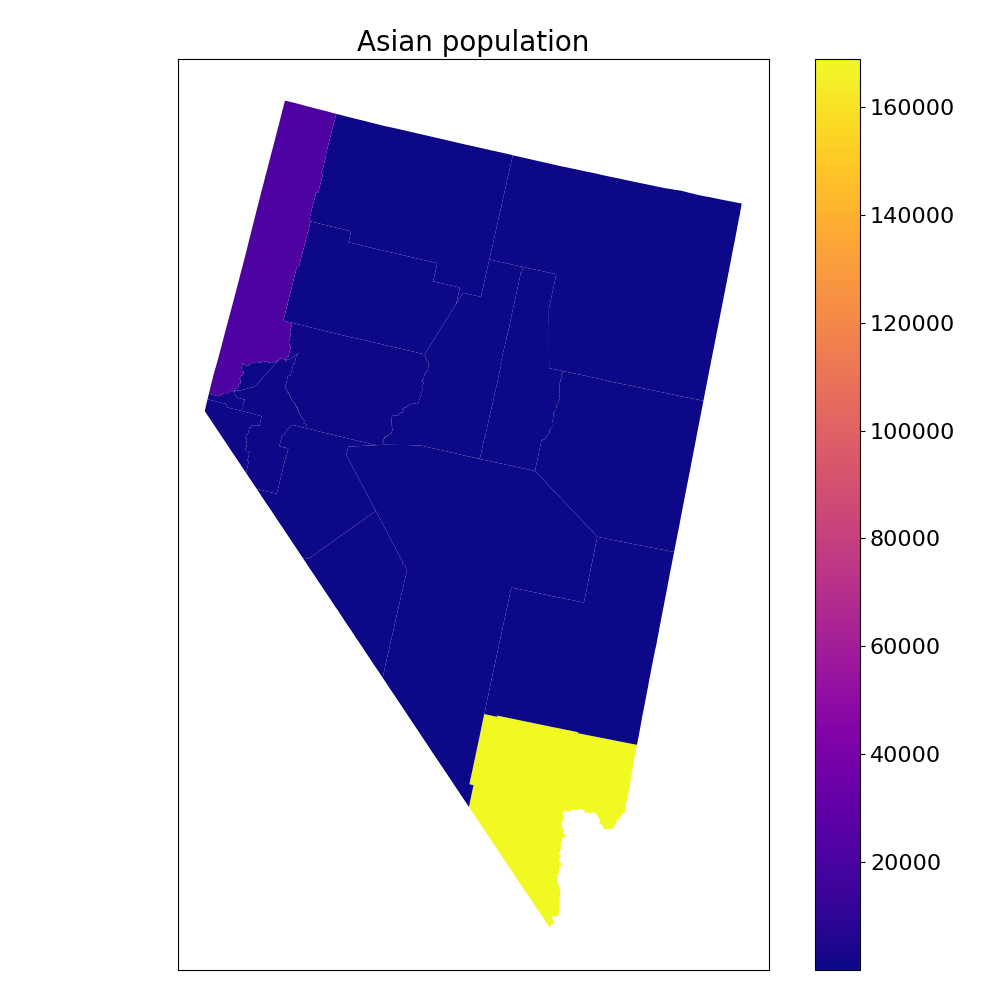}\\
        
        \rotatebox{90}{\hspace{20mm}Hispanic}&\includegraphics[width=0.22\textwidth]{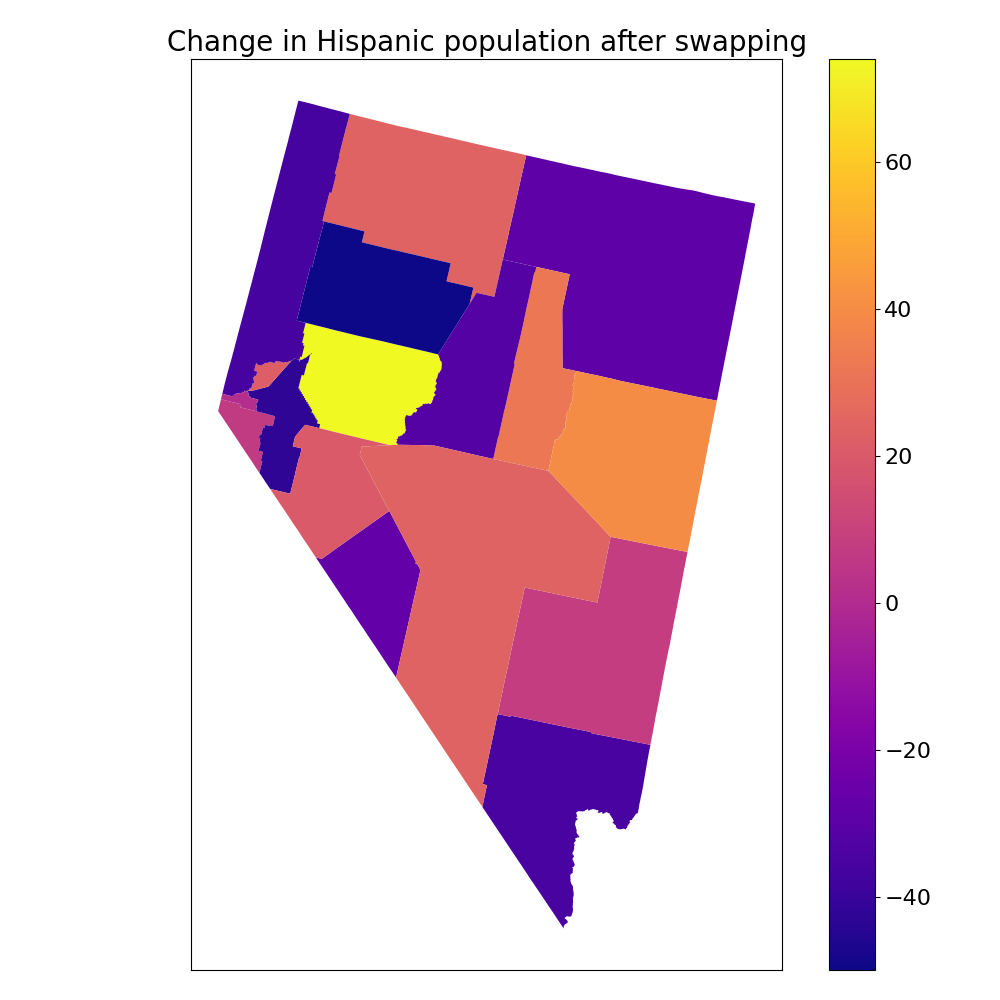}
                                             &\includegraphics[width=0.22\textwidth]{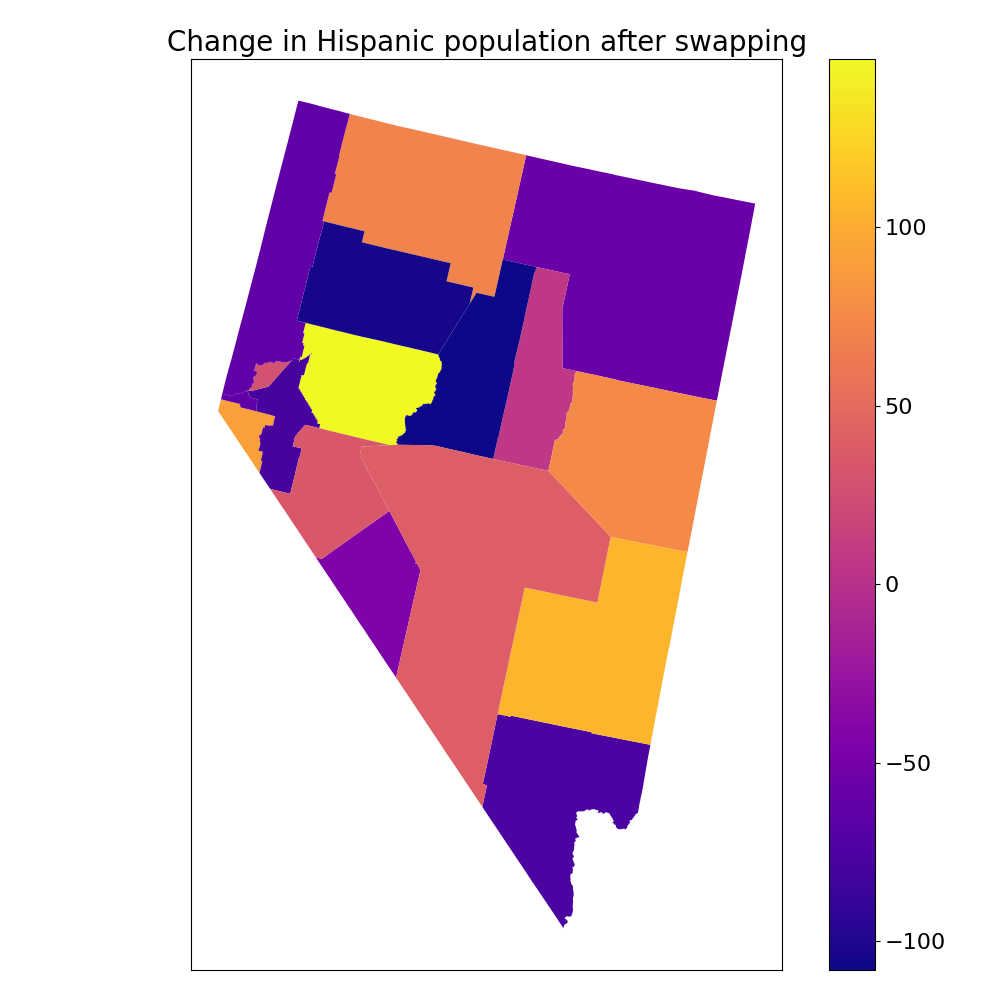}&\includegraphics[width=0.22\textwidth]{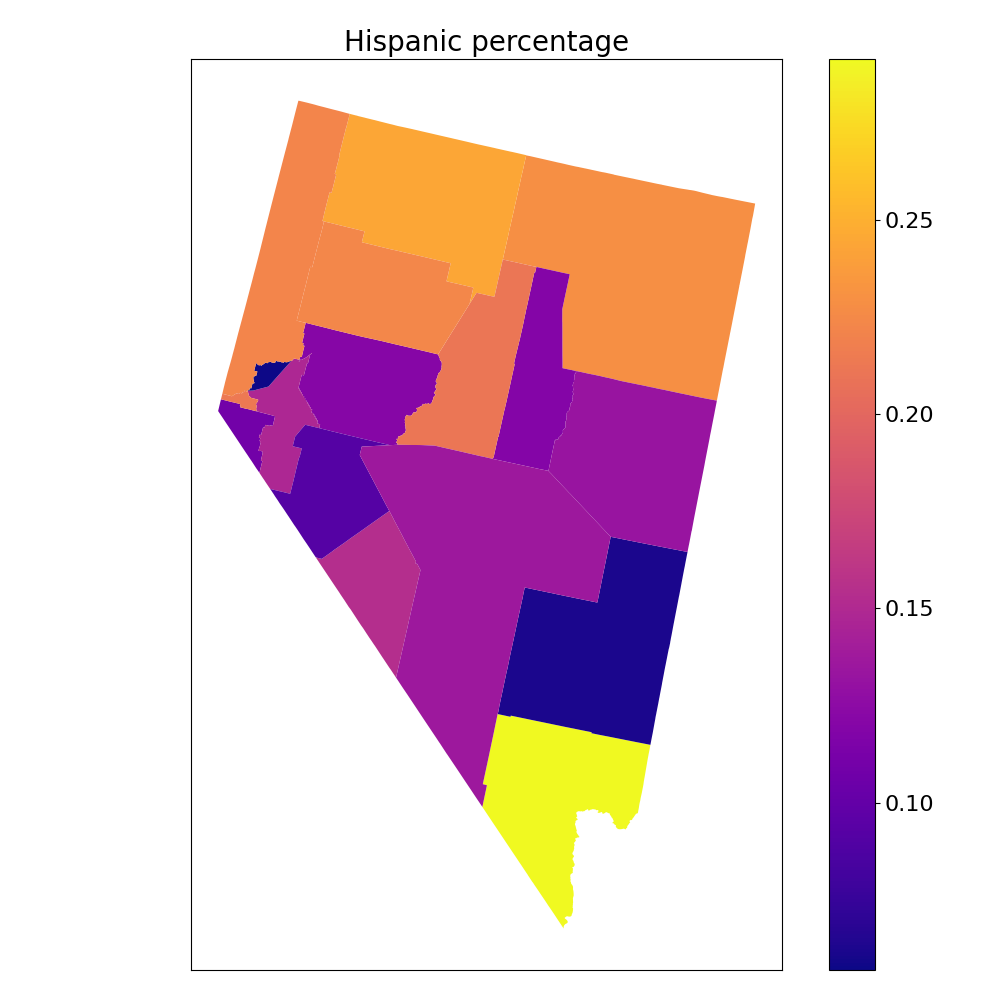}&\includegraphics[width=0.22\textwidth]{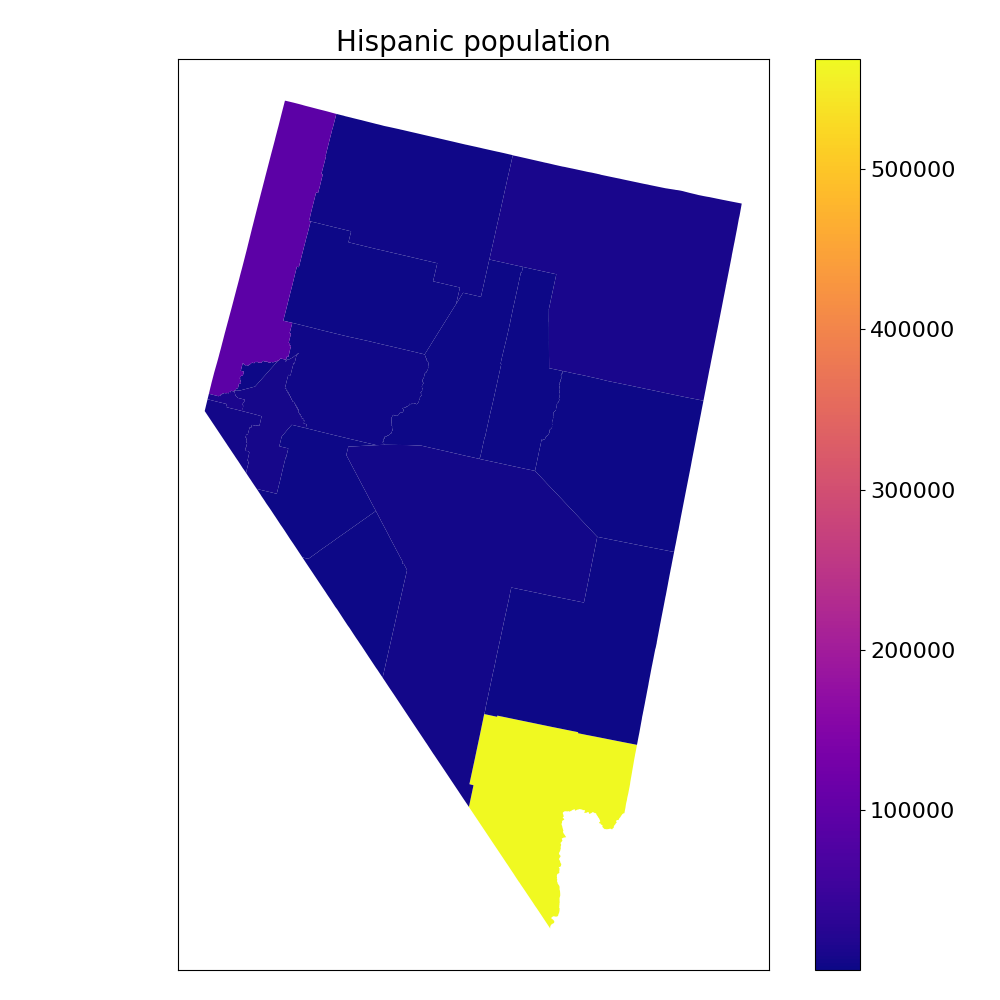}\\
        &2\% Swap Rate Effect & 10\% Swap Rate Effect & Percentage & Population
    \end{tabular}
    \caption{The effect of swapping on White and Hispanic population in Nevada, with the White/Hispanic percentage and White/Hispanic population shown for reference.}
    \label{fig:nv_full_maps}
\end{figure}

\begin{figure}
    \centering
    \begin{tabular}{lcccc}
        \rotatebox{90}{\hspace{20mm}White}&\includegraphics[width=0.22\textwidth]{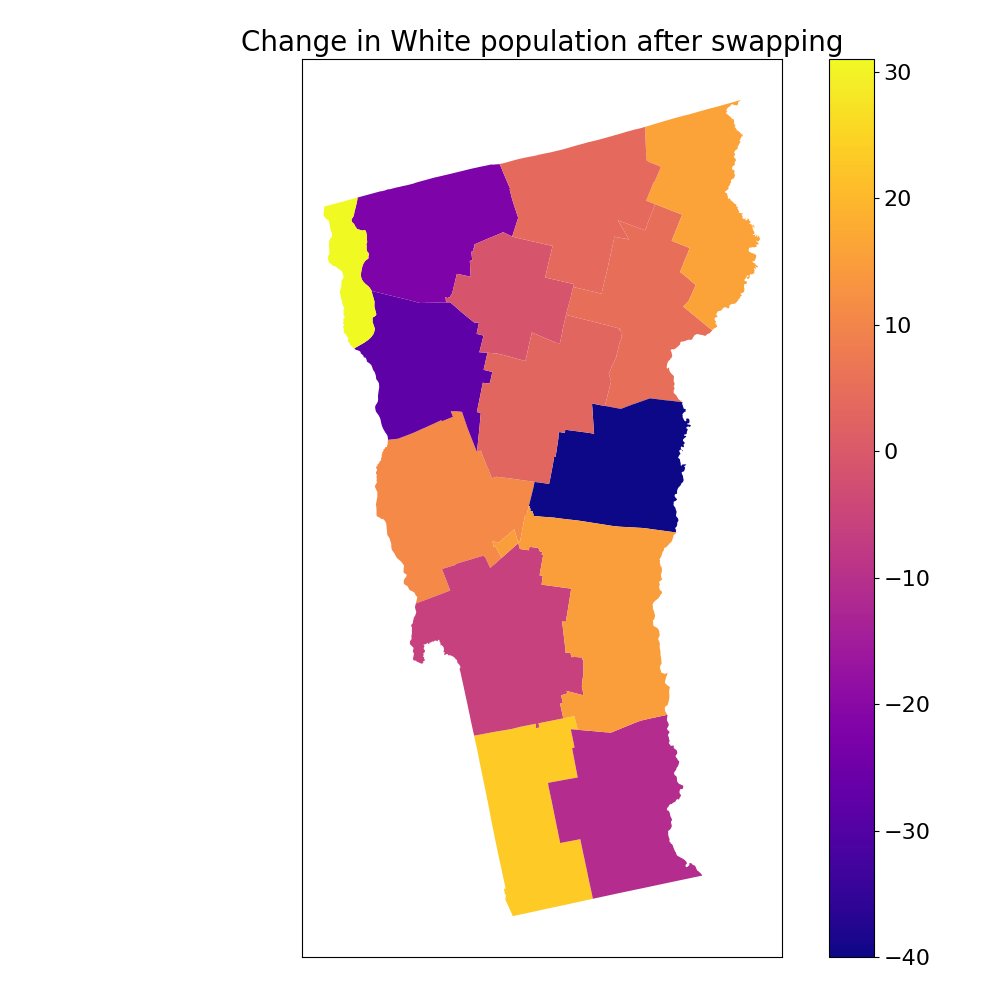}
                                          &\includegraphics[width=0.22\textwidth]{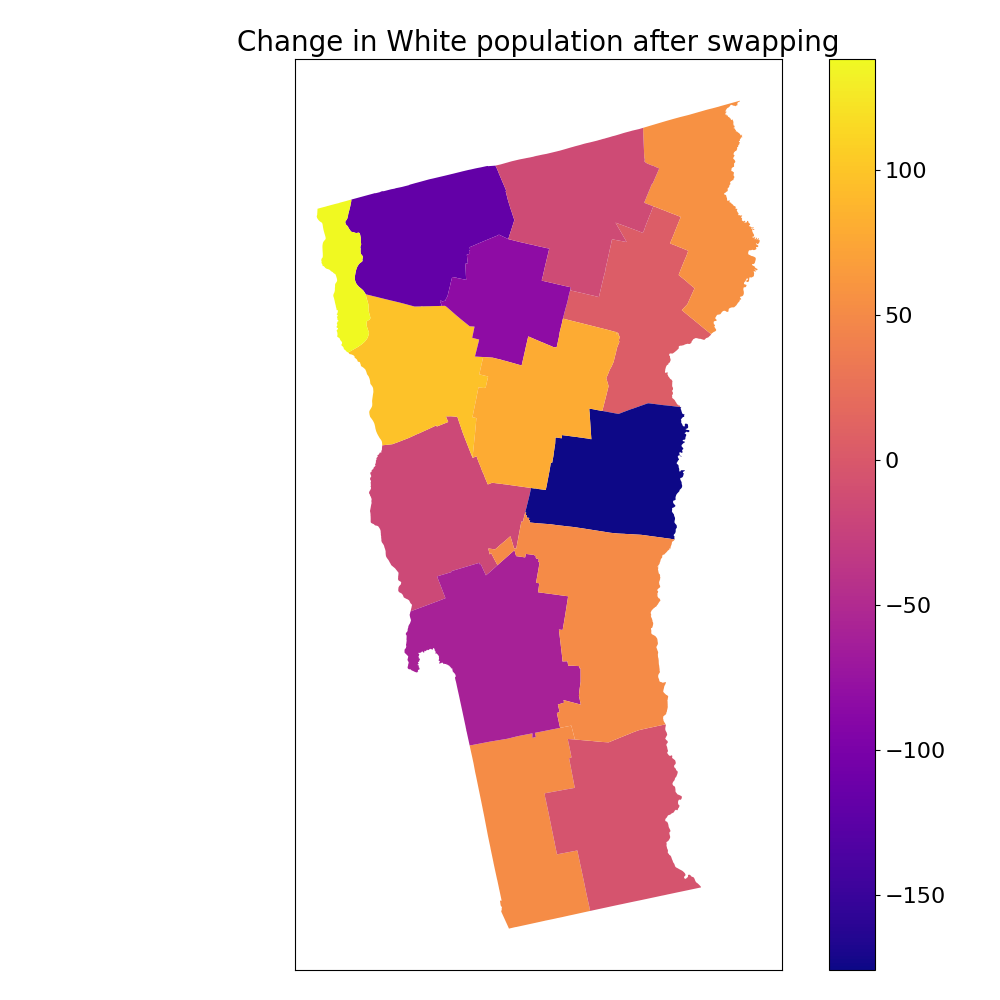}&\includegraphics[width=0.22\textwidth]{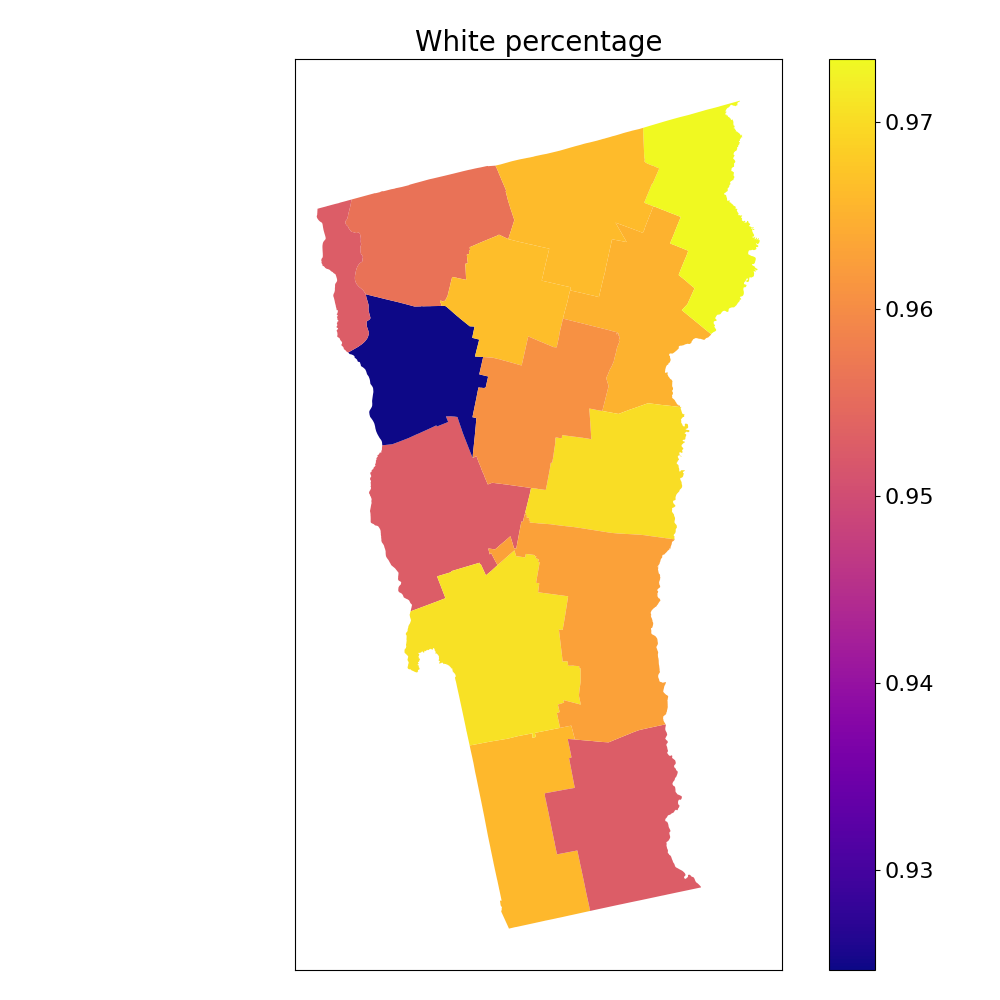}&\includegraphics[width=0.22\textwidth]{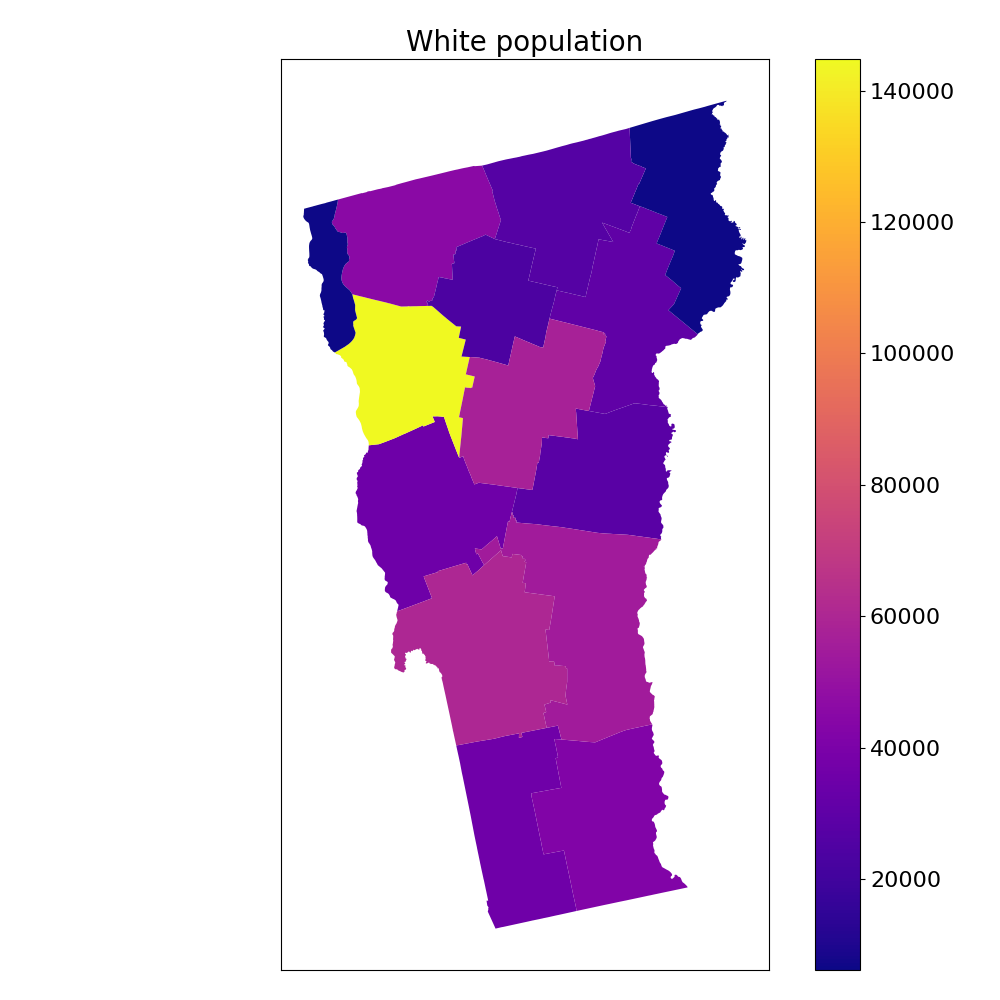}\\

        \rotatebox{90}{\hspace{20mm}Black}&\includegraphics[width=0.22\textwidth]{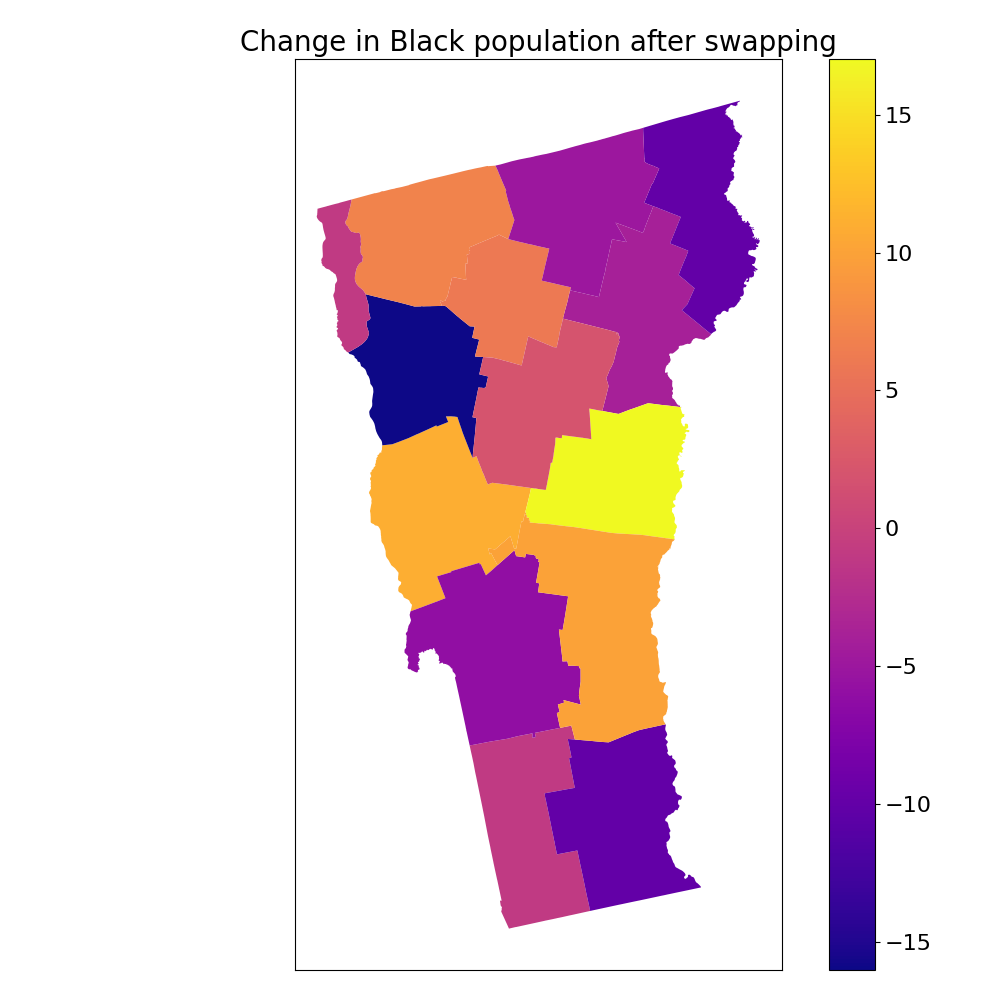}
                                          &\includegraphics[width=0.22\textwidth]{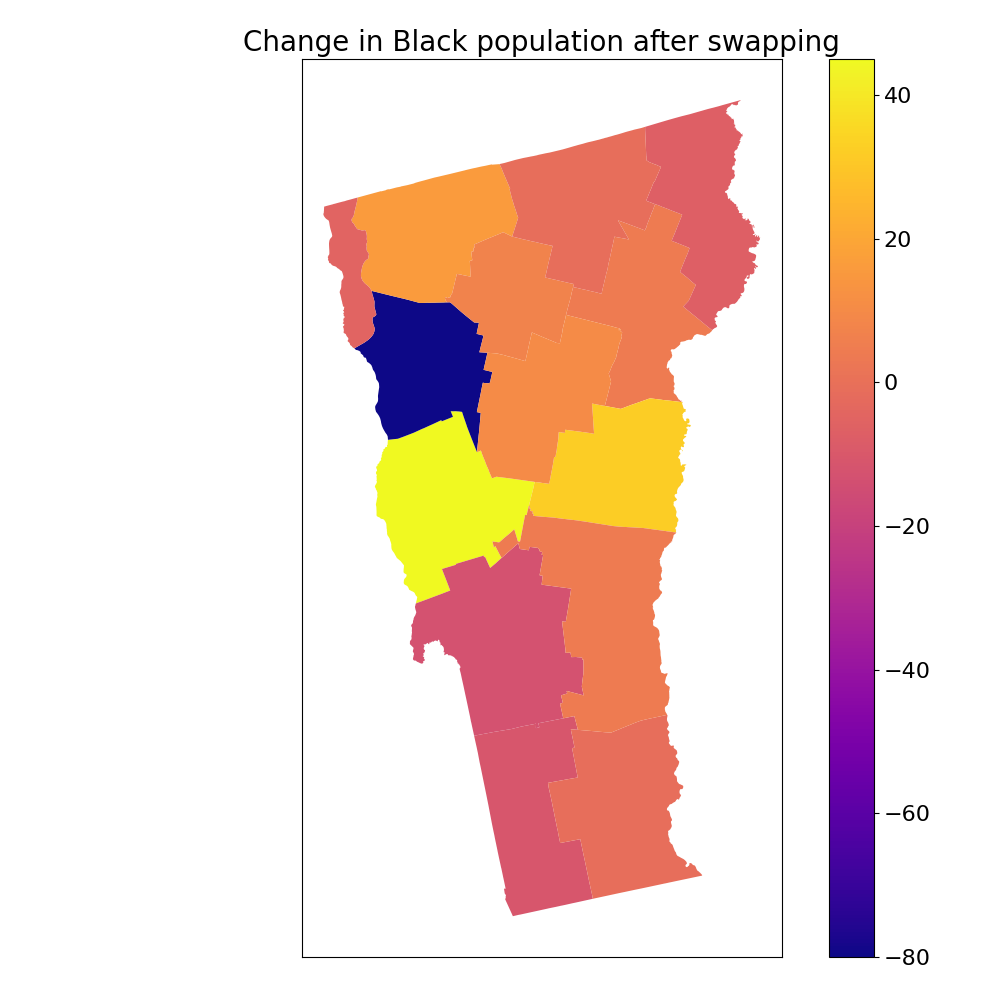}&\includegraphics[width=0.22\textwidth]{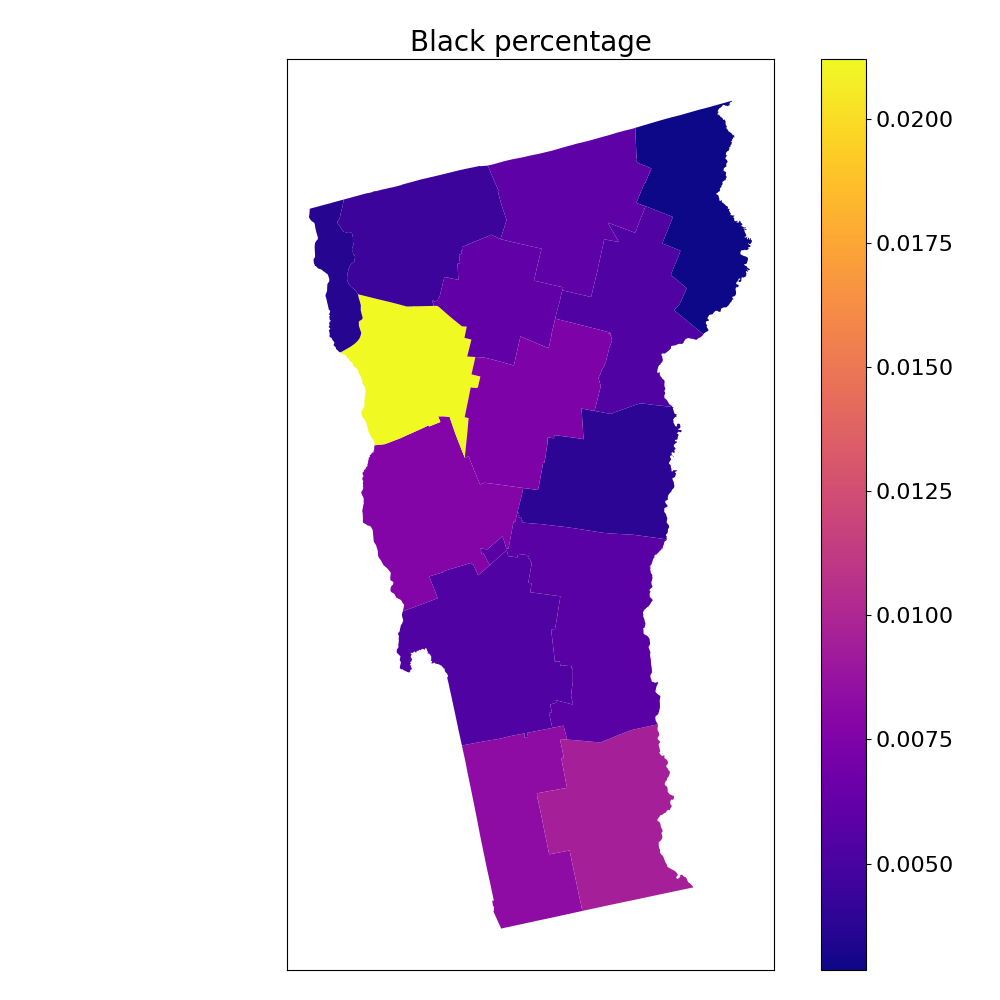}&\includegraphics[width=0.22\textwidth]{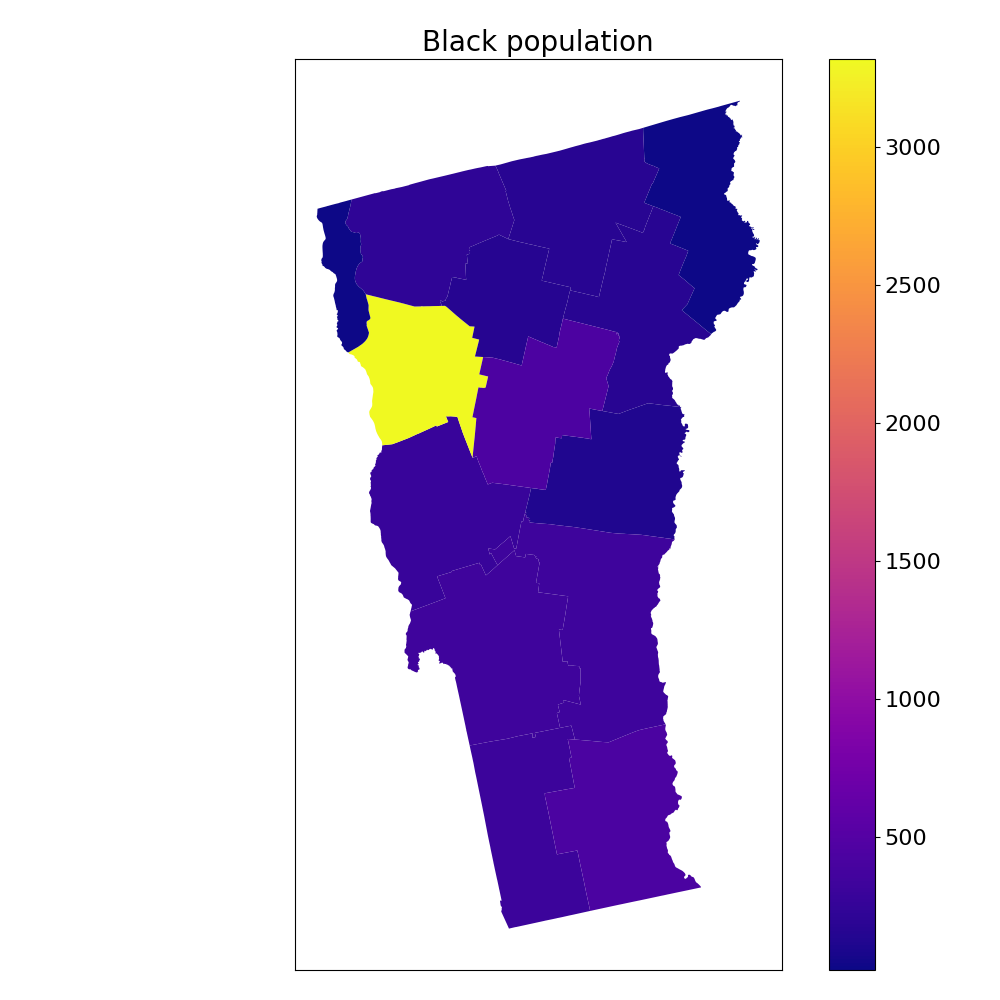}\\

        \rotatebox{90}{\hspace{20mm}Asian}&\includegraphics[width=0.22\textwidth]{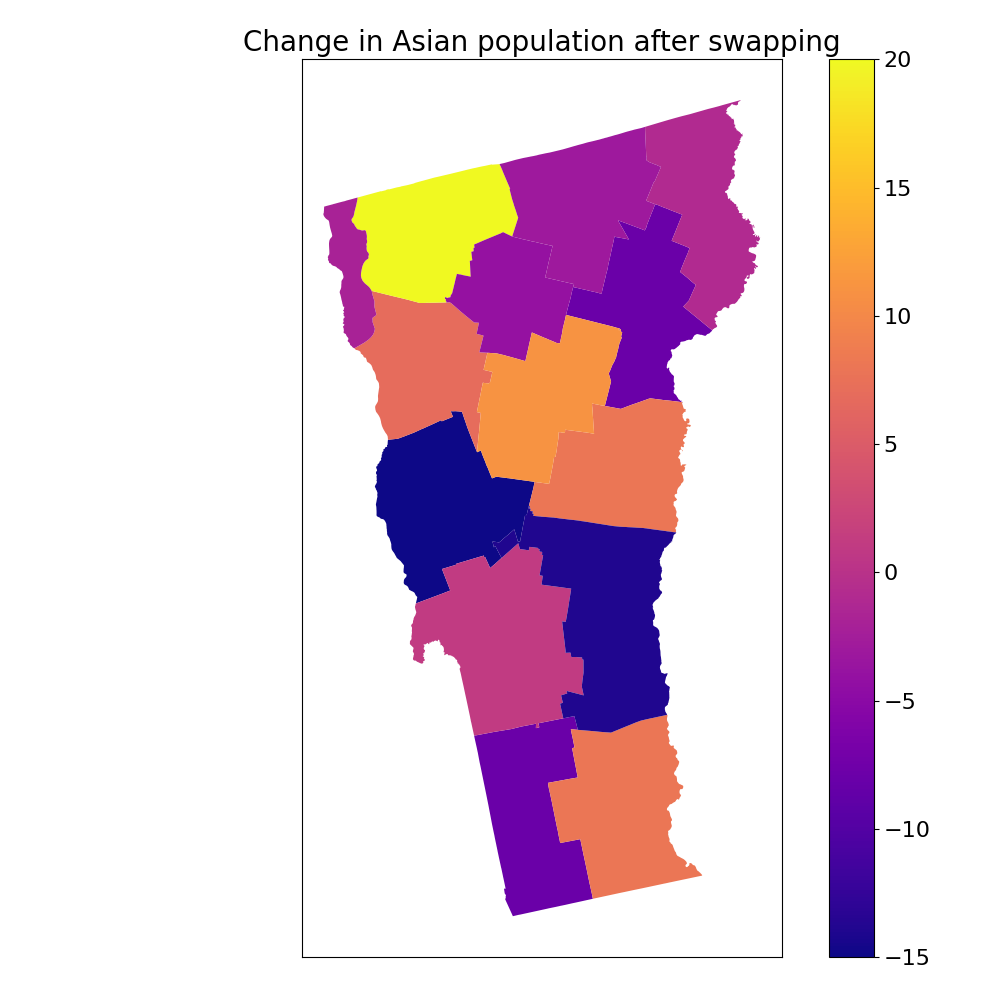}
                                          &\includegraphics[width=0.22\textwidth]{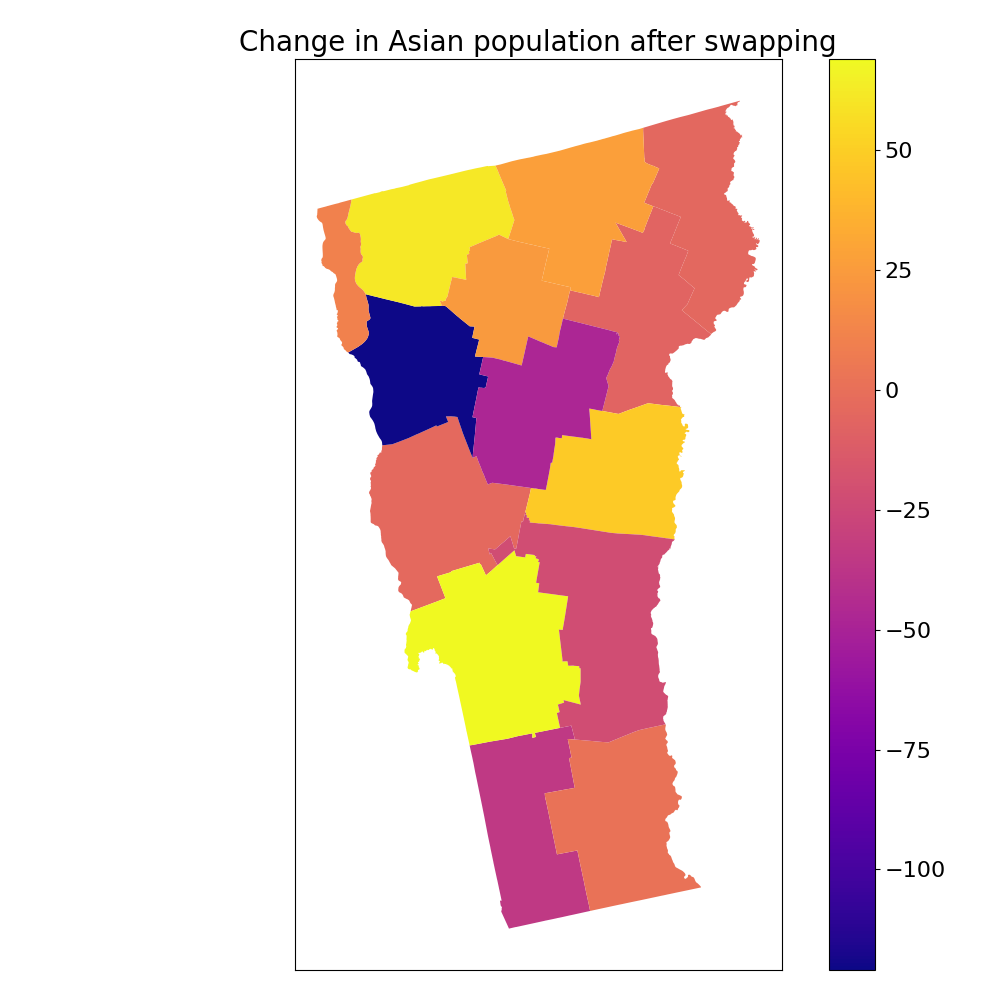}&\includegraphics[width=0.22\textwidth]{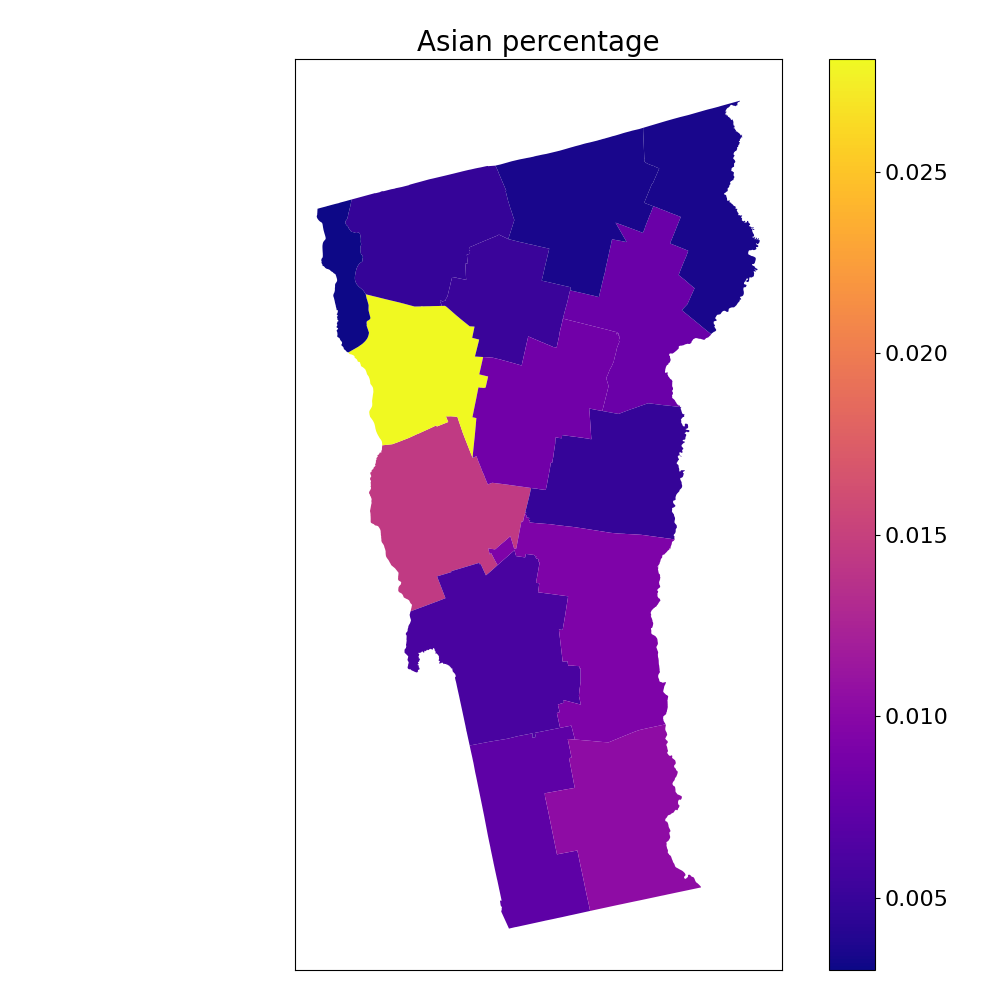}&\includegraphics[width=0.22\textwidth]{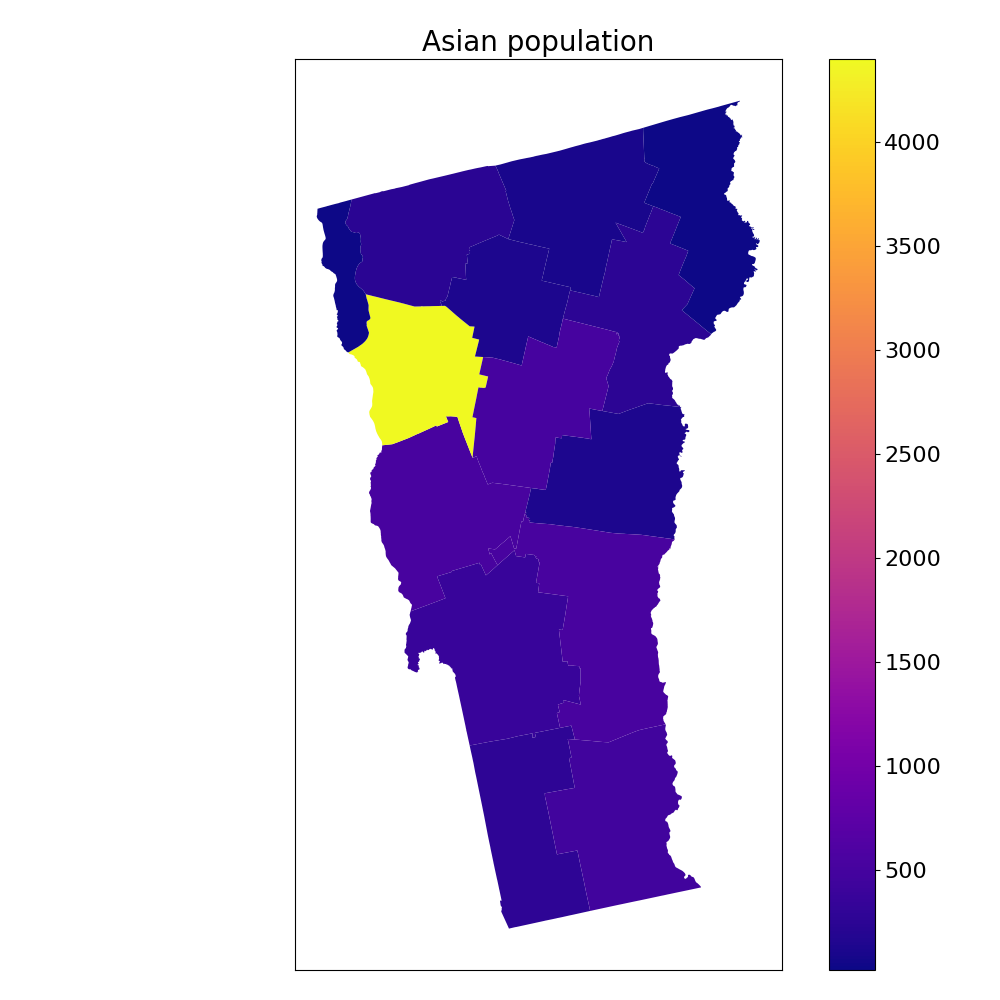}\\
        
        \rotatebox{90}{\hspace{20mm}Hispanic}&\includegraphics[width=0.22\textwidth]{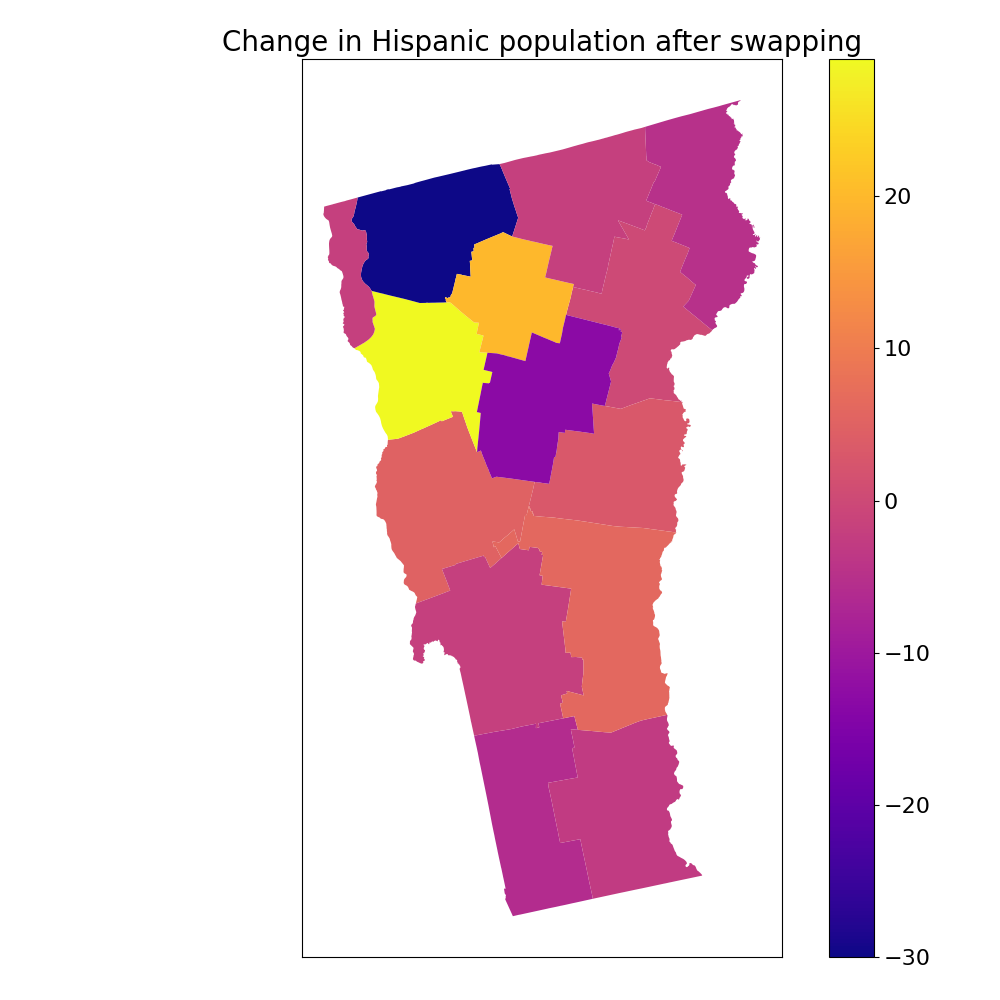}
                                             &\includegraphics[width=0.22\textwidth]{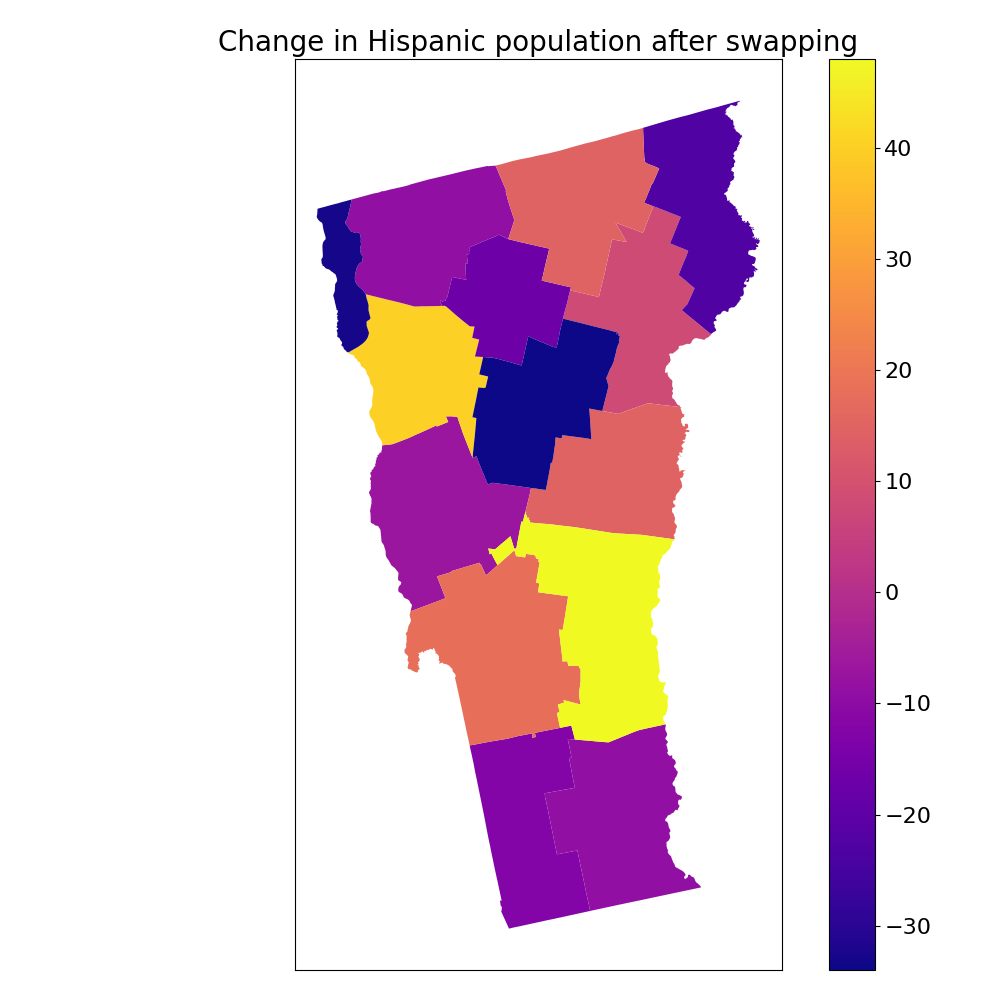}&\includegraphics[width=0.22\textwidth]{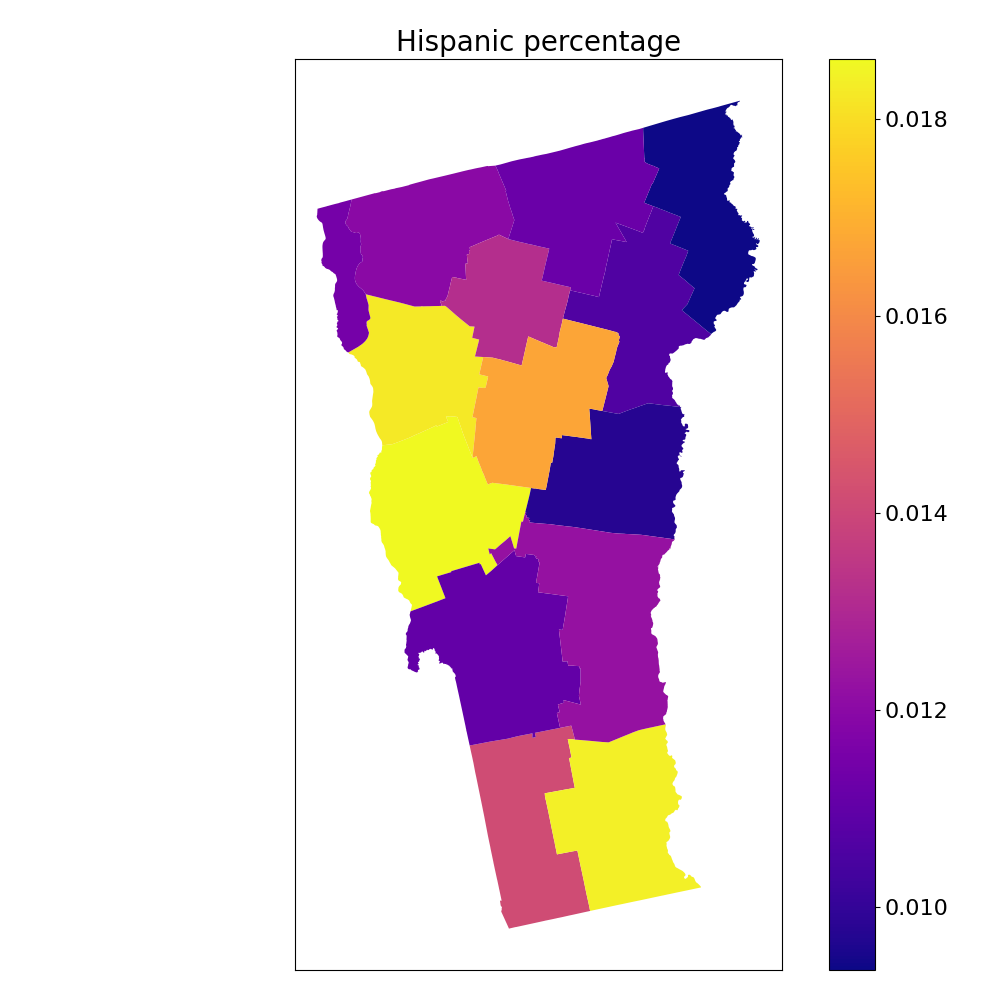}&\includegraphics[width=0.22\textwidth]{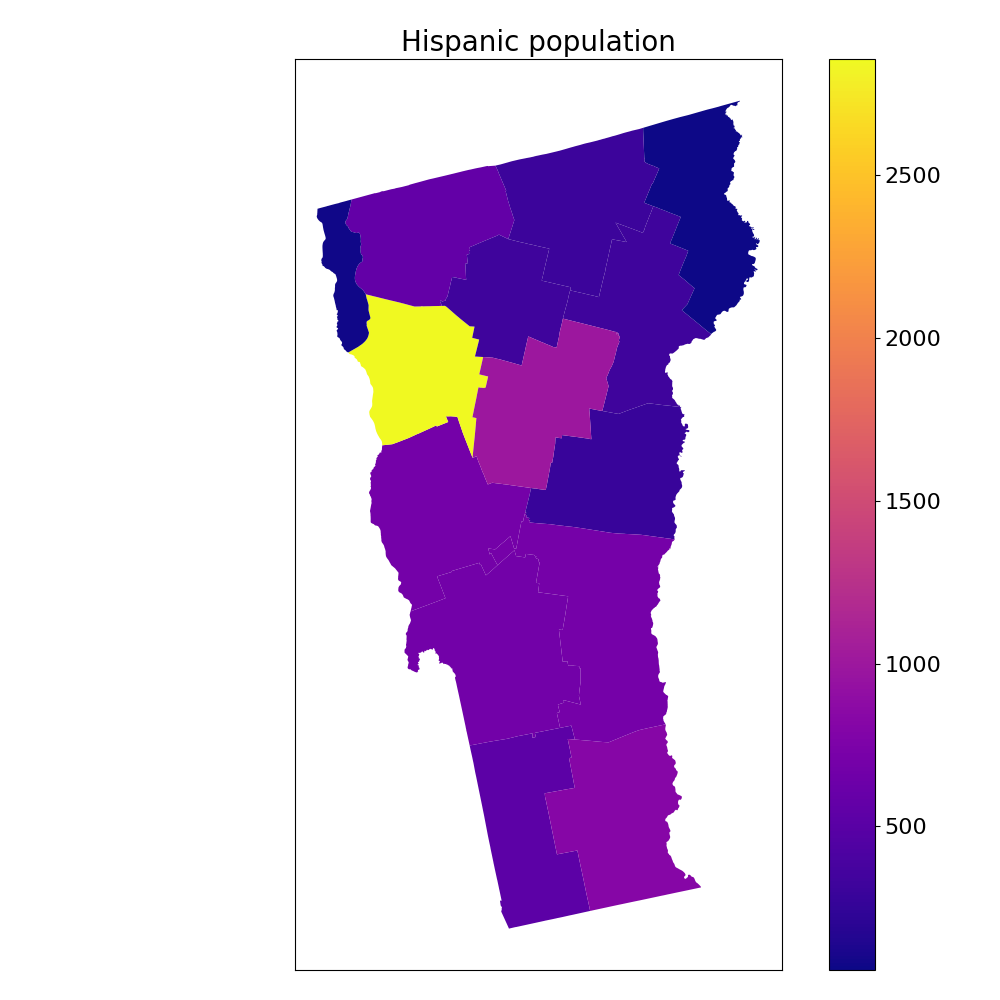}\\
        &2\% Swap Rate Effect & 10\% Swap Rate Effect & Percentage & Population
    \end{tabular}
    \caption{The effect of swapping on White and Hispanic population in Vermont, with the White/Hispanic percentage and White/Hispanic population shown for reference.}
    \label{fig:vt_full_maps}
\end{figure}

\begin{table}
    \centering
    Target Distribution for a 2\% Swap Rate
    
    \begin{tabular}{|r|c|c|c|c|c|c|c|c|}
    \hline
                  & W      & B      & AI/AN & AS    & H/PI  & OTH    & 2+    & Multiple Races \\ \hline
    \% overall    & 70.8\% & 23.9\% & 0.3\% & 0.7\% & 0.0\% & 1.1\%  & 0.6\% & 2.5\%          \\ \hline
    \% of targets & 21.2\% & 10.4\% & 4.3\% & 5.4\% & 0.6\% & 10.3\% & 8.8\% & 39.0\%         \\ \hline
    \end{tabular}

    \caption{This table shows the distribution of the race among targeted households, compared to the distribution of the race of all households.}
    \label{tab:target_distribution_2pct}
\end{table}

\begin{table}

\setlength\doublerulesep{5mm} 
    \centering
    Partner Distribution for a 2\% Swap Rate
    \begin{tabular}{|r|c|c|c|c|c|c|c|c|}
    \hline
    \diagbox[width=\dimexpr \textwidth/8+5\tabcolsep\relax, height=1cm]{Target}{Partner}& W      & B      & AI/AN & AS    & H/PI  & OTH   & 2+    & Multiple Races \\ \hline
    W              & 65.0 & 29.1 & 0.1 & 1.4 & 0.0 & 3.5 & 0.9 & 10.4         \\ \hline
    B              & 42.9 & 50.9 & 0.1 & 1.6 & 0.1 & 3.6 & 0.8 & 8.5          \\ \hline
    AI/AN          & 67.5 & 28.5 & 0.7 & 0.1 & 0.0 & 1.5 & 1.6 & 3.2          \\ \hline
    AS             & 67.0 & 26.5 & 0.0 & 2.5 & 0.0 & 3.2 & 0.7 & 3.1          \\ \hline
    H/PI           & 66.0 & 28.9 & 0.0 & 1.0 & 0.0 & 3.6 & 0.5 & 2.1          \\ \hline
    OTH            & 58.3 & 32.1 & 0.5 & 0.5 & 0.0 & 7.7 & 0.8 & 5.1          \\ \hline
    2+             & 65.5 & 30.5 & 0.5 & 0.5 & 0.0 & 2.2 & 0.8 & 3.7          \\ \hline
    Multiple Races & 66.1 & 29.4 & 0.5 & 0.9 & 0.1 & 2.5 & 0.6 & 7.3          \\ \hline\hline
    \% overall     & 70.8 & 23.9 & 0.3 & 0.7 & 0.0 & 1.1 & 0.6 & 2.5          \\ \hline
    \% of partners & 64.3 & 26.6 & 0.3 & 1.1 & 0.0 & 2.5 & 0.6 & 4.4          \\ \hline
    \end{tabular}
    \caption{For a swap target of a particular race, this table shows the distribution of the race of the partner household. The row labels are the race of the target household. Each row shows the distribution of the race of the partner household for that particular type of target household (each row is normalized to sum to 100). For reference, below is distribution of the household races among all households and among the households selected as partners.}
    \label{tab:partner_distribution_2pct}
\end{table}

\begin{table}
\centering
\begin{tabular}{|c|c|c|c|c|c|c|c|c|}
\hline
Household Size & W    & B    & AI/AN & AS   & H/PI & OTH  & 2+   & Multiple Races \\ \hline
1              & 26\% & 29\% & 40\%  & 23\% & 30\% & 15\% & 44\% & 0\%            \\ \hline
2              & 39\% & 27\% & 24\%  & 25\% & 24\% & 13\% & 23\% & 27\%           \\ \hline
3              & 14\% & 17\% & 14\%  & 18\% & 8\%  & 13\% & 12\% & 23\%           \\ \hline
4              & 13\% & 15\% & 13\%  & 19\% & 12\% & 17\% & 10\% & 21\%           \\ \hline
5              & 5\%  & 6\%  & 5\%   & 7\%  & 20\% & 15\% & 5\%  & 13\%           \\ \hline
6              & 1\%  & 2\%  & 2\%   & 3\%  & 3\%  & 10\% & 2\%  & 8\%            \\ \hline
7              & 0\%  & 1\%  & 1\%   & 1\%  & 2\%  & 4\%  & 1\%  & 4\%            \\ \hline
8              & 0\%  & 0\%  & 0\%   & 1\%  & 0\%  & 0\%  & 2\%  & 2\%            \\ \hline
9              & 0\%  & 0\%  & 0\%   & 0\%  & 2\%  & 14\% & 0\%  & 1\%            \\ \hline
10             & 0\%  & 0\%  & 0\%   & 1\%  & 0\%  & 0\%  & 0\%  & 1\%            \\ \hline
11             & 0\%  & 0\%  & 0\%   & 2\%  & 0\%  & 0\%  & 0\%  & 1\%            \\ \hline
12             & 0\%  & 0\%  & 0\%   & 0\%  & 0\%  & 0\%  & 0\%  & 1\%            \\ \hline
13             & 0\%  & 0\%  & 0\%   & 0\%  & 0\%  & 0\%  & 0\%  & 0\%            \\ \hline
14             & 0\%  & 1\%  & 0\%   & 0\%  & 0\%  & 0\%  & 0\%  & 0\%            \\ \hline
15             & 0\%  & 0\%  & 0\%   & 0\%  & 0\%  & 0\%  & 0\%  & 0\%            \\ \hline
16             & 0\%  & 0\%  & 0\%   & 0\%  & 0\%  & 0\%  & 0\%  & 0\%            \\ \hline
36             & 0\%  & 0\%  & 0\%   & 0\%  & 0\%  & 0\%  & 0\%  & 0\%            \\ \hline
\end{tabular}
\caption{The distribution over household sizes for each race (where houses with multiple races present are all in the ``Multiple Races'' column). This illustrates a correlation between household size and majority race.}
\label{tab:household_size_race_corr}
\end{table}

\begin{table}
\centering
\begin{tabular}{|l|l|l|l|}
\hline
State & \begin{tabular}[c]{@{}l@{}}Average tract entropy\\ before swapping\end{tabular} & \begin{tabular}[c]{@{}l@{}}Average tract entropy\\ after 2\% swapping\end{tabular} & \begin{tabular}[c]{@{}l@{}}Average tract entropy\\ after 10\% swapping\end{tabular} \\ \hline
AL    & 0.598                                                                           & 0.603                                                                              & 0.624                                                                               \\ \hline
WI    & 0.431                                                                           & 0.434                                                                              & 0.447                                                                               \\ \hline
TX    & 0.807                                                                           & 0.810                                                                              & 0.819                                                                               \\ \hline
NV    & 0.997                                                                           & 0.999                                                                              & 1.006                                                                               \\ \hline
VT    & 0.232                                                                           & 0.233                                                                              & 0.241                                                                               \\ \hline
\end{tabular}
\caption{Average racial entropy at the tract level before and after swapping.}
\label{tab:entropies}
\end{table}

\begin{figure}
    \centering
    \begin{tabular}{ccc}
        \rotatebox{90}{\hspace{17mm}2\% Swap Rate}&\includegraphics[scale=0.2335]{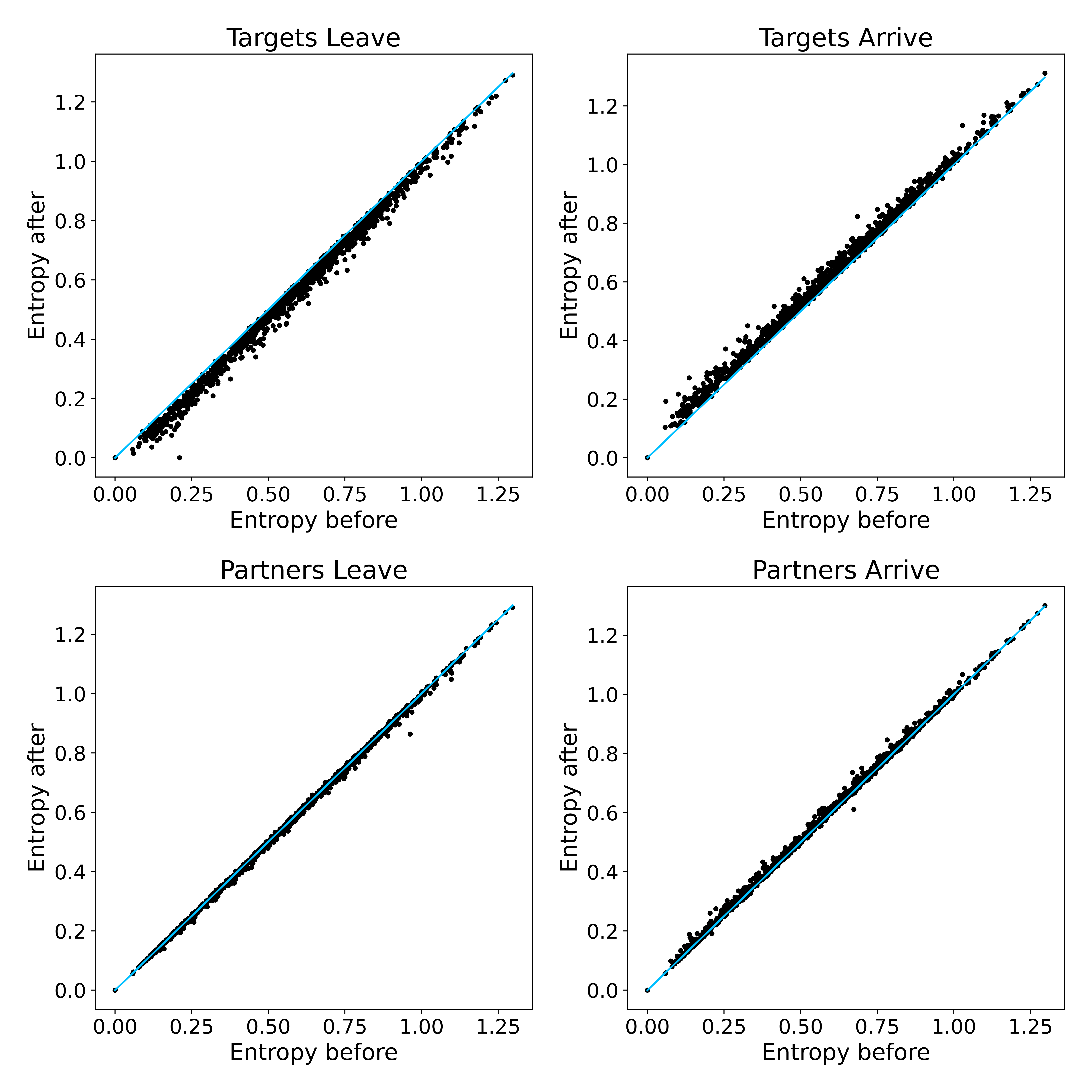}&\includegraphics[scale=0.28]{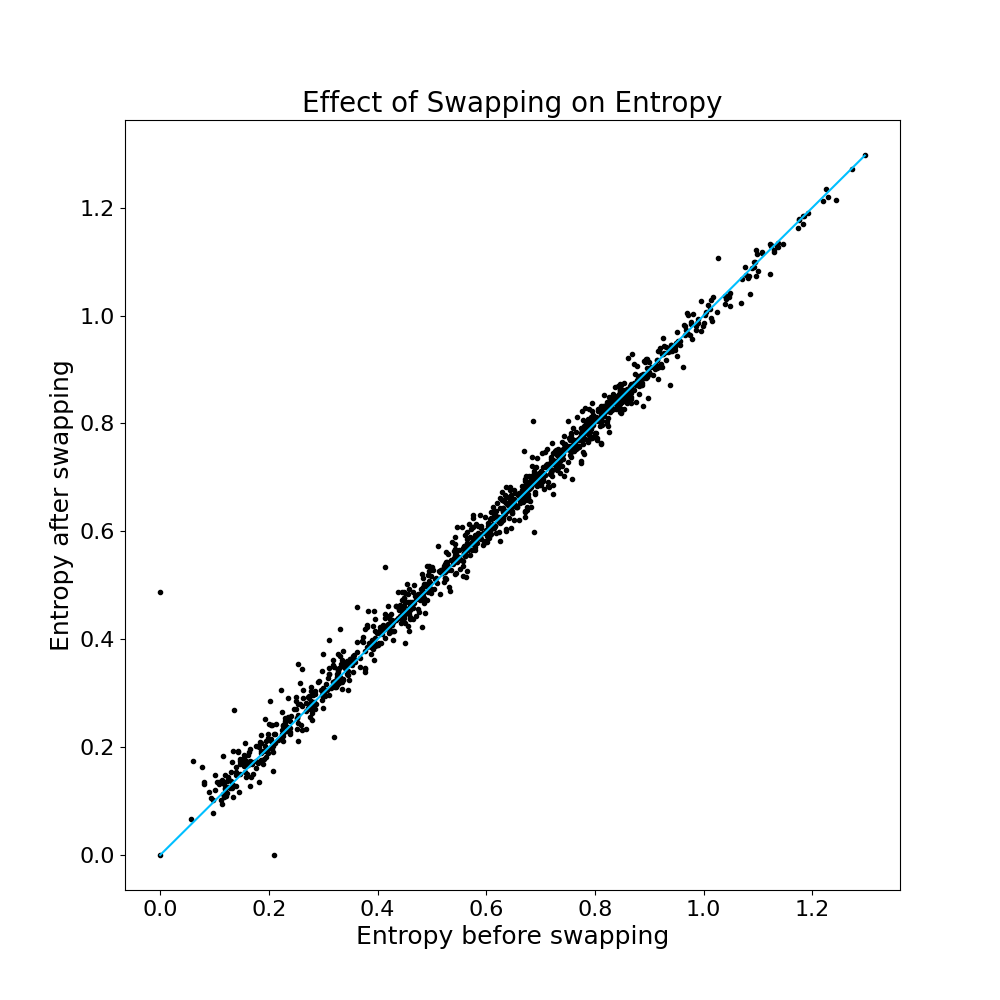}\\
        \rotatebox{90}{\hspace{17mm}10\% Swap Rate}&\includegraphics[scale=0.2335]{AL_figures/0.1_panels_bw.png}&\includegraphics[scale=0.28]{AL_figures/entropy_0.1_bw.png}
    \end{tabular}
    \caption{The effect of swapping on racial entropy at the tract level in Alabama.}
    \label{fig:entropy_al}
\end{figure}
\begin{figure}
    \centering
    \begin{tabular}{ccc}
        \rotatebox{90}{\hspace{17mm}2\% Swap Rate}&\includegraphics[scale=0.2335]{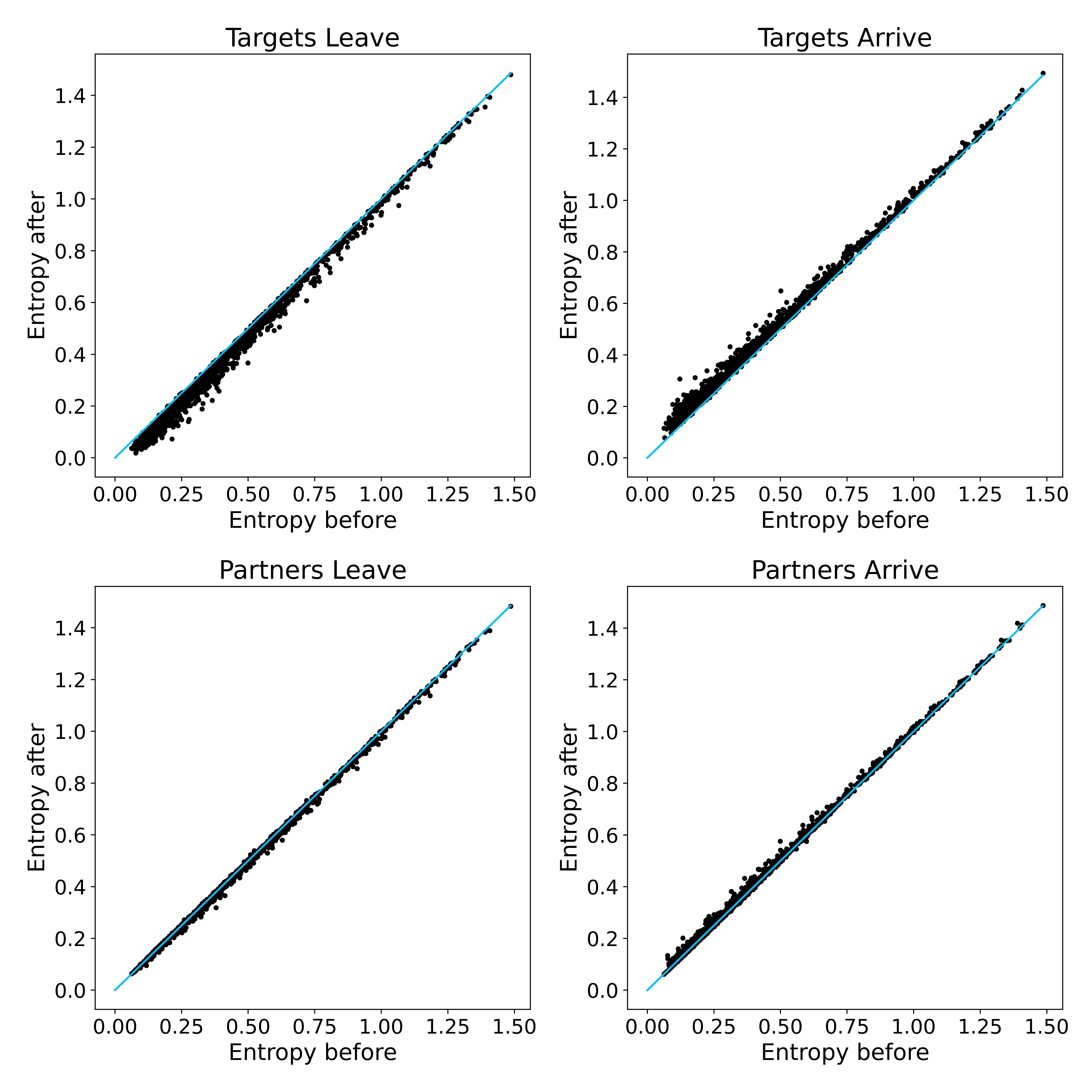}&\includegraphics[scale=0.28]{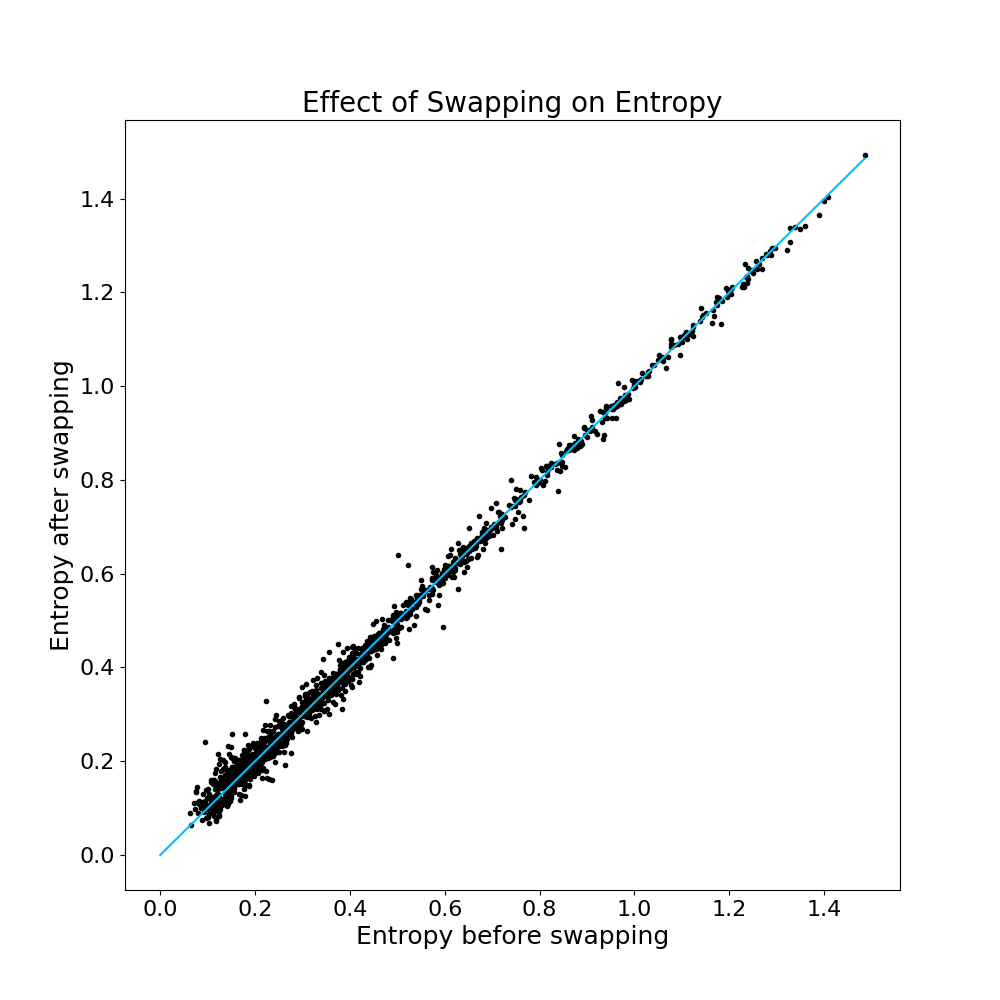}\\
        \rotatebox{90}{\hspace{17mm}10\% Swap Rate}&\includegraphics[scale=0.2335]{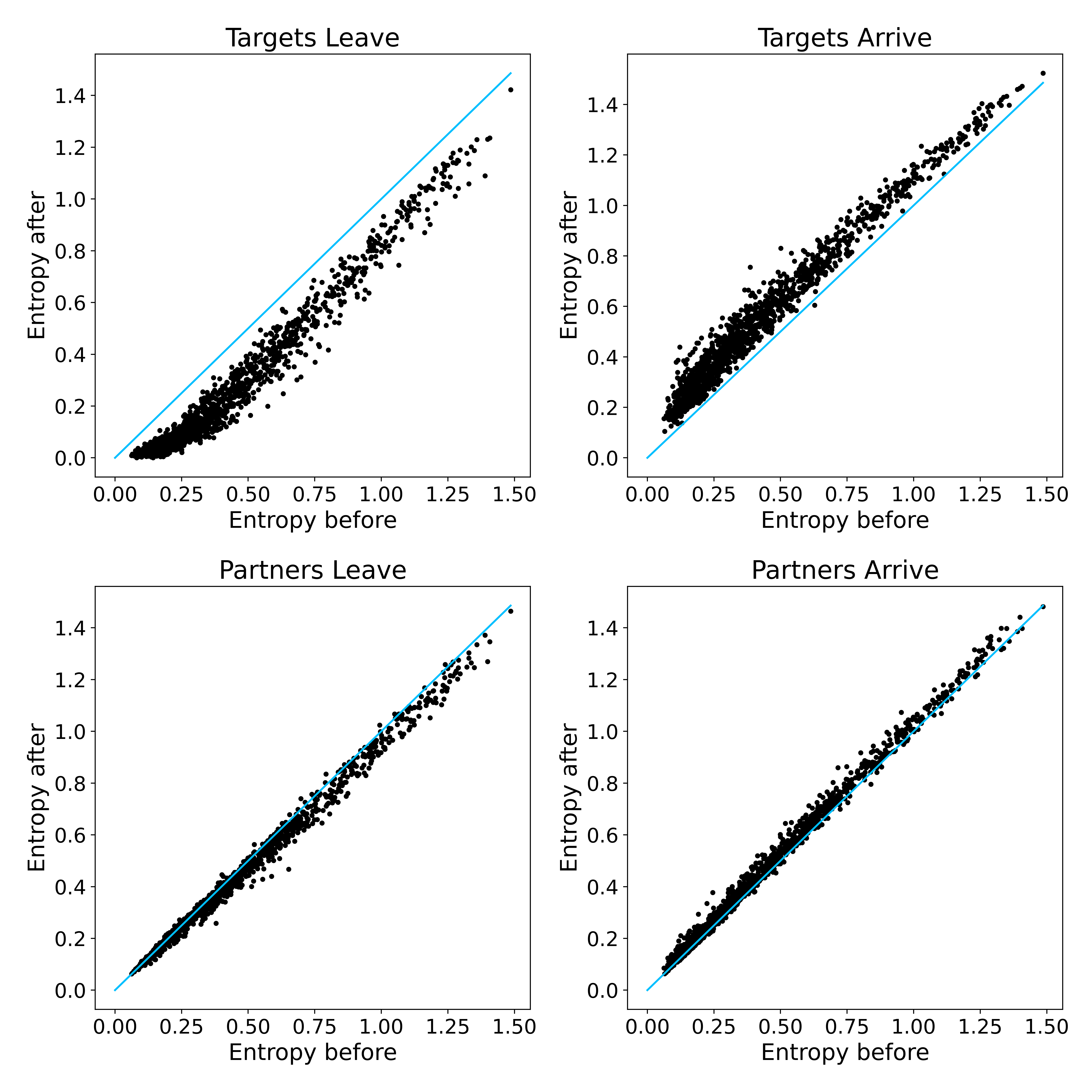}&\includegraphics[scale=0.28]{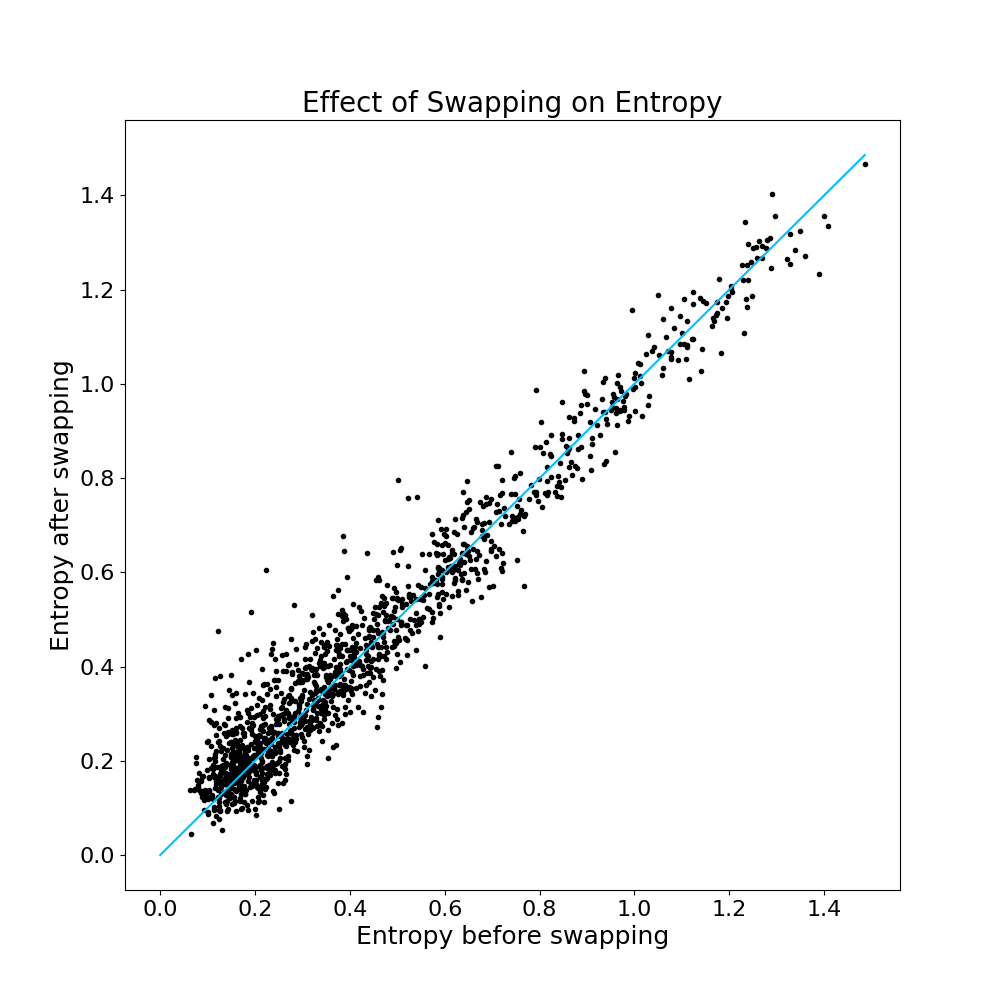}
    \end{tabular}
    \caption{The effect of swapping on racial entropy at the tract level in Wisconsin.}
    \label{fig:entropy_wi}
\end{figure}
\begin{figure}
    \centering
    \begin{tabular}{ccc}
        \rotatebox{90}{\hspace{17mm}2\% Swap Rate}&\includegraphics[scale=0.2335]{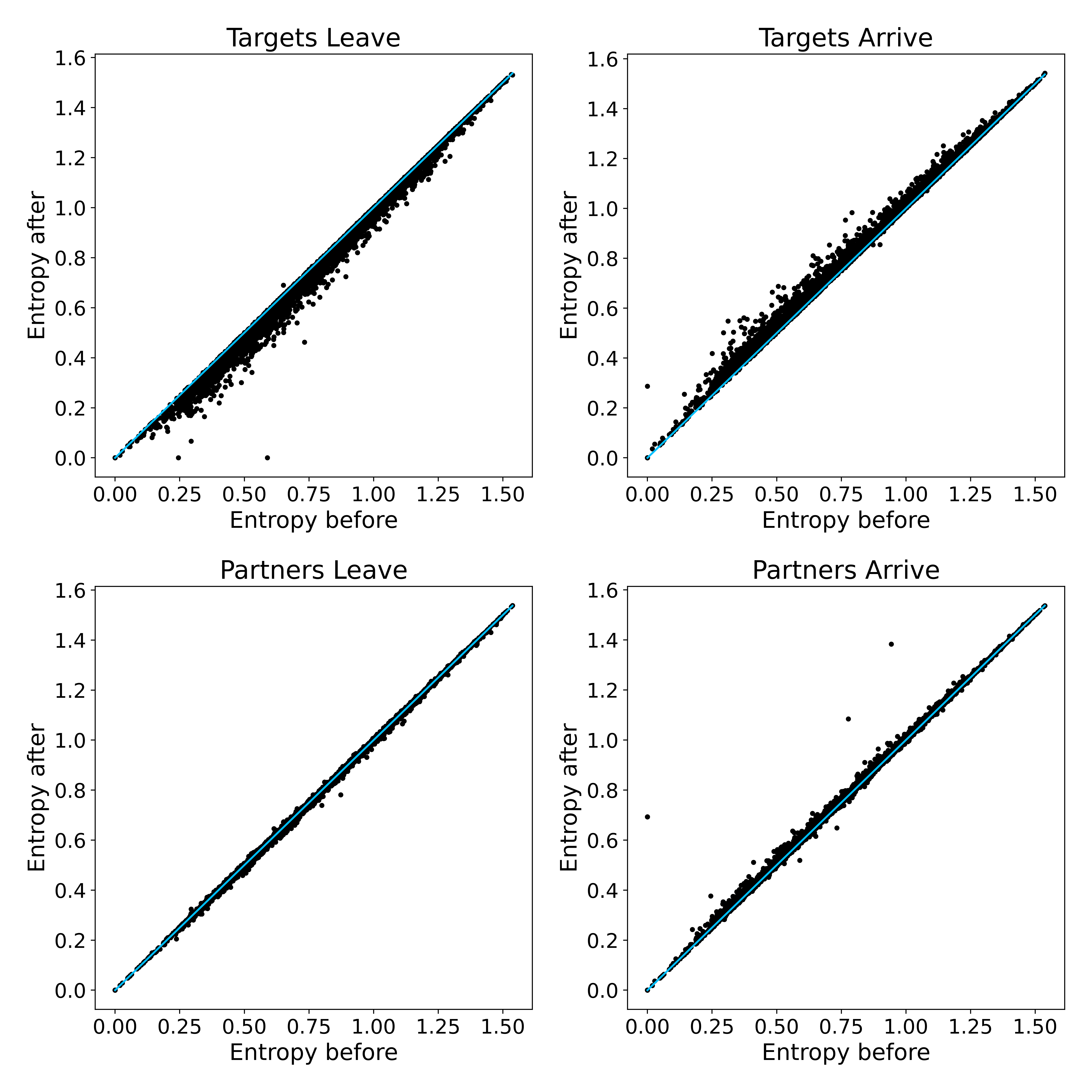}&\includegraphics[scale=0.28]{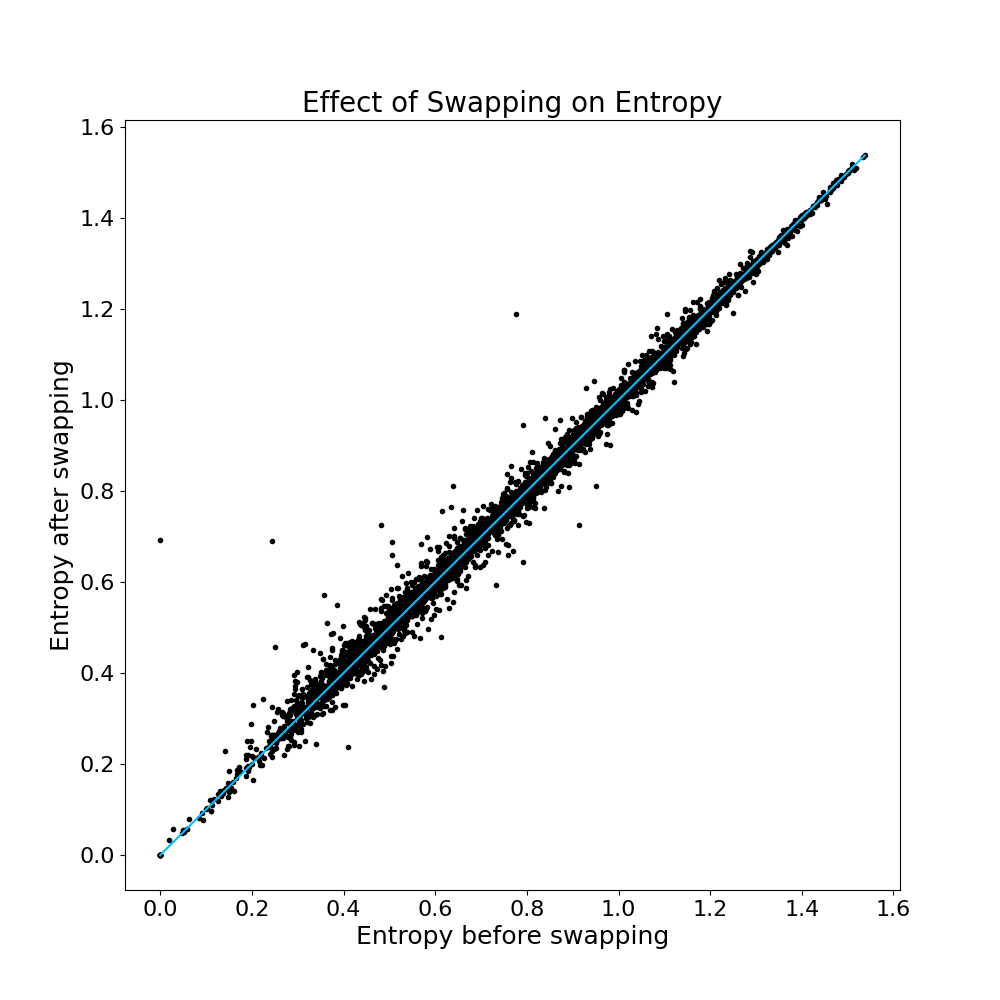}\\
        \rotatebox{90}{\hspace{17mm}10\% Swap Rate}&\includegraphics[scale=0.2335]{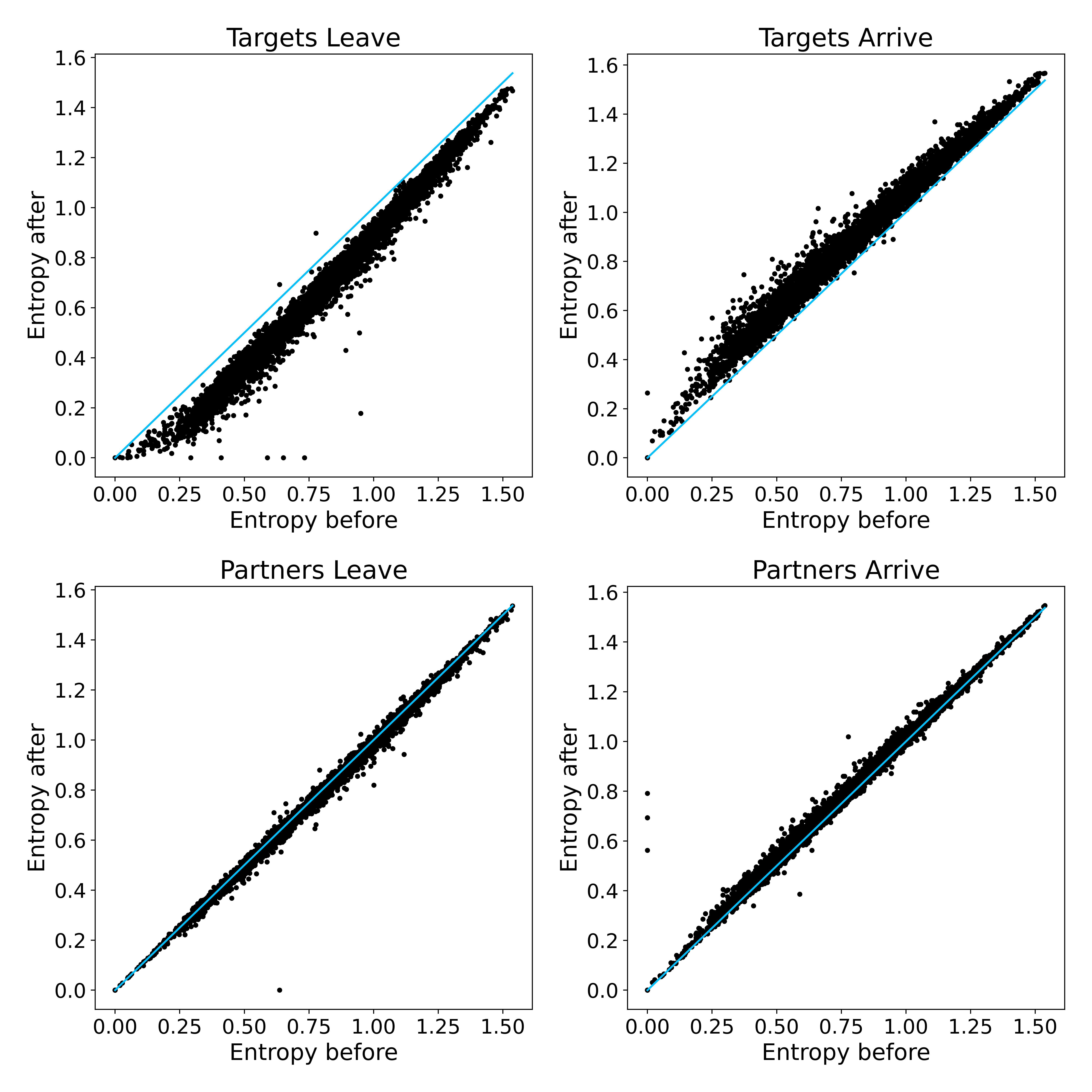}&\includegraphics[scale=0.28]{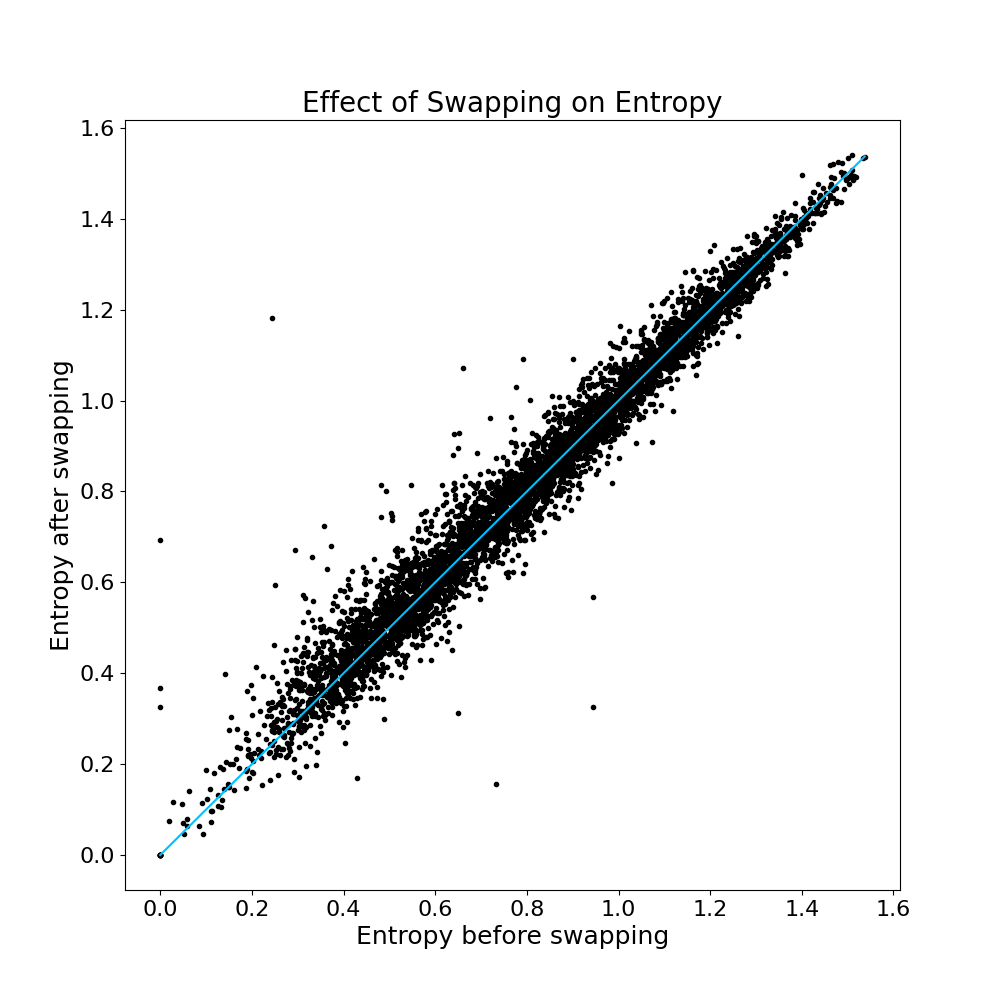}
    \end{tabular}
    \caption{The effect of swapping on racial entropy at the tract level in Texas.}
    \label{fig:entropy_tx}
\end{figure}
\begin{figure}
    \centering
    \begin{tabular}{ccc}
        \rotatebox{90}{\hspace{17mm}2\% Swap Rate}&\includegraphics[scale=0.2335]{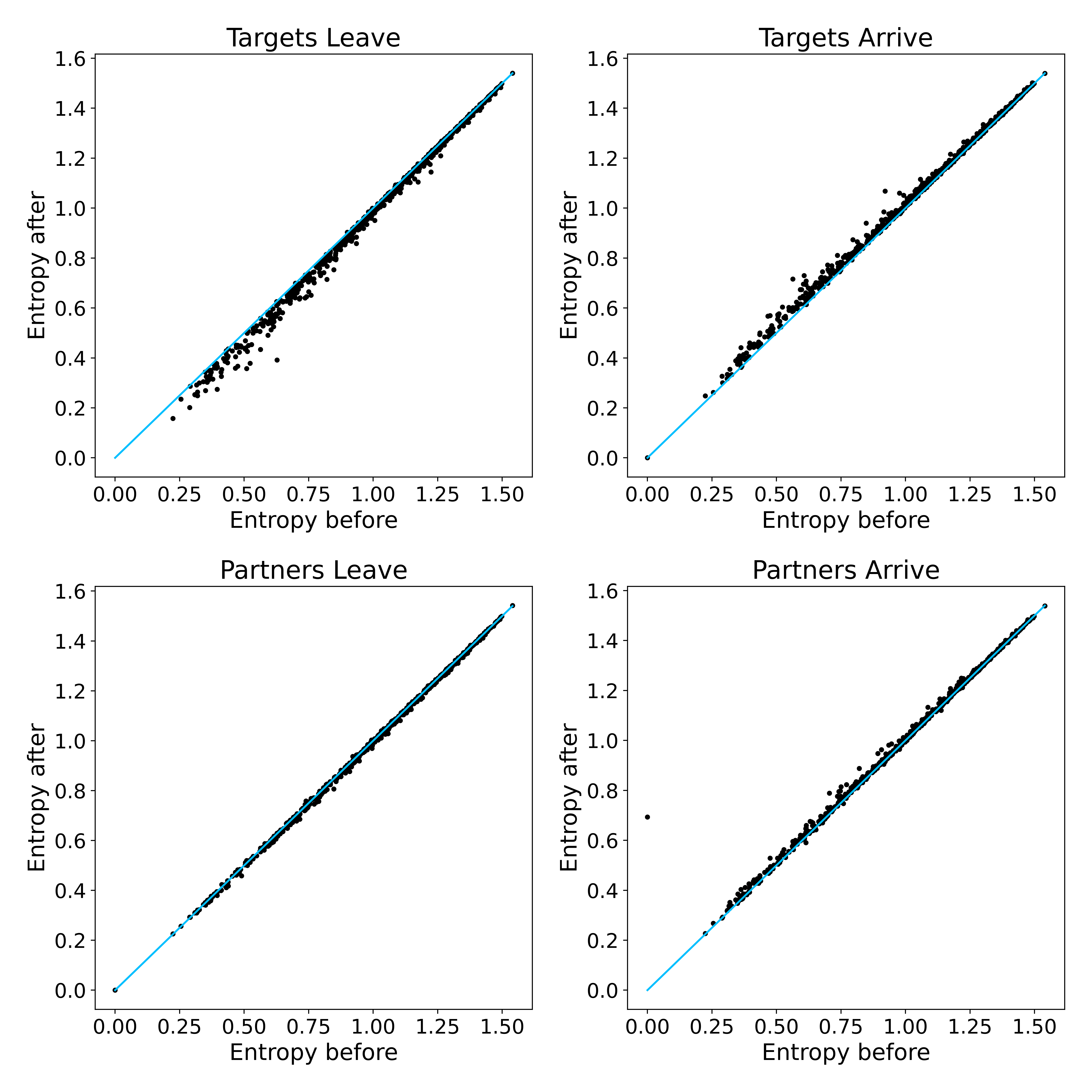}&\includegraphics[scale=0.28]{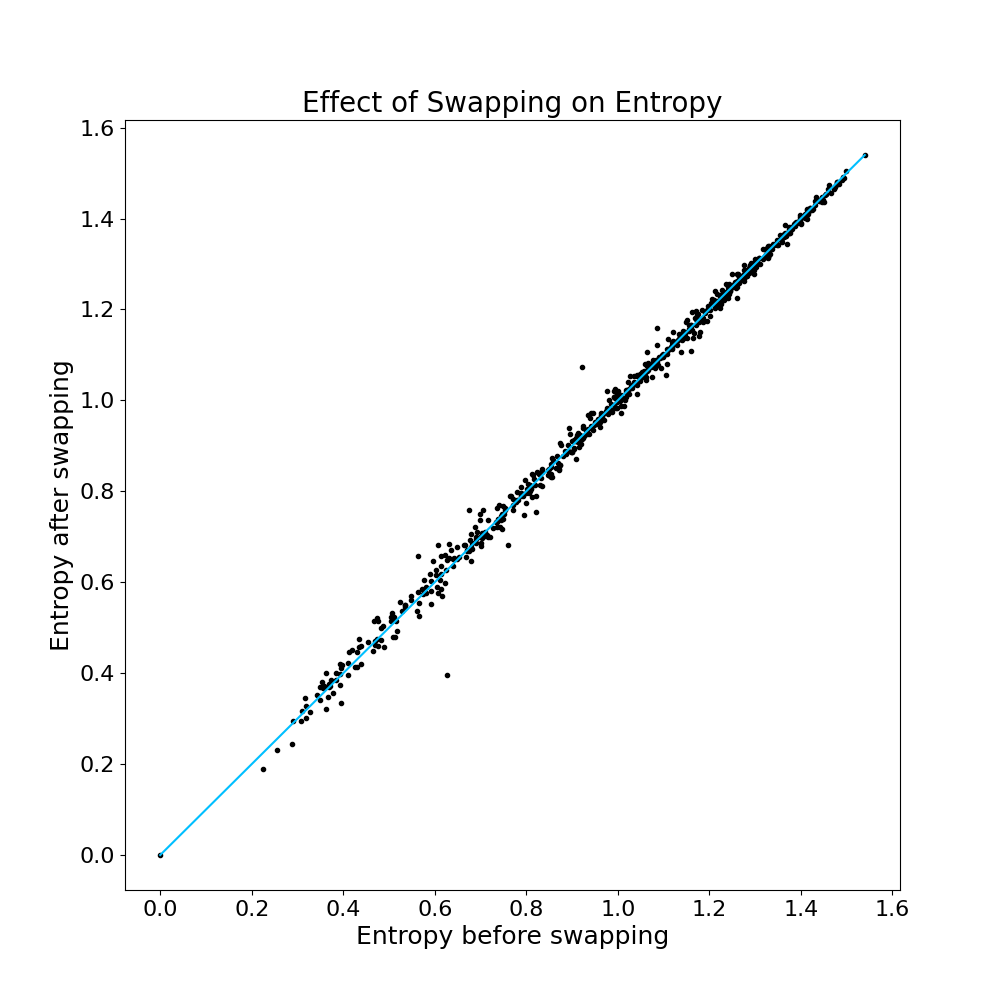}\\
        \rotatebox{90}{\hspace{17mm}10\% Swap Rate}&\includegraphics[scale=0.2335]{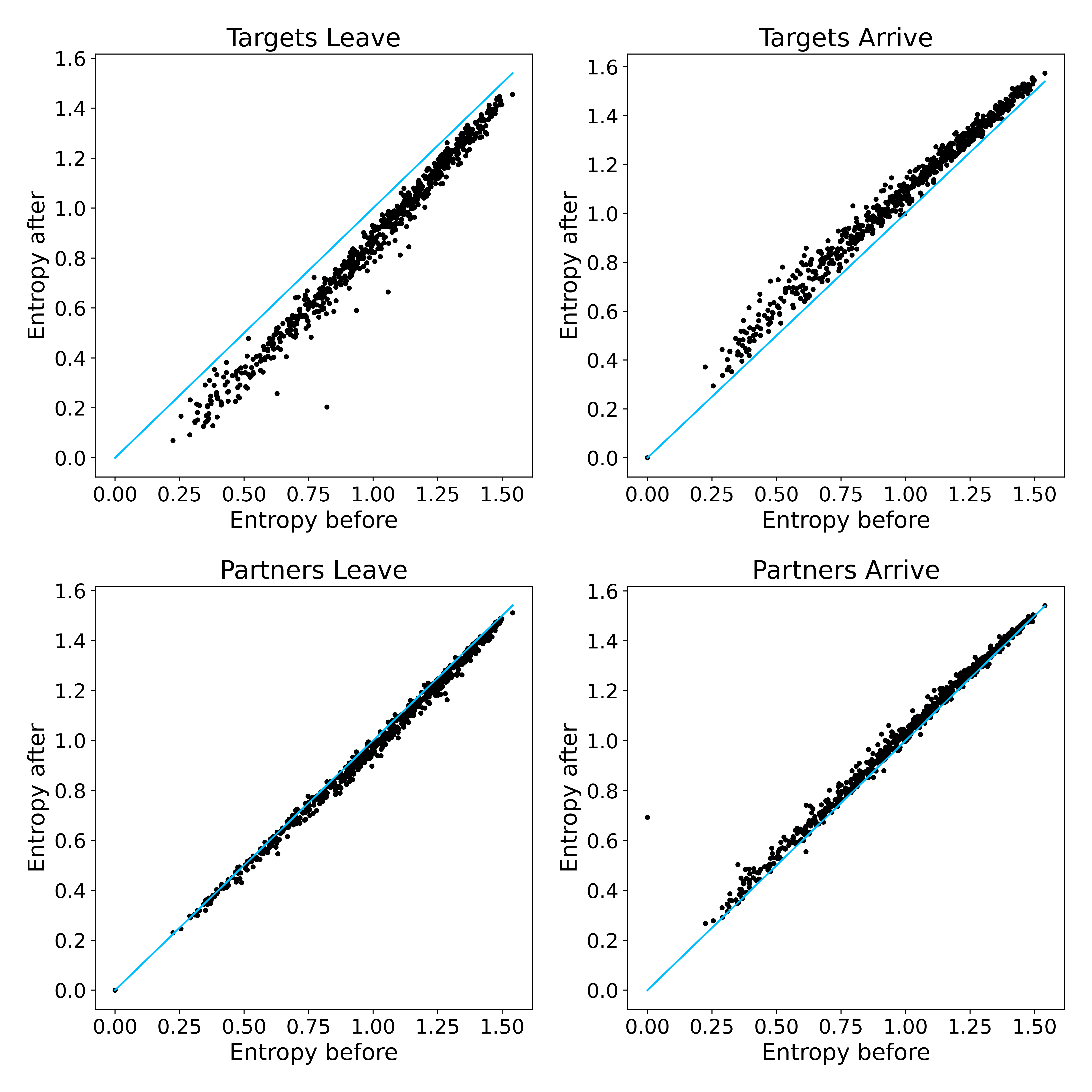}&\includegraphics[scale=0.28]{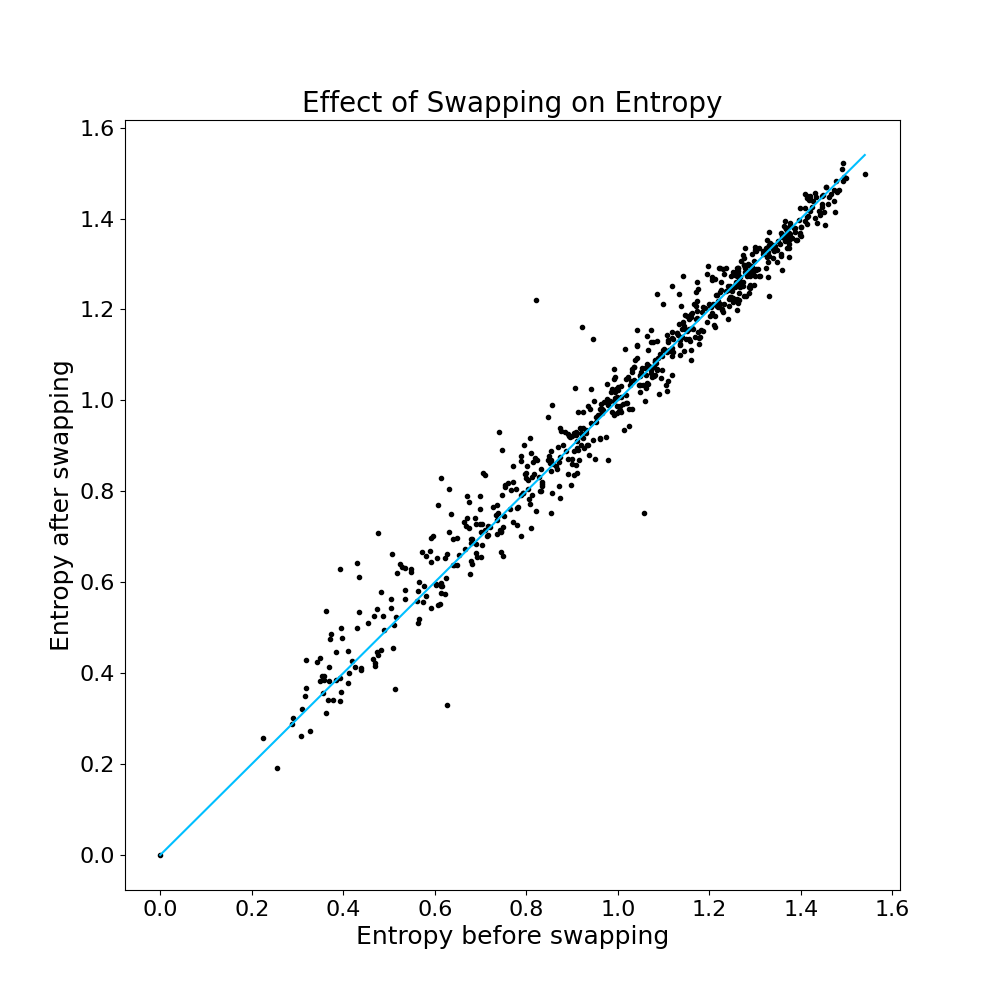}
    \end{tabular}
    \caption{The effect of swapping on racial entropy at the tract level in Nevada.}
    \label{fig:entropy_nv}
\end{figure}
\begin{figure}
    \centering
    \begin{tabular}{ccc}
        \rotatebox{90}{\hspace{17mm}2\% Swap Rate}&\includegraphics[scale=0.2335]{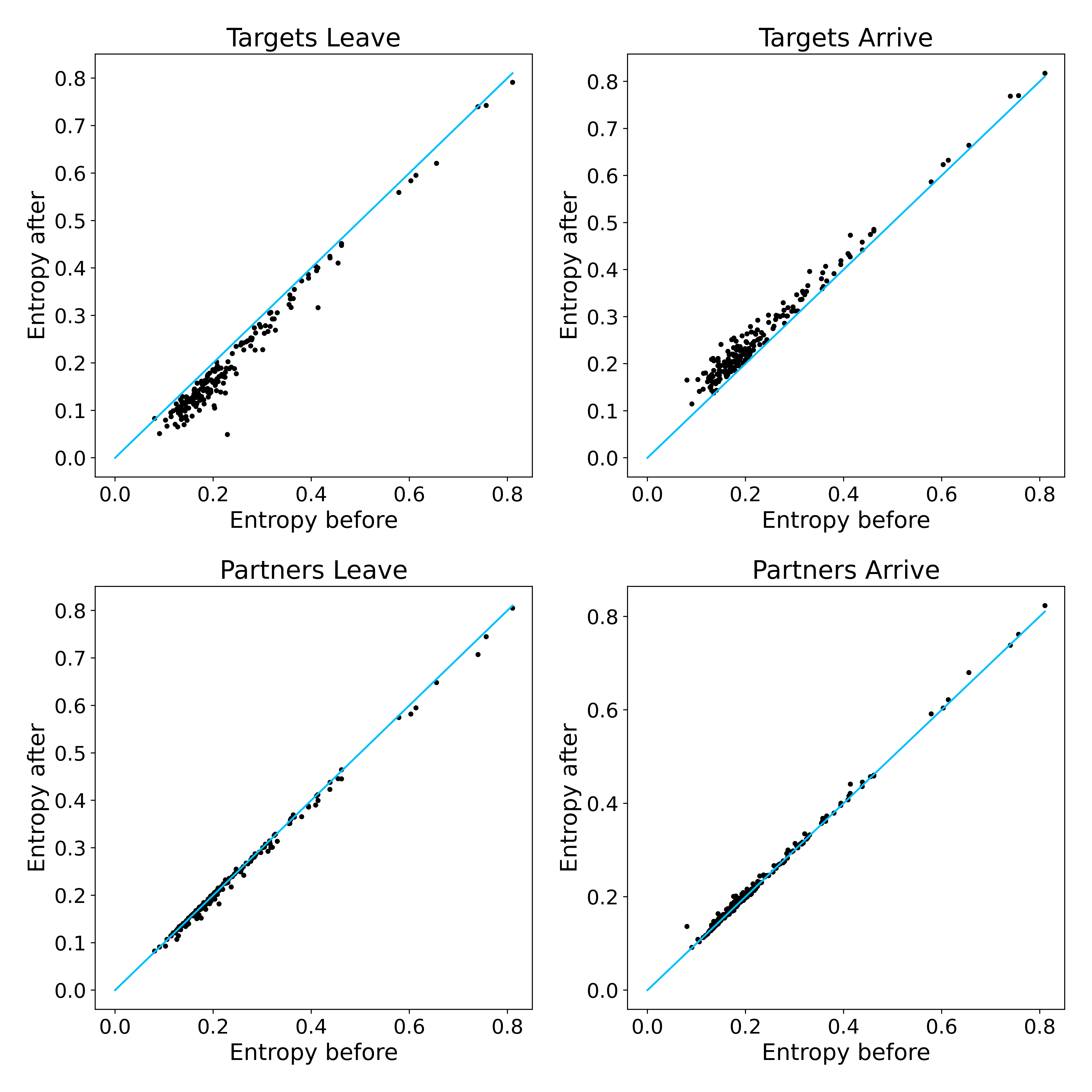}&\includegraphics[scale=0.28]{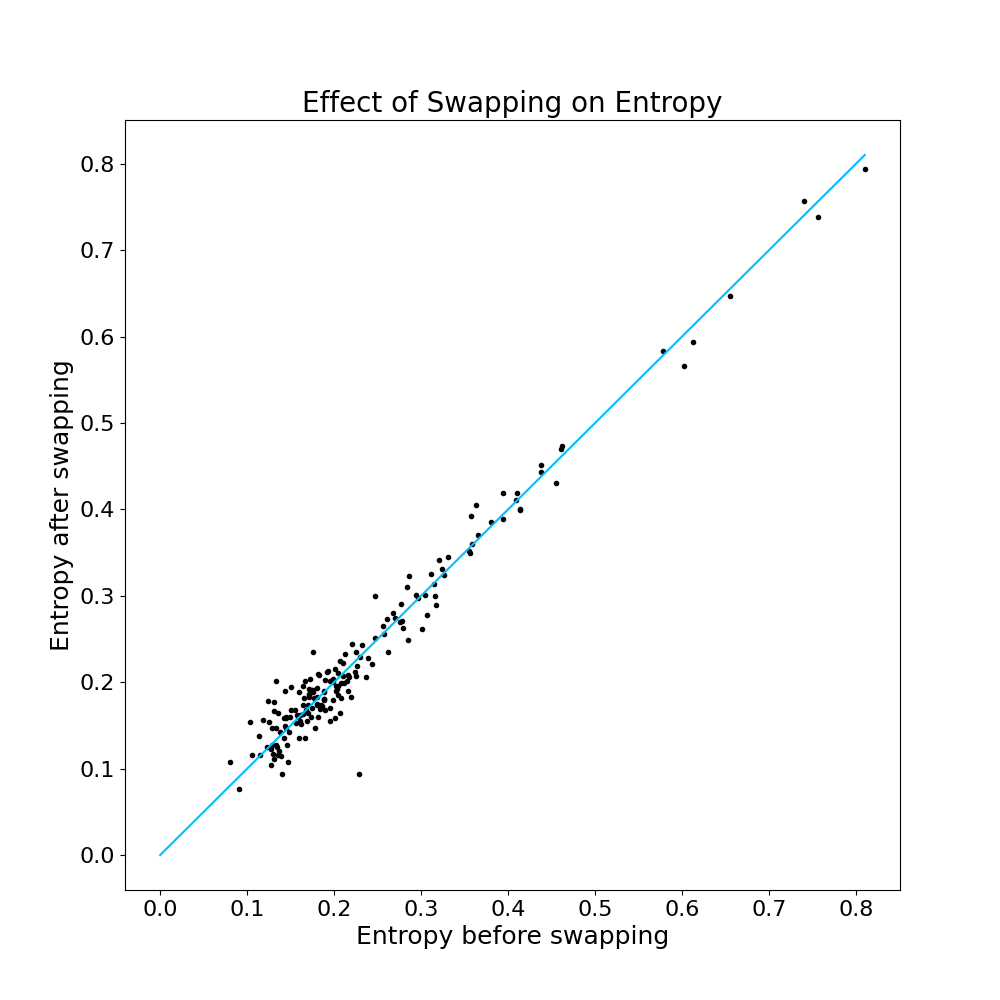}\\
        \rotatebox{90}{\hspace{17mm}10\% Swap Rate}&\includegraphics[scale=0.2335]{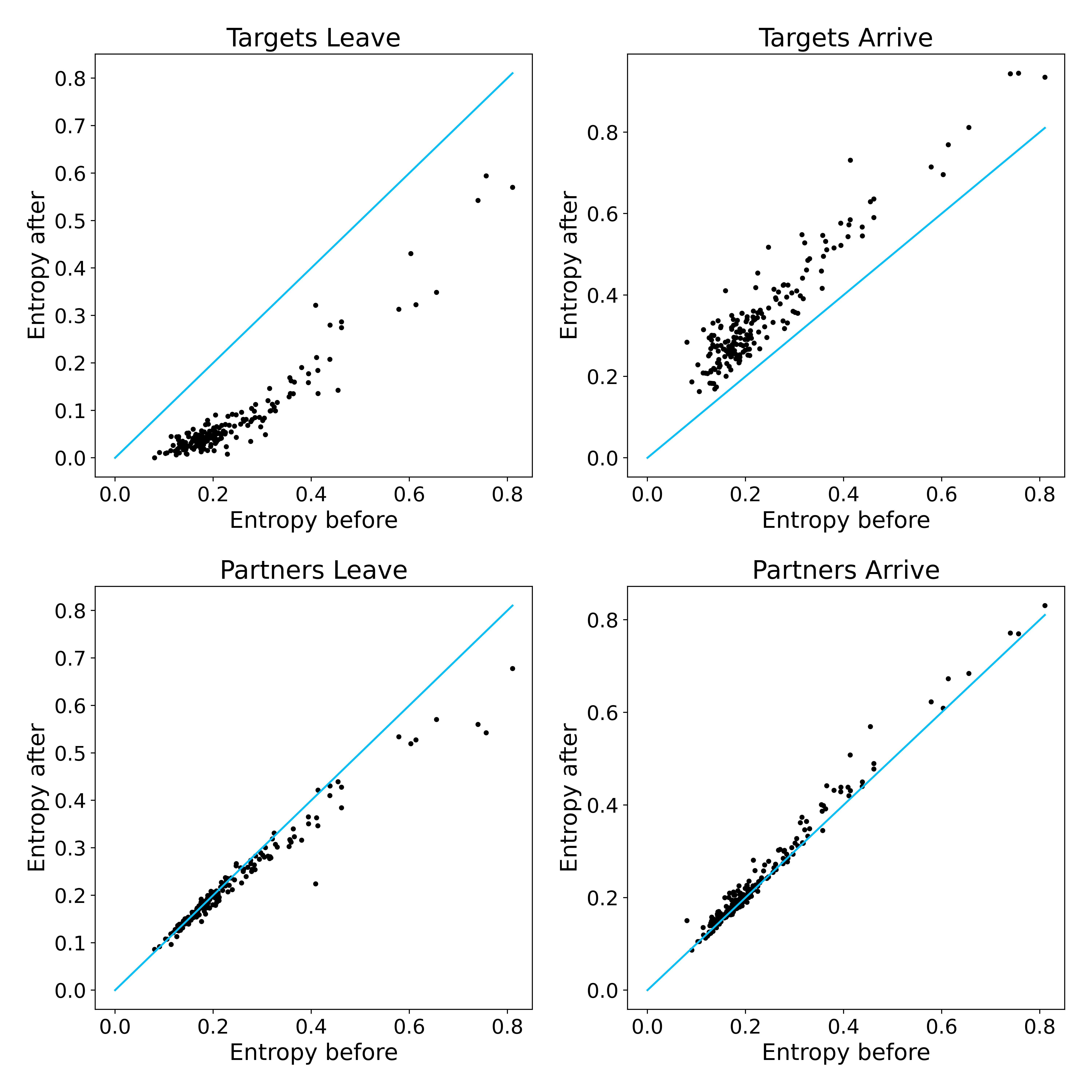}&\includegraphics[scale=0.28]{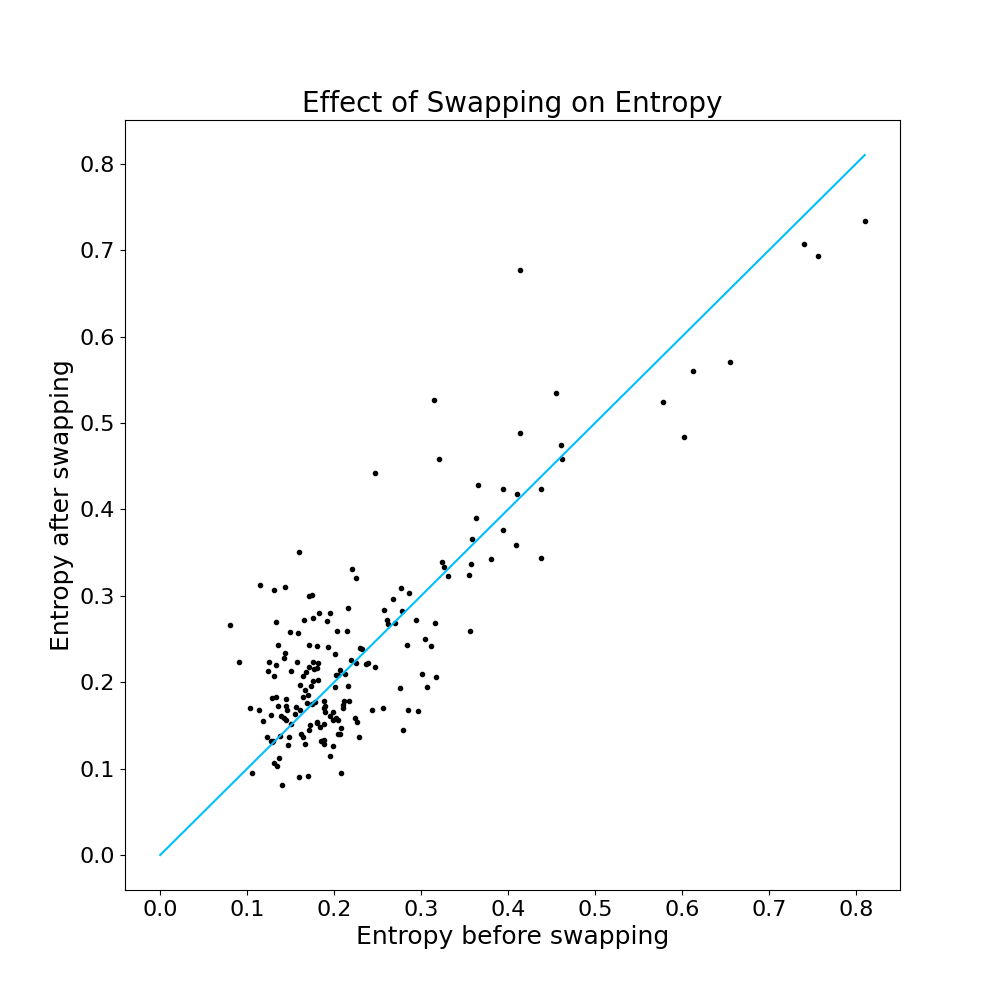}
    \end{tabular}
    \caption{The effect of swapping on racial entropy at the tract level in Vermont.}
    \label{fig:entropy_vt}
\end{figure}

\begin{figure}
    \centering
    \includegraphics[width=\textwidth]{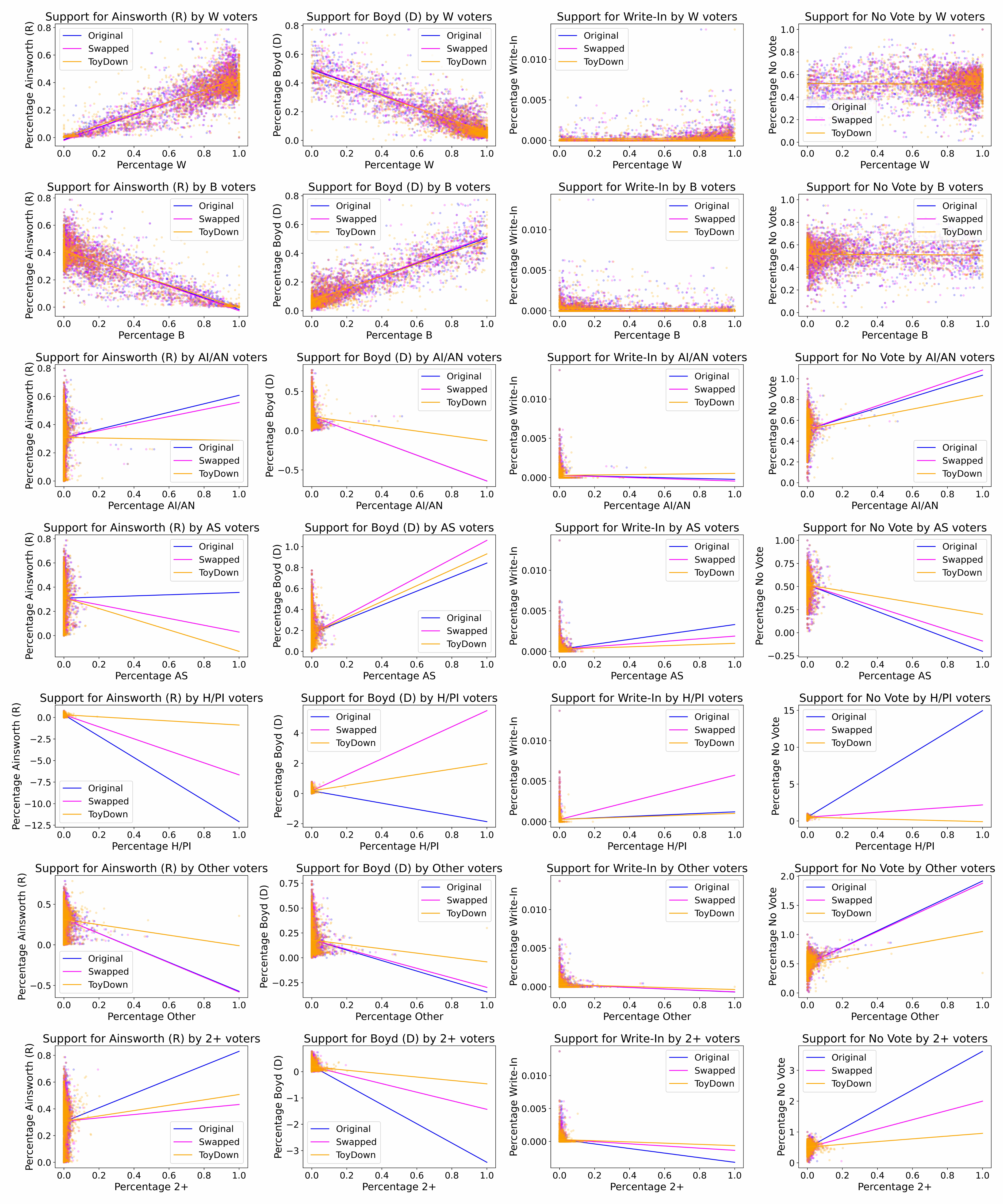}
    \caption{Ecological regression run on unswapped (in blue), swapped with a 10\% swap rate (in magenta), and ToyDown protected with $\epsilon=0.25$ (in orange) data. The outputs of ER (the estimates of supports of racial group $r$ for candidate $c$) are the intersections of the regression lines with the line $x=1$. In this figure, ER \emph{is not} weighted by population.}
    \label{fig:er}
\end{figure}

\begin{figure}
    \centering
    \includegraphics[width=\textwidth]{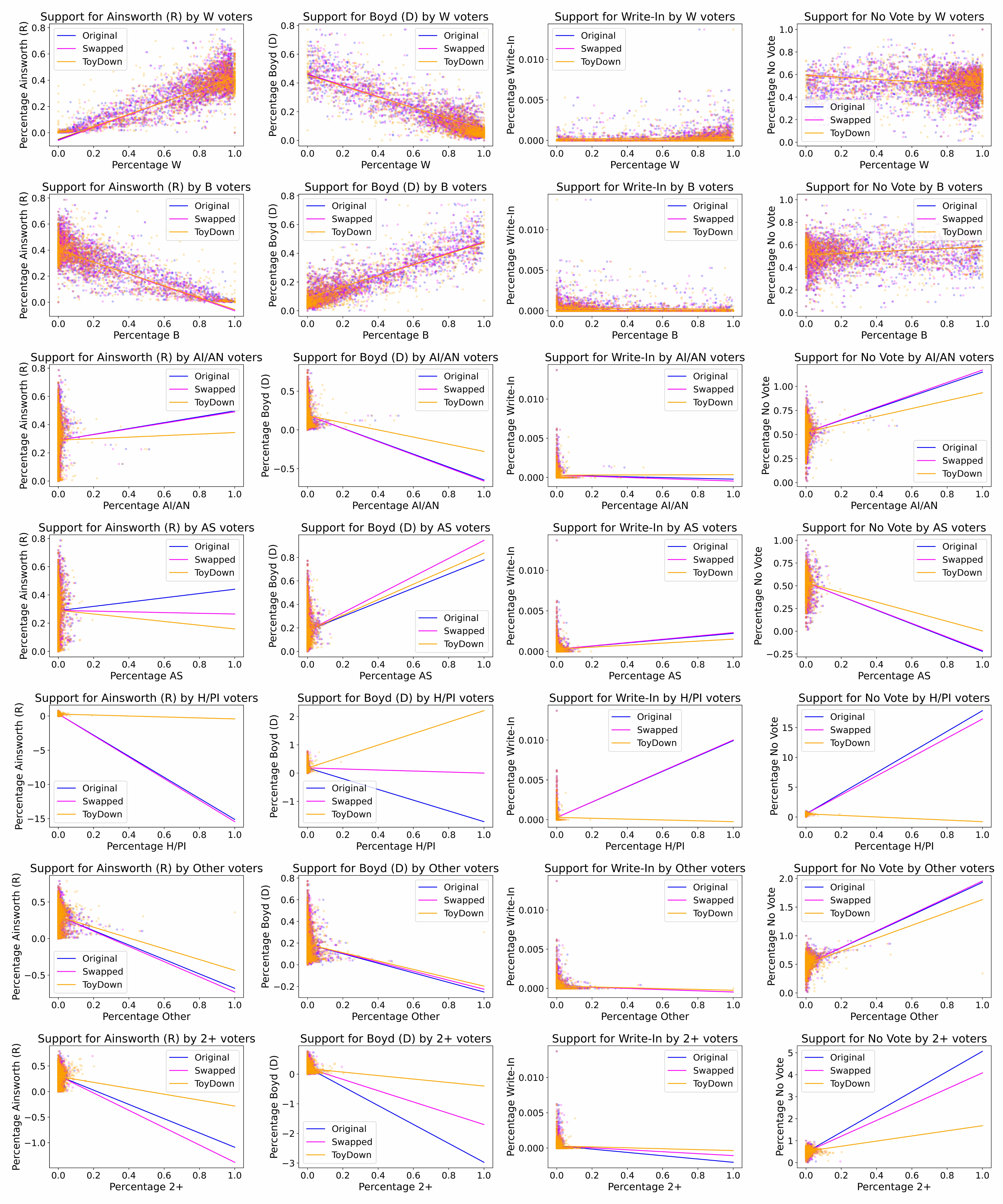}
    \caption{Ecological regression run on unswapped (in blue), swapped with a 10\% swap rate (in magenta), and ToyDown protected with $\epsilon=0.25$ (in orange) data. The outputs of ER (the estimates of supports of racial group $r$ for candidate $c$) are the intersections of the regression lines with the line $x=1$. In this figure, ER \emph{is} weighted by population.}
    \label{fig:er_weighted}
\end{figure}

\end{document}